\documentclass{aa}

\newcommand{\gsim}{\;\lower.6ex\hbox{$\sim$}\kern-7.75pt\raise.65ex\hbox{$>$}\;}
\newcommand{\lsim}{\;\lower.6ex\hbox{$\sim$}\kern-7.75pt\raise.65ex\hbox{$<$}\;}

\usepackage{graphicx}
\usepackage{longtable}
\usepackage{lscape}

\usepackage{color}

\begin{document}
\title{Variable stars in the bar of the Large Magellanic 
Cloud: the photometric catalogue\thanks{Based on data 
collected at the
European Southern Observatory, proposal numbers 62.N-0802, 66.A-0485,
and 68.D-0466} }

\author{Luca Di Fabrizio\inst{1,2}, 
Gisella Clementini\inst{1},
Marcella Maio\inst{1}, 
Angela Bragaglia\inst{1}, 
Eugenio Carretta\inst{1,3}, 
Raffaele Gratton\inst{3}, 
Paolo Montegriffo\inst{1},
and Manuela Zoccali\inst{4,5}}

\offprints{L. Di Fabrizio}

\institute{INAF - Osservatorio Astronomico di Bologna, Via Ranzani 1, 
  40127 Bologna, ITALY, (gisella.clementini, marcella.maio, angela.bragaglia,
  eugenio.carretta, paolo.montegriffo)@bo.astro.it
\and
INAF - Centro Galileo Galilei \& Telescopio Nazionale Galileo, 
PO Box 565, 
38700 S.Cruz de La Palma, Spain, difabrizio@tng.iac.es 
\and
INAF - Osservatorio Astronomico di Padova, Vicolo dell'Osservatorio 5, 35122
  Padova, ITALY, gratton@pd.astro.it
%\and European Southern Observatory, Karl-Schwarzschild-Strasse 2, D-85748 
%Garching bei M\"{u}nchen, Germany, 
\and Pontificia Universidad Cat\'olica de Chile, Departado de Astronom\'{\i}a 
y Astrof\'{\i}sica, Av. Vicuna Mackenna 4860, 782-0436, Macul, Santiago, Chile,
mzoccali@astro.puc.cl
\and 
Princeton University Observatory, Peyton Hall, 
Princeton NJ 08544, USA
}

%\email{difabrizio@tng.iac.es,gisella.clementini@bo.astro.it, marcella.maio@bo.astro.it,
%angela.bragaglia@bo.astro.it, 
%eugenio.carretta@bo.astro.it, gratton@pd.astro.it, paolo.montegriffo@bo.astro.it, mzoccali@eso.org}

\date{Received 14 July 2004/Accepted 27 September 2004}

\abstract{
The catalogue of the Johnson-Cousins $B,V$ and $I$ light curves 
obtained for 162 
variable stars (135 RR Lyrae, 4 candidate Anomalous Cepheids, 11 Classical Cepheids,
 11 eclipsing binaries and 1 $\delta$ 
Scuti star) in two areas close to the bar of the Large Magellanic Cloud is
presented along with coordinates, finding charts, periods, epochs, 
amplitudes, and mean 
quantities 
(intensity- and magnitude-averaged luminosities) of the variables with full 
coverage of the 
light variations. 
A star by star comparison is made with MACHO and OGLE II photometries based 
on both 
variable and constant stars in common, and the
transformation relationships to our photometry are provided.
The pulsation properties of the RR Lyrae stars in the sample are discussed in 
detail.
Parameters of the Fourier decomposition of the light curves are derived
for the fundamental 
mode RR Lyrae stars with complete and regular curves (29 stars). 
They are 
used to estimate metallicities, absolute magnitudes, 
intrinsic $(B-V)_0$ colours, and temperatures of the variable stars, according 
to Jurcsik and 
Kov\'acs  
(1996), and Kov\'acs and Walker (2001) method. Quantities derived 
from the Fourier parameters are compared with
the corresponding observed quantities. In particular,  
%,{\it and to discuss the luminosity-metallicity relation for RR Lyrae stars}.
the ``photometric" metallicities are compared with the spectroscopic 
metal abundances derived by Gratton et al. (2004) from low
resolution spectra obtained with FORS at the Very Large Telescope.
} 

\authorrunning{Di Fabrizio et al.}
\titlerunning{Variable stars in the LMC: the photometric catalogue}
   \maketitle

%\end{abstract}

%
%________________________________________________________________

\section{Introduction}
RR Lyrae stars and Cepheids are primary distance indicators and set the 
astronomical distance scale to the Large Magellanic Cloud and to the galaxies of the Local 
Group. Being from 2 to 6-7 magnitudes brighter than the 
RR Lyrae stars, Cepheids allow to reach galaxies as far as $\simeq$ 20 Mpc
%in the M100 cluster
%and beyond ({\it controllare qual'e' la galassia più lontana in cui
%sono state datte le Cefeidi con HST}),
%({\bf REFERENZA, deve essere un 
(see Freedman et al. 2001).
%o un Ferrarese et al. }).
Conspicuous samples of these variables have been discovered in the 
Large Magellanic Cloud (LMC) as a by-product 
of the microlensing surveys conducted by the MACHO collaboration (Alcock et 
al. 1996, hereinafter A96)
and by OGLE II (Udalski et al. 1997).  
A96 found more than 7,900 
RR Lyrae stars  in the $\sim$
39,000 arcmin$^2$ of the LMC they surveyed, among which  
181 double-mode pulsators (RRd's, Alcock et al. 1997, 2000),
as well as large 
numbers of Cepheids and eclipsing binary systems. Similar numbers are
reported by OGLE II (Soszy\'nski et al. 2003), who also increased to 230 
the number of double-mode RR Lyrae stars.
Calibrated photometry for the LMC RR Lyrae stars has been 
published by both the MACHO collaboration (Alcock et al. 2003a) and the OGLE II
team (Soszy\'nski et 
al. 2003). However, non-standard photometric passbands were used  
by MACHO, and  the RR Lyrae stars are near the
limiting magnitudes of these surveys, so  that the photometric accuracy 
of the individual light curves is reduced. This limits the use of
these samples in the derivation of very precise estimates of the LMC 
distance, or in the study and theoretical reproduction of the light curves  
(see for instance Marconi \& Clementini 2004).
%
%the determination of the parameters of the Fourier decomposition 
%of the light curves and the study of the    
%details of the light variation (such as humps or 
%bumps) are of interest.
Besides, in these experiments the variable stars were mainly observed in $V$ and $I$
and only to a lesser extent in the $B$ passband, thus limiting the 
comparison with most of the 
Galactic samples which instead  generally use $B$ and $V$.

We have obtained accurate multiband time series photometry reaching 
 $V \sim$ 23 (i.e. $\sim$ 3.5 mag fainter than the RR 
Lyrae stars in the LMC) of two 13$^{\prime} \times 13^{\prime}$ fields close to the bar of the LMC
and studied their variable stars (135 RR Lyrae, 4 candidate Anomalous 
Cepheids, 11 Classical Cepheids, 11 eclipsing binaries, and 1 $\delta$ Scuti).
The photometric data were complemented by spectroscopic observations
obtained with the 3.6 m and the VLT ESO telescopes in 1999 and 2001, respectively,
and used to derive
individual metallicities 
for 103
% 2 rrd fatte solo in B01 e 101 variabili, 98 RR e 3 candidate ACs in G04
 of the variables in the present sample, 
and the luminosity-metallicity relation
(M$_V(RR) -$ [Fe/H]) of the LMC
RR Lyrae stars (Bragaglia et al. 2001, Clementini et al. 2003a, hereinafter C03,
Gratton et al. 2004, hereinafter G04). 
A discussion of 
the astrophysical impact of the new data on the derivation of the $M_V(RR)-$[Fe/H]  
relationship and 
on the definition of the distance to the LMC has been presented
in C03. 

 In this paper we present 
the  catalogue of the $B,V,I$ light curves obtained for the 
162 short period variables we have identified in the two fields.
% which were  
% observed in the 
%standard Johnson-Cousins system defined by the $B$, $V$, and $I$
% filters ESO450, ESO451, ESO425, and the Landolt (1992) standard areas
%PG0231+51, SA95 and PG0918+029. 
In Section 2 we describe the acquisition, reduction and calibration of the 
data. Section 3 describes the identification, the period search procedures
and the characteristics of the variables. In Section 4 we present the star-by-star
comparison with MACHO and OGLE II photometries, based on both variable and constant
stars in common, and provide transformation relationships. The 
period distribution and the period amplitude relations followed by the RR Lyrae 
stars in our sample are discussed in Section 5. In Section 6 we discuss 
the metallicities, absolute
magnitudes, intrinsic $(B-V)_0$ colours, and effective temperatures 
derived from the the Fourier
decomposition of the light curve of the {\it ab}-type RR Lyrae stars 
with regular
light curves (29 stars) and compare them with the corresponding 
observed quantities.
%{\it Finally, in Section 7 we summarize our results and give some
%final remarks, la mettiamo?}.

\section{Observations and reductions}

The photometric observations presented in this paper were carried out at 
the 1.54 m Danish telescope
located in La Silla, Chile, on the nights 4-7 January 1999, UT, and 
23-24 January 2001, UT, respectively. 
The journal of the photometric observations is provided in 
Table~1 along 
with information about sky conditions 
during the observations.

\noindent
\begin{table*}[ht]
%\vskip 4 cm
\begin{center}
\caption{Journal of the photometric observations}
%\vspace*{5mm}
\begin{tabular}{c c c c c c c c c c}
\hline
\multicolumn{1}{c}{Observing date (UT)} &
\multicolumn{7}{c}{N.of Observations}&
\multicolumn{1}{c}{Photom. cond.}&
\multicolumn{1}{c}{Seeing}\\
&\multicolumn{3}{c}{Field A}&&\multicolumn{3}{c}{Field B}&\\
     &      $B$  &  $V$  &  $I$&&    $B$  &  $V$  &  $I$& & arcsec \\
\hline
Jan.~ 4, 1999&  11  &  20 &  --  && ~3 &  10 &  --  & clear & 1.4-1.8 \\
Jan.~ 5, 1999&  ~3  &  11 &  --  && 11 &  22 &  --  & clear & 1.4-1.9 \\
Jan.~ 6, 1999&  10  &  19 &  --  && ~3 &  ~9 &  --  & photometric&1.3-1.4 \\
Jan.~ 7, 1999&  ~3  &  ~8 &  --  && ~7 &  14 &  --  & poor-cirri&1.3-1.7 \\
Jan.~23, 2001&  ~7  &  ~7 &   7  && ~7 &  ~7 &   7  & photometric&0.9-1.6 \\
Jan.~24, 2001&  ~7  &  ~7 &   7  && ~7 &  ~8 &   7  & photometric&0.8-1.2 \\
             &      &     &      &&    &     &      &            &         \\
Total        &  41  &  72 &   14 && 38 &  70 &  14  &    --      &   --   \\
\hline
\end{tabular}
\end{center}
%\label{t:tab1}
\end{table*}
%The field selection was based on data from the MACHO collaboration (A96):
%that discovered about 7,900 RR Lyrae's  in the $\sim$
%39,000 arcmin$^2$ of the LMC surveyed, among which  73 RRd's (A97), and
%recently published on the MACHO web site  (http://wwmacho.mcmaster.ca) 
%coordinates, average luminosities, periods and calibrated light curves 
%for all of
%them.  
In both observing runs we centered our observations 
at two different positions, hereinafter 
called field A and B,  close to the bar of the LMC and contained in fields \#6
and \#13 of the MACHO microlensing experiment (see A96 and the 
MACHO web site at http://wwwmacho.mcmaster.ca).
Field A turned out also to have an about 40\% overlap with OGLE II field LMC\_SC21 (Udalsky et
al. 2000).
%, Acta Astronom. 50, 307). 
 The observed fields and their positions
with respect  to MACHO's map of the LMC are shown in Figure~1,
where the elongated rectangle indicates the position of the 
OGLE II field LMC\_SC21.

\begin{figure} 
\includegraphics[width=8.8cm]{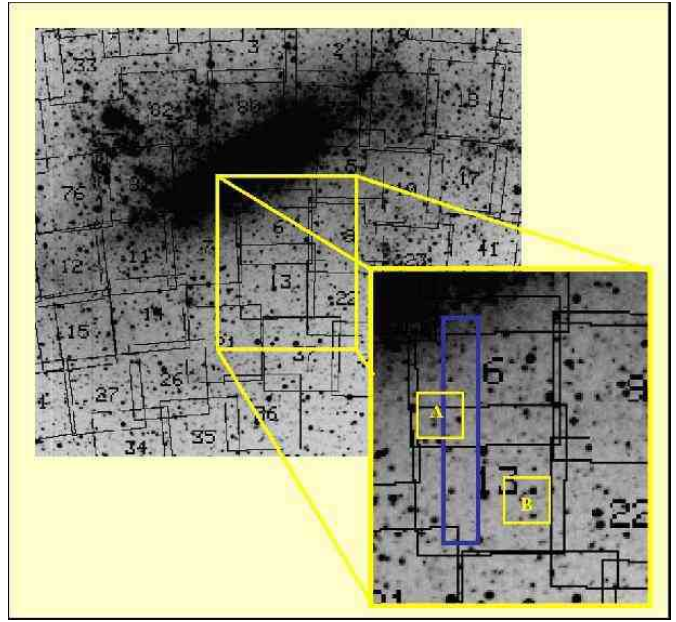}
\caption[]{The light squares indicate the approximate positions of 
our observed fields with 
respect to MACHO's fields \# 6 and \#13. The elongated rectangle 
identifies the position of the OGLE II field LMC\_SC21.
}
%\label{f:fig1}
\end{figure}

The two positions were chosen in order to maximize the number of
known RRd's observable with only two pointings of the 1.54 m Danish telescope,
since a major purpose of our study was to derive the
mass-metallicity relation for double mode pulsators (Bragaglia et al. 2001).
We expected to observe about 80 RR Lyrae's according to A96 average density of 
RR Lyr's in the LMC, among which 5 and 4 double mode RR Lyrae (RRd), in field A 
and B, respectively (Alcock et al. 1997, 
hereinafter A97). Coordinates
(epoch 2000) of the two centers are: $\alpha$ = 5:22:48.49, $\delta$ =
--70:34:06 (field A), and $\alpha$ = 5:17:35.7, $\delta$ = --71:00:13 (field
B). In both observing runs the telescope was equipped with the DFOSC focal
reducer. In 1999 data were acquired on a  Loral/Lesser 2052x2052 pixel
chip (CCD \#C1W7,  scale 0.4 arcsec/pix, field of view  of 13.7
arcmin$^2$), and a filter wheel mounting the Johnson standard system.
Observations were done in the Johnson-Bessel $B$ and $V$ filters (ESO 450, 
and 451), and we obtained 58 $V$
and 27 $B$ frames for field A, and 55 $V$ and 24 $B$ frames for field B. 
Seeing conditions were quite variable during each night and the whole 
observing run; typical
values were in the range 1.3-1.9 arcsec (see Table 1)
\footnote{These are the values measured from the FWHM of the observed stellar profiles.
Note that these  values likely overestimate the real seeing
FWHM, since it is now acknowledged that there was some photon diffusion 
on the Loral-Lesser CCD at the 1.54 m Danish telescope. This problem is not
present in the EEV chip used in the 2001 observations.}.

Exposure times varied
from   180 to 300 sec in $V$ and from 360 to 480 sec in $B$, depending
on weather/seeing conditions and hour angle. They  were chosen as an optimal
compromise between S/N and time resolution of the light variations of the RR
Lyrae variables. Eighteen stars from Landolt (1992) standard  fields were
observed during each night in order to secure the transformation  to the
standard Johnson photometric system. 

In the 2001 run, data were acquired on an EEV 42-80 CCD (2048x4096 pixels, 
scale of 0.39 arcsec/pix and  field of view  of 13.7 arcmin$^2$). The CCD
has pixel size of 15 $\mu$m and is  back-illuminated to increase its quantum
efficiency, particularly at shorter wavelengths. Due to the field of view
of the
DFOSC focal reducer, only half of the CCD is actually used to image
data.  Observations were done in the Johnson-Bessel $B$, $V$ and in the 
$i$-Gunn
filters\footnote{The $i$-Gunn observations can be reliably 
transformed to the standard $I$ of the Landolt-Cousins system}
  (ESO 450,  451, and 425) 
and we obtained 14 $V$, 14 $B$ and 14 $i$ frames for
field A, and 15 $V$, 14 $B$, and 14 $i$ frames for field B.  Exposure times
were of 360 sec in $B$, and 180 sec in $V$ and $i$.

Both nights of the 2001 run were fully photometric with good seeing conditions.
Transparency and  seeing  were better in the second night with most
frequent values of the  seeing  around 1.0 arcsec in $B$ and $V$, and 0.8 arcsec
in $i$. A large number of standard stars in Landolt (1992) - Stetson (2000)
standard fields PG0918+029, PG0231+051, PG1047+003, and SA98 
were observed several times  during both nights to estimate the
nightly  extinction and to tie the observations to the standard
Johnson-Cousins  photometric system (see Section 2.2).  Two exposures  of
different length were taken at any pointings of the standard fields, in order
to obtain well exposed measurements of both bright and faint standard stars. 

\subsection{Reductions}  

%%\begin{figure} 
%%%%%\includegraphics[width=8.8cm]{fig4_calib.ps}
%%%\includegraphics[width=8.8cm]{difabrizio.fig2.ps}
%%\caption[]{Calibration relations.}
%%%\label{f:fig2}
%%\end{figure}

%Final 
Reduction and analysis of the 1999  photometric data were done using the
package DoPHOT (Schechter, Mateo \& Saha 1993), which uses an elliptical
Gaussian PSF to evaluate instrumental magnitudes.  We used a PSF
varying with the position on the frame and
run DoPHOT independently on all frames, with a threshold for source detection
of 5 $\sigma$ above the local sky. The resulting tables were then aligned to 
the ``best" frame for
each field (i.e., to the one taken in best seeing and weather conditions, and
near meridian) and stars were counteridentified
using a private software written by P.
Montegriffo. Catalogues were produced, all
containing the same number of stars, and with a unique identifying number: this
helped in the following variability search and study. The number of objects
classified as stars in each frame is variable (from several thousands to about
30,000). The final 1999 catalogues, after counteridentification in $V$ and $B$, 
contain about 29,000 objects  for field A and about 23,000 for field B; this 
difference seems reasonable since field A is slightly closer to the LMC bar and thus more
crowded than field B.

Photometric reductions of the 2001 data were done using DAOPHOT/ALLSTAR II
(Stetson 1996) and ALLFRAME (Stetson 1994). DAOPHOT/ALLSTAR II  allows to
obtain very precise brightness estimates and astrometric positions for stellar
objects in individual  two-dimensional digital images starting from a rough 
initial estimate for the position and brightness of each star, and a model of
the PSF for each frame. We used a source detection threshold of  4 $\sigma$
above the local sky background, and a PSF which varied quadratically with
the position in the frame. Modelling of the PSF in each frame was obtained by
considering a set of about 100 stars. The resulting PSFs are hybrid models
consisting of an analytic function and a table of residuals, 
thus offering both the advantages of an analytic and of an empirical PSF.

Because of the high crowding of our LMC fields, in addition to 
DAOPHOT/ALLSTAR, reductions were executed with ALLFRAME, which  performed the
simultaneous consistent reductions of all the 2001 multicolour images of  our fields: 42
frames for field A, and 43 frames for field B, respectively. By combining
informations coming from all images it was thus possible  to obtain a better
precision in the identification and centering of the stars, and to 
resolve objects that appeared blended in frames with worse seeing conditions.

Aperture corrections were derived for the $B, ~V, ~I$ reference frames from
about 10 bright and relatively isolated stars in each frame. The choice of these
stars has been particularly difficult for field A, the more crowded one, for
which we also derived larger corrections. The mean differences between PSF and
aperture magnitudes were used to correct the PSF magnitudes of all other
objects. The $B, ~V, ~I$ corrections (aperture minus PSF) were:
--0.140, --0.073, --0.020 mag for field A, and --0.026, --0.035, --0.040 mag for field B
respectively.

Aperture magnitudes for the photometric standard stars were computed 
using PHOT in DAOPHOT, rejecting all saturated stars and all objects with 
less than 
1000 detected counts. The aperture radii for these stars were determined
from curves of growth. 

\subsection{Night extinction calculation and absolute photometric calibration}  
Only the third night (January 6, 1999) of the 1999 run was fully
photometric. Viceversa, both nights in the 2001 run were  photometric and 
with good seeing conditions. Since the 2001 run was definitely superior 
both for photometric quality and seeing, and since a much larger
number of standard stars were  observed, our
entire photometric data set has been tied to the standard Johnson-Cousins
photometric system through  the absolute photometric calibration of the 
2001 run. 

The extinction coefficients for the nights were computed from 
observations of the 
standard
stars in the  selected areas PG0918 and SA98 (Landolt 1992). We used 7 bright 
standard stars
in PG0918, with measurements at different airmasses ($1.180 <\sec z < 1.626$) 
to
estimate the extinction coefficients for the night of January 23, and
7 bright standard stars of SA98 with measurements at  $1.145 <\sec z < 2.028$,
to estimate the extinction for the night of January 24. The derived first
order  extinction coefficients are: $K_V = 0.142 \pm 0.008$,  $K_B = 0.240 \pm
0.020$, and $K_i = 0.071 \pm 0.006$ for January 23;  $K_V = 0.123\pm
0.005$,  $K_B = 0.220 \pm 0.009$, and $K_i = 0.052 \pm 0.010$ for January 24.
These extinction coefficients well compare to the average ones for La Silla,
as deduced from the relevant web pages.

Stetson (2000) has extended Landolt (1992) standard fields to a fainter
magnitude limit, reaching $V\sim$ 20 mag. To transform to the standard
Johnson-Cousins photometric  system, we used Stetson (2000) standard star
magnitudes, as available from   the web site
http://cadcwww.hia.nrc.ca/standards,   for a large number of standards in
Landolt's fields PG0918+029, PG0231+051, PG1047+003, and SA98. We have
verified that  Stetson (2000) standard system reproduces very well the
Johnson-Cousins standard system by Landolt (1992). 
In fact, if we restrict only to the original Landolt standards in each field,
and derive the calibrating equations using both Landolt's and Stetson's values,
the colour terms agree to the thousandth of magnitude both in $B$ and $V$.
%In fact colour terms of the 
%calibration equations  we derive using the sample of Landolt (1992) standard
%stars observed in the above areas and  Stetson (2000) magnitude values agree
%within a thousandth of magnitude with the colour terms  obtained by 
%using Landolt's magnitudes 
%of Landolt's (1992) standard stars, both in $V$ and $B$. 
In $I$ there
are two deviating stars, namely PG0231 for which Landolt's $I$ magnitude is
about 0.2 mag too bright, and SA98-1002 whose Landolt's $I$ magnitude is about
0.02-0.04 mag fainter. If these two stars are discarded, agreement to within a
thousandth of magnitude is found for the $I$ colour terms as well.

We measured magnitudes for 67 stars in these areas. However, since  most of the
new faint standard stars observed by Stetson (2000)  only have $V$
measurements, while the $B$ and $I$ database is still poor, only a subset of 27
stars with accurate standard magnitudes in all three photometric pass-bands of
our interest were actually used in the calibration procedure.  Aperture
photometry magnitudes of these stars measured in the two nights of the 2001
run, corrected for the extinction appropriate to each night, were  combined 
to derive
the following  calibration equations:
%$$ V = 1.003 (\pm 0.018) \times v - 5.220$$
$$  B-b = 0.111 (\pm 0.032) \times (b-v) - 5.472$$
$$  V-v = 0.021 (\pm 0.017) \times (b-v) - 5.175$$
$$  I-i = -0.023 (\pm 0.025) \times (v-i) - 6.507$$

\noindent
where $B, V, I$ are in the Johnson-Cousins system, while $b$, $v$, $i$ are the
instrumental magnitudes. The calibration relations are 
%shown in  Figure~2. They
are based on 127 measurements in the two nights of the 2001 run of the restricted sample of
27 standard stars with magnitude and colours in the ranges   $12.773 < V <
17.729$, $-0.273 < B-V < 1.936$, $-0.304< V-I < 2.142$.
Note that we adopted 
an iterative rejecting procedure, eliminating those objects  that deviated
 more than 2.5 $\sigma$ 
(where  $\sigma$ is the standard deviation of the residuals) from the least
square fit regression lines. Photometric zero points accuracies are of 0.02 
mag in  $V$
and 0.03 mag in $B$ and $I$,  respectively.

\subsection{Comparison between the 1999 and the 2001 photometries}

Figure~2 shows the instrumental colour magnitude diagrams (CMDs) 
obtained from the
photometric reductions of one $V$ and one $B$ frames of field A, from the 
1999 and 2001 data sets
respectively, using the various different packages employed in this study,
namely  DoPhot for the 1999 data set (left panel), 
DAOPHOT + ALLSTAR (central panel), 
and DAOPHOT + ALLSTAR + ALLFRAME (right panel) for the
2001 data set. 
\begin{figure} 
\includegraphics[width=8.8cm]{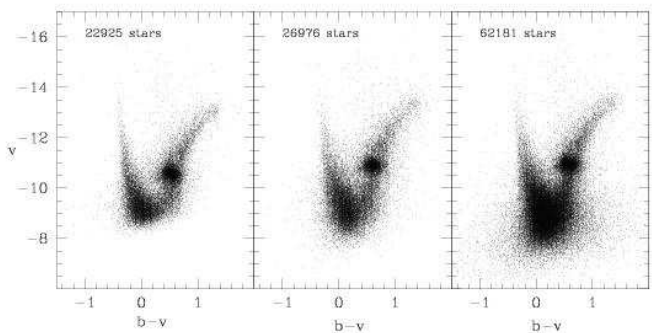}
\caption[]{Comparison between instrumental CMDs (all based on two
frames) 
of the 1999 and
2001 datasets. 
Left panel: DoPhot reductions of the 1999 data; 
central panel: DAOPHOT+ALLSTAR reductions of the 2001 data; 
right panel: DAOPHOT+ALLSTAR+ALLFRAME reductions of the 2001 data.}
%\label{f:fig1a}
\end{figure}
The figure very well illustrates the superiority of the 
2001 data and reduction procedures
 with respect to the 1999 ones. In particular, the increased number of 
objects and the fainter magnitude limit reached by the 2001 data in  the 
central panel of  Figure~2 is due predominantly to the better seeing and photometric conditions and
the improved sensitivity of the CCD in run 2001, and in part 
to the better performances of the DAOPHOT reduction package with
respect to DoPhot. The CMD in the right panel
demonstrates the efficiency and superiority of the  ALLFRAME package to resolve
and measure faint stellar objects in crowded fields: the number of stars in 
the right panel of the figure is more than doubled and reaches one magnitude fainter than data shown
in the other two panels. For these reasons all considerations about the
CMDs have been based on the ALLFRAME reductions of the 2001 data (see C03). 

\section{Identification of the variable stars}

Variable stars were identified on the 1999 $v$
and $b$ instrumental time-series independently, using the
program VARFIND, by P. Montegriffo. VARFIND performs the following actions:
(i) normalizes the files containing measures of the fitted stars to a 
reference frame, using 
all stars in 1.5 magnitude bins about 2 magnitudes brighter than the 
expected average level of the RR Lyrae variables to 
determine mean frame-to-frame offsets with respect to the reference
frames. As $v$ and $b$ reference frames we chose those taken 
in the best seeing and photometric conditions; (ii) computes the average magnitude of each star and its standard
deviation by combining all frames in a given filter, using the offsets 
determined in step (i); (iii) displays the scatter diagrams of the average 
measurements, namely the standard deviations vs. average $<v>$ and $<b>$ plots
from which candidate variables are identified thanks to their large 
{\it rms} and picked up interactively. In our scatter diagrams the 
RR Lyrae's and the Cepheids define very well distinct groups 
of stars with large {\it rms} values, respectively at  
18.6$<V<$19.8 mag and  
15.1$<V<$16.6 mag; 
(iv) extracts the time-series sequence of each candidate variable and of its
selected reference stars (see below).
%The identification of the variable stars was performed in the $v$
%and $b$  frames independently,  and 

The search procedure was repeated several
times, subsequently lowering the  detection threshold. Stars whose standard
deviations of the $v$ and $b$ measurements were larger than 3 $\sigma$, where
$\sigma$ is the rms of  {\it bona-fide} non-variable stars at same magnitude level,
were flagged as candidate variables and closely inspected for 
%Each  candidate object was then checked for 
variability using the program
GRATIS  (GRaphycal Analyzer of TIme Series) a private software developed at the
Bologna  Observatory by P. Montegriffo, G. Clementini and L. Di Fabrizio.
This code,  directly interfaced to VARFIND, allows to display the sequence
of differential measurements of the object with respect to the selected 
reference stable stars, as a function of the Heliocentric Julian day of 
observation, and to perform a period search on these data (see below).  
%Figure~4 shows the scatter diagram used to identify candidate variables from
%the $V$ time series data of field A: crosses mark the actual variable objects
%identified in this field (98 confirmed and 5 candidate variable stars). The region defined by the RR Lyrae
%variables is clearly visible at  $V \sim $19.4 mag, while variables around $V
%\sim$15.6 mag are Cepheids.
%%%\begin{figure} 
%%%%\includegraphics[width=8.8cm]{clementini.fig7_new.ps}
%%%caption[]{Scatter diagram in the $V$ band
%%%for stars in field A, produced by 
%%%the program VARFIND. A total number of 28,638 objects are displayed in 
%%%the full plot. Crosses 
%%%mark 
%%%the variable stars identified in this field (103 objects). The region at 
%%%18.6$<V<$19.8 mag corresponds to the RR Lyrae variables, while variables 
%%%fwith 15.1$<V<$16.6 mag are Cepheids. Dots indicate all the objects checked for 
%%%variability in this field (1165 stars).}
%%%\label{f:fig1a}
%%%\end{figure}
A total number of 1165 and 747 objects were checked for variability in fields A
and B, respectively. We are confident that our  identification of the RR
Lyrae stars is rather complete, and
% as far as the RRab variables in  field B are  
%concerned;
%given the larger crowding, identification may be less
%complete in field A. Moreover, we cannot
%totally rule out that some  RRc's may have escaped detection in both fields, 
%due to their smaller amplitudes. 
we will come back to this point in Sections 3.2 and 5. 
%As an alternative approach we also searched for variables using the following 
%procedure: (i) we selected all stars in a 1.5 magnitude bin bracketing the 
%expected range 
%for RR Lyrae variables on a reference $V$ frame; 
%(ii) we determined the mean frame-to-frame offset with respect to the reference
%frame by iteratively eliminating all stars deviating more than 2.5 $\sigma$ from
%the average. These mean offsets are finally based on about 2500 ``constant" 
%stars having a $V$ magnitude similar to that of the RR Lyrae's;
%(iii) we then determined the average $V$ magnitude and its {\it rms} for all
%stars by combining all $V$ frames, using the offsets determined in step (ii); 
%(iv) all stars with {\it rms} larger than three times the typical value 
%for this magnitude range (again determined iteratively) were closely examined 
% for variability looking for possible periods in the range 0.2-1 days;
%(v) the same procedure was repeated using the B frames.
%On the whole, results obtained using the two procedures agree very well.

Variable stars were then counteridentified on the 2001 frames using
private software by P. Montegriffo. A few further variables originally missed
by the search on the 1999 data were recovered in the comparison with MACHO and
OGLE II datasets (see Section 4). In the end  
the two fields were found to contain a total number of 
162 short period variable stars (P$<$ 7 days), mainly of RR Lyrae type 
(125 single-mode and 10 double-mode, one of which not previously 
known from A97; see Section 5.1), and an additional 8 candidate
variable objects: 5 possible binary systems, 1 possible {\it ab-}type RR Lyrae, 
and 2 other variables that we were not able to classify.  

The number of variables divided by type and field
is given in Table~2.
\noindent
\begin{table}[ht]
%\vskip 4 cm
\begin{center}
\caption{Number and type of variables identified in the two fields}
%\vspace{0.5cm}
\begin{tabular}{lccc}
\hline
Type&\multicolumn{1}{c}{Field A}&\multicolumn{1}{c}{Field B}&\multicolumn{1}{c}{Total}\\
\hline
RRab& 52& 35& 87\\
RRc& 20& 18& 38\\
RRd& 6& 4& 10\\
Anomalous Cepheid & 3& 1& 4\\
Cepheids& 10& 1& 11\\
Binaries& 6& 5& 11\\
$\delta$ Scuti& 1& ---& 1\\
Total&98&64&162\\
Candidate variables& 5 & 3&8 \\
\hline
\end{tabular}
\end{center}
%\label{t:tab2}
\end{table}
Finding charts for all the variables are provided
in Figures 3 to 10, where each field is divided in 4 quadrants  
6.8$^{\prime} \times 6.8^{\prime}$ large (subfields A1, A2, A3, A4,
and  B1, B2, B3, B4, respectively), which correspond to the
pre-imaging fields of our spectroscopic study with FORS1 at the VLT 
(see G04). 
There is some overposition at the centre of the each set of 4 quadrants 
and a few objects appear twice. 
RR Lyrae stars are marked by red open circles in the electronic version
of the finding charts, the other variables are in blue.
%Different colours are used to identify the 
%various types of variables in the electronic version of these
%finding charts. Colour coding is: red for RRab and RRc, yellow for RRd,
%light blue for Cepheids, magenta for candidate Anomalous Cepheids, blue for
%binaries, and green for the $\delta$ Scuti star.
Two RR Lyrae stars fall outside the FORS fields and are shown 
separately in 
Figure~11. 

\subsection{Period search and average quantities}

All variables were studied using their differential photometry with respect to
two stable, well isolated objects used as reference stars, whose constancy was
carefully checked on the full 1999-2001 data set.  Coordinates and calibrated
magnitudes of the  reference stars from the 2001 photometry 
are given in Table~3. Errors quoted in the table include both  the internal
error contribution given by ALLFRAME (about 0.005 mag in $V$ and $I$, and  0.004 
mag in $B$), and
the  systematic errors in the transformation to the standard system (which include 
uncertainties of the aperture corrections: about 0.02 mag in $V$ and $I$ and 
0.03 mag in $B$, and the zero points of the photometric calibration:  $\pm
0.02$ mag in $V$, and $\pm 0.03$ mag in $B$ and $I$,  see Section 2.2).

Note that in a preliminary analysis,  variables were studied using their 
differential
photometry with respect to a larger number of comparison stars selected   in
each field (namely four stars per field). However, since results were  very
much the same in the final  study we used just one star per field, namely
in each field the star with most accurate magnitude determinations and with 
colours better
matching the RR Lyrae's average colour.
This procedure minimize any  colour effect
on the differential light curves and amplitudes of the variable stars, due to 
the colour of the comparison stars and the different colour response of the
detectors used in the two runs. 

\begin{table*}[ht]
\begin{center}
%\vskip 4 cm
\caption{Coordinates and magnitudes of the comparison stars}
%\vspace{0.5cm}
%\scriptsize
\begin{tabular}{rcccccccc}
\hline
\multicolumn{1}{c}{Id} &\multicolumn{1}{c}{$\alpha_{2000}$}&\multicolumn{1}{c}
{$\delta_{2000}$}&\multicolumn{1}{c}{$V$}&\multicolumn{1}{c}{n$_{V}$}&
\multicolumn{1}{c}{$B$}&\multicolumn{1}{c}{n$_{B}$}&
\multicolumn{1}{c}{$I$}&\multicolumn{1}{c}{n$_{I}$}\\
\hline
\multicolumn{9}{c}{Field A}\\
1253~~~& 5 22 57.93 & $-$70 31 31.96 &16.889$\pm$0.026 & 14&17.575$\pm0.045$&
14&16.102$\pm$0.025&14\\
%1550~~~& 5 23 08.42 & $-$70 28 50.59 &16.78$\pm$0.03 & 5&18.45$\pm0.04$&5&&\\
% 879~~~& 5 22 50.71 & $-$70 34 46.50 &17.07$\pm$0.03 & 5&18.47$\pm0.04$&5&&\\
%9101~~~& 5 22 53.17 & $-$70 29 50.30 &19.43$\pm$0.03 & 5&20.38$\pm0.04$&5&&\\
\multicolumn{9}{c}{Field B}\\
 128~~~& 5 16 29.75 & $-$71 01 46.62 &16.194$\pm$0.023 & ~15&16.888$\pm0.037$
&14&15.410$\pm$0.035&14\\
% 346~~~& 5 17 28.94 & $-$71 04 34.74 &16.96$\pm$0.03 & ~5&18.39$\pm0.05$&~3&&\\
% 531~~~& 5 17 25.04 & $-$71 02 38.26 &18.15$\pm$0.03 & ~5&19.19$\pm0.05$&~3&&\\
%4340~~~& 5 18 08.17 & $-$71 00 41.17 &19.30$\pm$0.03 & ~5&20.22$\pm0.05$&~3&&\\
\hline
\end{tabular}

%Note: ({\bf Qui vanno controllate le coordinate - MARCELLA})
\end{center} 
%\label{t:tab5}
\end{table*}

In order to define the periodicities we run GRATIS
on the instrumental differential photometry of the 
 variable stars. GRATIS performs a period search 
according to two different algorithms: (a) the Lomb periodogram  (Lomb 1976,
Scargle 1982) and (b) the best-fit of the  data with a truncated Fourier series
(Barning 1962). We first performed the
Lomb analysis on a wide period interval. Then the Fourier algorithm was used to
refine the period definition and to find the best fitting model from which  to
measure the amplitude and average luminosity of each variable. The period
search employed each of the complete (1999+2001) $\Delta b$, $\Delta v$, and
$\Delta i$ data-sets. 
We derived periods and epochs accurate to the third-fourth 
decimal place for all the
variable in our sample, 
well sampled the $B$ and $V$ light curves 
for about 95\% of the
RR Lyrae stars, and detected the Blazhko modulation of the light
curve (Blazhko 1907) in about  17\% of the RRab's and 5.3\% of
the RRc's (see Section 5).
Complete coverage of the light variation was also obtained for 
%Thanks to the two years base line we were able to: (i)
%derive periods and epochs for all the 156 variables  accurate the accuracy depends on the light curve data  sampling,
%which, in the best cases is 72 $V$, 41 $B$, 14 $I$ data points for  variables 
%in field A, and 70 $V$, 38 $B$, 14 $I$ data points for variables in field B;
%(ii) identify ; and  finally (iii) 
%to
%Beside the RR Lyrae stars, we also fully covered  the light curve of 
4 candidate Anomalous
Cepheids (see Section 3.2), for 9 eclipsing binaries with short
 orbital period 
(P$<$1.4 days),  and for 6 of the Cepheids.
%, and reasonably well sampled the light 
%curves  of the remaining 5 Cepheids. 
GRATIS also performs a
search for multiple periodicities, and was run on the data of the 10
double-mode variables falling in our two  fields, 9 in A97 and 1 newly
discovered. However, our data sampling for  these stars is 
inadequate to allow
a very  accurate derivation of the double-mode periodicities: on this
particular aspect, the very extensive data set collected by MACHO and
OGLE II are clearly 
superior to ours.

Best fitting models of the light variation were computed for all variables with
full light curve coverage, using GRATIS. These models are based on 
Fourier series, with the
number of harmonics  generally varying from 1 to 5 for the {\it c}-type RR
Lyrae's, and from 4 to 12 for the {\it ab}-type variables. 
%Models were further
%hand-edited whenever needed. This actually occurred with the $I$ light curves
%due to the small number of   observations in this band, and  in a few {\it
%ab}-type variables, where the skewness of the light curves could be reproduced
%with difficulty, only using a large number of harmonics and at the cost of 
%disturbing unphysical oscillations of a model curve based only on 
%Fourier terms at other phases.  
Intensity-average differential
$<\Delta v>$,  $<\Delta b>$, and $<\Delta i>$ magnitudes were derived for 
all the variables with complete light curves as the integral over the 
entire pulsation cycle of the models best fitting the observed data.
By adding the instrumental magnitudes of the reference stars, we obtained the
$b$, $v$, $i$ mean instrumental magnitudes of the variables, and the mean 
$B$, $V$, $I$
magnitudes in the Johnson-Cousins system 
were calculated using the calibration equations given in Section 2.2 and the
aperture corrections in Section 2.1.

Average residuals from the best fitting models for RR Lyrae's with well sampled
light curves are 0.02-0.03 mag in $V$ and 0.03-0.04 mag in $B$ for the
single-mode, non Blazhko variables, and 0.05-0.10 in $V$ and 0.06-0.12 in  $B$
for the  double-mode stars. The lower accuracy of the $B$ light curves is
because the RR Lyrae stars are intrinsically fainter 
in this passband. 

The individual $B,V,I$ photometric 
measurements of the variables are provided in Table~4.
%, where stars
%are devided by field (Field A: first part of Table~4, 
%Field B: second portion of Table~4) and within each part of
%the table variables are grouped per type: 
%{\it ab-}, {\it c-}, {\it d-}type RR Lyrae, $\delta$ Scuti, candidate Anomalous Cepheids, Cepheids, 
%eclipsing binaries.
For each star we indicate the star identification number, the field where
the star is located, 
the variable type, Heliocentric Julian Day of observations and
corresponding $V$, $B$, $I$ magnitudes. 

\begin{table}[ht]
%\begin{center}
%\vskip 4 cm
\caption{$V,B,I$ photometry of the variable stars}
%\vspace{0.5cm}
\scriptsize
\begin{tabular}{cccccc}
\hline
\multicolumn{6}{c}{Star \#2525 - Field A - RRab}\\
\hline
\multicolumn{1}{c}{HJD}&\multicolumn{1}{c}{$V$}
&\multicolumn{1}{c}{HJD}&\multicolumn{1}{c}{$B$}&
\multicolumn{1}{c}{HJD}&\multicolumn{1}{c}{$I$}\\
\multicolumn{1}{c}{($-$2451183)}&&\multicolumn{1}{c}{($-$2451183)}&&
\multicolumn{1}{c}{($-$2451933)}\\
  0.623172& 19.708  & 0.626309& 20.227& 0.580303  & 18.666 \\
  0.630545& 19.741  & 0.634897& 20.243& 0.608358  & 18.591 \\
  0.660672& 19.738  & 0.666204& 20.047& 0.633786  & 18.731 \\
  0.670556& 19.558  & 0.685707& 19.517& 0.683115  & 18.799 \\
  0.681100& 19.231  & 0.704341& 19.193& 0.708370  & 18.712 \\
  0.690070& 19.120  & 0.722720& 19.010& 0.757143  & 19.024 \\
  0.699977& 19.006  & 0.747280& 19.057& 0.784249  & 19.127 \\
  0.708693& 18.863  & 0.766320& 19.206& 1.574978  & 18.922 \\
  0.718368& 18.831  & 0.785521& 19.278& 1.600522  & 18.922 \\
  0.727072& 18.759  & 0.807245& 19.430& 1.625198  & 18.970 \\
\hline
\end{tabular}
%\normalsize

A portion of Table 4 is shown here for guidance regarding its form
and content. The entire catalogue is available only electronically
%at: {\it http://www.bo.astro.it/$\sim$gisella
at CDS.
%\end{center} 
%\label{t:tab4}
\end{table}
%\begin{table}[ht]
%\begin{center}
%\vskip 4 cm
%\caption{$V,B,I$ photometry of the variable stars}
%\vspace{0.5cm}
%\scriptsize
%\begin{tabular}{rccrccrc}
%\hline
%\multicolumn{8}{c}{Star \#1408 - Field B - RRab}\\
%\hline
%\multicolumn{1}{c}{~~~HJD$-$2451183}&\multicolumn{1}{c}{$V$}
%&&\multicolumn{1}{c}{~~~HJD$-$2451183}&\multicolumn{1}{c}{$B$}&&
%\multicolumn{1}{c}{~~~HJD$-$2451933}&\multicolumn{1}{c}{$I$}\\
%  0.646828& 19.502  &&  0.651180& 19.989&& 0.593738&  18.786\\
%  0.675914& 19.759  &&  1.633004& 19.273&& 0.621226&  18.891\\
%  0.732152& 19.925  &&  1.653004& 19.375&& 0.646214&  18.780\\
%  0.756781& 19.746  &&  1.671511& 19.445&& 0.695879&  18.881\\
%  0.775797& 19.919  &&  1.689891& 19.539&& 0.720694&  18.845\\
%  0.795763& 19.425  &&  1.713559& 19.577&& 0.771342&  18.847\\
%  0.818054& 19.796  &&  1.731326& 19.573&& 0.796747&  18.938\\
%  1.628432& 18.958  &&  1.749381& 19.584&& 1.587742&  18.595\\
%  1.637356& 18.963  &&  1.786881& 19.725&& 1.612892&  18.565\\
%  1.648652& 19.054  &&  1.838339& 19.787&& 1.637777&  18.637\\
%\hline
%\end{tabular}
%%\end{center} 
%\label{t:tab5}
%\end{table}
In Table~5 and 6 we summarize the main characteristics of the
variables for stars in field A and B, separately. Namely we list: identifier, 
coordinates
($\alpha$ and $\delta$) at the 2000 equinox, variable star type, period, 
heliocentric 
Julian day (HJD) of maximum light for the pulsating variables (RR Lyrae's, Cepheids and
$\delta$ Scuti) and of the primary (deeper) minimum light for the 
eclipsing binaries, number of data-points on the $V,B,I$ light curves,
$V,B,I$ 
%intensity and magnitude-averaged 
mean magnitudes and amplitudes of
the light curves, computed as the difference between maximum and 
minimum of the best fitting models, for the variable stars with complete 
coverage of the light variation. At the bottom of each table we also give 
informations on the candidate variables. 
The atlas of light curves is presented in the Appendix. 
%({\bf MARCELLA: la stella 28539 del campo A non ha magnitudini medie in 
%tabella ma non ha neanche una nota che spieghi perche', puoi controllare sul 
%quaderno con lo studio delle curve di luce? Grazie.}). 
%Examples of light curves of an {\it ab-}, a {\it c-} and a {\it d-}type RR 
%Lyrae, of a candidate Anomalous Cepheid, a Classical Cepheid and an
%eclipsing binary are shown in Figure~14, where the photometric data 
%are folded according to the ephemerides given 
%in Table~5 and 6.
%({\bf Discutere qui la luminosita' media delle RR Lyrae nei due campi,
%ricalcolata con le 65 o 66 RRab con curva di luce completa e senza
%shifts del campo A e quelle del campo B e rimandare al C03 per le
%implicazioni astrofisiche.})
%where variables are divided per field and type, and within each 
%group are ordered by increasing period. 

The average apparent luminosities of the RR Lyrae stars with full coverage
of the light curve and without shifts between the 1999 and 2001 photometry
are 
%({\bf QUESTE MEDIE VANNO RICALCOLATE AGGIUNGENDO DELLE 7 RR LYRAE 
%RECUPERATE DA
%MACHO QULLE CON CURVA COMPLETA E NON SOVRALUMINOSE}):
$<V>=19.417 \pm 0.019$ ($\sigma$ =0.154, 67 stars), 
$<B>=19.816 \pm 0.021$ ($\sigma$ =0.171, 67 stars) in field A, and
$<V>=19.318 \pm 0.022$ ($\sigma$ =0.157, 49 stars), 
$<B>=19.678 \pm 0.023$ ($\sigma$ =0.159, 49 stars) in field B.
These values (the $V$ average luminosities in particular)
 are fully consistent with those presented in C03. We refer to this
 paper for an in-depth discussion of their implications on the distance
 to the LMC and related issues.
 We also recall that our average luminosities for the field LMC 
 RR Lyrae stars are in very good agreement with Walker (1992) mean
 apparent luminosity of the RR Lyrae stars in the LMC globular clusters
 (see Section 6 of C03).

 It has often been argued on the better way to compute the 
average magnitude of a variable star and on the colour that better 
represents the temperature of an RR Lyrae star (Sandage 1990, 1993; Carney, 
Storm \& Jones 1992; Bono, Caputo \& Stellingwerf 1995).  
 The average magnitudes of the variable stars in Tables~5 and 6 were computed 
in two different ways, as intensity-averaged means (Columns 8,9,10) and
as magnitude-averaged means (Columns 11,12,13).  Based on theoretical
 grounds it has been claimed that large differences may exist between
these two different types of averages, and that for RR Lyrae stars 
the difference may be as large as 0.1-0.2 mag in $V$ and $B$, respectively   
(Bono et al. 1995).
In Figure~12 we plot the differences between the two types of
averages for star in Field A and B separately. Magnitude-averaged mean 
magnitudes are generally fainter than the intensity-averaged
mean magnitudes, and the differences increase for fainter magnitudes.
However, they are generally small and only in a few cases exceed 
0.1 mag. At the luminosity level of the RR Lyrae stars the average 
differences are $<V_{mag} - V_{int}>$=0.020, $<B_{mag} - B_{int}>$=0.035 and
$<I_{mag} - I_{int}>$=0.010 for stars in Field A, and 
0.022, 0.042, and 0.011 mag for stars in Field B.
%({\bf RICALCOLARE QUESTE DIFFERENZE AGGIUNGENDO LE VARIABILI RECUPERATE DA MACHO})

Figures~13 and ~14 show the position of the various types of
variables in the $V, B-V$ CMDs of Field A and B.

\begin{figure} 
\includegraphics[width=8.8cm, clip=true]{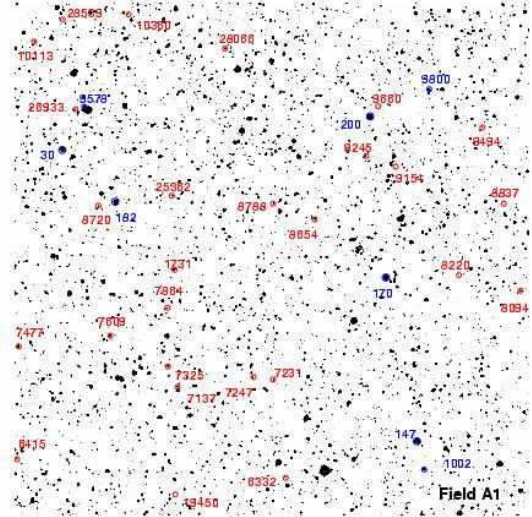}
\caption[]{LMC sub-field A1 ($6.8^{\prime} \times 6.8^{\prime}$), North-East
quadrant. North is up and East is left.
Variables are marked by open circles. Identification numbers are as in Table~5.}
%\label{f:fig3a}
\end{figure}

\begin{figure} 
\includegraphics[width=8.8cm, clip=true]{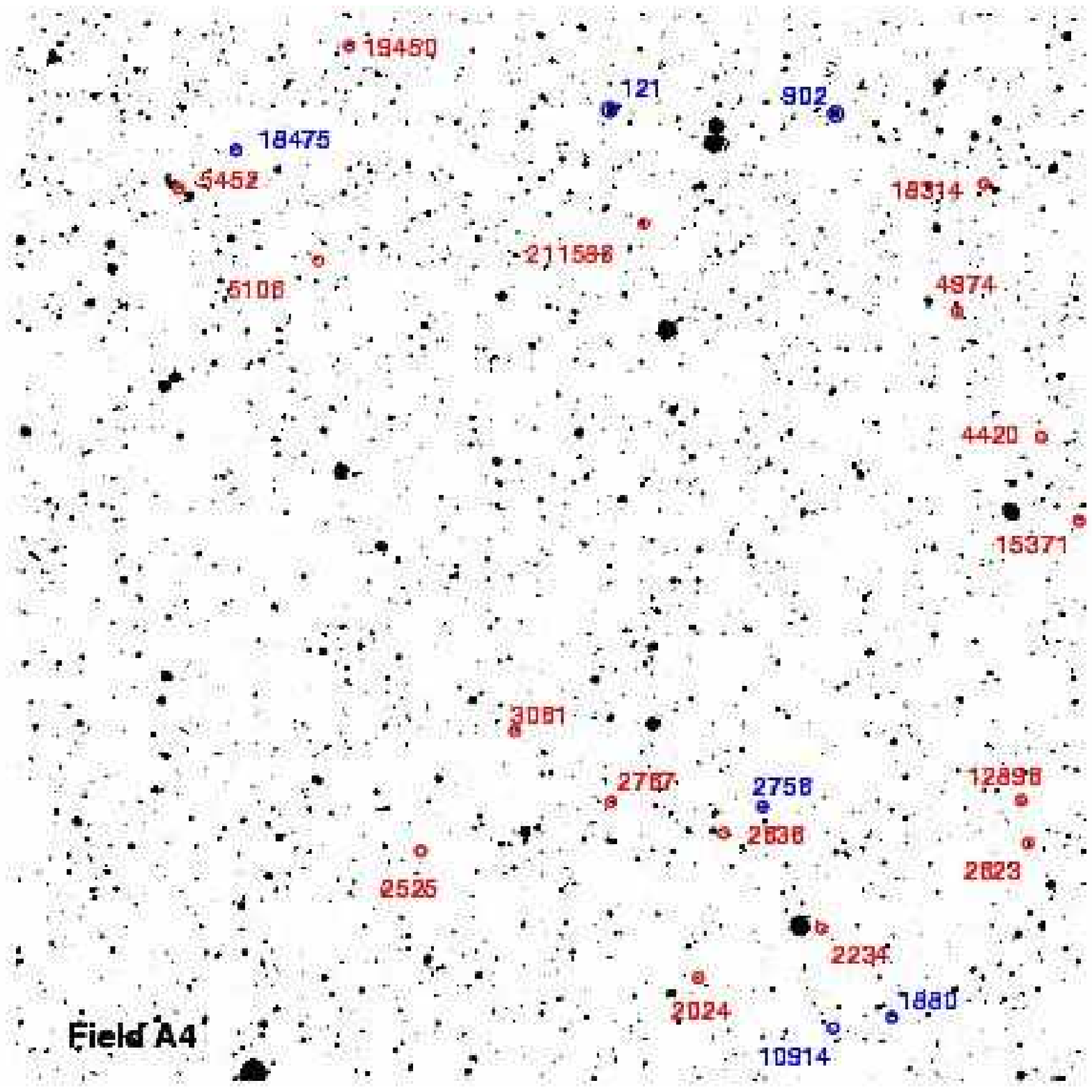}
\caption[]{FORS1 LMC sub-field A4 ($6.8^{\prime} \times 6.8^{\prime}$), 
South-East quadrant. North is up and East is left.
Variables are marked by open circles.
Identification numbers are as in Table~5.
}
%\label{f:fig3d}
\end{figure}

\begin{figure} 
\includegraphics[width=8.8cm, clip=true]{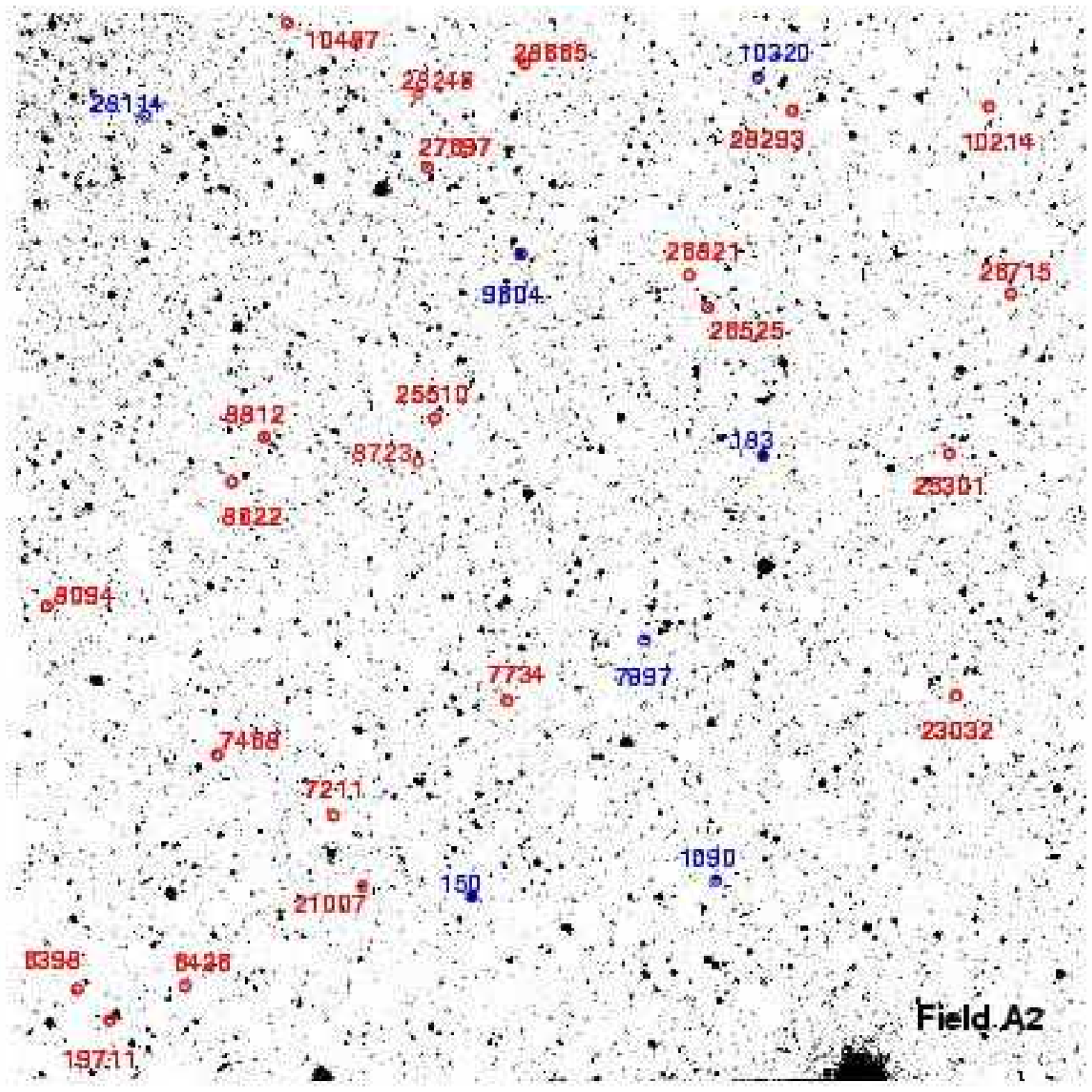}
\caption[]{FORS1 LMC sub-field A2 ($6.8^{\prime} \times 6.8^{\prime}$), 
North-West quadrant.
North is up and East is left. Variables are marked by open circles.
Identification numbers are as in Table~5.
}
%\label{f:fig3b}
\end{figure}

\begin{figure} 
\includegraphics[width=8.8cm, clip=true]{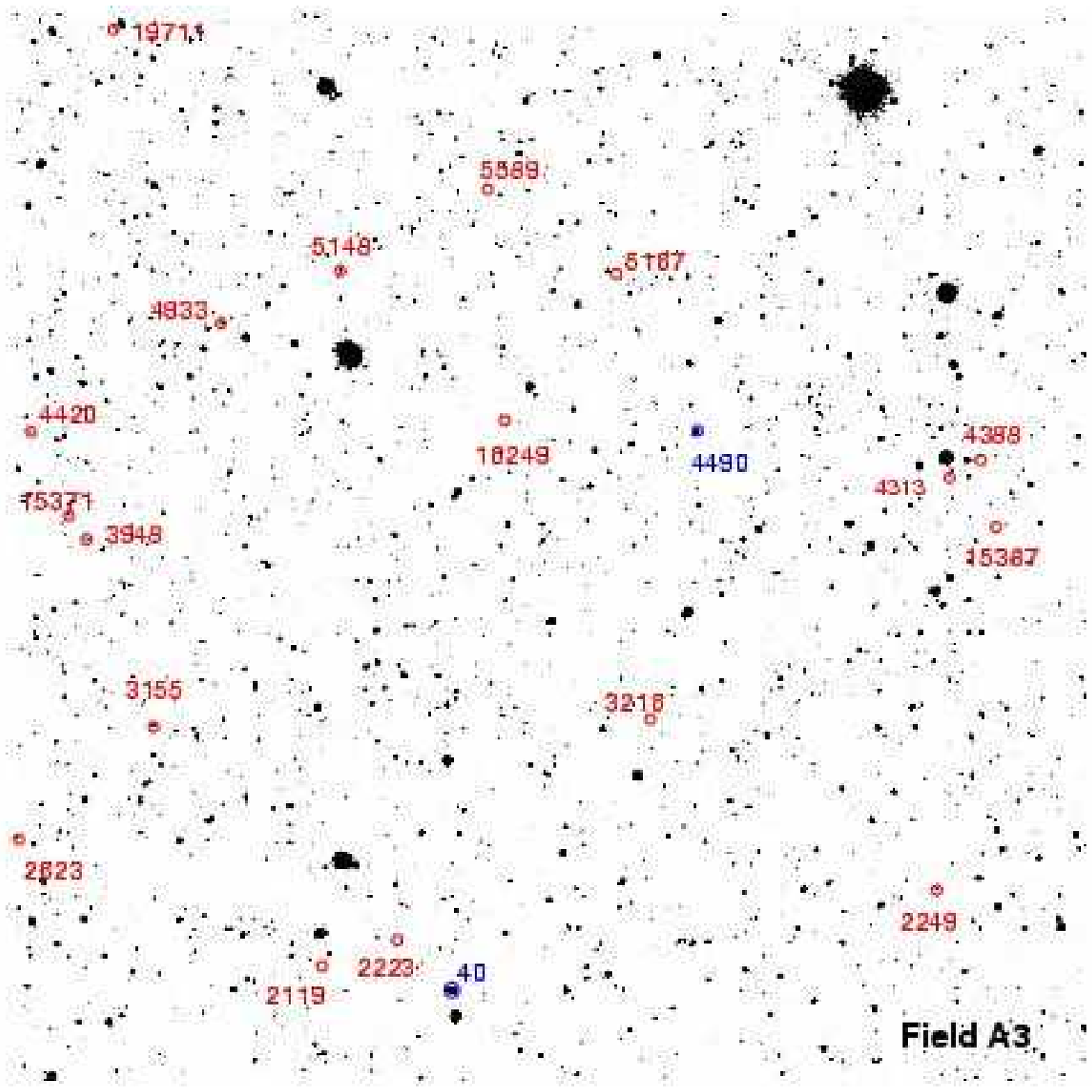}
\caption[]{FORS1 LMC sub-field A3 ($6.8^{\prime} \times 6.8^{\prime}$), 
South-West quadrant.
North is up and East is left. Variables are marked by open circles.
Identification numbers are as in Table~5.
}
%\label{f:fig3c}
\end{figure}

\begin{figure} 
\includegraphics[width=8.8cm, clip=true]{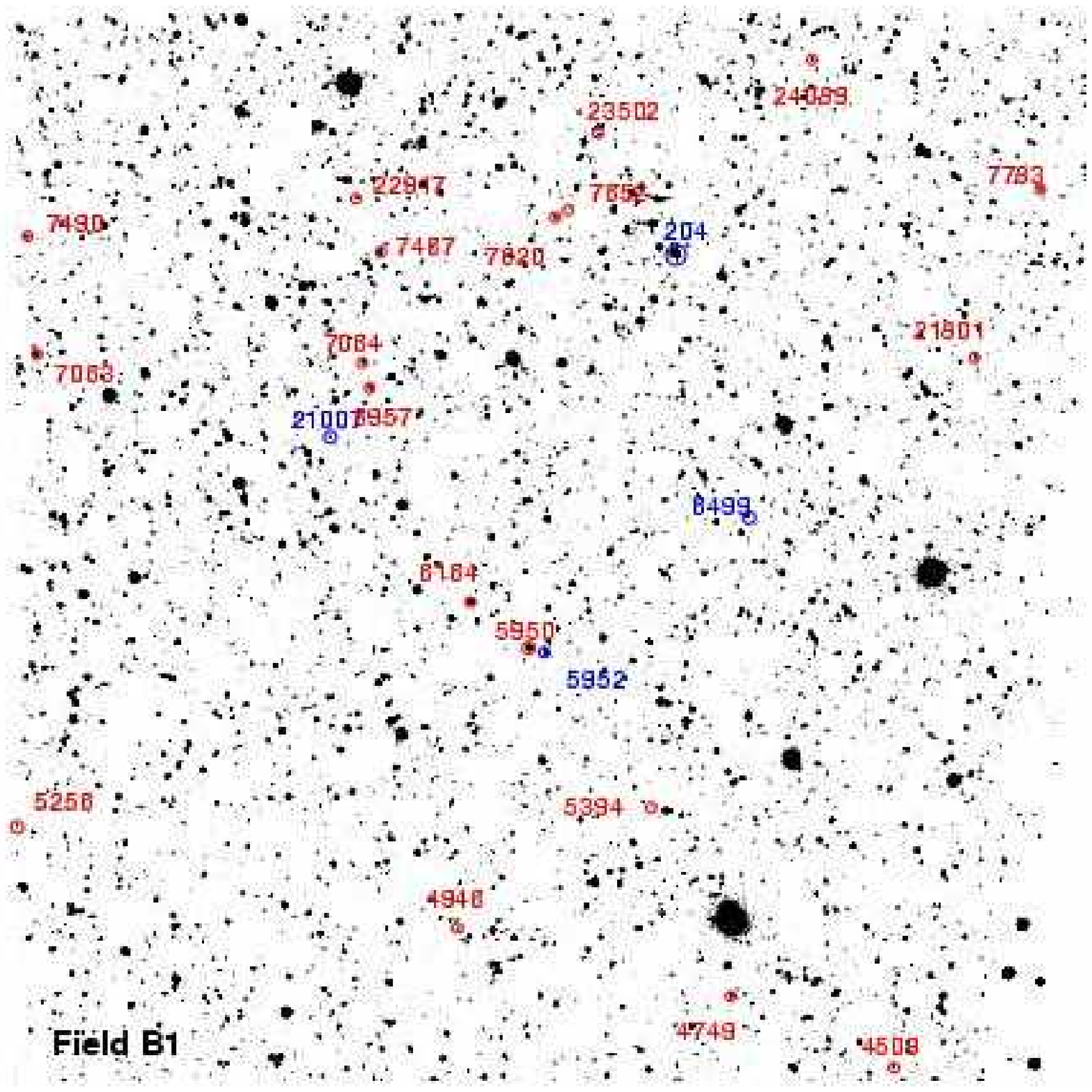}
\caption[]{FORS1 LMC sub-field B1 ($6.8^{\prime} \times 6.8^{\prime}$), 
North-East quadrant.
North is up and East is left. Variables are marked by open circles.
Identification numbers are as in Table~6.
}
%\label{f:fig4a}
\end{figure}

\begin{figure} 
\includegraphics[width=8.8cm, clip=true]{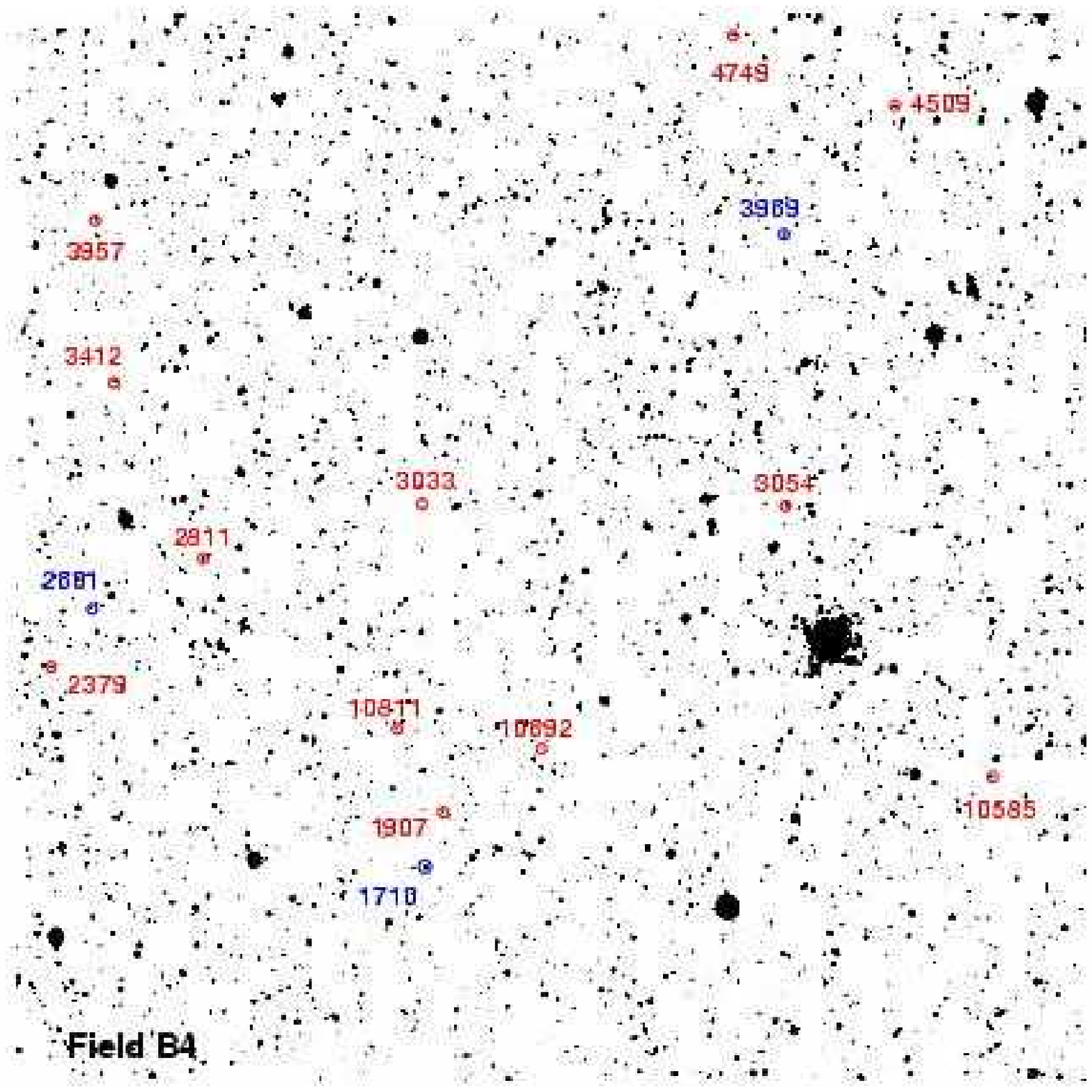}
\caption[]{FORS1 LMC sub-field B4 ($6.8^{\prime} \times 6.8^{\prime}$), 
South-East quadrant.
North is up and East is left. Variables are marked by open circles.
Identification numbers are as in Table~6.
}
%\label{f:fig4d}
\end{figure}

\begin{figure} 
\includegraphics[width=8.8cm, clip=true]{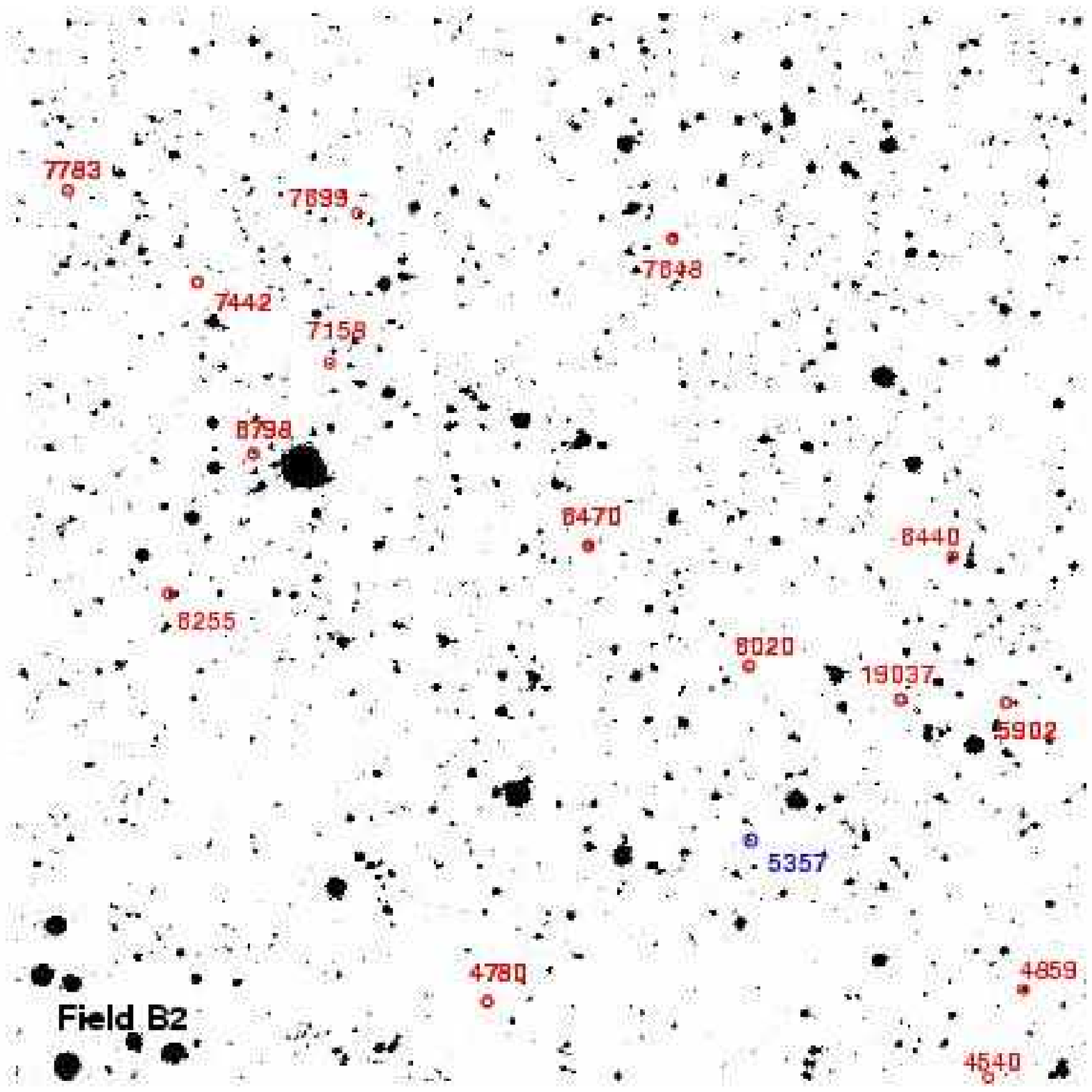}
\caption[]{FORS1 LMC sub-field B2 ($6.8^{\prime} \times 6.8^{\prime}$), 
North-West quadrant.
North is up and East is left. Variables are marked by open circles.
Identification numbers are as in Table~6.
}
%\label{f:fig4c}
\end{figure}

\begin{figure} 
\includegraphics[width=8.8cm, clip=true]{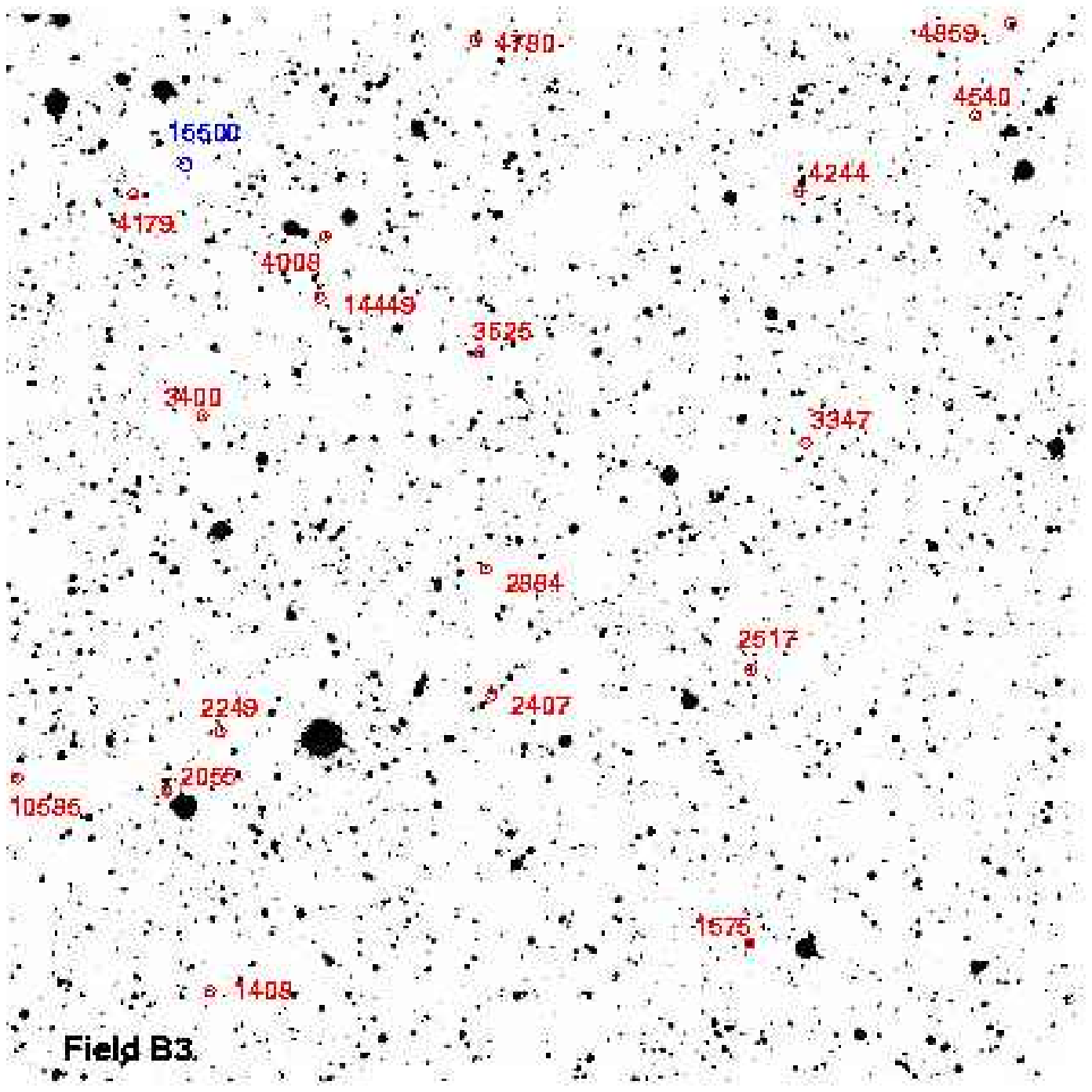}
\caption[]{FORS1 LMC sub-field B3 ($6.8^{\prime} \times 6.8^{\prime}$), 
South-West quadrant.
North is up and East is left. Variables are marked by open circles.
Identification numbers are as in Table~6.
}
%\label{f:fig3b}
\end{figure}

\begin{figure} 
\includegraphics[width=4.3cm]{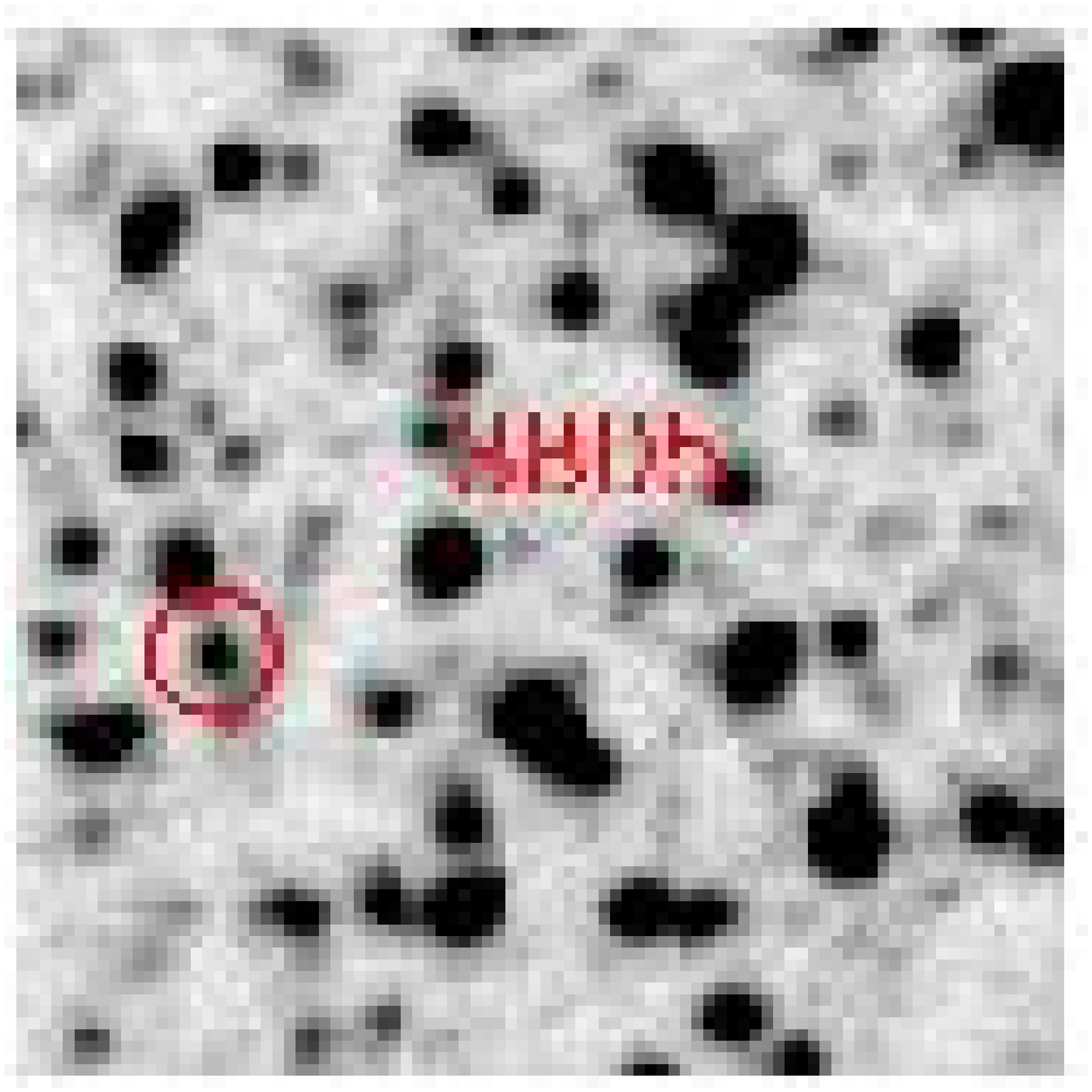}
\includegraphics[width=4.3cm]{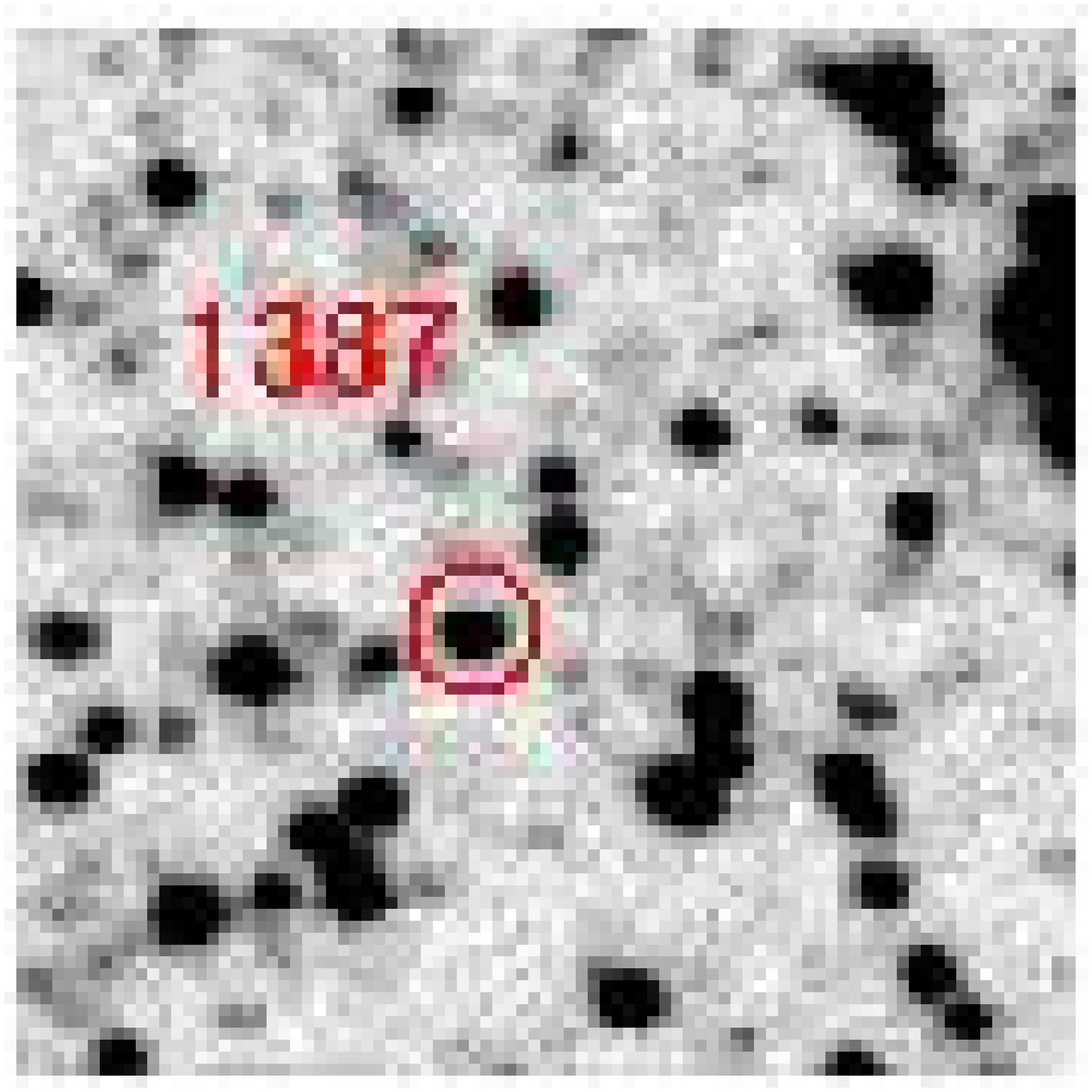}
\caption[]{Left panel: finding chart of the {\it ab-}type RR Lyrae star \# 3805, which is 
located slightly outside
sub-field A4 in the East direction. 
Right panel: finding chart of the {\it c-}type RR Lyrae star \# 1387, which is located slightly 
outside sub-field B4 in the South/East direction.
Both maps show a 
40 $\times$ 40 arcsec area, 
North is up and East is left.
}
%\label{f:fig3bis}
\end{figure}

%\begin{figure} 
%%\includegraphics[width=4cm]{1387_new.ps}
%\caption[]{Finding chart for variable \# 1387, which is located slightly 
%outside sub-field B$_4$ in the XXXX {\bf (CHECK-ANGELA)} direction. 
%The map shows a 39{\bf (CHECK-ANGELA)}$\times$ 
%39{\bf CHECK} arcsec area, 
%North is up and East is left.
%({\bf Mettiamo il Nord ed East sulla figura?})
%}
%\label{f:fig4bis}
%\end{figure}

%\begin{figure} 
%%\includegraphics[width=9cm]{RR_ab_ca_page_4_one.ps}
%%\includegraphics[width=9cm]{RR_c_cb_page_1_one.ps}
%%\includegraphics[width=9cm]{RR_dm_cb_page_1_one.ps}
%%\includegraphics[width=9cm]{Cef_ca_page_1_one.ps}
%%\includegraphics[width=9cm]{Cef_ca_page_1_two.ps}
%%\includegraphics[width=8.8cm]{fig14.ps}
%\caption[]{Examples of light curves. From top to bottom:
%an {\it ab-}, a {\it c-},
%a {\it d-} type RR Lyrae, a candidate Anomalous Cepheid, a Classical Cepheid, 
%and a binary
%system.}
%\label{f:fig1a}
%\end{figure}
 
%\begin{figure} 
%%\includegraphics[width=9cm]{Cef_ca_page_1_one.ps}
%%\includegraphics[width=9cm]{Cef_ca_page_1_two.ps}
%%\includegraphics[width=9cm]{Bin_cb_page_1_one.ps}
%\caption[]{Examples of light curves. From top to bottom:
%an Anomalous Cepheid, a Classical Cepheid, and a binary
%system. ({\bf Marcella rifai questa
%figura con queste stelle ma la macro usata per il tuo 
%talk di Lampedusa}).}
%\label{f:fig1a}
%\end{figure}

%\begin{table*}
\begin{landscape}
\begin{table}
%\small
\caption{Informations and 
average quantities for the variable stars in field A.}
%\begin{center}
\tiny
\begin{tabular}{rcccllrccccccllll}
\hline
       Id~~ &  $\alpha$  &  $\delta$    & Type &     ~~~~P       &     ~~~Epoch        &  Np~~~  &   $<{\rm V_{int}}>$&$<{\rm B_{int}}>$&$<{\rm I_{int}}>$&$<{\rm V_{mag}}>$&$<{\rm B_{mag}}>$&$<{\rm I_{mag}}>$&~A$_{\rm V}$&~A$_{\rm B}$&~A$_{\rm I}$&~~Notes\\
          &    (2000)  &    (2000)    &      &   (days)  &    ($-$2400000)  & (V,B,I)  &   (mag)  &   (mag) &	  (mag) & (mag)    &  (mag)   &   (mag)  &  (mag)  &(mag)    &(mag)   &\\
\hline
%\caption{Informations and average quantities for the single-mode RR Lyrae variables}
%\hline
%&&&&&&&&&&&&&&&&\\
   1731   & 5:23:38.69 & -70:31:08.17 &  ab  &  0.58245  &  51183.63490  &   6,14,$-$  &   $-$    &    $-$   &   $-$	&   $-$    &   $-$    &   $-$   &   ~~$-$   &  1.100: &  ~~$-$   & Incomplete\\
   2525   & 5:23:32.39 & -70:39:15.34 &  ab  &  0.61615  &  51933.57692  &  69,41,$-$  &  19.340  &  19.764  &   $-$   &  19.376  &  19.826  &   $-$	&  0.991  &  1.272  &  ~~$-$   &\\
   2767   & 5:23:17.70 & -70:38:55.90 &  ab  &  0.53106  &  51183.52271  &  64,33,$-$  &  19.467  &  19.874  &  19.074  &  19.517  &  19.962  &  19.089  &  1.091  &  1.363  &  ~~$-$   &\\
   3061   & 5:23:25.13 & -70:38:28.94 &  ab  &  0.47622  &  51182.69038  &  66,41,11   &  19.631  &  20.037  &  19.220  &  19.679  &  20.121  &  19.239  &  0.809  &  1.093  &  0.765 &   Blazhko\\
   3805   & 5:24:04.69 & -70:37:18.42 &  ab  &  0.62740  &  51185.78000  &  60,37,10   &  19.402  &  19.866  &  18.850  &  19.415  &  19.889  &  18.848  &  0.623  &  0.805  &  0.345 &\\
   3948   & 5:22:40.34 & -70:37:16.96 &  ab  &  0.66656  &  51182.61069  &  69,40,14   &  19.292  &  19.686  &  18.628  &  19.331  &  19.757  &  18.647  &  0.959  &  1.287  &  0.654 &   Blazhko?,(a)\\
   4313   & 5:21:33.88 & -70:36:52.65 &  ab  &  0.64222  &  51933.55000  &  54,33,9    &  19.270  &  19.779  &  18.451  &  19.276  &  19.791  &  18.488  &  0.356  &  0.454  &  $-$ &   \\
   4933   & 5:22:29.99 & -70:35:53.61 &  ab  &  0.61350  &  51182.85143  &  65,32,13   &  19.103  &  19.531  &  18.542  &  19.127  &  19.572  &  18.546  &  0.793  &  1.044  &  0.442 &   Blazhko?\\
   4974   & 5:22:51.21 & -70:35:47.69 &  ab  &  0.58069  &  51933.57692  &  69,41,11   &  19.384  &  19.809  &  18.778  &  19.406  &  19.850  &  18.801  &  0.764  &  1.039  &  0.482 &\\
   5106   & 5:22:14.40 & -70:37:43.54 &  ab  &  0.56476  &  51183.84000  &  46,29,12   &  18.820  &  18.939  &  18.587  & 18.827   &  18.952  &  18.596  &  0.391:  &  0.565:  &  0.455:: &  Blend\\
   5148   & 5:22:20.87 & -70:35:34.08 &  ab  &  0.56862  &  51186.68716  &  71,39,12   &   $-$    &    $-$   &   $-$	&   $-$    &   $-$    &   $-$	 &   ~~$-$   &   ~~$-$   & ~~$-$   &  ~~~~(b)\\
   5167   & 5:21:59.65 & -70:35:34.99 &  ab  &  0.63023  &  51934.57159  &  64,33,11   &  19.359  &  19.837  &  18.808  &  19.373  &  19.864  &  18.822  &  0.570  &  0.765  &  0.550:&\\
   5331   & 5:23:15.30 & -70:35:14.46 &  ab  &  0.58234  &  51185.61000  &  32,21,14   &  19.673  &  20.079  &  19.103  &  19.696  &  20.133  &  19.107  &0.932::: &  1.192  &  0.322 &\\
   5452   & 5:23:51.03 & -70:35:01.06 &  ab  &  0.67849  &  51933.62602  &  67,38,12   &  19.296  &  19.799  &   $-$	&  19.309  &  19.812  &   $-$	 &  0.589  &  0.569  &  ~~$-$   &\\
   5589   & 5:22:09.54 & -70:35:02.50 &  ab  &  0.63648  &  51934.60800  &  59,33,11   &  19.574  &  20.079  &  18.942  &  19.578  &  20.089  &  18.949  &  0.364  &  0.417  &  0.391:&   Blazhko?\\
   6398   & 5:22:40.71 & -70:33:50.18 &  ab  &  0.56026  &  51182.78609  &  70,41,12   &  19.317  &  19.745  &  18.730  &  19.347  &  19.802  &  18.736  &  0.883  &  1.156  &  0.561 &\\
   6426   & 5:22:32.47 & -70:33:48.73 &  ab  &  0.66224  &  51182.78094  &  70,41,14   &  19.185  &  19.584  &  18.555  &  19.227  &  19.660  &  18.569  &  1.045  &  1.312  &  0.707 &\\
   7211   & 5:22:21.12 & -70:32:43.96 &  ab  &  0.51978  &  51182.97180  &  67,36,14   &   $-$    &    $-$   &   $-$	&   $-$    &   $-$    &   $-$   & 0.875:: & 1.377:: &0.655:::& Incomplete\\
   7247   & 5:23:25.53 & -70:32:33.45 &  ab  &  0.56171  &  51182.56088  &  70,40,14   &  19.408  &  19.795  &  18.857  &  19.429  &  19.835  &  18.852  &  0.714  &  0.865  &  0.422 &\\
   7325   & 5:23:39.08 & -70:32:24.81 &  ab  &  0.48677  &  51183.68571  &  68,40,13   &  19.435  &  19.845  &  18.893  &  19.479  &  19.927  &  18.906  &  1.131  &  1.449  &  0.677 &\\
   7468   & 5:22:30.00 & -70:32:20.56 &  ab  &  0.63550  &  51182.79217  &  62,37,12   &  19.615  &  20.126  &  18.850  &  19.622  &  20.142  &  18.855  &  0.492  &  0.692  &  0.304:&\\
   7477   & 5:24:02.92 & -70:32:08.60 &  ab  &  0.65641  &  51183.08070  &  69,40,14   &  19.183  &  19.552  &  18.670  &  19.249  &  19.628  &  18.695  &  1.108  &  1.371  &  0.730 &\\
   7609   & 5:23:48.34 & -70:32:00.33 &  ab  &  0.57336  &  51182.80975  &  65,38,14   &  19.313  &  19.699  &    $-$	&  19.340  &  19.751  &   $-$	 &  0.788  &  1.004  &  ~~$-$   &\\
   7734   & 5:22:07.82 & -70:31:59.83 &  ab  &  0.61699  &  51183.25568  &  53,24,$-$  &    $-$   &    $-$   &   $-$	&   $-$    &   $-$    &   $-$	 &  0.502  &  0.661  &  ~~$-$   &  ~~~~(c)\\
	  & 	       &	      &      &           &	         &  13,14,10   &    $-$   &    $-$   &   $-$	&   $-$    &   $-$    &   $-$	 &  0.519  &  0.673  &  0.311 &\\
   8094   & 5:22:43.00 & -70:31:23.70 &  ab  &  0.74575  &  51182.23076  &  60,40,13   &  19.353  &  19.891  &  18.676  &  19.360  &  19.908  &  18.682  &  0.452  &  0.644  &  0.323 &\\
   8220   & 5:22:52.79 & -70:31:11.11 &  ab  &  0.67684  &  51182.92564  &  70,41,12   &  19.469  &  19.920  &  18.770  &  19.483  &  19.949  &  18.780  &  0.608  &  0.829  &  0.382 &\\
   8720   & 5:23:50.14 & -70:30:16.73 &  ab  &  0.65767  &  51934.68800  &  66,38,14   &  19.129  &  19.489  &  18.590  &  19.185  &  19.584  &  18.607  &  1.163  &  1.447  &  0.760 &\\
   8788   & 5:23:22.35 & -70:30:14.64 &  ab  &  0.55960  &  51183.01548  &  70,41,14   &  19.444  &  19.844  &   $-$	&  19.482  &  19.916  &   $-$	 &  0.939  &  1.240  & 0.830:::&   Blazhko\\
   9154   & 5:23:02.88 & -70:29:44.63 &  ab  &  0.61981  &  51933.61768  &  67,40,11   &  19.552  &  20.032  &  18.931  &  19.569  &  20.059  &  18.933  &  0.662  &  0.755  &  0.255  &Blazhko,(d)\\
          & 	       &	      &      &           &	         &  53,26,$-$  &    $-$   &   $-$    &   $-$    &   $-$    &   $-$    &   $-$    &  0.759  &  0.913  &  ~~$-$  &\\
	  & 	       &	      &      &           &	         &  14,14,11   &    $-$   &    $-$   &   $-$	&   $-$    &   $-$    &   $-$	 &  0.384  &  0.478  & 0.263::&\\
   9245   & 5:23:07.61 & -70:29:36.50 &  ab  &  0.55980  &  51182.81277  &  65,38,12   &    $-$   &    $-$   &   $-$	&   $-$    &   $-$    &   $-$	 &  0.706  &  0.948  & 0.402::&   ~~~~(e)\\
   9494   & 5:22:49.20 & -70:29:13.50 &  ab  &  0.57860  &  51185.74736  &  69,40,12   &  19.217  &  19.584  &   $-$	&  19.266  &  19.674  &   $-$	 &  1.150  &  1.410  &  ~~$-$   &\\
   9660   & 5:23:05.71 & -70:28:56.83 &  ab  &  0.62181  &  51183.68200  &  67,41,14   &  19.392  &  19.862  &  18.795  &  19.407  &  19.890  &  18.796  &  0.669  &  0.811  &  0.312 &\\
  10214   & 5:21:31.10 & -70:28:12.01 &  ab  &  0.59994  &  51934.74600  &  60,32,10   &  19.204  &  19.639  &    $-$	&  19.217  &  19.665  &   $-$	 &  0.633  &  0.824  &  ~~$-$   &\\
  10487   & 5:22:24.55 & -70:27:40.52 &  ab  &  0.58957  &  51182.40810  &  54,38,12   &  19.569  &  20.022  &  18.886  &  19.603  &  20.073  &  18.896  &  0.913  &  1.106  &  0.603 &\\
  12896   & 5:22:46.09 & -70:38:54.95 &  ab  &  0.57368  &  51185.65865  &  71,37,12   &  19.589  &  20.027  &  18.955  &  19.620  &  20.087  &  18.964  &  0.911  &  1.248  &  0.691 &\\
  15371   & 5:22:41.56 & -70:37:07.14 &  ab  &  0.58712  &  51185.66370  &  31,28,12   &  19.460  &  19.874  &  18.557  &  19.480  &  19.936  &  18.605  &  0.929  &  1.201  &  0.957 &\\
  15387   & 5:21:30.38 & -70:37:11.30 &  ab  &  0.55983  &  51933.58300  &  58,30,5    &  19.612  &  20.043  &   $-$	&  19.630  &  20.075  &   $-$	 &  0.705  &  0.839: &  ~~$-$   &\\
  16249   & 5:22:08.22 & -70:36:31.00 &  ab  &  0.60475  &  51183.72300  &  71,41,13   &  19.378  &  19.759  &   18.844 &  19.430  &  19.837  &  18.847  &  1.118  &  1.403  &  0.360 &Blazhko,(d)\\
          & 	       &	      &      &           &	         &  57,27,$-$  &    $-$   &    $-$   &   $-$	&   $-$    &   $-$    &   $-$	 &  1.252  &  1.500  &  ~~$-$ &\\
	  & 	       &	      &      &           &	         &  14,14,13   &    $-$   &    $-$   &   $-$	&   $-$    &   $-$    &   $-$	 &  0.764  &  1.090  &  0.337:&\\
  18314   & 5:22:49.08 & -70:34:59.12 &  ab  &  0.58711  &  51183.17420  &  70,41,13   &  19.410  &  19.790  &  18.815  &  19.459  &  19.793  &  18.835  &  1.120  &  1.426  &  0.719 &\\
  19450   & 5:23:37.89 & -70:34:06.71 &  ab  &  0.39792  &  51182.55700  &  70,41,14   &  19.662  &  19.983  &  19.286  &  19.737  &  20.116  &  19.324  &  1.344  &  1.709  &  1.098 &\\
  19711   & 5:22:38.14 & -70:34:02.02 &  ab  &  0.55296  &  51181.77241  &  29,18,14   &  19.200  &  19.535  &  18.607  &  19.244  &  19.606  &  18.617  &  0.961  &  1.148  &  0.756 &\\
  21007   & 5:22:18.85 & -70:33:10.84 &  ab  &  0.75730  &  51933.80000  &  14,13,8    &  19.319  &  19.841  &  18.487  &  19.323  &  19.851  &  18.488  &  0.340  &  0.456  &  0.162 &   \\
  25301   & 5:21:33.95 & -70:30:24.47 &  ab  &  0.56059  &  51182.71139  &  63,32,12   &  19.766  &  20.237  &   $-$	&  19.805  &  20.317  &   $-$	 &  1.005  &  1.359  &  ~~$-$   &\\
  25362   & 5:23:38.48 & -70:30:08.55 &  ab  &  0.57746  &  51182.73348  &  70,41,14   &  19.443  &  19.816  &  18.764  &  19.488  &  19.903  &  18.785  &  1.078  &  1.466  & 0.671::&\\
  25510   & 5:22:13.37 & -70:30:11.50 &  ab  &  0.64956  &  51183.78150  &  64,38,11   &  19.150  &  19.614  &  18.554  &  19.160  &  19.631  &  18.561  &  0.609  &  0.707  &  0.417 &   Blazhko?\\
  26525   & 5:21:52.45 & -70:29:28.68 &  ab  &  0.52288  &  51186.67057  &  66,37,14   &  19.473  &  19.913  &    $-$	&  19.507  &  19.991  &   $-$	 &  0.863  &  1.269  &  ~~$-$   &\\
  26821   & 5:21:53.90 & -70:29:17.47 &  ab  &  0.58755  &  51185.75937  &  65,39,13   &  19.624  &  20.097  &  19.097  &  19.644  &  20.132  &  19.104  &  0.752  &  0.989  &  0.473 &\\
  26933   & 5:23:53.78 & -70:28:59.71 &  ab  &  0.48829  &  51186.66446  &  59,35,14   &  19.295  &  19.577  &  18.750  &  19.355  &  19.679  &  18.754  &  1.188  &  1.490  &  0.791 &\\
  28066   & 5:23:30.05 & -70:28:11.07 &  ab  &  0.59975  &  51933.60857  &  51,24,$-$  &    $-$   &    $-$   &   $-$	&   $-$    &   $-$    &   $-$	 &  0.496  &  0.666  &  ~~$-$   &   ~~~~(f)\\
	  & 	       &	      &      &           &	         &  13,14,12   &    $-$   &    $-$   &   $-$	&   $-$    &   $-$    &   $-$	 &  0.503  &  0.629  &  0.326 &\\
  28246   & 5:22:14.65 & -70:28:06.71 &  ab  &  0.47777  &  51934.62544  &  15,32,10   &  19.605  &  20.030  &  19.144  &  19.660  &  20.106  &  19.182  &  1.270  &  1.543  &  1.023 &   \\
  28293   & 5:21:46.08 & -70:28:13.20 &  ab  &  0.66148  &  51186.76300  &  56,24,7    &  19.520  &  20.053  &	$-$     &  19.524  &  20.062  &	  $-$    &  0.403  &  0.553  &0.472:::&	~~~~(g)\\
  28539   & 5:23:55.73 & -70:27:48.61 &  ab  &  0.61388  &  51934.75195  &  68,36,11   &  19.533  &  19.916  &  18.796  &  19.592  &  20.028  &  18.815  &  1.149  &  1.495  &  0.707 &\\
% 211586   & 5:23:15.30 & -70:35:14.46 &  ab  &  0.54890  &  51932.83990  &  14,13,11   &  19.649  &  20.040  &  19.022  &  19.673  &  20.082  &  19.026  &$>$0.596 &$>$0.966 &$>$0.307&\\
% 227671   & 5:22:14.65 & -70:28:06.71 &  ab  &  0.47674  &  2451934.618  &  14,14,10   &  19.603  &  20.015  &  19.159  &  19.653  &  20.115  &  19.198  &  1.222  &  1.569  &  1.013 &\\
\hline                                                                           														 
&&&&&&&&&&&&&&&&\\
   2024  & 5:23:10.96  &  -70:40:03.33 &  c  &  0.36008  &  51933.70166  &  72,40, 8  & 19.500  &  19.876  &  19.165  &  19.513 &  19.893  &  19.177 &  0.509  &  0.606  &  0.455  &\\
   2119  & 5:22:22.20  &  -70:39:59.98 &  c  &  0.26526  &  51934.60511  &  64,39,10  & 19.659  &  19.986  &  19.407  &  19.663 &  19.990  &  19.421 &  0.297  &  0.354  &0.479::::&\\
   2223  & 5:22:16.48  &  -70:39:50.18 &  c  &  0.28784  &  51934.58000  &  67,36,10  & 19.556  &  19.836  &  19.136  &  19.568 &  19.856  &  19.145 &  0.493  &  0.604  &  0.499  &\\
   2234  & 5:23:01.41  &  -70:39:44.47 &  c  &  0.32280  &  51933.74938  &  38,26,$-$ &    $-$  &   $-$    &	$-$   &    $-$  &    $-$   &	$-$  &  0.531  &  0.664  &  ~~$-$    &   ~~~~(f)\\
	 & 	       &	       &     &           &	         &  14,14,12  &    $-$  &   $-$    &	$-$   &    $-$  &    $-$   &	$-$  &  0.425  &  0.584  &  0.389  &\\
   2623  & 5:22:45.59  &  -70:39:11.37 &  c  &  0.29130  &  51183.62631  &  65,40,13  & 19.368  &  19.631  &  19.046  &  19.379 &  19.649  &  19.057 &  0.441  &  0.595  &  0.454  &\\
   2636  & 5:23:09.26  &  -70:39:08.61 &  c  &  0.31611  &  51934.76812  &  69,40,7   & 19.595  &  19.896  &  19.080  &  19.605 &  19.919  &  19.083 &  0.464  &  0.633  &  0.234  &\\
   3216  & 5:21:57.05  &  -70:38:25.85 &  c  &  0.21824  &  51185.76536  &  67,39,$-$ &   $-$	&    $-$   &   $-$    &    $-$  &    $-$   &	$-$  &  0.407  &  0.515  &  ~~$-$    &   ~~~~(h)\\
	 & 	       &	       &     &           &	         &  53,26,$-$ &    $-$  &   $-$    &   $-$    &    $-$  &    $-$   &	$-$  &  0.411  &  0.526  &  ~~$-$    &\\
	 & 	       &	       &     &           &	         &  14,13,14  &    $-$  &   $-$    &   $-$    &    $-$  &    $-$   &	$-$  &  0.392  &  0.513  &  0.405  &\\
   4388  & 5:21:31.63  &  -70:36:46.15 &  c  &  0.34194  &  51185.64493  &  63,30, 9  & 19.427  &  19.758  &   $-$    &  19.431 &  19.764  &	$-$  &  0.305  &  0.363  &  0.200::&\\
   6332  & 5:23:20.39  &  -70:33:53.53 &  c  &  0.25047  &  51933.59613  &  68,41,14  & 19.433  &  19.753  &  19.005  &  19.439 &  19.766  &  19.007 &  0.374  &  0.527  &  0.300  &	Blazhko,(i)\\
	 &	       &	       &     &  0.24961  &  51933.57400  &  14,14,14  & 19.444  &  19.750  &  19.023  &  19.452 &  19.764  &  19.027 &  0.383  &  0.531  &  0.298  &   \\
\hline										      
\end{tabular}
%\end{center}
\normalsize
\end{table}
\end{landscape}
%%%%%%%%%%%%%%%%%%%%%%%%%%%%%%%%%%%%%%%%
%\pagestyle{empty}
\begin{landscape}
\begin{table}
%\small
%\caption{Informations and average quantities for the variable stars in field A.}\\
%\begin{center}
%\scriptsize
%\small
Table 5: - continued -\\

%\caption{Informations and average quantities for the single-mode RR Lyrae 
%variables}
\tiny
\begin{tabular}{rcccllrccccccllll}
\hline
       Id~~ &  $\alpha$  &  $\delta$    & Type &     ~~~~P       &     ~~~Epoch        &  Np~~~  &   $<{\rm V_{int}}>$&$<{\rm B_{int}}>$&$<{\rm I_{int}}>$&$<{\rm V_{mag}}>$&$<{\rm B_{mag}}>$&$<{\rm I_{mag}}>$&~A$_{\rm V}$&~A$_{\rm B}$&~A$_{\rm I}$&~~Notes\\
%       Id  &  $\alpha$  &  $\delta$    & Type &     P       &     Epoch        &  Np  &$<{\rm V_{int}}>$&$<{\rm B_{int}}>$&$<{\rm I_{int}}>$&$<{\rm V_{mag}}>$&$<{\rm B_{mag}}>$&$<{\rm I_{mag}}>$&A$_{\rm V}$&A$_{\rm B}$&A$_{\rm I}$&Note\\
          &    (2000)  &    (2000)    &      &   (days)  &    ($-$2400000)  & (V,B,I)  &    (mag)  &   (mag) &     (mag) & (mag)    &  (mag)   &   (mag)  &  (mag)  &(mag)    &(mag)   &\\
\hline
&&&&&&&&&&&&&&&&\\
%   6332  & 5:23:20.39  &  -70:33:53.53 &  c  &  0.25047  &  51933.59613  &  68,41,14  & 19.433  &  19.753  &  19.005  &  19.439 &  19.766  &  19.007 &  0.374  &  0.527  &  0.300  &    Blazhko,(i)\\
%         &             &               &     &  0.24961  &  51933.57400  &  14,14,14  & 19.444  &  19.750  &  19.023  &  19.452 &  19.764  &  19.027 &  0.383  &  0.531  &  0.298  &	\\
   6415  & 5:24:03.13  &  -70:33:38.44 &  c  &  0.44299  &  51186.74600  &  64,38,12  & 19.206  &  19.573  &  18.713  &  19.215 &  19.583  &  18.715 &  0.438  &  0.473  &  0.229  &\\
   7231  & 5:23:22.37  &  -70:32:35.39 &  c  &  0.32349  &  51183.11007  &  66,36,14  & 19.322  &  19.643  &  18.861  &  19.331 &  19.664  &  18.866 &  0.413  &  0.607  &  0.287  &\\
   7864  & 5:23:39.20  &  -70:31:38.15 &  c  &  0.31347  &  51182.78235  &  67,39,14  & 19.464  &  19.774  &  19.047  &  19.475 &  19.792  &  19.050 &  0.433  &  0.545  &  0.233  &\\
   8622  & 5:22:28.87  &  -70:30:35.86 &  c  &  0.32082  &  51934.69553  &  66,39,14   & 19.542  &  19.868  &  19.103  &  19.552 &  19.887  &  19.108 &  0.429  &  0.628  &  0.307  &\\
   8812  & 5:22:26.38  &  -70:30:19.05 &  c  &  0.35485  &  51182.82224  &  69,41,14   & 19.397  &  19.767  &  18.821  &  19.410 &  19.791  &  18.829 &  0.515  &  0.626  &  0.396  &\\
   8837  & 5:22:45.64  &  -70:30:14.33 &  c  &  0.31629  &  51182.66999  &  64,41,14   & 19.566  &  19.905  &  19.057  &  19.580 &  19.927  &  19.061 &  0.501  &  0.631  &  0.328  &\\
  10113  & 5:24:00.31  &  -70:28:06.22 &  c  &  0.35231  &  51186.78595  &  69,40,14   & 19.486  &  19.878  &  18.891  &  19.494 &  19.894  &  18.895 &  0.415  &  0.535  &  0.321  &\\
  10360  & 5:23:45.34  &  -70:27:44.11 &  c  &  0.27926  &  51934.58000  &  53,26,$-$  &    $-$  &    $-$   &	$-$    &    $-$  &    $-$   &	 $-$  &  0.431  &  0.541  &   ~~$-$   &	~~~~(f)\\
	 & 	       &	       &     &           &	         &  13,13,13   &    $-$  &    $-$   &  19.199  &    $-$  &    $-$   &  19.201 &  0.363  &  0.507  &  0.203  &\\
  26715  & 5:21:29.21  &  -70:29:23.32 &  c  &  0.35646  &  51182.69358  &  52,25,$-$  & 19.378  &  19.725  &	$-$    &  19.388 &  19.746  &	 $-$  &  0.486  &  0.678  &    ~~$-$  &	~~~~(l)\\
  27697  & 5:22:13.98  &  -70:28:34.98 &  c  &  0.38293  &  51184.64328  &  66,39,$-$  & 19.166  &  19.541  &  18.744  &  19.174 &  19.554  &  18.746 &  0.396  &  0.471  &  ~~$-$    &	Blazhko?,(m)\\
	 & 	       &	       &     &           &	         &  53,25,$-$  &    $-$  &    $-$   &	$-$    &    $-$  &    $-$   &	 $-$  &  0.380  &  0.486  &    ~~$-$  &\\
	 & 	       &	       &     &           &	         &  13,14,11   &    $-$  &    $-$   &	$-$    &    $-$  &    $-$   &	 $-$  &  0.485  &  0.513  &  0.229: &\\
  28665  & 5:22:06.49  &  -70:27:55.55 &  c  &  0.30047  &  51183.77067  &  64,38,$-$  &    $-$  &    $-$   &	$-$    &    $-$  &    $-$   &	 $-$  &  0.348  &  0.536  &   ~~$-$   &	~~~~(n)\\
	 &	       &	       &     &           &		 &  52,25,$-$  &    $-$  &    $-$   &	$-$    &    $-$  &    $-$   &	 $-$  &  0.336  &  0.553  &   ~~$-$   &\\
	 &	       &	       &     &           &		 &  12,13,12   &    $-$  &    $-$   &  19.264  &    $-$  &    $-$   &  19.270 &  0.422  &  0.564  &  0.342  &\\
\hline										        		      
&&&&&&&&&&&&&&&&\\
%&&&&&&&&&&&&&&&&&&&&&\\ 
   2249  & 5:21:33.31  & -70:39:51.28  &  d  &  0.30731  &  51182.99235  &  70,41,12&  19.372  &  19.704  &  18.878  &  19.389 &  19.729  &  18.903 &  0.600  &  0.709  &  0.744::&\\
   3155  & 5:22:35.25  & -70:38:28.39  &  d  &  0.38161  &  51934.66700  &  66,40,14&  19.209  &  19.577  &  18.792  &  19.218 &  19.598  &  18.803 &  0.419  &  0.664  &  0.496  &\\
         & 	       & 	       &     &  0.38141  &               &          &          &          &          &         &          &         &         &         &         &\\
   4420  & 5:22:44.66  & -70:36:35.68  &  d  &  0.35989  &  51182.65973  &  72,41,14&  19.409  &  19.726  &  18.784  &  19.417 &  19.740  &  18.794 &  0.419  &  0.554  &  0.432  &\\
   7137  & 5:23:37.64  & -70:32:41.42  &  d  &  0.34301  &  51934.60052  &  72,40,14&  19.413  &  19.736  &   $-$    &  19.420 &  19.750  &    $-$  &  0.413  &  0.557  &  0.424  &\\
   8654  & 5:23:15.72  & -70:30:27.24  &  d  &  0.34544  &  51183.77084  &  66,35,12&  19.269  &  19.651  &  18.848  &  19.275 &  19.658  &  18.849 &  0.475  &  0.817  &  0.113  &\\
  23032  & 5:21:33.40  & -70:31:57.24  &  d  &  0.34226  &  51182.93313  &  71,39,13&  19.597  &  19.993  &   $-$    &  19.682 &  20.081  &    $-$  &  0.693  &  0.935  &  ~~$-$    &\\

\hline
%&&&&&&&&&&&&&&&&\\
&&&&&&&&&&&&&&&&\\
%     DELTA SCUTI
 28114   &  5:22:35.51  &  -70:28:15.66  &  $\delta$S  &  0.11268  &  51183.63220  & 69,40,11 & 19.940  &  20.273 &   $-$	 &  19.943 &  20.280  &    $-$  & 0.280 & 0.388 & 0.185 &\\
\hline

%\end{tabular}
%\end{landscape}
%%%%%%%%%%%%%%%%%%%%%%%%%%%%%%%%%%%%%%%%
%\pagestyle{empty}
%\begin{landscape}
%\caption{Informations and average quantities for the single-mode RR Lyrae variables}
%\small
%Tabella 6.2: Ceph.eidi Anomale, Ceph.eidi, Binarie, Delta Scuti
%\tiny
%\newline
%\begin{tabular}{rcccllccccccccccccccccccc}
%\hline
%\hline
%N&$\alpha$    &$\delta$       &Tipo&P          &Epoca            &Np&Na&Res.&Np&Na&Res.&Np&Na&Res.&$<{\rm V_{int}}>$&$<{\rm B_{int}}>$&$<{\rm I_{int}}>$&$<{\rm V_{mag}}>$&$<{\rm B_{mag}}>$&$<{\rm I_{mag}}>$&A$_{\rm V}$&A$_{\rm B}$&A$_{\rm I}$&Note          \\
%        &       (2000)     &      (2000)        &           &(giorni)      &($-$2400000)      & (V) & (V)   &(mag)    & (B)  & (B)    &  (mag) & (I)  & (I)         &(mag)      &(mag)  &(mag) &(mag) & (mag) &(mag) &(mag) & (mag) &(mag)      &(mag)     & \\
%\hline

&&&&&&&&&&&&&&&&\\
%    Ceph.EIDI ANOMALE
   9578  &  5:23:52.36  &  -70:28:57.87  &  AC  &  0.54758  &  51186.81780  & 62,34,12&  18.626  &  19.277  &  17.789  &  18.620 &  19.293  &  17.790  &  0.307  &  0.576 &  0.205  &\\
   9604  &  5:22:07.01  &  -70:29:07.41  &  AC  &  0.61569  &  51182.38306  & 62,29,12&  18.932  &  19.234  &  18.550  &  18.947 &  19.253  &  18.558  &  0.655  &  0.774 & 0.532 &\\
  10320  &  5:21:48.72  &  -70:28:00.82  &  AC  &  0.29177  &  51185.76536  & 66,37,14&  18.655  &  19.236  &  $-$     &  18.658 &  19.244  &	 $-$   &  0.264  &  0.419 &  ~~$-$  &\\
\hline
%&&&&&&&&&&&&&&&&\\
&&&&&&&&&&&&&&&&\\
%     Ceph.EIDI
    30   &  5:23:55.92  &  -70:29:31.92  &  Ceph. &  3.66050  &  51933.77646  & 67,41,14& 15.396  &  15.980  &	$-$   &  15.403 &  15.992  &	$-$  & 0.392 & 0.532 &  ~~$-$  & \\
    40	 &  5:22:12.23  &  -70:40:09.81  &  Ceph. &  2.39797  &  51934.66602  & 69,40,13& 15.753  &  16.237  &  15.212  &  15.765 &  16.260  &  15.225 & 0.501 & 0.691 & 0.513 & \\
   121	 &  5:23:17.88  &  -70:34:30.81  &  Ceph. &  2.13218  &  51180.99088  & 66,41,12& 15.975  &  16.532  &  15.288  &  15.980 &  16.543  &  15.290 & 0.344 & 0.472 &0.205::& \\
   147	 &  5:22:59.43  &  -70:33:24.15  &  Ceph. &  4.69248  &  51181.61046  & 67,37,14& 15.467  &  16.099  &	$-$   &  15.497 &  16.170  &	$-$  & 0.830 & 1.218 &  ~~$-$  & \\
   150	 &  5:22:10.50  &  -70:33:14.98  &  Ceph. &  3.13782  &  51181.70621  & 56,36,14& 16.251  &  16.837  &  15.524  &  16.257 &  16.852  &  15.523 & 0.415 & 0.574 &  ~~$-$  & \\
   170	 &  5:23:04.43  &  -70:31:13.83  &  Ceph. &  6.66000  &  51933.30000  & 64,35,12& 15.277  &  16.105  &	$-$   &  15.284 &  16.117  &	$-$  & 0.394 & 0.568 &  ~~$-$  & \\
   182	 &  5:23:47.50  &  -70:30:13.46  &  Ceph. &  2.80545  &  51184.80502  & 61,38,14& 15.820  &  16.389  &	$-$   &  15.862 &  16.479  &	$-$  & 0.940 & 1.354 &  ~~$-$  & \\
   183	 &  5:21:48.21  &  -70:30:25.86  &  Ceph. &  2.49288  &  51182.99807  & 68,39,10& 16.259  &  16.830  &  15.566  &  16.283 &  16.879  &  15.567 & 0.689 & 0.976 &0.193:::&\\
   200	 &  5:23:07.02  &  -70:29:05.06  &  Ceph. &  2.73228  &  51179.44316  & 61,41,14& 15.568  &  16.142  &  14.920  &  15.575 &  16.156  &  14.921 &0.424: & 0.571 & 0.208 & \\
   902	 &  5:23:00.61  &  -70:34:32.31  &  Ceph. &  1.17104  &  51182.99969  & 69,41,13& 16.936  &  17.460  &	$-$   &  16.941 &  17.469  &	$-$  & 0.368 & 0.487 &  ~~$-$  & \\
\hline
&&&&&&&&&&&&&&&&\\
%&&&&&&&&&&&&&&&&\\
%    BINARIE
  1880	 &  5:22:56.07  &  -70:40:17.73  &  EB  &  2.18677  &  51183.76190  & 69,40,13 &  18.997  &  19.139 &	$-$    &  18.921 &  19.078  &	 $-$  &  ~~$-$  &  ~~$-$  &  ~~$-$  & \\
  2756	 &  5:23:06.01  &  -70:38:57.82  &  EB  &  1.18211  &  51183.81265  & 70,39,14 &  19.327  &  19.595 &	$-$    &  19.342 &  19.612  &	 $-$  & 0.733 & 0.781 &  ~~$-$  & \\
  4490	 &  5:21:53.47  &  -70:36:35.13  &  EB  &  1.38051  &  51184.69877  & 70,39,10 &  19.016  &  19.011 &  19.017  &  19.022 &  19.015  &  19.020 & 0.727 & 0.523 &  ~~$-$  & \\
  9800	 &  5:22:57.60  &  -70:28:43.46  &  EB  &  0.59749  &  51184.78793  & 66,41,14 &  18.590  &  18.627 &  18.504  &  18.602 &  18.638  &  18.518 & 0.617 & 0.626 & 0.592 & \\
 10914	 &  5:23:00.60  &  -70:40:22.44  &  EB  &  0.56184  &  51183.79300  & 65,39,13 &  19.767  &  20.041 &	$-$    &  19.784 &  20.058  &	 $-$  & 0.689 & 0.647 &0.615::& \\
 18475	 &  5:23:46.56  &  -70:34:46.31  &  EB  &  0.80928  &  51185.66801  & 60,41,12 &  19.821  &  19.872 &	$-$    &  19.826 &  19.880  &	 $-$  & 0.452 & 0.570 &  ~~$-$  & \\
\hline
%	CANDIDATE
&&&&&&&&&&&&&&&&\\
 1002	 &  5:22:58.16  &  -70:33:46.41  &  EB  &  ~~~~$-$  & ~~~~~$-$  	   &    $-$    &    $-$   &   $-$   &   $-$    &   $-$   &    $-$   & $-$ &$\sim$0.29  &$\sim$0.32 &$>$0.2&\\
 1090	 &  5:21:51.82  &  -70:33:08.41  &  EB  &  ~~~~$-$  & ~~~~~$-$  	   &    $-$    &    $-$	  &   $-$   &   $-$    &   $-$   &    $-$   & $-$  &$\sim$0.23  &$\sim$0.34 &$>$0.3&\\
 3276	 &  5:24:04.96  &  -70:38:06.35  & ?    &  ~~~~$-$  & ~~~~~$-$  	   & 70,38,14  &  19.040  &  19.268 &  18.862  &  19.051 &  19.279  &  18.877 &$>$0.5  &$>$0.6 &$>$0.6&\\
 7997	 &  5:21:57.21  &  -70:31:36.27  & ~EB?  &  ~~~~$-$  & ~~~~~$-$  	   &    $-$    &    $-$   &   $-$   &   $-$    &   $-$   &    $-$   & $-$ &$\sim$0.14  &$\sim$0.46 &$>$0.5&\\
 8723    &  5:22:14.60  &  -70:30:27.49  &  EB	&$\sim$1.16 & 51183.78600  & 51,21,14  &  19.152  &  19.859 &   $-$    &  19.161 &  19.862  &  18.455 &  0.789  &~~$-$  &  ~~$-$ &   \\
%28002	 &     ???      &     ???        & Cand.&  $-$      &     $-$       &58,27,$-$ &   $-$    &  $-$    &	$-$    &   $-$   &    $-$   &    $-$  &  ~$-$  &  ~$-$  &  ~$-$  &\\
\hline
\end{tabular}
%%%%%%%%%%%%%%%%%%%%%%%%%%%%%%%%%%%%%%%	
%\medskip
Notes: \\
(a) The minimum light in 2001 is systematically brighter than in 1999
 both in $B$ and $V$, possibly indicating a Blazhko modulation.
 
(b) No reliable average magnitudes and amplitudes are available since in 1999 the star occasionally fell 
    on a CCD bad column.
    
(c) The 2001 light curves are systematically fainter than the 1999 ones. The star could be an unresolved 
    blend in the 1999 photometry. Amplitudes and average luminosities are 
    provided for the 1999 (upper line) and 2001 data (lower line), separately.
    
(d) Amplitudes and average luminosities are provided for the combined 1999 + 2001 data (upper line), and for the
    1999 (middle line) and 2001 data 
    (lower line), separately.
    
(e) The 2001 light curves are systematically fainter than the 1999 ones. The star could either be
    an unresolved blend in the 1999 photometry, or could be affected by Blazhko effect.  
    Amplitudes correspond to the 1999 data-set.
    
(f) The 2001 light curves are systematically brighter than the 1999 ones. 
    Amplitudes and average luminosities are 
    provided for the 1999 (upper line) and 2001 data (lower line), separately.
    
(g) Light curves are very noisy possibly indicating the presence of secondary periodicities.

(h) The 2001 light curves are systematically slightly brighter than in 1999, particularly in $B$.
    Amplitudes are 
    provided for the combined 1999 + 2001 data (upper line), and for the 1999 (middle line) and
    the 2001 data (lower line), separately.
    
(i) Period and shape of the 2001 light curves are slightly different than in 1999.
    Amplitudes and average luminosities are 
    provided for the combined 1999+2001 data (upper line) and for the 2001 data (lower line), separately.

(l) The star was not observed in 2001.
(m) Amplitudes and average luminosities are provided separately for the combined
1999 + 2001 data (upper line), for the 1999 data (middle line), and for the 2001 data (lower line).
(n) The 2001 light curves are systematically fainter than the 1999 ones. Amplitudes are provided 
separately for the combined 1999 + 2001 data (upper line), for the 1999 data (middle line), and for the 2001 data (lower line).
%(r) Amplitudes are lower limits.\\    
%\label{t:tab1}                                                                                                                
%\end{tabular}
%\end{center}
\normalsize
\end{table}
\end{landscape}
%\noindent
%%%%%%%%%%
%CAMPO A %
%%%%%%%%%%
%\end {table*}                                                                                                                             

%\begin{table*}

\begin{landscape}
\begin{table}
%\small
\caption{Informations and 
average quantities for the variable stars in field B.}
%\begin{center}

\tiny
\begin{tabular}{rcccllrccccccllll}
%\begin{tabular}{rcccllrcccccccccccllll}
\hline
       Id~~ &  $\alpha$  &  $\delta$    & Type &     ~~~~P       &     ~~~Epoch        &  Np~~~  &   $<{\rm V_{int}}>$&$<{\rm B_{int}}>$&$<{\rm I_{int}}>$&$<{\rm V_{mag}}>$&$<{\rm B_{mag}}>$&$<{\rm I_{mag}}>$&~A$_{\rm V}$&~A$_{\rm B}$&~A$_{\rm I}$&~~Notes\\
%       Id &  $\alpha$  &  $\delta$    & Type &     P       &     Epoch        &  Np  &   $<{\rm V_{int}}>$&$<{\rm B_{int}}>$&$<{\rm I_{int}}>$&$<{\rm V_{mag}}>$&$<{\rm B_{mag}}>$&$<{\rm I_{mag}}>$&A$_{\rm V}$&A$_{\rm B}$&A$_{\rm I}$&Note\\
          &    (2000)  &    (2000)    &      &   (days)  &    ($-$2400000)  & (V,B,I)  &   (mag)  &   (mag) &	  (mag) & (mag)    &  (mag)   &   (mag)  &  (mag)  &(mag)    &(mag)   &\\
\hline
%CAMPO B
%
%%%%%%%%%% 
% RRab    %
%%%%%%%%%%
%
&&&&&&&&&&&&&&&&\\
    1408   & 5:17:13.79 &  -71:06:06.91 &  ab & 0.62600 & 51182.73500 & 52,31,14  & 19.343  &  19.772  &  18.777  &  19.369  &  19.813  &  18.778  &  0.812  &  0.978  &  0.376: &\\
    1575   & 5:16:31.27 &  -71:05:48.49 &  ab & 0.67389 & 51182.40483 & 70,37,14  & 19.250  &  19.666  &  18.690  &  19.290  &  19.733  &  18.703  &  1.029  &  1.284  &  0.740  &\\
    1907   & 5:18:12.30 &  -71:04:59.49 &  ab & 0.58048 & 51182.92634 & 55,23,$-$ & $-$     &   $-$    &    $-$   &	$-$  &   $-$    &   $-$    &  0.658  &  0.796  &   ~~$-$   &Blazhko,(a)\\
	   & 	        &		&     &         &	      & 70,37,13  & 19.287  &  19.724  &  18.726  &  19.306  &  19.762  &  18.739  &   ~~$-$   &   ~~$-$   &   ~~$-$   &\\
    2055   & 5:17:17.39 &  -71:04:50.18 &  ab & 0.52295 & 51182.41018 & 47,34,12  &  $-$    &  $-$     &    $-$   &   $-$    &	$-$     &   $-$    & 0.749:::&  1.304  &  0.534::&\\
    2249   & 5:17:13.01 &  -71:04:27.10 &  ab & 0.61044 & 51933.59040 & 69,35,14  & 19.346  &  19.775  &  18.689  &  19.371  &  19.823  &  18.728  &  0.747  &  0.987  &  0.634  &\\
    2379   & 5:18:43.16 &  -71:04:03.24 &  ab & 0.48854 & 51182.14762 & 66,36,13  &  $-$    &	$-$    &    $-$   &   $-$    &   $-$	&   $-$       & 0.931:: &1.292::  &0.482::::& Incomplete\\
    2407   & 5:16:51.61 &  -71:04:13.40 &  ab & 0.67004 & 51182.54647 & 68,36,14  & 19.186  &  19.613  &  18.531  &  19.201  &  19.639  &  18.539  &  0.652  &  0.818  &  0.466  &Blazhko?\\
    2884   & 5:16:52.13 &  -71:03:25.18 &  ab & 0.61943 & 51183.71406 & 70,36,11  & 19.217  &  19.630  &  18.655  &  19.249  &  19.689  &  18.639  &  0.869  &  1.182  &  0.475::&\\
    3033   & 5:18:13.98 &  -71:03:00.56 &  ab & 0.49938 & 51933.59874 & 69,37,14  &  $-$    &	$-$    &    $-$   &   $-$    &   $-$	&   $-$    &  1.157: & 1.254:  &0.371::::& Incomplete\\
    3054   & 5:17:45.36 &  -71:03:01.45 &  ab & 0.50798 & 51183.21819 & 68,35,14  & 19.066  &  19.440  &    $-$   &  19.089  &  19.486	&   $-$    &  0.902  &  1.149  &  ~~$-$   &\\
    3400   & 5:17:14.46 &  -71:02:26.58 &  ab & 0.48616 & 51182.21055 & 67,37,14  & 19.469  &  19.805  &  18.824  &  19.530  &  19.921  & 18.851   &  1.263  &  1.619  &  0.836  &\\
    3412   & 5:18:38.21 &  -71:02:14.47 &  ab & 0.53020 & 51182.91258 & 68,36,13  & 19.425  &  19.834  &  18.856  &  19.460  &  19.902  & 18.876   &  0.834  &  1.159  &  0.707  &\\
    4244   & 5:16:27.60 &  -71:00:59.85 &  ab & 0.55621 & 51182.84333 & 70,37,14  & 19.260  &  19.633  &  18.782  &  19.297  &  19.702  & 18.808   &  0.944  &  1.159  &  ~~$-$	 &Blazhko?,(b)\\
	   &		&	        &     &         &	      & 55,23,$-$ &  $-$    &   $-$    &    $-$   &   $-$    &   $-$    &   $-$    &  0.881  &  1.179  &  ~~$-$    &\\
	   &		&	        &     &         &	      & 15,14,14  &   $-$   &	$-$    &    $-$   &   $-$    &   $-$	&   $-$    &  1.122  &  1.399  &  0.800  &\\
    4540   & 5:16:13.67 & -71:00:28.34  &  ab & 0.56892 & 51182.52598 & 61,31, 8  & 19.414  &  19.801  &  18.858  &  19.450  &  19.866  & 18.878   &  0.880  &  1.208  &  0.668  &~~~~(c)\\
    4780   & 5:16:53.00 & -71:00:02.53  &  ab & 0.61757 & 51934.63778 & 69,35,14  & 19.396  &  19.860  &  18.707  &  19.411  &  19.895  & 18.718   &  0.595  &  0.972  &   ~~$-$   &Blazhko?,(b)\\
	   &		&	        &     &         &	      & 54,21,$-$ &  $-$    &   $-$    &    $-$   &   $-$    &   $-$    &   $-$    &  0.592  &  0.930  &  ~~$-$    &\\
	   &		&	        &     &         &	      & 15,14,14  &   $-$   &	$-$    &    $-$   &   $-$    &   $-$	&   $-$    &0.466::::&0.673::: &  0.499  &\\
    4859   & 5:16:10.87 & -70:59:54.34  &  ab & 0.52336 & 51184.68400 & 46,21,$-$ & 19.240  &  19.617  &    $-$   &  19.294  &  19.687	&   $-$    &  1.061  &  1.182  &   ~~$-$   &~~~~(c)\\
    5394   & 5:17:15.73 & -71:04:11.37  &  ab & 0.50993 & 51182.86900 & 64,31,11  & 19.463: &  19.957: &  19.117: &  19.499: &  19.976: & 19.121:  &$>$0.643 &$>$0.735 &  0.318    &  Incomplete\\
    5902   & 5:16:12.23 & -70:58:04.86  &  ab & 0.56975 & 51182.87903 & 54,23,$-$ & 19.121  &  19.472  &    $-$   &  19.165  &  19.571  &   $-$    &  1.015  &  1.282  &   ~~$-$   &~~~~(c)\\
    5950   & 5:18:06.56 & -70:57:48.52  &  ab & 0.49957 & 51182.07579 & 55,24,14  &   $-$   &   $-$    &    $-$   &   $-$    &   $-$    &   $-$    &  $>$0.4 & $>$0.5  &   ~~$-$   & Incomplete\\
    6020   & 5:16:32.44 & -70:57:53.68  &  ab & 0.61428 & 51183.70139 & 53,22,$-$ &   $-$   &	$-$    &    $-$   &   $-$    &   $-$	&   $-$    &  0.858  &  0.973  &   ~~$-$   &~~~~(d)\\
	   & 	        &	        &     &         &	      & 15,14,14  &   $-$   &	$-$    &    $-$   &   $-$    &   $-$	&   $-$    &  0.931  &  1.157  &  0.534  &\\
    6440   & 5:16:16.68 & -70:57:11.56  &  ab & 0.49482 & 51934.75800 & 43,23,13  & 19.247  &  19.612  &  18.865  &  19.280  &  19.664	& 18.876   &  0.875  &  1.053  &  0.592  &\\
    6798   & 5:17:11.33 & -70:56:32.65  &  ab & 0.58405 & 51182.44080 & 53,22,$-$ & 19.253  &  19.658  &  18.711  &  19.291  &  19.730  & 18.720   &  1.035  &  1.376  &   ~~$-$   &Blazhko,(e)\\
	   & 	        &	        &     &         &	      & 15,14,14  &   $-$   &	$-$    &    $-$   &   $-$    &   $-$	&   $-$    &  0.760  &  1.043  &  0.499  &\\
    7063   & 5:18:43.98 & -70:55:55.77  &  ab & 0.65428 & 51183.79572 & 68,36,13  & 19.195  &  19.629  &  18.549  &  19.213  &  19.654  &  18.569  &  0.633  &  0.686  &  0.457  &Blazhko?\\
    7158   & 5:16:45.89 & -71:03:27.94  &  ab & 0.80192 & 51182.91400 & 54,28,12  & 19.198  &  19.610  &  18.452  &  19.212  &  19.636  &  18.462  &  0.679  &  0.830  &  0.433  &\\
    7442   & 5:17:15.68 & -70:55:26.77  &  ab & 0.57795 & 51933.63581 & 70,36,14  & 19.426  &  19.835  &  18.855  &  19.446  &  19.876  &  18.877  &  0.650  &  0.917  &  0.655  &\\
    7620   & 5:18:03.53 & -70:55:03.12  &  ab & 0.65616 & 51182.75690 & 70,37,14  & 19.079  &  19.409  &  18.461  &  19.133  &  19.489  &  18.471  &  1.071  &  1.366  &  0.710  &\\
    7652   & 5:17:32.87 & -71:03:13.94  &  ab & 0.50700 & 51934.78000 & 61,36,12  & 19.426  &  19.763  &  19.191  &  19.492  &  19.888  &  19.207  &  1.270  &  1.629  &  0.811  &\\
   10692   & 5:18:04.54 & -71:04:34.91  &  ab & 0.55094 & 51182.70020 & 68,37,14  & 19.548  &  19.956  &  18.918  &  19.574  &  20.009  &  18.938  &  0.862  &  1.109  &  0.747: &Blazhko?\\
   10811   & 5:18:15.95 & -71:04:27.07  &  ab & 0.47637 & 51182.81620 & 69,36,14  & 19.431  &  19.797  &   $-$    &  19.492  &  19.893	&   $-$    &  1.166  &  1.392  &  0.452  &\\
   14449   & 5:17:05.32 & -71:01:40.85  &  ab & 0.58413 & 51934.57998 & 64,36,13  & 19.514  &  19.851  &  18.727  &  19.450  &  19.906  &  18.734  &  0.804  &  1.016  &  0.329::&\\
   19037   & 5:16:20.66 & -70:58:06.31  &  ab & 0.41128 & 51933.76200 & 67,33,14  & 19.702  &  20.023  &  19.293  &  19.784  &  20.174  &  19.332  &  1.466  &  1.821  &  0.917  &\\
   21801   & 5:16:36.29 & -70:58:21.72  &  ab & 0.50723 & 51182.88136 & 68,34,14  & 19.598  &  20.013  &  19.216  &  19.627  &  20.067  &  19.235  &  0.874  &  1.198  &  0.653::&\\
   22917   & 5:18:19.05 & -70:54:56.03  &  ab & 0.56468 & 51182.63629 & 60,35,14  & 19.426  &  19.797  &  18.887  &  19.462  &  19.863  &  18.893  &  0.957  &  1.283  &  0.729  &\\
   23502   & 5:18:01.58 & -70:54:31.81  &  ab & 0.47217 & 51934.61759 & 46,31,9   & 19.385  &  19.662  &   $-$    &  19.446  &  19.792  &   $-$	   &  1.296  &  1.638  &1.300::: &~~~~(c)\\
   24089   & 5:17:44.64 & -70:54:03.08  &  ab & 0.55977 & 51183.74000 & 55,21,$-$ & 19.365  &  19.798  &    $-$   &  19.370  &  19.797  &   $-$    &  0.371  &  0.485  &  1.386  &~~~~(f)\\
           & 	        &	        &     & 	&             &  $-$,14,8 &	  $-$      & $-$	  &   $-$    &   $-$	&   $-$    &	$-$     &  1.339  &  ~~$-$ & ~~$-$	&\\
\hline         
% 
%%%%%%%%%% 
% RRc    %
%%%%%%%%%%
%
%&&&&&&&&&&&&&&&&&&&&&\\
%&&&&&&&&&&&&&&&&&&&&&\\
%&&&&&&&&&&&&&&&&&&&&&\\
%\begin{tabular}{rcccllccccccccccccccccccc}
%\hline
%\hline
%N&$\alpha$    &$\delta$       &Tipo&P          &Epoca            &Np&Na&Res.&Np&Na&Res.&Np&Na&Res.&$<{\rm V_{int}}>$&$<{\rm B_{int}}>$&$<{\rm I_{int}}>$&$<{\rm V_{mag}}>$&$<{\rm B_{mag}}>$&$<{\rm I_{mag}}>$&A$_{\rm V}$&A$_{\rm B}$&A$_{\rm I}$&Note          \\
%        &  (2000)          &       (2000)       &           &(giorni)      &($-$2400000)      & (V) & (V)   &(mag)    & (B)  & (B)    &  (mag) & (I)  & (I)         &(mag)      &(mag)  & (mag) & (mag) & (mag)  & (mag) & (mag) & (mag)       &(mag)      &(mag)     &               \\
%\hline
&&&&&&&&&&&&&&&&\\
    1387   & 5:18:47.52 & -71:05:58.35 &  c  & 0.36448 & 51185.77393 & 69,36,14  & 19.252  &  19.561  &  18.737  &  19.263  &  19.580  &  18.741  &  0.443  &  0.587  &  0.316  &\\
    2517   & 5:16:31.25 & -71:04:03.32 &  c  & 0.23176 & 51934.70774 & 69,36,12  & 19.695  &  19.925  &  19.374  &  19.706  &  19.943  &  19.379  &  0.452  &  0.555  &  0.312  &\\
    2811   & 5:18:31.15 & -71:03:21.74 &  c  & 0.27757 & 51183.04260 & 70,35,13  & 19.384  &  19.691  &  18.956  &  19.389  &  19.699  &  18.958  &  0.313  &  0.371  &  0.242  &\\
    3625   & 5:16:52.66 & -71:02:01.87 &  c  & 0.27654 & 51182.93263 & 61,35,14  &    $-$  &  $-$     &   $-$	 &   $-$    &	$-$    &   $-$    &  0.292  &  0.379  &   ~~$-$	&~~~~(g)\\
           & 	        &	       &     &         &	     & 46,21,$-$ &    $-$  &  $-$     &   $-$	 &   $-$    &	$-$    &   $-$    &  0.305  &  0.343  &   ~~$-$	&\\
           & 	        &	       &     &         &	     & 15,14,14  &    $-$  &  $-$     &   $-$	 &   $-$    &	$-$    &   $-$    &  0.322  &  0.378  &  0.194  &\\
           & 	        &	       &     & 0.27995 & 51182.89758 & 59,35,$-$ &    $-$  &  $-$     &   $-$	 &   $-$    &	$-$    &   $-$    &  0.297  &  0.403  &   ~~$-$	&\\
           & 	        &	       &     &         &	     & 44,21,$-$ &    $-$  &  $-$     &   $-$	 &   $-$    &	$-$    &   $-$    &  0.293  &  0.360  &   ~~$-$	&\\
           & 	        &	       &     &         &	     & 15,14,14  &    $-$  &  $-$     &   $-$	 &   $-$    &	$-$    &   $-$    &  0.308  &  0.379  &  0.197  &\\
    3957   & 5:18:39.60 & -71:01:12.61 &  c  & 0.34232 & 51182.70003 & 68,35,13  & 19.533  &  19.932  &  19.038  &  19.543  &  19.948  &  19.050  &  0.427  &  0.509  &  0.218  &\\
    4008   & 5:17:04.72 & -71:01:17.28 &  c  & 0.28474 & 51934.75471 & 66,34,14  & 19.281  &  19.559  &  18.855  &  19.290  &  19.574  &  18.859  &  0.421  &  0.507  &  0.282  &\\
    4179   & 5:17:19.90 & -71:01:02.08 &  c  & 0.35587 & 51186.69270 & 50,29,14  & 19.173  &  19.502  &  18.640  &  19.183  &  19.520  &  18.644  &  0.438  &  0.560  &  0.298  &\\
    4749   & 5:17:49.67 & -71:00:01.38 &  c  & 0.32703 & 51184.82256 & 63,35,13  & 19.314  &  19.653  &  18.834  &  19.321  &  19.667  &  18.838  &  0.394  &  0.512  &  0.306  &\\
    4946   & 5:18:11.02 & -70:59:35.60 &  c  & 0.31275 & 51934.63003 & 68,34,12  & 19.432  &  19.739  &  18.929  &  19.443  &  19.757  &  18.933  &  0.433  &  0.561  &  0.298  &\\
    5256   & 5:18:45.57 & -70:58:56.79 &  c  & 0.34248 & 51182.67690 & 70,36,14  & 19.259  &  19.614  &  18.795  &  19.268  &  19.628  &  18.796  &  0.417  &  0.527  &  0.210  &\\
    6164   & 5:18:10.12 & -70:57:30.75 &  c  & 0.37487 & 51934.68304 & 70,36,14  & 19.057  &  19.353  &  18.604  &  19.068  &  19.374  &  18.613  &  0.451  &  0.615  &  0.389  &\\
    6255   & 5:17:17.83 & -70:57:26.43 &  c  & 0.35239 & 51933.61784 & 58,33,11  & 19.264  &  19.600  &  18.807  &  19.272  &  19.614  &  18.810  &  0.380  &  0.539  &  0.223  &\\
    6957   & 5:18:18.03 & -70:56:08.75 &  c  & 0.40567 & 51182.81240 & 70,37,13  & 19.197  &  19.532  &  18.673  &  19.205  &  19.549  &  18.680  &  0.396  &  0.568  &  0.370  &\\
    7064   & 5:18:18.59 & -70:55:58.64 &  c  & 0.40070 & 51934.75034 & 70,37,13  & 19.122  &  19.433  &  18.644  &  19.135  &  19.454  &  18.655  &  0.474  &  0.607  &  0.451  &\\
    7490   & 5:18:44.67 & -70:55:10.81 &  c  & 0.30481 & 51186.72128 & 70,37,13  &    $-$  &   $-$    &    $-$   &   $-$    &	$-$    &   $-$    &  0.505  &  0.637  &   ~~$-$	&~~~~(h)\\
           & 	        &	       &     &         &	     & 55,22,$-$ &    $-$  &   $-$    &    $-$   &   $-$    &	$-$    &   $-$    &  0.483  &  0.528  &   ~~$-$	&\\
           & 	        &	       &     &         &	     & 15,14,13  &    $-$  &   $-$    &    $-$   &   $-$    &	$-$    &   $-$    &  0.568  &  0.709  &  0.430  &\\
    7648   & 5:16:38.53 & -70:55:09.52 &  c  & 0.34268 & 51186.66994 & 69,35,14  & 19.384  &  19.688  &  18.947  &  19.399  &  19.707  &  18.951  &  0.467  &  0.598  &  0.339  &\\
    7783   & 5:17:25.70 & -70:54:51.95 &  c  & 0.34634 & 51933.69588 & 70,35,12  & 19.279  &  19.564  &  18.828  &  19.298  &  19.596  &  18.837  &  0.553  &  0.742  &  0.417  &\\
   10585   & 5:17:29.07 & -71:04:45.16 &  c  & 0.26954 & 51934.63450 & 70,37,12  & 19.628  &  19.934  &  19.158  &  19.641  &  19.959  &  19.167  &  0.478  &  0.657  &  0.407: &\\
\hline										      
\end{tabular}
%\end{center}
\normalsize
\end{table}
\end{landscape}
%%%%%%%%%%%%%%%%%%%%%%%%%%%%%%%%%%%%%%%%
%\pagestyle{empty}
\begin{landscape}
\begin{table}
%\small
%\caption{Informations and average quantities for the variable stars in field A.}\\
%\begin{center}
%\scriptsize
%\small
Table 6: - continued -\\

\tiny
\begin{tabular}{rcccllrccccccllll}
\hline
%\begin{tabular}{rcccllrccccccccccccccc}
%\hline
       Id~~ &  $\alpha$  &  $\delta$    & Type &     ~~~~P       &     ~~~Epoch        &  Np~~~  &   $<{\rm V_{int}}>$&$<{\rm B_{int}}>$&$<{\rm I_{int}}>$&$<{\rm V_{mag}}>$&$<{\rm B_{mag}}>$&$<{\rm I_{mag}}>$&~A$_{\rm V}$&~A$_{\rm B}$&~A$_{\rm I}$&~~Notes\\
            &    (2000)   &    (2000)     &      &   (days)    &    ($-$2400000)  & (V,B,I)  &             (mag)  &             (mag) &	  (mag)        & (mag)            &          (mag)   &           (mag)  &     (mag)  &   (mag)    &(mag)      &\\
\hline 
%
%%%%%%%%%% 
% RRd    %
%%%%%%%%%%
%
&&&&&&&&&&&&&&&&\\
    3347   & 5:16:26.98 & -71:02:36.58 &  d  & 0.36040 & 51182.84429 & 70,37,14&  19.204  &  19.547  &  18.762  &  19.211  &  19.558  &  18.765  &  0.372  &  0.452  &  0.256 &\\
    4509   & 5:17:36.92 & -71:00:28.22 &  d  & 0.37130 & 51934.77500 & 69,37,14&  19.462  &  19.823  &  18.717  &  19.468  &  19.833  &  18.725  &  0.333  &  0.429  &  0.397 &\\
    6470   & 5:16:45.06 & -70:57:07.67 &  d  & 0.36979 & 51934.63003 & 70,37,14&  19.207  &  19.571  &  18.831  &  19.218  &  19.583  &  18.840  &  0.448  &  0.480  &  0.258 &\\
    7467   & 5:18:17.14 & -70:55:16.22 &  d  & 0.35775 & 51183.81805 & 70,37,14&  19.043  &  19.372  &  18.597  &  19.055  &  19.389  &  18.599  &  0.487  &  0.560  &  0.252 &\\

\hline                                                                                                                        
%\end{tabular}                                                                                                                 
%			Ceph.EIDE ANOMALA
%
%\begin{tabular}{rcccllccccccccccccccccccc}
%\hline
%\hline
%N&$\alpha$    &$\delta$       &Tipo&P          &Epoca            &Np&Na&Res.&Np&Na&Res.&Np&Na&Res.&$<{\rm V_{int}}>$&$<{\rm B_{int}}>$&$<{\rm I_{int}}>$&$<{\rm V_{mag}}>$&$<{\rm B_{mag}}>$&$<{\rm I_{mag}}>$&A$_{\rm V}$&A$_{\rm B}$&A$_{\rm I}$&Note          \\
%        &  (2000)          &       (2000)       &           &(giorni)      &($-$2400000)      & (V) & (V)   &(mag)    & (B)  & (B)    &  (mag) & (I)  & (I)         & (mag)  & (mag)  & (mag) & (mag) & (mag)   & (mag) & (mag) & (mag) &(mag)      &(mag)     &               \\
%\hline
&&&&&&&&&&&&&&&&\\
    5952   &  5:18:04.30  &  -70:57:50.13  &  AC &  0.63300  &  51183.69855  &  64,35,12 &  18.459  & 19.120 &  17.648  &  18.463  &  19.135  &  17.631  &  0.325 & 0.601 &  0.274  &\\
\hline                                                                                                                        
%
%			Ceph.EIDE
%
&&&&&&&&&&&&&&&&\\
     204    & 5:17:54.15 & -70:55:16.84 &  Ceph.   &  3.11055  &   51182.93750  &	68,36,13  &16.009 & 16.620 &   $-$  &  16.038  &  16.679  &	$-$   & 0.765: & 1.058: & ~~$-$ &\\
%			BINARIE
%						    
\hline                                                                                                                        
%&&&&&&&&&&&&&&&&&&&&&\\
%&&&&&&&&&&&&&&&&&&&&&\\
%						    
&&&&&&&&&&&&&&&&\\
    1710   & 5:18:13.72  & -71:05:19.83  &  EB   &  0.73439  &  51186.68150  &  70,34,14 &  19.402  & 19.501  &  19.240  &  19.416  &  19.518  &  19.232 &  0.643  &  0.650  &  0.682  &\\
    5357   & 5:16:32.35  & -70:59:00.82  &  EB   &  1.02684  &  51183.35399  &  70,37,14 &  19.288: & 19.424: &   $-$    &  19.066: &  19.226: &   $-$	 &1.015::: &1.039::: &  ~~$-$    &\\
    6499   & 5:17:48.42  & -70:56:57.99  &  EB   &  0.60784  &  51184.77200  &  68,34,14 &  19.376  & 19.500  &   $-$    &  19.383  &  19.509  &   $-$   &  0.401  &  0.433  &  0.329  &\\
   15500   & 5:17:15.83  & -71:00:50.06  &  EB   &  0.78568  &  51184.71789  &  66,37,14 &  19.792  & 19.884  &   $-$    &  19.805  &  19.899  &   $-$   &  0.620  &  0.704  &  ~~$-$    &\\
   21007   & 5:18:21.08  & -70:56:27.16  &  EB   &  1.35753  &  51186.69270  &  68,34,14 &  20.010  & 20.207  &   $-$    &  20.024  &  20.226  &   $-$   &  0.639  &  0.790  &  0.503  &\\
			
%			
%
%			CANDIDATE
\hline
&&&&&&&&&&&&&&&&\\
%   1147   &     ????   &      ????       & Cand. &    $-$    &    $-$        &  48,23,11 &    $-$   &  $-$    &  $-$     &   $-$    &  $-$     &  $-$    &  0.110: & 0.150:  &  $-$    &\\   
    2601   & 5:18:39.98  & -71:03:41.03  & EB &    ~~$-$    &     ~~~~~$-$       & $-$  &   $-$    &  $-$  &  $-$     &   $-$    &   $-$   &  $-$   &  $\sim$0.61 & $\sim$0.63  &  $>$0.3    &\\   
    3969   & 5:17:45.59  & -71:01:17.38  & ?  &    9.90::    &    ~~~~~$-$        & $-$  &   $-$   &   $-$   &   $-$   &  $-$   &   $-$    &   $-$   & $\sim$0.21& $\sim$0.23  &  $>$0.15   &\\  
%   6596   &     ????   &      ????       & Cand. &    $-$    &    $-$        &  40,23,14 &    $-$   &  $-$    &  $-$     &   $-$    &  $-$     &  $-$    &  0.740: & 0.640:  &  0.200::&\\
    7699   & 5:17:03.10    &  -70:55:00.65   & RR?  &    ~~$-$    &   ~~~~~$-$        &  14,5,$-$ &    $-$   &  $-$    &  $-$     &   $-$    &  $-$     &  $-$    &  0.600: & 0.500:  & ~~ $-$    &\\  
%  24079   &     ????   &      ????       & Cand. &    $-$    &    $-$        &  18,13,$-$&    $-$   &  $-$    &  $-$     &   $-$    &  $-$     &  $-$    &  $-$    &  $-$    &  $-$    &\\
%
\hline
\end{tabular}
%%%%%%%%%%%%%%%%%%%%%%%%%%%%%%%%%%%%%%%	
\medskip
																					       
Notes: \\

%({\bf QUESTE NOTE VANNO TRADOTTE E SNELLITE})\\
%{\bf AGGIUNGERE QUI IN FONDO LE INFO. SULLE CANDIDATE VARIABILI: 
%5 STELLE NEL CAMPO B}\\
(a) Amplitudes are from the 1999 data set (upper line), average luminosities are from the combined 1999+2001 data set (lower line).

(b) Amplitudes and average luminosities are 
    provided separately for the combined 1999 + 2001 data (upper line), for the 1999 data (middle line), and
    for the 2001 data (lower line).
    
(c) The star was not observed in 2001.

(d) The 2001 light curves are systematically fainter than the 1999 ones. 
    Amplitudes and average luminosities are 
    provided for the 1999 (upper line) and 2001 data (lower line), separately.
    
(e) Amplitudes and average luminosities are 
    provided for the 1999 (upper line) and 2001 data (lower line), separately.
    
(f) The star does not have $V$ observations in 2001. The 1999 $B$ light curve has a different shape and amplitude than the
    2001 one. Amplitudes and average luminosities for the 1999 data are provided in the fist line, the $B$ amplitude 
    of the 2001 data in the second line.
    
(g) The 2001 light curves are systematically brighter and have smaller amplitudes than the 1999 ones. 
    Both in 1999 and 2001 the V data provide a slightly different period than the B ones.
    Amplitudes and average luminosities are provided separately for the combined 1999 + 2001 data (upper line), 
    for the 1999 data (middle line), and for the 2001 data (lower line), and for each of the two periodicities.

(h) The 2001 light curves are systematically fainter than the 1999 ones. 
    Amplitudes are 
    provided for the combined 1999 + 2001 data (upper line), for the 1999 data (middle line), and
    for the 2001 data (lower line), separately.
    
%\label{t:tab2}                                                                                                                
%\end{tabular}
%\end{center}
\normalsize
\end{table}
\end{landscape}
%\noindent
%%%%%%%%%%
%CAMPO A %
%%%%%%%%%%
%\end {table*}                                                                                                                             

%It has often been argued on the better way to compute the 
%average magnitude of a variable star and on the colour that better 
%represents the temperature of an RR Lyrae star (Sandage 1990, 1993; Carney, 
%Storm \& Jones 1992; Bono, Caputo \& Stellingwerf 1995).  
% The average magnitudes of the variable stars in Tables~5 and 6 were computed 
%in two different ways, as intensity-averaged means (Columns 8,9,10) and
%as magnitude-averaged means (Columns 11,12,13).  Based on theoretical
% grounds it has been claimed that large differences may exist between
%these two different types of averages, and that for RR Lyrae stars 
%the difference may be as large as 0.1-0.2 mag in $V$ and $B$, respectively   
%(Bono et al. 1995).
\begin{figure}
%\resizebox{4.3cm}{!}{
%%%\includegraphics[width=8.8cm]{intmagA2.ps}
\includegraphics[width=8.8cm]{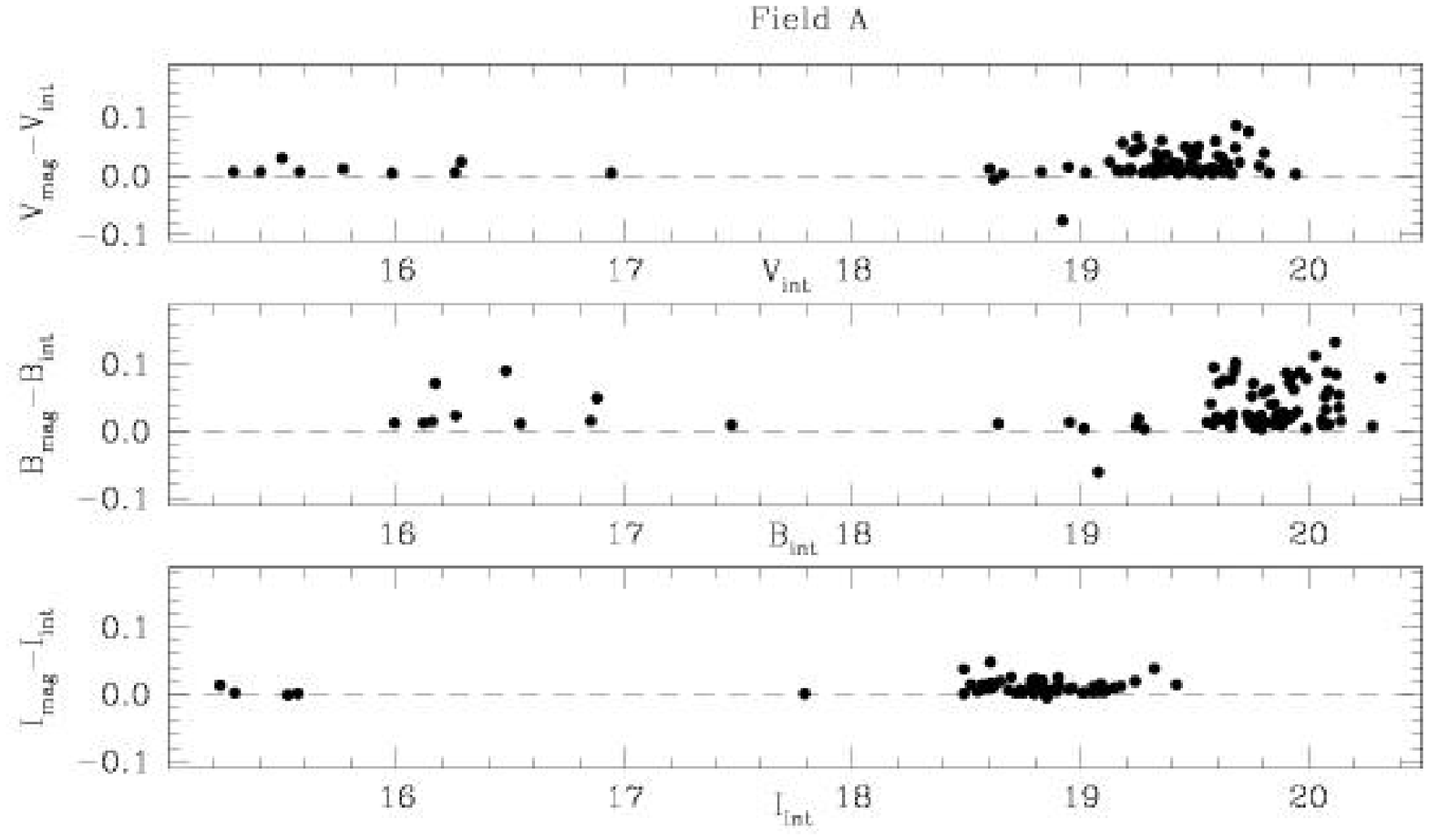}
%\includegraphics[bb=18 180 592 506, width=8.8cm, draft=true]{difabrizio.fig12a.ps}
%}
%\resizebox{4.3cm}{!}{
%%%\includegraphics[width=8.8cm]{intmagB2.ps}
\includegraphics[width=8.8cm]{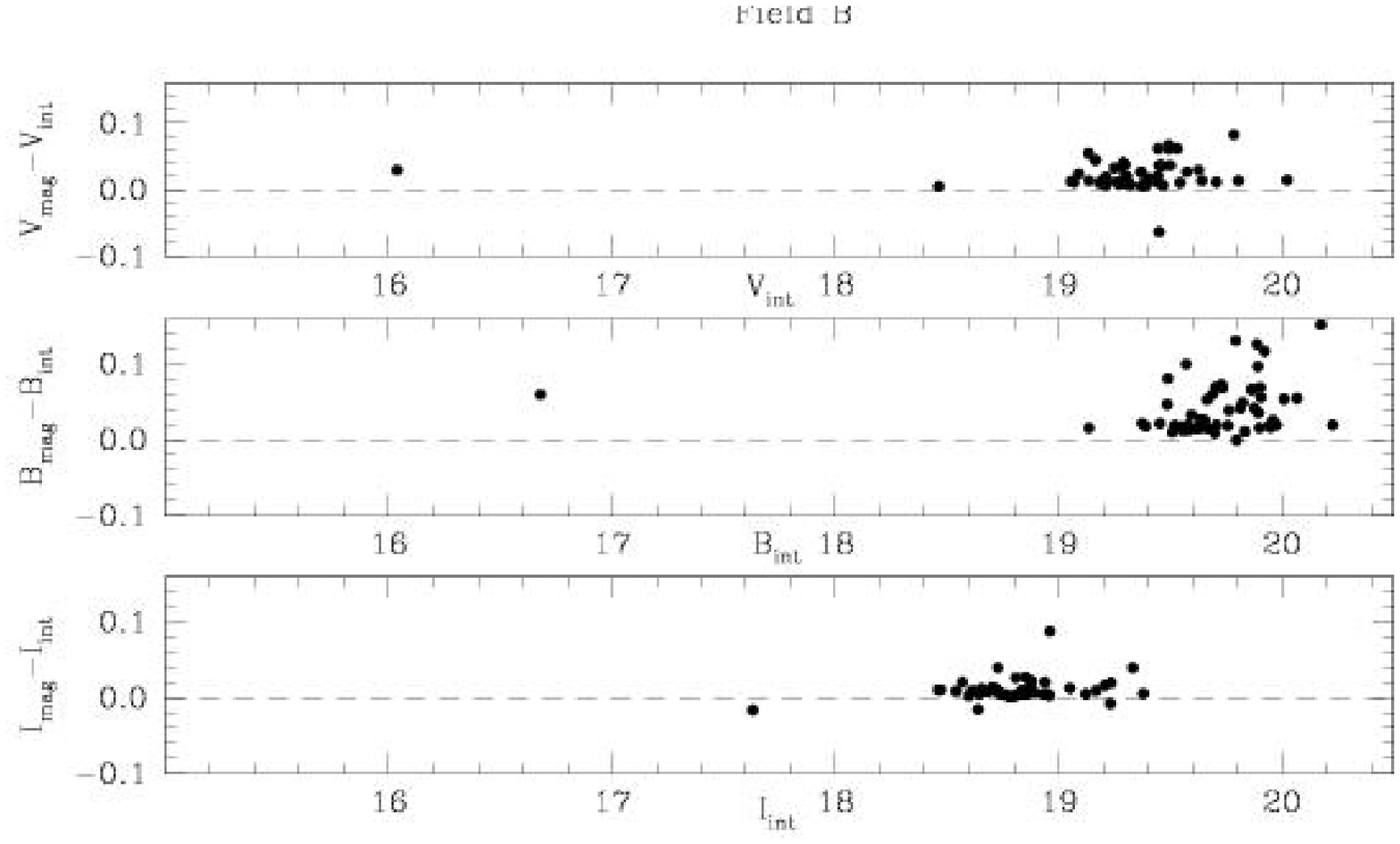}
%\includegraphics[bb=18 180 592 506, width=8.8cm, draft=true]{difabrizio.fig12b.ps}
%}
%%\includegraphics[width=8.8cm]{intmagA2.ps}
%\centerline{\hbox{\psfig{figure=intmagA.ps,width=15.0cm,clip=}}}
%\medskip
\caption{Differences between magnitude-averaged and intensity-averaged
mean magnitudes for the variable stars in field A (upper panels) and B 
(lower panels), separately.}
%({\bf AGGIUNGERE LE VARIABILI RECUPERATE DA MACHO})
% Black filled circles: $V_{mag} - V_{int}$;
%blue filled circles: $B_{mag} - B_{int}$; red filled circles:
%$I_{mag} - I_{int}$. 
%({\it Questa figura va rifatta in modo che sia
%leggibile anche se non a colouri, 
%va divisa in 3 pannelli separati per B, V ed I come nelle
%figure del confronto con Ogle, le label sugli assi devono
%essere  $V_{mag} - V_{int}$, $B_{mag} - B_{int}$ e 
%$I_{mag} - I_{int}$, per l'asse y, e $V_{int}$, $B_{int}$ e
%$I_{int}$ sull'asse x})}
%i.e. simboli diversi e 
%scrivere bene gli assi etc.})}
%\label{f:intmagA}
\end{figure}
%\begin{figure}
%%\includegraphics[width=8.8cm]{intmagB2.ps}
%\centerline{\hbox{\psfig{figure=intmagA.ps,width=15.0cm,clip=}}}
%\medskip
%\caption{Differences between intensity-averaged and magnitude-averaged
%mean magnitudes for the variable stars in field B.}
% Black filled circles: $V_{mag} - V_{int}$;
%blue filled circles: $B_{mag} - B_{int}$; red filled circles:
%$I_{mag} - I_{int}$. 
%({\it Questa figura va rifatta in modo che sia
%leggibile anche se non a colouri vedi commenti su figura precedente})}
%\label{f:intmagB}
%\end{figure}
\begin{figure}
\includegraphics[width=8.8cm]{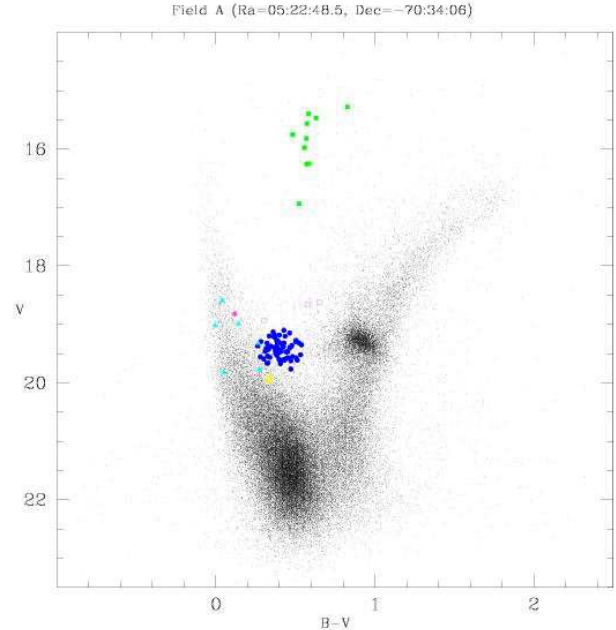}
%\includegraphics[bb= 18 144 592 718, width=8.8cm, draft=true]{difabrizio.fig13.ps}
%\centerline{\hbox{\psfig{figure=intmagA2.ps,width=15.0cm,clip=}}}
%\medskip
\caption{Position of the variable stars on the 
 $V$ {\it vs} $B-V$ colour - magnitude diagram of field A.
%, from 
%C03.
%The box outlines the clump stars of this field.
Different symbols are used for the various type of variables 
(RR Lyrae stars: filled circles; candidate Anomalous Cepheids: open squares; 
blended variables: asterisks; Cepheids: 
filled squares; binaries: filled triangles; crosses: $\delta$ Scuti) 
which are 
plotted according to their intensity average magnitudes and colours.
}
%\label{f:intmagB}
\end{figure}
\begin{figure}
\includegraphics[width=8.8cm]{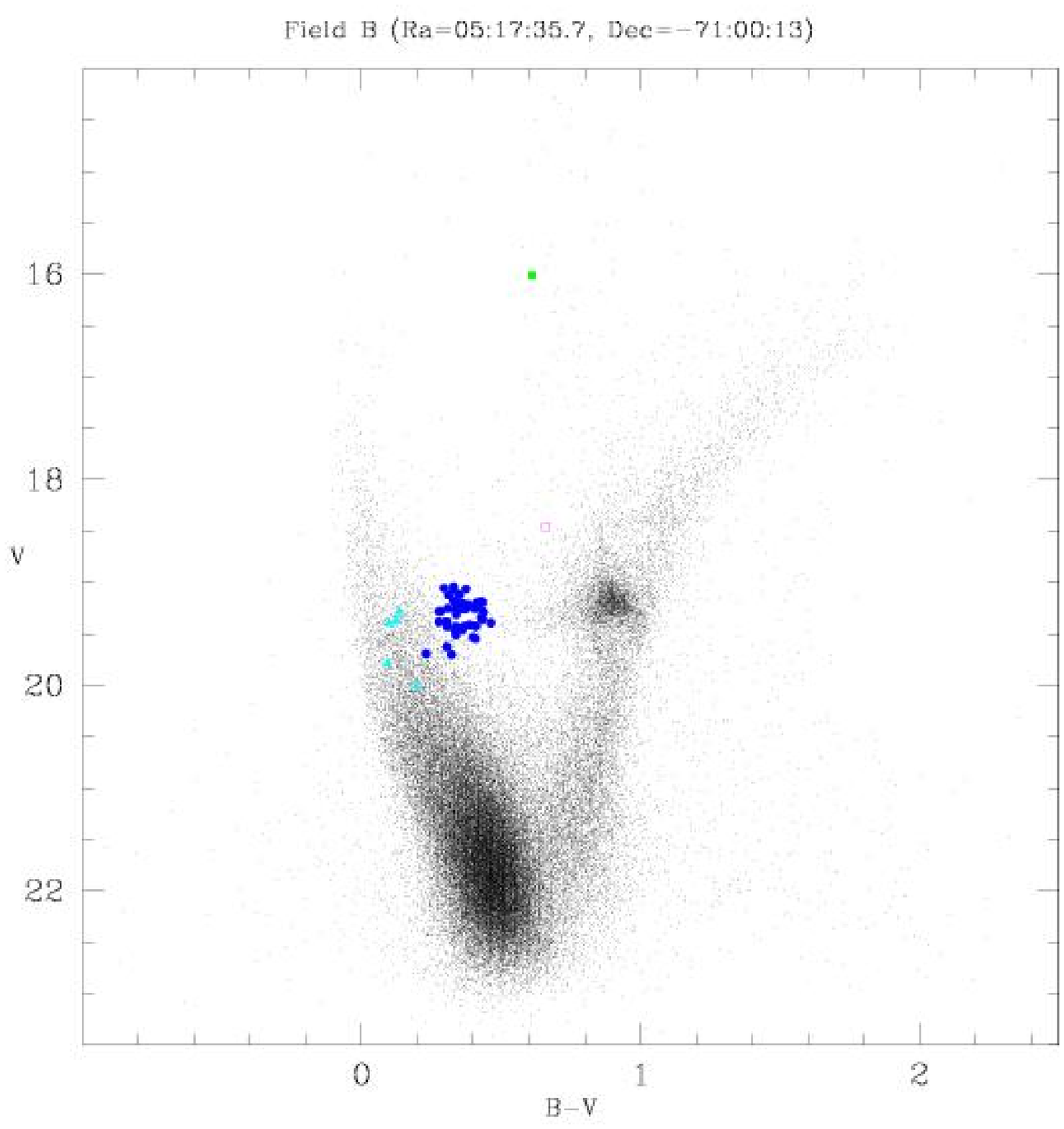}
%\includegraphics[bb= 18 144 592 718, width=8.8cm, draft=true]{difabrizio.fig14.ps}
%\centerline{\hbox{\psfig{figure=intmagA.ps,width=15.0cm,clip=}}}
%\medskip
\caption{Same as Figure~13 for the variable stars in Field B.
}
%\label{f:intmagB}
\end{figure}

%%%Figures~14 and ~15 show the position of the various types of
%%%variables in the $V, B-V$ CMDs of Field A and B.
%taken from C03
%({\it Rimettiamo le stesse figure, e chiediamo il permesso ad AJ?}). 
Variable stars are plotted according to their intensity-averaged magnitudes 
and colours, and with different symbols corresponding to different types. 
$B,V,I$ magnitudes and coordinates (in pixels) of all the 
stars shown in these figures 
(63409 in field A and 58556 in field B) 
are available in electronic form 
%from {\it http://www.bo.astro.it/$\sim$gisella
%{\bf upon request from the authors}.
at CDS.
%({\it Le rendiamo disponibili? in che formato diamo le coordinate? in
%pixel o le trasformiamo allo stesso sistema delle coordinate delle
%variabili, mettiamo una porzione di questa tabella nel paper?)}.

\subsection{The variable stars just above the horizontal branch}

Our sample contains 5 variables (star \#5106, \#9578, 
\#9604 and \#10320 in Field A, and
star \#5952 in field B) with periods in the range 
from 0.29 to 0.63 days, which is typical of RR Lyrae stars, but 
with $V$ average magnitudes 
from 0.5 to about 0.9 mag brighter 
%({\bf CHECK PER LA 5106}) 
than the average luminosities of the RR Lyrae in the same 
fields 
(see open squares and asterisk 
in Figures~13 and 14). They also have amplitudes
generally smaller than the RR Lyrae of similar period.
Average luminosities and amplitudes of these stars are
summarized in Table~7 where, in columns 7 and 8, we also list 
the difference in
magnitude with respect to the average luminosity of the
RR Lyrae stars in the same field (see Section 3.1).

\begin{table*}[ht]
%\vskip 4 cm
\begin{center}
\caption{Characteristcs of the 5 variables above the HB}
\scriptsize
%\vspace*{5mm}
\begin{tabular}{r c c c c c c  c c c}
\hline
Id~~~&Field&$<V>$&$<B>$& A$_{V}$&A$_{B}$&$\Delta V$&$\Delta B$&$<SHARP_V>$&$<SHARP_B>$\\
%     &     &(int-averaged)&(int-averaged)& & & & \\
\hline
 5106 & A & 18.820 & 18.939 & 0.391 & 0.565 & 0.597 & 0.877 & +0.286 & +0.337\\
 9578 & A & 18.626 & 19.277 & 0.307 & 0.576 & 0.787 & 0.535 & +0.264 & +0.220\\
 9604 & A & 18.932 & 19.234 & 0.655 & 0.774 & 0.481 & 0.578 & +0.099 & +0.110\\
10320 & A & 18.655 & 19.236 & 0.264 & 0.419 & 0.758 & 0.576 & $-$0.209 & $-$0.340\\
 5952 & B & 18.459 & 19.120 & 0.325 & 0.601 & 0.862 & 0.560 & +0.144 & +0.092\\
\hline
\end{tabular}
\end{center}
%\small	

Notes: $\Delta V$=$<V_{RR}> - <V_{*}>$, $\Delta B$=$<B_{RR}> - <B_{*}>$
%\label{t:tab1}
\normalsize
\end{table*} 
These objects could be RR Lyrae variables blended  
with stars of comparable
luminosity on the red and blue sides of the horizontal branch (HB) of the old
stellar population in the LMC, namely 
clump and/or young main sequence stars. Indeed, star
\# 5952 in field B is considered the blend of an
RR Lyrae and a red giant in MACHO web catalogue
of variable stars (Alcock et al. 2003a, see Section 4.2.1). 
Table~8 shows schematically how the luminosities and amplitudes of a typical
RR Lyrae in field A (namely the {\it ab-}type 
RR Lyrae \# 2525) are expected to change, during the
pulsation cycle, were the star blended to a red giant with 
luminosity equal to the average magnitude of the clump stars 
in the same field: $<V_{Clump~A}>$=19.304, and $<B_{Clump~A}>$=20.215 mag,
according to C03. The comparison between light curves of resolved and
blended variable is shown in Figure~15.
 \begin{table}[ht]
%\vskip 4 cm
\begin{center}
\caption{Blend of an {\it ab-}type RR Lyrae and a clump star in field A}
%\scriptsize
%\vspace*{5mm}
\begin{tabular}{c c c c c c}
\hline
\multicolumn{6}{c}{$<V_{RR}>$=19.326~~~$<B_{RR}>$=19.757}\\
\multicolumn{6}{c}{A$_{V}$(RR)=0.882~~~A$_{B}$(RR)=1.177}\\
\multicolumn{6}{c}{$<V_{Clump~A}>$=19.304~~~$<B_{Clump~A}>$=20.215}\\
\hline
Phase&$V_{RR}$&$B_{RR}$&&$V_{RR+Clump}$&$B_{RR+Clump}$\\
\hline
0.00 & 18.798 & 19.026 & & 18.269 & 18.713\\
0.10 & 19.024 & 19.331 & & 18.402 & 18.933\\
0.20 & 19.191 & 19.625 & & 18.493 & 19.128\\
0.30 & 19.318 & 19.819 & & 18.558 & 19.246\\
0.40 & 19.477 & 20.019 & & 18.634 & 19.360\\
0.50 & 19.542 & 20.012 & & 18.664 & 19.356\\
0.60 & 19.583 & 20.114 & & 18.682 & 19.411\\
0.70 & 19.580 & 20.156 & & 18.681 & 19.433\\
0.80 & 19.680 & 20.203 & & 18.723 & 19.456\\
0.90 & 19.592 & 19.994 & & 18.686 & 19.346\\
1.00 & 18.798 & 19.026 & & 18.269 & 18.713\\
\hline
\multicolumn{6}{c}{$<V_{RR+Clump}>$=18.551~~~$<B_{RR+Clump}>$=19.190}\\
\multicolumn{6}{c}{A$_{V}$(RR+Clump)=0.454~~~A$_{B}$(RR+Clump)=0.743}\\
%     &        &        & &        &        \\
\multicolumn{6}{c}{$\Delta V$=$<V_{RR+Clump}> - <V_{RR}>$ = 0.775}\\
\multicolumn{6}{c}{$\Delta B$=$<B_{RR+Clump}> - <B_{RR}>$ = 0.567}\\         
\hline
\end{tabular}
\end{center}
\end{table}
%\small	
\begin{figure*} 
\begin{center}
\includegraphics[width=15.0cm]{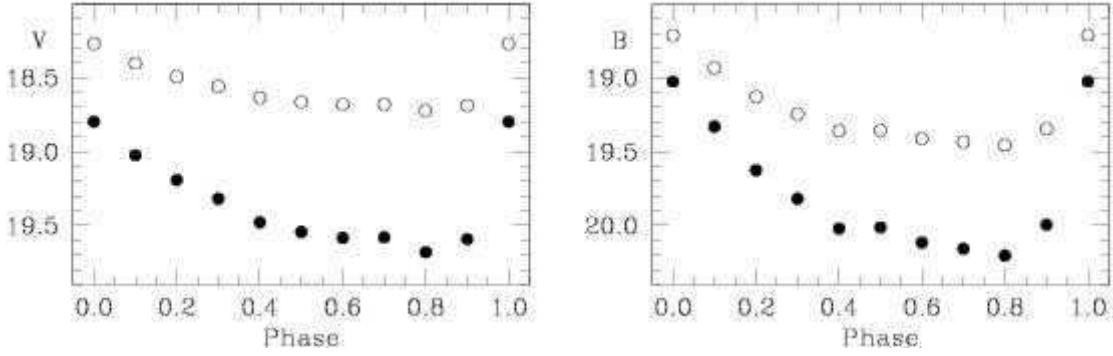}
\caption[]{Schematic light curves of a resolved 
{\it ab-}type RR Lyrae star in field A (filled circles) and of its 
blend with a clump star in the same field (open circles).
}
%\label{f:fig5b}
\end{center}
\end{figure*}

%({\bf AGGIUSTARE IL PEZZO CHE SEGUE PER INCLUDERE LA 5106 CHE SEMBRA DAVVERO
%ESSERE UN BLEND}) 
This exercise shows that as the result of the blend the variable star would
appear about 0.8 and 0.6 mag brighter than
its average $V$ and $B$ luminosities of RR Lyrae star,
its $V$ amplitude would be reduced by
about 50\% and the $B$ amplitude by about 37\%.  These numbers are 
very similar to the $\Delta V$ and $\Delta B$ value and amplitudes listed in
Table~7, thus showing that blending is a plausible cause of the
overluminosities of these 5 variable stars.  
In order to further investigate the blending hypothesis  we checked the 
frames. 
Stars \# 9578 and \#9604 appear to be rather
isolated. Stars \# 10320, \#5952 and \#5106 instead have
faint companions that may occasionally fall within the PSF of the
primary star in bad seeing conditions.  This should produce an
increased scatter of the light curves as is indeed the case for star 
\# 5106
% (see Figure~17) 
which also is 
rather blue ($B-V$=0.119) indicating that this RR Lyrae is likely blended 
with a main sequence star. The other 
4 objects in Table~7  
have instead all rather clean light curves (star \# 5952 in particular) 
and show no shifts between the 1999
and 2001 light curves that might hint they could be unresolved blends in
our 1999 photometry, which was taken in less favourable seeing conditions.
\begin{figure*} 
\begin{center}
\includegraphics[width=15.0cm]{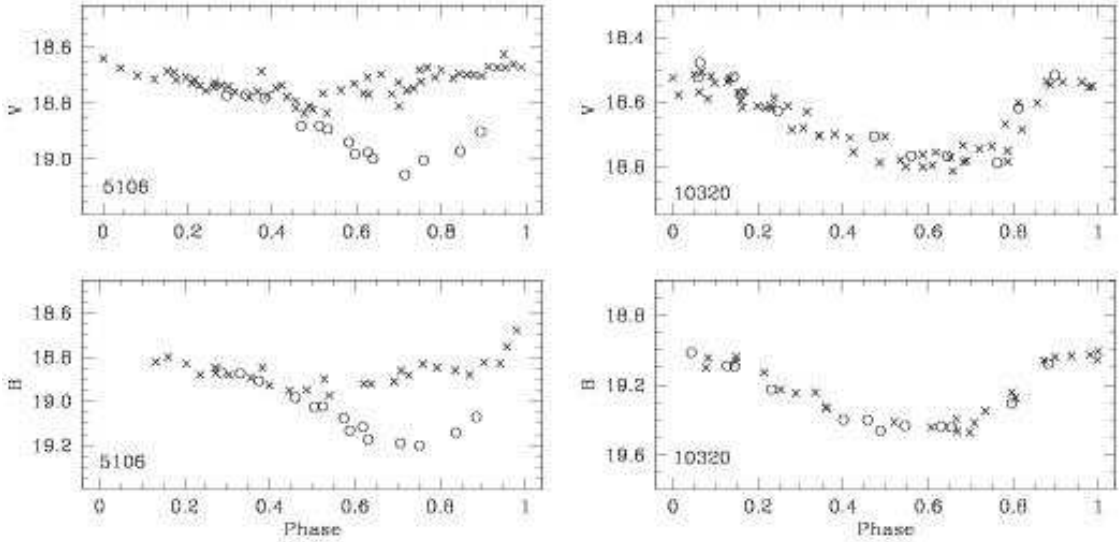}
\caption[]{Light curves of the variable stars \# 5106 
(left panels) and \# 10320 (right panels), open circles and crosses correspond to the 
2001 and 1999 data, respectively. 
}
%\label{f:fig5b}
\end{center}
\end{figure*}
Figure~16 shows the B,V light curves of star \# 5106 (left panels) and 
\# 10320 (right panels).
%, different symbols are used for the 
%2001 and 1999 data, respectively.
The 1999 light curves of \# 5106 are overluminous, particularly at minimum 
light, and have smaller amplitudes
compared to the 2001 ones, as if the star was an unresolved blend in the 1999
photometry, those of star \# 10320 do not show
any systematic difference between the two datasets.   

For each photometrized object DAOPHOT returns a shape 
defining parameter called {\it
SHARP}, which is related to the intrinsic angular size of the object 
image and  
measures the regularity and symmetry of the PSF
stellar profile. According to DAOPHOT user manual 
%({\bf REFERENZA-MARCELLA})
% {\it SHARP} is a zero-th order estimate of the 
%square of the quantity: $$SHARP^2 \sim \sigma^{2}_{observed} - 
%\sim \sigma^{2}_{PSF}$$ O
objects with values of  $SHARP >>$ 0 are 
galaxies and blended doubles, objects with values of  $SHARP <<$ 0 are
cosmic rays and image defects. In our 2001 photometry stars at the luminosity level of the
HB generally have: $\mid SHARP \mid <$ 0.10$-$0.20.
Average $SHARP$ values 
%in the 2001 $V,B$ data 
for the 5 overluminous variables
are given in Columns 9 and 10 of Table~7. 
Stars \# 5952 and \# 9604 have very good $SHARP$ values, $SHARP$ of  
star \# 9578 is worse but still acceptable.
 Star \# 10320 has negative values of $SHARP$ reflecting the fact that 
is at the frame edge 
 where there are geometric distorsions.  
Finally, \# 5106 has large positive values of  $SHARP$ possibly indicating 
that the star is double. In conclusion, 
star \# 5106 is likely a blended variable, while 
 if the other four   
stars are actually blends, the two components must be
completely unresolved, so to appear as just one single object 
within the PSF profile.

Tests with artificial stars performed to evaluate the completness of our
photometry in field A show that at the luminosity level of
the RR Lyrae and clump stars (19.20 $\leq V \leq $ 19.40 mag) 
our photometry is complete to 96.5 \%. Since there are 78 RR Lyrae stars
in field A we thus estimate that about 2-3
%esattamente 2.73
of this type of variables may be lost 
due to incompletness/blending, and, roughly scaling down to the smaller number
of RR Lyrae stars and lower crowding, 
%about 1.8 
less than 2 in field B. 
These estimates are reasonably consistent with the number of variables detected
just above the HB in each field.

G04 obtained spectra with FORS1 at the Very Large Telescope (VLT) 
and measured 
the metallicity of 3 of the overluminous variables.
 All of them appear as single
objects in the FORS1 slit.
% ({\bf ANGELA E' VERO?}). 
The derived metallicities are: [Fe/H]=$-1.96\pm 0.16$ for \#9604, 
[Fe/H]=$-1.66\pm 0.09$ for \#10320, and [Fe/H]=$-1.59\pm 0.03$ for \#5952,
for an average value of [Fe/H]=$-1.74\pm 0.11$.
The spectra of these 
3 objects are shown in figures 9 and 21 of G04, along with those of LMC 
RR Lyrae and clump stars, and of Anomalous
Cepheids (ACs) in $\omega$ Cen (see G04 figure 20), taken with the same 
instrumental set-up.
The 3 stars have spectra very similar to the ACs in $\omega$ Cen.
No clear evidence of spectral features due to secondary unresolved 
componens are seen, however star \# 5952 
has a prominent G-band similar to that observed 
in the spectrum of the clump star shown in figure 9 of  G04.

The 5 overluminous variables were observed by MACHO and classified respectively 
as: 
{\it ab-}type RR Lyrae stars (\# 5106 and \# 9604), an RRab 
blended with a red giant (\# 5952), and eclipsing binaries (\# 9578 and
10320; see Table~9).
 The average $V$ magnitudes of stars \# 5106 and 5952 agree with ours 
 within 0.05 mag, with our values being systematically fainter.
 Stars \# 10320, \# 9604 and 9578 are instead brighter in our 
 photometry, by 0.14, 0.17 and 0.27 mag, respectively. Nevertheless, 
 even in MACHO photometry they lie above the HB. 

Finally, we note that stars \# 9604 and \# 10320 were also observed
by OGLE II (see Section 4.3 and Table~14) and classified {\it ab-} and 
{\it c-}type RR Lyrae, respectively.
OGLE II average luminosities and light curves of star \# 9604 agree within
0.1 mag, with our values being slightly brighter (by 0.04 mag in $B$ and
0.11 mag in $I$, see Table~14). Similarly, OGLE II $B$ data for star
\# 10320 agree within 0.03 mag to our value, being 0.03 mag 
fainter (we do not have $I$ photometry for this
star). However, OGLE II $V$ average luminosities are respectively 0.79
and 0.71 mag fainter than ours, causing these two variables to have
rather unlikely colours for RR Lyrae stars: ($B-V$)$_{9604}$=$-$0.45, 
($V-I$)$_{9604}$=1.21 mag, and ($B-V$)$_{10320}$=$-$0.17, 
($V-I$)$_{10320}$=1.53 mag in OGLE II photometry. Indeed, the OGLE II $V$ 
light curves of these objects are very poor. No actual $V$ light 
variation is seen for \# 10320, possibly indicating a mismatched
$B,V,I$ counteridentification.

In conclusion, based on the available observational evidences star
\# 5106 is likely to be the blend of an {\it ab-}type RR Lyrae with a
young main sequence star.   
Instead, it is not possible to definitely assign a classification to the
%confirm or 
%rule out whether the 
other 4 overluminous variables.
%could be blended objects.
Sub-arcsec photometry
% as that achievable with the Hubble Space Telescope
would be needed to shed some light on this issue.

On the other hand, given the complex stellar
population in the LMC, we should also consider whether these 4 
objects could be
pulsating variables intrinsically brighter that the RR Lyrae stars, such as
the  Anomalous Cepheids (ACs) 
commonly found in dwarf
Spheroidal galaxies (Pritzl et al. 2002 and references 
therein), or the low luminosity (LL) Cepheids (Clementini et
al. 2003b) and  the short period Classical Cepheids (SPCs) found in a 
number of dwarf Irregular galaxies (Smith et al. 1992, Gallart et al. 1999, 
 2004,
Dolphin et al. 2002).

Anomalous Cepheids are metal-poor (Population II) helium burning stars in the
instability strip, from about 0.5 up to about 2 mag (Bono et al. 1997) 
brighter than the 
HB of the old stars. They
generally have periods in the range 0.3-2 days, but are too luminous for their
periods to be Population II Cepheids (Wallerstein \& Cox 1984). 
The high luminosity can be accounted for if
they are more massive than normal old HB stars,
%. These
%larger masses are thought to occur because these stars 
as if they formed from the
coalescence of a close binary (originally a blue straggler), although in some
cases they may result from the evolution of younger, single massive stars. At
low metallicities (Z$\leq$0.0004, i.e. [Fe/H]$\leq -1.7$), a hook in the HB is
predicted, the so called ``HB turnover'' (see Caputo 1998, and references
therein), so that stars with masses larger than $\sim 1.3 M_\odot$ may cross
the instability strip. Thus, there is a limiting metallicity above
which no Anomalous Cepheid should be generated (Bono et
al. 1997, Marconi et al. 2004). This limit in metallicity should be about
[Fe/H]$\sim -1.7$ for variables around $\sim 1.3 M_\odot$ and [Fe/H]$\sim -2.3$
for variables around $\sim 1.8 M_\odot$. 
While very common in dwarf Spheroidal
galaxies, Anomalous Cepheids are very rare in globular clusters: only one is
known in the very metal-poor cluster NGC\,5466 (Zinn \& Dahn 1976,
[Fe/H]=$-2.22$\ according to Harris 1996) and two suspected ones are found 
in 
$\omega$~Cen (Nemec et al. 1994, Kaluzny et al. 1997), a cluster spanning
a wide range in metallicity
% with at least three separate enrichment peaks
(Norris, Freeman, \& Mighell 1996, Suntzeff \& Kraft 1996, Pancino et al.
2002) and suspected of being the 
% and suggested to possibly be the 
remnant of a disrupted dwarf galaxy.

The short period Cepheids are blue loop stars, i.e. stars that have ignited the
helium in non degenerate cores ($M \geq 2.5 M_\odot$), and have periods shorter
than 10 days. They fall on the extension to short periods of the Classical
Cepheids P/L relations (see Smith et al. 1992, Gallart et al. 1999, 2004,
Dolphin et al. 2002).

Observed for the first time in NGC~6822 dwarf Irregular galaxy 
(Clementini et
al 2003b), the LL Cepheids have small amplitudes, luminosities just
above the HB, and are fainter and have shorter periods than the 
short period Cepheids.  

%{\bf CONTROLLARE CHE QUESTI PEZZI NON SIANO TROPPO UGUALI ALLO G04}).
It is not possible to decide to which of the above classes these four variables 
brighter than the  
HB more likely belong, based on 
the period-luminosity (P/L) relations, since at their short
periods the P/L relations of Anomalous and Classical Cepheids
merge and are almost indistinguishable. Indeed, in the  P/L plane stars \#9604, 
\#5952 and \#9578 fall on the extension to short periods of the fundamental mode
Anomalous and Classical Cepheids, while star \#10320 lies on the extension to
short periods of the first overtone P/L relations (see Figure 2 of Baldacci et
al. 2004).
%
%Gratton et al. (2004) measured the metal abundance 
%([Fe/H]) for three of these stars,
%namely: [Fe/H]=$-1.96\pm
%0.16$ for \#9604, [Fe/H]=$-1.66\pm 0.09$ for \#10320, and [Fe/H]=$-1.59:\pm
%0.03$ for \#5952, with a corresponding average value of [Fe/H]=$-1.74\pm
%0.11$. 
Knowledge of the metallicity may allow to break the degeneracy in the
P/L relation, since short period Classical Cepheids and ACs are expected
to have different metallicities, similar to those of their respective Population I
 and II parent populations.
%there is 
%a limiting metallicity above
%which no Anomalous Cepheid should be generated (Bono et
%al. 1997, Marconi et al. 2003)\footnote{This limit should be
%[Fe/H]$\sim -1.7$ for variables around $\sim 1.3 M_\odot$ and 
%[Fe/H]$\sim -2.3$
%for variables around $\sim 1.8 M_\odot$.}, and 
%Anomalous Cepheids are expected to have low metal abundances, 
%similar to
%or lower than the metallicity of the oldest populations in the host system,
%while short period Classical Cepheids should have metallicities similar to
%those of the Population I component in the system.
%Since both individual and average metallicities 
%() 
%of the three variables are
%well below the average metal abundance of the LMC RR Lyrae stars 
%([Fe/H]=$-1.48\pm 0.03$; G04) they 
Based on the individual and average metallicities G04 conclude   
that the three overluminous variables they analyzed would more likely be
ACs with masses $M \sim 1.3
M_\odot$ rather than the short period tail of the LMC Classical Cepheids. 
Star \#9578 lacks a metallicity estimate, hence its possible 
classification as AC is more uncertain.

\section{A star by star comparison with MACHO and OGLE II photometries}
\subsection{Introduction}
Fields A and B 
are contained in MACHO's fields \#6 and \#13, respectively, and
there is a 42.1\% overlap between field A and OGLE II field LMC\_SC21.
Both MACHO and OGLE II 
catalogues are available on line.
In particular, the MACHO collaboration has made available on web
(see http://wwwmacho.mcmaster.ca/Data/MachoData.html) 
coordinates and instrumental photometry for about 9 milion
LMC stars, and instrumental time-series  for all the variables
they  have identified in the LMC. For the variables they also
publish calibrated average magnitudes\footnote{MACHO instrumental
time-series and the calibrated average magnitudes of the LMC
variable stars are also available 
at the CDS at Strasburg.
%on the VizieR On-line Data
%Catalog, II/247 at 
%http://cdsweb.u-strasbg.fr/viz-bin/VizieR-2?-source=II/247. 
}.
Calibrated photometric maps (including time-series data of the
variable stars) for all the LMC fields observed by OGLE II are
instead available on OGLE II web page at
http://www.astrouw.edu.pl/$\sim$ogle/ogle2/rrlyr\_lmc.html.
It was thus possible  to make a detailed comparison between our
and  MACHO and OGLE II photometries, for both 
variables and constant stars in common.

%
%({\it This is a very important point because ....
%qui va detto che i cataloghi di MACHO e OGLE II (MACHO) in particolare per
%le stelle variabili sono una fonte ricchissima e se quindi si riesce
%ad agganciare bene anche la calibrazione fotometrica ad un sistema standard
%ben definito questo ne aumenta l'utilizzabilita'}).
%

Before going into the details of this comparison
we note that two major differences exist between our, MACHO, and 
OGLE II databases:
(i) observing strategy, exposures and time resolution of
our photometric observations were specifically designed to achieve a very 
accurate definition of the average luminosity level of the RR Lyrae
stars in the bar of the LMC, 
and provide a valuable counterpart to Walker (1992) study of the RR Lyrae
stars in the LMC globular clusters.
RR Lyrae's are instead by-products close to  the 
limiting magnitude of MACHO and OGLE surveys, whose main target 
was the detection of microlensing events in the LMC;
(ii) although we used DoPhot to reduce the 1999 time series, the final 
photometry and calibration of our full dataset was handled by 
DAOPHOT+ALLFRAME, while both MACHO and OGLE II photometries used
the DoPhot package\footnote{Actually MACHO used SoDoPhot (Son of DoPhot), a 
revised package based on DoPhot algorithms but optimized
to MACHO image data.}.
These packages may give similar results when crowding is not too severe;
however DAOPHOT+ALLFRAME 
is much more efficient than 
DoPhot to resolve and measure faint stellar objects in crowded fields.
This is clearly shown in Figure~2, where, thanks to ALLFRAME, we reach 
about 1-1.5
mag fainter and resolve almost twice the number of stars as with DoPhot.
Moreover, 
 DoPhot is reported to give systematically 
brighter magnitudes for faint stars in crowded regions than DAOPHOT
due to its sky fitting procedure (Alcock et al. 1999, hereinafter A99). These differences 
should be kept in mind to 
interpret the results of the comparisons discussed 
in the next subsections.
  
\subsection{Comparison with MACHO photometry}

The MACHO collaboration has published calibrated photometry,  
namely magnitude-averaged mean magnitudes (Alcock et al. 2003a),   
only for the LMC
variable stars.
%, for which 
%can be found either 
%at wwwmacho.mcmaster.ca/Data/MachoData.htm or 
%on the VizieR On-line
%Data Catalog, II/247 at 
%http://cdsweb.u-strasbg.fr/viz-bin/VizieR-2?-source=II/247.
%({\bf Qui va controllato che  
%i valori pubblicati sulle due web pages siano gli stessi - MARCELLA)}.
A99 provide a detailed 
description of the photometric 
calibration to the Kron-Cousins $V$ and $R$ system of 
the twenty top-priority MACHO fields of the LMC which include 
fields \#6 and \#13. They quote an internal precision of 
%the MACHO photometry calibration of 
$\sigma_V$=0.021 mag (based on 20,000 stars 
with $V \lsim$ 18 mag) and, from the 
comparison 
with other published measurements, they estimate a 
mean offset between MACHO and all the other data of $\Delta V$=$-$0.035 mag
(see fig. 7 of their paper).
A99 calibration is referred to as  
version 9903018 in following publications of the MACHO team (e.g.
Alcock et al. 2004).
However, 
the calibrated average magnitudes
available on MACHO web pages (which, at the time this paper 
is being written, correspond to the last update of
April 18th 2002)  
are  
based on a different 
version of A99 photometric calibration (see Alcock et al. 2004).
%{\bf (CHECK, 
%questa \'e la versione corrispondente al 
%paper 
%Alcock et al. 
%2003a, VizieR On-line
%Data Catalog, II/247,   che 
% nell'Alcock et al. 2004
%dicono e' basata su una versione diversa della calibrazione fotometrica
%dei dati, ma non \'e detto a che versione della calibrazione A99 corrisponde
%e sulla pagina web non siamo riuscite a trovare il numero della versione}).
MACHO catalogue is undoubtedly an invaluable
inventory of the LMC variable star content; however, 
because of the non-standard passbands,  
%used by MACHO, 
the severe ``blending"
problems in the fields close to the LMC bar, and the complexity 
of the calibration procedures (see A99 for details), 
the absolute photometric calibration is a 
major concern. As 
a matter of fact different versions of the MACHO calibrated 
light curves exist,
 and it would be very important to know which version most closely
matches the standard system in order to be able to fully exploit 
the catalogue.
%s often used in the study of the variable
%stars. 
While working at the
present paper we discussed this issue with 
members of the MACHO team who were working on  
the 
calibration procedures and/or were using the MACHO variable star catalogues (namely Dr.s D. Alves, C. Clement, 
and G. Kov\'acs). 
%either at Meetings or  
%via e-mail, 
We exchanged datasets and made comparisons between our
photometry 
and data 
based on different versions of the MACHO photometric calibrations.
% they
%made available to us.
In the following we report results 
%that were kindly made available to us.
% 
%
%to compare our data with.  
%
%with members of the
%MACHO team 
%who % 
%asked us to make comparison between MACHO and our photometry
%for the variable stars in common by 
%However, since none of these revised calibrations has been 
%published yet ({\bf CHECK, sia su pagina web che sui papers
%relativi, astro-ph/03100281 in particolare ed il mail della Clement, in base al
%quale i dati dell'astro-ph/0310281 were transformed to the 
%Kron-Cousins V and R systems using the equations derived by A99, there 
%in a MACHO project internal report Dave Alves designed as version 9903018})
%and so far no further up-date of the 
%MACHO web page values has been done yet, 
%Since a number of different versions of the MACHO calibrated photometric data 
%exist, 
%these comparisons
%with MACHO's data 
%which are 
based on 4 different
datasets of MACHO's photometry,
namely:
\begin{enumerate}
\item  MACHO's magnitude-averaged 
 mean values for the variables in common (77 and 54 variables 
   in field A and B, respectively) as published 
on MACHO web pages.
%, and are
%based on 
%A99 calibration version XXXXXX. 
This comparison is described in Section 4.2.1.

\item MACHO's time-series photometry for 42 RR Lyrae stars (25 in Field A and 
 and 17 in field B, respectively), kindly made available by 
 G. Kov\'acs. This point-to-point comparison of the light curves 
 is described in Section 4.2.2.
% {\it (vedere se questi valori sono stati pubblicati e 
%    come vanno citati, e su che versione della calibrazione sono 
%basati, chiedere a Kov\'acs, se sono i dati usati per l'Alcock et al. 2003b.
%ApJ 598, 597, in questo paper dicono che sono basati sulla calibrazione in
%A99)}
%({\bf manda mail a Kov\'acs per ref. calibrazione, GISELLA})
\item MACHO's magnitude-averaged  mean magnitudes for 7 {\it c-}type 
RR Lyrae stars (3 in Field A and 
 and 4 in field B, respectively) whose data
 were sent us by C. Clement (see Section 4.2.3).

\item MACHO's photometry for 18 RR Lyrae stars (9 in each field) 
and for the non-variable stars in 4$^{\prime} \times 4^{\prime}$
%  ({\bf Quanto grandi sono
%  in media i patches, ci vogliono le dimensioni, MARCELLA?})
areas surrounding the variables, whose photometric data were kindly 
made available by D. Alves. 
%({\bf manda mail a Alves per ref. calibrazione, GISELLA})
%
%{\bf (vedere se questi dati sono stati pubblicati e come
%vanno  citati, chiedere ad Alves su che versione della calibrazione
%sono basati}). 
These comparisons are described in Sections 4.2.4 and
4.2.5, respectively.

\end{enumerate}

\subsubsection{Comparison with MACHO photometry for the variable stars 
in common: the web catalogue}
We have retrieved from the MACHO web archive
 coordinates and magnitude-averaged mean magnitudes for all the variables
 identified by MACHO in our 
 %the photometric data ....
%({\it mettere l'indirizzo web e poi dire che abbiamo controidentificato le 
%variabili, dire che software e' stato usato per fare la controidentificazione,
%seguire la traccia di quanto e' stato scritto per il confronto con OGLE II}).
%corresponding to our 
fields A and B and counteridentified the variable stars
in common by coordinates
% for the coordinates, 
using private software by P. Montegriffo.
Counteridentifications between our and MACHO identification numbers 
are provided in Table~9 where we also give the classifications.

\begin{table}[ht]
\begin{center}
%\vskip 4 cm
\caption{Counteridentification between MACHO and us for
the variable stars in common in field A and B, separately}
%\vspace{0.5cm}
%\scriptsize
\begin{tabular}{ccll}
\hline
\multicolumn{4}{c}{Field A}\\
\hline
\multicolumn{1}{c}{Id$_{MACHO}$}&\multicolumn{1}{c}{Id$_{this~paper}$}
&\multicolumn{1}{c}{Type$_{MACHO}$}&\multicolumn{1}{c}{Type$_{this~paper}$}\\
\hline
%rr.txt in /dati3/gisella/lmc/marcella
%campo A
%id_macho id_noi  class_macho   class_noi
~6.7055.7    & ~~30    & ~~Ceph.1st & ~~~~Ceph    \\
6.6810.11    & ~~40    & ~~Ceph.1st & ~~~~Ceph    \\	
6.6931.37    & ~~57    & ~~EB       & ~~~~~~~?       \\  
6.6932.22    & ~121    & ~~Ceph.1st & ~~~~Ceph    \\
6.6933.19    & ~147    & ~~Cep.Fun  & ~~~~Ceph    \\
6.6812.27    & ~150    & ~~Cep.Fun  & ~~~~Ceph    \\
6.6933.11    & ~170    & ~~Cep.Fun  & ~~~~Ceph    \\
6.7054.10    & ~182    & ~~Cep.Fun  & ~~~~Ceph    \\
6.6934.10    & ~200    & ~~EB	    & ~~~~Ceph    \\
6.6810.79    & ~242    & ~~EB       & ~~~~~~~?       \\  
\hline
\end{tabular}
\end{center}

Table 9 is presented in its entirety in the electronic edition
of the Journal. A portion is shown here for guidance regarding its form
and content.
%({\bf Qui va ricontrollato se ci sono anche la 4313 e la 21007 ed eventualmente
%vanno aggiunte in coda a questa tabella nella parte che non viene
%stampata}).
%\end{center} 
%\label{t:tab4}
\end{table}

\begin{figure} 
\includegraphics[width=8.8cm]{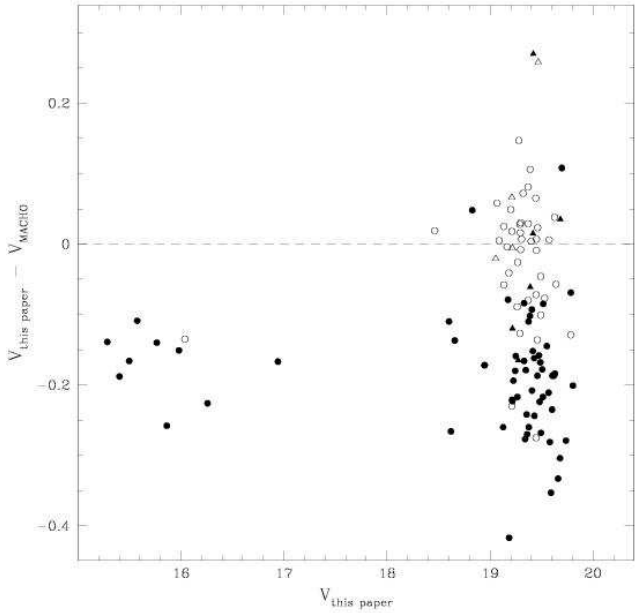}
\caption[]{Comparison between our and MACHO 
mean $V$ magnitudes for the variable stars in common.
Residuals are: this paper $-$ MACHO. 
Filled and open symbols are used for variable stars in field A and 
B, respectively. Triangles are the double mode RR Lyrae stars. 
%({\bf AGGIUNGERE IN QUESTA FIGURA TUTTE LE VARIABILI RECUPERATE DAL
%CONFRONTO CON MACHO})
%({\bf Questa figura va rifatta aggiundengo le stelle 28246, 28539 ed 
%eventualmente anche la 4313 e la 21007})
}
%%\label{f:fig1a}
\end{figure}
MACHO detected 85
%87 sono 86 perche' la 6.6691.70 e' fuori campo 
 variables in the portion of their field \# 6 in common with our field A. 
 We
have counteridentified all of them. 
Three of these stars (MACHO numbers: 6.6810.67, 6. 7052.518, and  
6.7054.463, corresponding to our stars: \# 354, 3394 and 17341)
are not found to significantly vary in our photometry. 
Other 5 variables classified eclipsing binaries  by MACHO, some of which 
with very long period  (P$>$ 60 day), have  
%The first two 
%have A$_V$ amplitudes of 0.02-0.03 mag in MACHO photometry, the latter
%{\bf (MARCELLA QUI BISOGNA VEDERE COM'E' LA CURVA DI LUCE DELLA TERZA IN MACHO,
%E VEDERE DI TUTTE E 3 LE AMPIEZZE DA NOI}.
%Indeed, MACHO light curves 
%for these stars are XXXX poor ?????? 
%({\bf Potrebbero essere 72 se ci sono anche la 4313 e la 21007, vanno controllate - MARCELLA}). 
%For other 5 variables, all classified eclipsing binaries by MACHO do 
%   
%({\bf MARCELLA QUESTI 16 OGGETTI NON CI SONO PROPRIO NELLA NOSTRA FOTOMETRIA O
%I CORRISPONDENTI OGGETTI DA NOI N ON VARIANO?)}, 
%are 
%({\bf Potrebbero essere 14 se due delle 4 RR mancanti sono la 4313 e la 21007})
%classified by MACHO as eclipsing binaries, 5 of them with very long  periods
%(P$>$ 10$^d$),  and all 11 
small amplitudes, sometime rather dubious in our photometry.
%, hence
%they may have been not detectable in our 1999 data
%({\bf QUI ANDRA' AGGIUSTATO DOPO CHE LE HO GUARDATE}).
% The remaining 4 ({\bf QUI VA CAMBIATO SOLO UNA PER NOI E' RUSCO LA
% 3394})
%({\bf Potrebbero essere 2 se due delle 4 RR mancanti sono la 4313 e la 21007}) 
%missing objects are all classified  as RRab's; however 3  of them %({\bf CHECK})
%have amplitudes too small and are rather dubious. 
On the other hand, we have
identified 26 additional variables apparently missed by MACHO; 
they
include 18 RR Lyraes (10 RRab's and 8 RRc's), 5 eclipsing  binaries, 
1 Cepheid, 1 $\delta$ Scuti, and 1 candidate variable of
unknown type. Thus we have about
34\% more short period variables than MACHO in field A.

57 variables have been found by  MACHO in the area in  common with field B.
We have counteridentified 56 of them.
%  (52 confirmed and 1 candidate
%variable stars in our list and three objects that have very small amplitudes). 
The missing object is at the very edge of
our field B and its photometry is not reliable.
% Questo oggetto e nella striscia dei frames del 2001 che non siamo risusciti
% a ridurre nel 1999 non e' stato controllato ma e' comunque al bordo dei
% frames 
Two of the variables in common, classified by MACHO as eclipsing 
binaries, have rather small and dubious amplitudes in both 
photometries. 
%
%, are all classified by
%MACHO as eclipsing binaries and 2 of them have very long periods  (P$>$ 10$^d$).
%One further eclipsing binary (star \# 2601 has larger amplitudes, XXXX
%but our photometry is too sparse to derive any period, the light curve 
%adopting MACHO period ($\sim$ 14.816 day) is OK.
%small amplitudes, hence they may have been not detectable in our 1999 data.
%Thus we may have actually  lost 4 of the MACHO variables in this field, 
%all classified  as RRab's, 3 with  period around half a day, and one with
%P$\sim$0.$^d$9 and rather bright.
In field B we have identified 13 additional variables that were
not detected by MACHO; they include 9 RR Lyraes (3 RRab's and 6 RRc's) and
4 eclipsing  binaries. Thus we have about 24\% more short period 
variables than MACHO in field B.

We also noticed that MACHO classification of some of the variable stars in common
does not match ours (see Columns 3 and 4 of Table 9). 
In particular, there are 6 variables
classified as eclipsing binaries by MACHO that we classify as 
RR Lyrae stars (3) and Cepheids (1 Classical and 2 candidate ACs), 2 RR
Lyrae for MACHO that we classify as an eclipsing binary and the
blend of an RRab and a main sequence star, and an RR Lyrae + giant branch
star for
MACHO that we classify as candidate Anomalous Cepheid. Finally we assign a different
pulsation mode to 13 other variables,  classified as RR Lyrae stars in both
photometries.

\begin{figure} 
\includegraphics[width=8.8cm]{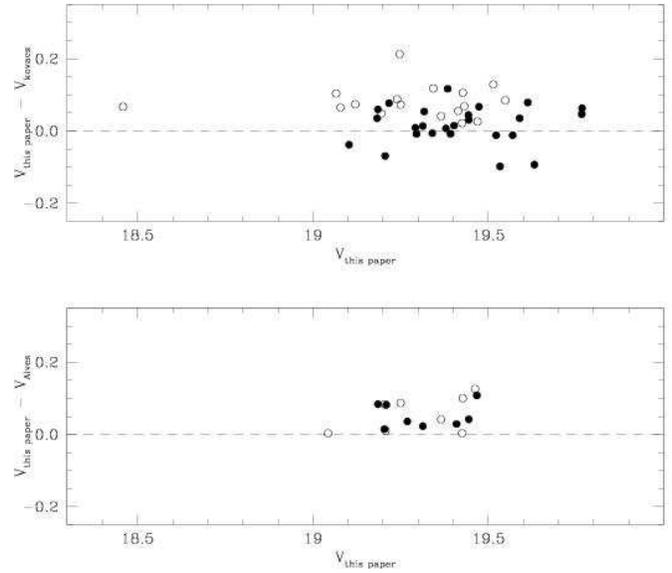}
\caption[]{Comparison between our and Kov\'acs (top panel) and
Alves (bottom panel) 
mean $V$ magnitudes for the variable stars in common.
Residuals are: this paper $-$ others. 
Filled and open symbols are used for variable stars in field A and 
B, respectively. 
%({\bf AGGIUNGERE IN QUESTA FIGURA TUTTE LE VARIABILI RECUPERATE DAL
%CONFRONTO CON MACHO})
%({\bf Questa figura va rifatta aggiundengo le stelle 28246, 28539 ed 
%eventualmente anche la 4313 e la 21007})
}
%%\label{f:fig1a}
\end{figure}

\begin{figure*} 
\begin{center}
\includegraphics[width=15.0cm]{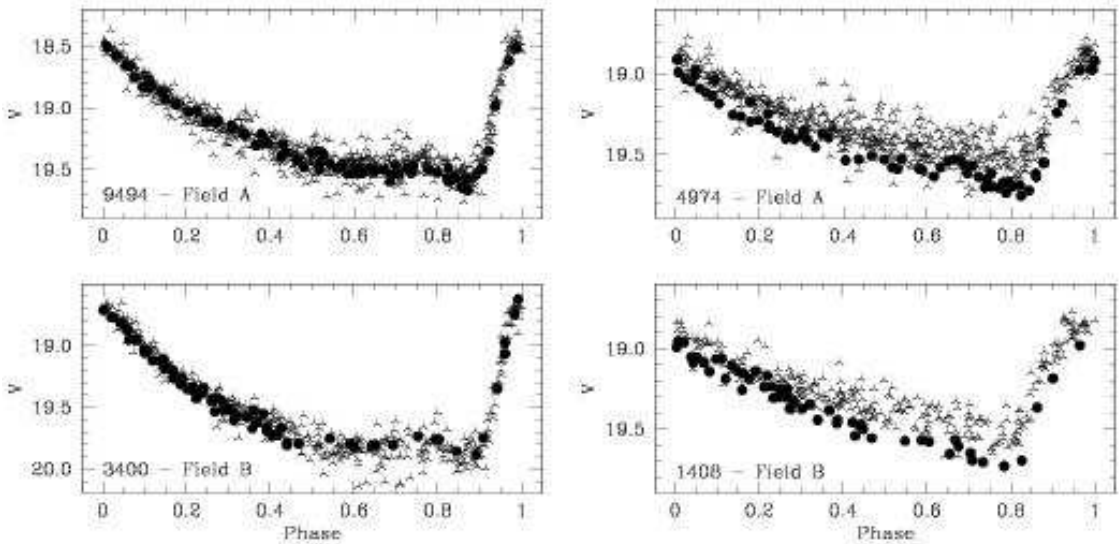}
\caption[]{Point-to-point comparison of the light curves for {\it ab-}type 
RR Lyrae stars in common with Kov\'acs subsample. Filled dots: our photometry,
three arms crosses: MACHO photometry. These represent the best (left) and
worst (right) cases.}
%\label{f:fig5b}
\end{center}
\end{figure*}

%
%tabella 
%
\noindent
\begin{table*}[ht]
%\vskip 4 cm
\begin{center}
\caption{Comparison with MACHO photometry for the variable stars in common:
 Kov\'acs subsample}
\scriptsize
%\vspace*{5mm}
\begin{tabular}{r r c c c c r r  c}
\hline
Id$_{MACHO}$~~~&Id$_{this~paper}$&$<V_{MACHO}>$&$<V_{this~paper}>$&
$<V_{Kovacs}>$&$<V_{this~paper}>$&$\Delta 1$~~~&$\Delta 2$~~~&Field\\
               &                 &(mag-averaged)&(mag-averaged)   &
 (int-averaged)&(int-averaged)    &              &                 & \\

\hline
  6.6931.603~~&  2525~~~~ & 19.636 &  19.376 &  19.346  & 19.340 & $-$0.260  &  $-$0.006 &  A\\
  6.6932.948~~&  4974~~~~ & 19.499 &  19.406 &  19.267  & 19.384 & $-$0.093  &     0.117 &  A\\
  6.6811.882~~&  6398~~~~ & 19.526 &  19.347 &  19.263  & 19.317 & $-$0.179  &     0.054 &  A\\
  6.6811.736~~&  6426~~~~ & 19.421 &  19.227 &  19.125  & 19.185 & $-$0.194  &     0.060 &  A\\
  6.6813.923~~&  9494~~~~ & 19.483 &  19.266 &  19.140  & 19.217 & $-$0.217  &     0.077 &  A\\
 6.6934.1136~~&  9660~~~~ & 19.615 &  19.407 &  19.400  & 19.392 & $-$0.208  &  $-$0.008 &  A\\
  6.6810.635~~& 12896~~~~ & 19.807 &  19.620 &  19.554  & 19.589 & $-$0.187  &     0.035 &  A\\
 6.6691.1079~~& 25301~~~~ & 20.006 &  19.805 &  19.719  & 19.766 & $-$0.201  &     0.047 &  A\\
  6.7054.801~~& 25362~~~~ & 19.656 &  19.488 &  19.399  & 19.443 & $-$0.168  &     0.044 &  A\\
 6.6692.1042~~& 26525~~~~ & 19.685 &  19.507 &  19.406  & 19.473 & $-$0.178  &     0.067 &  A\\
 13.5956.410~~&  1408~~~~ & 19.288 &  19.369 &  19.225  & 19.343 &    0.081  &     0.118 &  B\\
 13.5835.395~~&  1575~~~~ & 19.262 &  19.290 &  19.177  & 19.250 &    0.028  &     0.073 &  B\\
 13.6078.524~~&  3054~~~~ & 19.084 &  19.089 &  18.962  & 19.066 &    0.005  &     0.104 &  B\\
  6.6931.817~~&  3061~~~~ & 19.983 &  19.679 &  19.724  & 19.631 & $-$0.304  &  $-$0.093 &  A\\
 13.5957.489~~&  3400~~~~ & 19.607 &  19.530 &  19.443  & 19.469 & $-$0.077  &     0.026 &  B\\
 13.6199.527~~&  3412~~~~ & 19.596 &  19.460 &  19.403  & 19.425 & $-$0.136  &     0.022 &  B\\
  6.7053.758~~&  3805~~~~ & 19.567 &  19.415 &  19.387  & 19.402 & $-$0.152  &     0.015 &  A\\
  6.6811.752~~&  3948~~~~ & 19.497 &  19.331 &  19.283  & 19.292 & $-$0.166  &     0.009 &  A\\
 13.5837.629~~&  4540~~~~ & 19.443 &  19.450 &  19.358  & 19.414 &    0.007  &     0.056 &  B\\
 13.5837.382~~&  4859~~~~ & 19.278 &  19.294 &  19.152  & 19.240 &    0.016  &     0.088 &  B\\
  6.6811.591~~&  4933~~~~ & 19.387 &  19.127 &  19.141  & 19.103 & $-$0.260  &  $-$0.038 &  A\\
 13.5837.566~~&  5902~~~~ & 19.169 &  19.165 &  19.047  & 19.121 & $-$0.004  &     0.074 &  B\\
 13.6079.125~~&  5952~~~~ & 18.444 &  18.463 &  18.392  & 18.459 &    0.019  &     0.067 &  B\\
% 13.5837.447~~&  6020~~~~ & 19.236 &  19.170 &  19.149  &   $-$  & $-$0.066  &    $-$~~  &  B\\
  6.7054.713~~&  6415~~~~ & 19.436 &  19.215 &  19.275  & 19.206 & $-$0.221  &  $-$0.069 &  A\\
 13.5838.576~~&  6440~~~~ & 19.133 &  19.280 &  19.038  & 19.247 &    0.147  &     0.213 &  B\\
% 13.6078.438  &  1907	& 19.299 &    $-$  &  19.186  &   $-$  &    $-$  &   $-$  &  b\\
% 13.6199.646  &  2379	& 19.576 &    $-$  &  19.298  &   $-$  &    $-$  &   $-$  &  b\\
%%\hline
%%\hline
%%\end{tabular}
%%\end{center}
%\label{t:tab1}
%%\end{table}
%
%
%%\noindent
%%\begin{table}[ht]
%%\scriptsize
%\vskip 4 cm
%%\begin{center}
%%\caption{Comparison with MACHO light curves - Kov\'acs - Continued}
%\vspace*{5mm}
%%\begin{tabular}{c c c c c c c c c}
%%\hline
%%\hline
%%  (1)   &      (2)  &   (3)  &   (4)   &   (5)  &  (6)   &   (7)  &  (8)  &  (9)\\
 13.6201.449~~&  7063~~~~ & 19.195 &  19.213 &  19.147  & 19.195 &  0.018    & 0.048 &  B\\
%  6.6812.1107 &  7211	& 19.718 &    $-$  &  19.449  &   $-$  &    $-$  &   $-$  &  a\\
  6.7054.582~~&  7477~~~~ & 19.408 &  19.249 &  19.148  & 19.183 &  $-$0.159 & 0.035 &  A\\
  6.7054.373~~&  7609~~~~ & 19.617 &  19.340 &  19.299  & 19.313 &  $-$0.277 & 0.014 &  A\\
 13.6080.435~~&  7620~~~~ & 19.108 &  19.133 &  19.014  & 19.079 &  0.025    & 0.065 &  B\\
% 13.5959.722  &  7699	& 19.718 &    $-$  &    $-$   &   $-$  &    $-$  &   $-$  &  b\\
%  6.6812.881  &  7734	& 19.551 &    $-$  &  19.348  &   $-$  &    $-$  &   $-$  &  a\\
 6.6933.1036~~&  8788~~~~ & 19.706 &  19.482 &  19.413  & 19.444 &  $-$0.224 & 0.031 &  A\\
  6.6933.953~~&  9154~~~~ & 19.780 &  19.569   & 19.534 & 19.522 &  $-$0.211 & $-$0.012 &  A\\
%  6.6934.1067 &  9245	& 19.603 &    $-$  &  19.320  &   $-$  &    $-$  &   $-$  &  a\\
 6.6813.1071~~& 10487~~~~ & 19.838 &  19.603 &  19.581  & 19.569 &  $-$0.235 & $-$0.012 &  A\\
 13.6078.615~~& 10692~~~~ & 19.568 &  19.574 &  19.463  & 19.548 &  0.006    & 0.085 	&  B\\
 13.6078.672~~& 10811~~~~ & 19.538 &  19.492 &  19.363  & 19.431 &  $-$0.046 & 0.068 	&  B\\
  6.6931.779~~& 10914~~~~ & 19.853 &  19.784 &  19.704  & 19.767 &  $-$0.069 & 0.063 	&  A\\
 13.5957.581~~& 14449~~~~ & 19.459 &  19.450 &  19.385  & 19.514 &  $-$0.009 & 0.129 	&  B\\
  6.6690.904~~& 15387~~~~ & 19.814 &  19.630 &  19.533  & 19.612 &  $-$0.184 & 0.079 	&  A\\
  6.6811.969~~& 16249~~~~ & 19.674 &  19.430 &  19.372  & 19.379 &  $-$0.244 & 0.007    &  A\\
 13.6080.645~~& 22917~~~~ & 19.439 &  19.462 &  19.321  & 19.427 &     0.023 & 0.106    &  B\\
 13.6080.584~~& 24089~~~~ & 19.450 &  19.370 &  19.324  & 19.365 &  $-$0.080 & 0.041    &  B\\
  6.7055.830~~& 26933~~~~ & 19.597 &  19.355 &  19.303  & 19.295 &  $-$0.242 & $-$0.008 &  A\\
%  6.6934.1055 & 28066	& 19.642 &    $-$  &  19.413  &   $-$  &    $-$  &   $-$  &  a\\
  6.7055.1045 & 28539~~~~ & 19.945 &  19.592 &  19.631  & 19.533 &  $-$0.353 & $-$0.098 &  A\\
\hline
\end{tabular}
\end{center}
%\small	

%Notes: Values in Columns 3 and 4 are magnitude-averaged mean magnitudes 
%from MACHO's web pages  and our values, respectively; values in Columns 5 and 6 are 
%intensity-averaged mean magnitudes from Kov\'acs and our photometry, respectively.\\
%$\Delta 1$=$<V_{this~paper}> - <V_{MACHO}>$ using the magnitude means
%in Columns 3 and 4: $<\Delta 1>$=$-$0.123 mag ($\sigma$=0.120, 42 stars).\\
%$\Delta 2$=$<V_{this~paper}> - <V_{Kovacs}>$ using the intensity means
%in Column 5 and 6: $<\Delta 2>$=0.043 mag ($\sigma$=0.059, 42 stars).
%({\bf MARCELLA qui per la 9154 e la 16249 che sono Blazhko ho adottato i 
%valori del file rr\_Acalpulito3.out
%per le intensity-averaged che vengono dalla curva media tra 1999 e 2001 credo, occorre
%calcolare da queste curve medie gli equivalenti valori magnitude-averaged, vanno quindi rifatte
%le differenze e calcolati i delta medi e relativi sigma.}.\\
%\label{t:tab1}
\normalsize
\end{table*}
%, this comparison is again based on MACHO web page informations as
%updated at the XXXXX {\it (sempre la stessa data di tutti gli altri confronti)}.  
The comparison between MACHO mean V magnitudes
% (directly drawn 
%from the on line variable star catalogue) 
and our 
%as updated at the 
%at October 15th, 2003, which 
%corresponds 
%to the last web page update of April 18th 2002), and our 
magnitude-averaged values (see column 11 of tables 5 and 6)
for variables in common with full coverage of the light curve and
without systematic shifts between the 1999 and the 2001 photometries
is shown in Figure~17, where  
filled and open symbols are used for variables in field A 
and B, respectively, and triangles mark the double mode RR Lyrae stars.
The average $V$ difference, present photometry minus MACHO, is $-0.170$ mag
($\sigma=0.106$, 66 stars) 
%({\bf Questi numeri e i relativi sigma e numero di stelle 
%vanno tutti ricontrollati e ricalcolati - MARCELLA perche' la differenza e calcolato solo su 58 stelle
%in comune?})
%PERCHE' LE ALTRE NON HANNO IL VALOR MEDIO NELLE TABELLONE, QUINDI CURVA DI LUCE NON COMPLETA
 in field A, and $-0.013$ mag ($\sigma=0.099$, 44 stars) in field B. 
While there is very good agreement for stars in field B, there is a large systematic shift for the
variables in field A, with MACHO web luminosities being  on average {\bf fainter} than ours
by 0.170 mag. 
%({\bf RICALCOLARE 	QUESTI SHIFTS AGGIUNGENDO LE VARIABILI
%RECUPERATE DA MACHO}) . 
%{\it Il pezzo che segue in italico e' stato messo pari pari nel C03, che si fa?
%lo togliamo? cambiamo le parole? 
%The 
%quite large systematic shift found for variable stars in field A, 
%with MACHO web luminosities being {\bf fainter} than ours, is 
%rather surprising 
%and of no obvious explanation, since the different treatment of the 
%background in DoPhot is expected to produce {\bf brighter} magnitudes 
%than DAOPHOT+ALLFRAME photometry (A99)}. 
%The agreement for the variable stars in common in field B is much more 
%satisfying.
% 
%%({\it Quello che segue e' il pezzo vecchio che e' gia' stato
%%pubblicato in C03
% Moreover, we notice that 
%(i) a shift by +0.174 mag of field A photometry would result in an average 
%luminosity of the RR Lyrae stars in this area about 
%0.24 mag fainter than A00 median
%luminosity of the RRab's in the LMC,  and (ii) given the good agreement 
%existing in field B, 
%would imply a 
%rather unlikely difference of about 0.27 mag in the average luminosity 
%of the RR Lyrae stars in the two fields.})
% 
%
We thus suspect that there may be calibration problems, namely disalignements
and photometric shift between different fields,  affecting the individual
average magnitudes published on MACHO web catalogue for the LMC variable stars. 
%As a confirmation of this suspect 
On the other hand we also note that Alcock et al. (2000: hereinafter 
A00) median luminosity of
a sub-sample of 680 RRab's in the LMC ($<V>$=19.45 mag)
%, which is based on different
%version of A99 photometric calibration ({\bf CHECK, bisogna vedere
%se questo e' vero, qui bisogna scoprire su quale calibrazione e' basata
%la fotometria usata in A00}) 
is in good agreement, within the respective error bars, with the average
luminosity of the RR Lyrae stars in the LMC drawn from the present
photometry (see discussion in Section 6 of C03 and their Table~5).

\subsubsection{Comparison with MACHO photometry for the variable stars: 
Kov\'acs subsample}

MACHO time series calibrated data for a subsample  
%From the comparison between our light's curves for a subsample 
of 42 variables in common with our photometry 
(39 RRab, 1 RRc, 1 AC and 1 eclipsing binary, according to our 
classification; 41 RRab and 1 RRL+GB according to MACHO) were 
kindly provided to us by Dr. G. Kov\'acs.

This photometry is based on A99 calibration. 
The comparison between mean magnitudes 
is shown in the top panel of Figure~18. Individual values are 
provided in Table~10,
where we list MACHO's web page magnitude-averaged values (Column 3), the
present paper magnitude-averaged values (Column 4), and the
intensity-averaged values  from our photometry and Kov\'acs dataset in
Column 5 and 6, respectively. Finally, in Columns 7 and 8 we list the
corresponding residuals this paper  minus MACHO web  ($\Delta$1), and this
paper minus Kov\'acs ($\Delta$2).  The agreement with  Kov\'acs dataset is
generally good and without  apparent offsets between field A and B.
The average difference  $\Delta 2$=$<V_{this~paper}> - <V_{Kovacs}>$ is 0.043
mag ($\sigma$=0.059, 42 stars), to compare with 
%({\bf CHECK QUESTI NUMERI})
 $<\Delta1>$=$<V_{this~paper}> - <V_{MACHO~web}>$ = $-$0.123 mag 
($\sigma$=0.120, 42 stars). %({\bf CHECK QUESTI NUMERI})
Our average magnitudes are generally {\bf fainter}  than Kov\'acs' as
expected on the basis of the different reduction procedures (see dicussion in Section 4.1). 
Figure~19 shows the
point-to-point comparison of the light curves of 4 {\it ab-}type RR Lyrae
stars (two per each of our fields) representing respectively the best  (left
panels) and the worst (right panels) comparison between the two samples. The
two variables shown in the right panels of the figure  are systematically
brighter in MACHO photometry.
% possibly due to 
%blending.

\noindent
\begin{table*}[ht]
%\vskip 4 cm
\begin{center}
\caption{Comparison with MACHO photometry for the variable stars in common: Alves 
subsample}
\scriptsize
%\vspace*{5mm}
\begin{tabular}{r r c c c c l l }
\hline
Id$_{MACHO}$~~~&Id$_{this~paper}$&$<V_{MACHO}>$&$<V_{this~paper}>$&
$<V_{Alves}>$&$<V_{this~paper}>$&Type$_{MACHO}$&Type$_{this~paper}$\\
               &                 &(mag-averaged)&(mag-averaged)   &
 (int-averaged)&(int-averaged)    &              &                  \\

\hline
%Field A
   6.6931.650~~&  2767~~~~ &  19.602 &  19.517  & 19.359 &  19.467  & ~~~ RRab  & ~~~~RRab \\ 
   6.6810.428~~&  3155~~~~ &  19.338 &  19.218  & 19.127 &  19.209  & ~~~ RRc	& ~~~~RRd  \\ 
 13.6691.4052~~&  4420~~~~ &  19.402 &  19.417  & 19.380 &  19.409  & ~~~~~ $-$ & ~~~~RRd  \\ 
   6.6811.736~~&  6426~~~~ &  19.421 &  19.227  & 19.101 &  19.185  & ~~~ RRab  & ~~~~RRab \\  
 13.7054.2970~~&  7137~~~~ &  19.150 &   $-$  & 19.309 &  19.413  & ~~~ RRd	& ~~~~RRd  \\ 
   6.7054.373~~&  7609~~~~ &  19.617 &  19.340  & 19.290 &  19.313  & ~~~ RRab  & ~~~~RRab \\  
   6.6933.939~~&  8654~~~~ &  19.440 &  19.275  & 19.233 &  19.269  & ~~~ RRe	& ~~~~RRd  \\ 
  6.6933.1036~~&  8788~~~~ &  19.706 &  19.482  & 19.402 &  19.444  & ~~~ RRab  & ~~~~RRab \\  
   6.6692.853~~& 10214~~~~ &  19.440 &  19.217  & 19.189 &  19.204  & ~~~ RRab  & ~~~~RRab \\ 
\hline
%Field B
 13.5835.395~~&  1575~~~~ &  19.262 &  19.290  & 19.163 &  19.250  & ~~~ RRab  & ~~~~RRab \\  
 13.6078.524~~&  3054~~~~ &  19.084 &    $-$  & 18.953 &  19.066  & ~~~ RRab  & ~~~~RRab \\  
 13.5836.525~~&  3347~~~~ &  19.145 &  19.211  & 19.120 &  19.204  & ~~~ RRc   & ~~~~RRd  \\ 
 13.6199.527~~&  3412~~~~ &  19.596 &  19.460  & 19.422 &  19.425  & ~~~ RRab  & ~~~~RRab \\  
 13.5958.518~~&  4509~~~~ &  19.210 &  19.468  & 19.336 &  19.462  & ~~~ RRd   & ~~~~RRd  \\ 
 13.5838.497~~&  6470~~~~ &  19.224 &  19.218  & 19.197 &  19.207  & ~~~ RRc   & ~~~~RRd  \\ 
 13.6080.591~~&  7467~~~~ &  19.076 &  19.055  & 19.040 &  19.043  & ~~~ RRc   & ~~~~RRd  \\ 
 13.6080.645~~& 22917~~~~ &  19.439 &  19.462  & 19.327 &  19.427  & ~~~ RRab  & ~~~~RRab \\  
 13.6080.584~~& 24089~~~~ &  19.450 &  19.370  & 19.323 &  19.365  & ~~~ RRab  & ~~~~RRab \\  
\hline
\end{tabular}
\end{center}
%
%\scriptsize
% (1) MACHO ID\\								     
% (2) Our ID\\							     
% (3)  

Notes: 
%Values in Columns 3 and 4 are  magnitude-averaged mean magnitudes 
%from MACHO's web pages and our photometry, respectively; values in Column 5 and 6 
%are  
%intensity-averaged mean magnitudes  from MACHO's and our photometry, respectively.\\
Stars \# 7137 and \# 4509 do not appear in the MACHO on-line catalogue,
values in Column 3 for these stars are taken from A97 (see their Table 1).
%
%\label{t:tab1}
\normalsize
\end{table*}

\subsubsection{Comparison with MACHO photometry for the variable stars in 
common: Clement subsample}
%
%tanto per cambiare, per alcune stelle, i valori pubblicati
%nell'astro-ph/0310281 sono diversi da quelli che ci aveva dato la
%Clement...
%Qui faccio riferimento al loro articolo, e non ai dati che avevamo noi.
%

%Likewise, we compared the magnitudes of 7 RRc stars, M5-like variables (see
Alcock et al. (2004) discuss the properties of 330
first-overtone M5-like RR Lyrae variables contained in 16 LMC MACHO fields
including fields \#6 and \# 13.
These restricted sample includes MACHO ``best-fit" {\it c-}type RR Lyrae 
with $-0.56 < \log P < -0.4$, amplitudes $A_V> 0.3$ and amplitude
ratios in the range $ 0.75 < A_R/A_V < 0.85 $ (C. Clement private 
communication, Alcock et al. 2004).
Photometry of these stars is
based on version 9903018 of A99 calibration (Alcock et al. 2004).
% (Clement 2003, private 
%comunication). 
Seven of these RRc's are in our sample:
%These period and 
%amplitude cuts
%were made so that we would not include any RR2 variables or any 
%longer period variables that have anomalous light curves. We restricted
%the amplitude ratio so that we would avoid stars with incomplete
%phase coverage or with faint unresolved companions. 
%which includes a total of 333 stars contained in 16 LMC MACHO fields
%including fields \#6 and 13 (Clement, private comunication)
%and whose magnitudes were transformed to the Kron-Cousins V and
%R system using the equations derived by Alcock et al (1999, PASP 111,
%1539). In a MACHO project internal report, Dave Alves designated
%it as and the calibration version 9903018.
we find that 
MACHO's mean magnitudes are on average 0.07 mag {\bf brighter} than ours
(see Table~12 by Alcock et al. 2004), again 
as expected on the basis of the different reduction procedures.
This shift is totally consistent with that found from the larger sample of
newly calibrated MACHO light curves provided us by D. Alves (see following 
Section 4.2.4), 
but at odds with the results from the comparison with the MACHO web values. 
%The corresponding comparison table has been published by Alcock et al. (2004; 
%see their Table 12).
We explicitely notice that this is indeed a small sample, since it was 
selected as described above, but as discussed in Section 4.2.1, and
contrary to what stated by Alves (2004), we have
a much larger number of variable stars in common with MACHO database.
%{\it  Quando di si parla di questo confronto va corretta l'affermazione
%   di Alves sulla review fatta a Sidney e apparsa come astro-ph0310673.
%   NON ABBIAMO solo 7 stelle in comune, quelle sono solo le 7 c in comune
%  con le C mandateci dalla Clement che ha estratto solo il sottocampione 
%   citato nel mail di cclement su astbo3 del 23 Aprile 2003.}

%
%\newpage

\subsubsection{Comparison with MACHO photometry for the variable stars in common:
Alves subsample}

Dr. D. Alves kindly made available to us time series data for
a subsample of 18 RR Lyrae variables in common with our database (9 for 
each field, 10 RRab and 8 RRd 
according to our classification; 10 RRab, 4 RRc, 2 RRd, 1 RRe and 1 variable of 
unknown type according to MACHO, but classified RRd by A97) along with photometry for the non-variable 
stars falling in $\sim 4^{\prime} \times 4^{\prime}$ patches 
 surrounding these RR Lyrae stars. 
These photometric data are calibrated 
according to A99 and Alcock et al. (2004) calibrations (Alves 2004, private
communication).
 Counteridentifications and average magnitudes of these 18 stars
 are given in Table~11. 
The comparison between    
intensity-averaged magnitudes is generally good, (see 
columns 5 and 6 of Table~11 and bottom panel of Figure~18), with Alves values being 
0.061 mag {\bf brighter} ($\sigma= 0.042$, 18 stars) than ours and  without
significant differences between field A and B.
The corresponding comparison using the magnitude-averaged
luminosities of these RR Lyrae stars available on MACHO
web pages leads to a different result: MACHO web values 
are on average 0.067 mag
{\bf fainter} than ours (see Columns 3 and 4 of Table~11). 

\subsubsection{Comparison with MACHO photometry for the non-variable stars: 
Alves subsample}

\begin{figure} 
\begin{center}
\includegraphics[width=8.8cm]{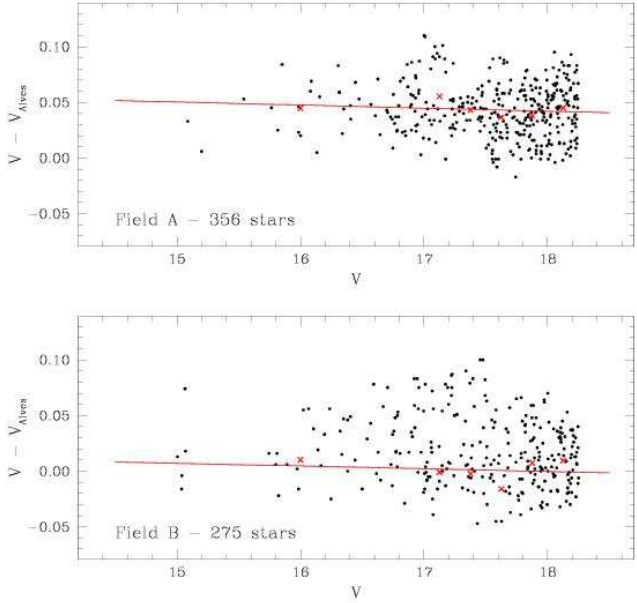}
\caption[]{Comparison with MACHO photometry (Alves subsample) for 
non-variable stars brighter than 18.25 mag. Residuals are this paper
minus MACHO. Lines indicate the linear fits of the average residuals of all
bins.
% $V_{this~paper} - V_{Alves}$ 
%per magnitude bin with a 2 $\sigma$ rejection 
%for stars 
%({\bf Nelle x e y labels togliere "this paper" - MARCELLA}) 
}
%\label{f:fig5b}
\end{center}
\end{figure}
\begin{figure} 
\begin{center}
\includegraphics[width=8.8cm]{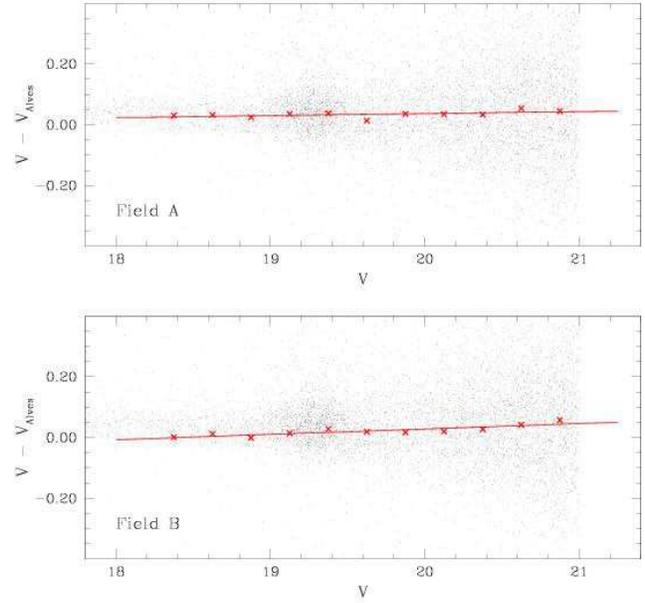}
\caption[]{Same as Figure 20 for non-variable with $18.25 <V< 21$ mag.
%({\bf Nelle x e y labels togliere "this paper" - MARCELLA}) 
%Comparison with MACHO photometry (Alves subsample) for 
%non-variable stars 
%fainter than 18.25 mag.
%({\bf Nelle x e y labels MACHO e our vanno come pedici senza le parentesi e 
%invece di "our" metter "this paper", fare i punti in modo che si 
%vedano meglio e le scritte un po' piu' grandi})
}
%\label{f:fig5b}
\end{center}
\end{figure}
%

%\newpage
%
%tabella 1
%
%\noindent
\begin{table}[ht]
%\vskip 4 cm
\begin{center}
\caption{Comparison with MACHO photometry for the non-variable 
stars with $V < 18.25$, from Alves subsample}
\scriptsize
%\vspace*{5mm}
\begin{tabular}{c c c c}
\hline
    Bin  &  $<\Delta V>$ &  $\sigma$ & N\\
\hline
%Field A
\multicolumn{4}{c}{Field A}\\
 15.00 $-$ 17.00 &  0.045  &  0.021  &   64\\ 
 17.00 $-$ 17.25 &  0.055  &  0.031  &   34\\ 
 17.25 $-$ 17.50 &  0.043  &  0.011  &   31\\ 
 17.50 $-$ 17.75 &  0.037  &  0.027  &   58\\ 
 17.75 $-$ 18.00 &  0.038  &  0.021  &   81\\ 
 18.00 $-$ 18.25 &  0.045  &  0.025  &   88\\ 
\hline
%Field B
\multicolumn{4}{c}{Field B}\\
 15.00 $-$ 17.00 &  0.026  &  0.029 &  61\\ 
 17.00 $-$ 17.25 &  0.022  &  0.037 &  36\\ 
 17.25 $-$ 17.50 &  0.027  &  0.041 &  29\\ 
 17.50 $-$ 17.75 &  0.011  &  0.031 &  30\\ 
 17.75 $-$ 18.00 &  0.012  &  0.029 &  57\\ 
 18.00 $-$ 18.25 &  0.010  &  0.022 &  62\\ 
\hline
\end{tabular}
\end{center}

Notes: $<\Delta V>$ = $<V_{this~paper}-V_{Alves}>$ 
%\label{t:tab1}
\end{table}
%
%tabella 2
%
\noindent
\begin{table}[ht]
%\vskip 4 cm
\begin{center}
\caption{Comparison with MACHO photometry 
for the non-variable 
stars with $18.25<V<21$, from Alves subsample}
\scriptsize
%\vspace*{5mm}
\begin{tabular}{c c r c}
\hline
Bin &$<\Delta V>$  &  $\sigma$ & N\\
\hline
%Field A
\multicolumn{4}{c}{Field A}\\
18.25 $-$ 18.50  &  0.030  &  0.027  &  123\\ 
18.50 $-$ 18.75  &  0.032  &  0.030  &  133\\ 
18.75 $-$ 19.00  &  0.024  &  0.031  &  188\\ 
19.00 $-$ 19.25  &  0.035  &  0.037  &  502\\ 
19.25 $-$ 19.50  &  0.037  &  0.042  &  673\\ 
19.50 $-$ 19.75  &  0.013  &  0.049  &  329\\ 
19.75 $-$ 20.00  &  0.035  &  0.063  &  327\\ 
20.00 $-$ 20.25  &  0.035  &  0.065  &  445\\ 
20.25 $-$ 20.50  &  0.033  &  0.098  &  603\\ 
20.50 $-$ 20.75  &  0.054  &  0.110  &  753\\ 
20.75 $-$ 21.00  &  0.045  &  0.129  &  826\\ 
\hline	 					      
%Field B
\multicolumn{4}{c}{Field B}\\
18.25 $-$ 18.50 &  0.001  &  0.030  &	 98    \\ 
18.50 $-$ 18.75 &  0.012  &  0.034  &	115    \\ 
18.75 $-$ 19.00 &$-0.001$ &  0.032  &   164   \\ 
19.00 $-$ 19.25 &  0.014  &  0.040  &	438   \\ 
19.25 $-$ 19.50 &  0.028  &  0.055  &	426   \\ 
19.50 $-$ 19.75 &  0.019  &  0.060  &	268   \\ 
19.75 $-$ 20.00 &  0.017  &  0.057  &	280   \\ 
20.00 $-$ 20.25 &  0.020  &  0.074  &	373   \\ 
20.25 $-$ 20.50 &  0.027  &  0.087  &	466   \\ 
20.50 $-$ 20.75 &  0.041  &  0.119  &	647   \\ 
20.75 $-$ 21.00 &  0.057  &  0.116  &	694   \\ 
\hline
\end{tabular}
\end{center}

Notes: $<\Delta V>$ = $<V_{this~paper}-V_{Alves}>$ 

%\label{t:tab1}
\end{table}
%
%tabella 3
%
%
%({\it Quanto segue va descritto meglio seguendo lo schema fatto nel pezzo 
%sul confronto con OGLE II}).
The non-variable stars in common 
%between our fields and patches of Fields \#6 and \#13 in the MACHO template,
%provided us by D. Alves 
were counteridentified by coordinates.
They correspond to a total number of 18996 stars 
(10467 in field A, and 8529 in field B, respectively).

%
%provided by D. Alves.
Comparison between the two photometries was done dividing 
the stars into a bright and a faint sample corresponding respectively 
to objects with 
$V < 18.25$ mag (356 stars in field A, and 275 in field B) and objects with 
%(470XXX stars in field A and 351XXX in field B) and  
$18.25<V<21$ mag (4902 stars in field A, and 3969 in field B). 
Within each subsample stars where further divided into
magnitude bins 0.25 mag wide. 
Average residuals were computed adopting a $\sigma$ rejection
procedure that discarded objects deviating more than 
2 $\sigma$ from the average in the bin.
%We compared our  photometry of the non-variable stars with MACHO data,
%using stars with magnitude brighter than 18.25 mag,   These
%data were kindly made available to us by D. Alves.  The comparison in
%field A was initially done on the satisfying these conditions.
%
%({\it Quanto segue va descritto meglio 
%decendo che le stelle sono state divise in bins in magnitudine coe mostrato
%dalle successive tabelle numero XXX e XXXX, abbiamo calcolato la differenza 
%media in ogni
%bin e fatta la regressione tra questi valori medi})>
%
In Table~12  we list the mean differences $V_{this~paper}$
$- V_{Alves}$ of the stars in the bright subsample (for objects in field A and
B separately), 
with their respective $\sigma$ and number of stars per magnitude bin.
Transformation equations between the two photometries  were then 
computed as the linear
fit of the average residuals of all bins. They are:
$$V_{this~paper}-V_{Alves} = -0.0028\times V_{this~paper} + 0.0916$$ 
in field A (356 objects) and:
$$V_{this~paper}-V_{Alves}= -0.0025\times V_{this~paper} + 0.0438$$
in field B (275 stars).
These linear fits are shown in Figure~20. 
%The transformation
%relations between MACHO and our photometry are then:
%$$V_{this~paper}=0.9972\times V_{Alves}+0.0913$$
%in field A, and:
%$$V_{this~paper}=0.9975\times V_{Alves}+0.0437$$

The same comparison done on the stars with $18.25<V<21$ mag
is 
provided in Table~13 and shown in Figure~21. 
The transformation equations in this magnitude range using 
a linear fit with a 2 $\sigma$ rejection are:
$$V_{this~paper}-V_{Alves} = 0.0067\times V_{this~paper} - 0.0953$$ 
in field A (4902 objects) and:
$$V_{this~paper}-V_{Alves}= 0.0177\times V_{this~paper} - 0.3251$$
in field B (3969 stars).

\subsection{Comparison with OGLE II photometry}
%{\it Qui va alleggerito tutto il discorso che segue e poi va controllato se
%Alcock et al. 2004 fanno il confronto con OGLE o no}
%\medskip
%\medskip
%
%{\it So far comparisons between OGLE II photometry and other literature data were 
%either based i) on bright variables stars as in  Udalski et al. (1998, Acta
%Astron. 48,1) who compared their LMC photometry  of Cepheids in NGC 1850 with
%that of Sebo and Wood (1995),  
%or in  Gronewegen (2000, astro-ph/0010298) who compared the  I photometry of
%LMC Cepheids in common between DENIS and OGLE;   or ii) on small numbers of
%objects  as in Udalski et al. (2000, Acta Astron. 50, 307 - astroph-0010151)
%where 6 objects in common were used to compare OGLEII and
%Walker's (1993) photometry of NGC 1835);  or iii) on average values drawn from 
%small samples of RR Lyrae stars (as in  Udalski et al. (2000, Acta Astron. 50,
%307 astroph-0010151)  that compared their average luminosity of the  RR Lyraes
%in field LMC\_SC21 to values we published in a preliminary version of the
%present paper.}
%\medskip 

%The release of the OGLE II photometric maps for the LMC (Udalski et al. 2000,
%astro-ph/0010150) and 
The partial overlap of our field A with 
OGLE II field LMC\_SC21 (Udalski et al. 2000) gave us the possibility to make a detailed 
comparison between 
the two photometries based on 
a large number of stars covering a wide magnitude  range.
We have retrieved from the OGLE archive
\footnote{\tiny ftp://sirius.astrouw.edu.pl/ogle/ogle2/maps/lmc%
\normalsize} the photometric data corresponding to field LMC\_SC21. %
%\footnote{\tiny The limits coordinates inside this file are (RA;DEC-RA;DEC)
%5:19:50.46 ; $-$71:01:31.4 $-$ 5:22:38.55 ;$-$70:05:03.5.%
%\normalsize} 
%(sirius.astrouw.edu.pl/\~ogle/ogle2/maps/lmc) the photometric data 
%corresponding to
%field LMC\_SC21. 
The overlapping region 
%between our  field A and OGLE II field LMC\_SC21
corresponds to 42.25\% and 9.84\% of our and OGLE II fields, respectively. This
region is located at roughly 5:21:29.9 $<\alpha<$ 5:22:38.6 
and $-$70:41:00.9 $<\delta<-$ 70:27:18.4, corresponding
to 1218.95 $<X<$ 2047.44 and 2976.78 $<Y<$ 4967.58 in OGLE II coordinate
system. Inside this area OGLE II has $B$,$V$,$I$ photometry for 
15524, 17067 and 17582 stars, respectively, 
to compare with our 21524 objects. 
Our limiting magnitude is about 1.5 mag fainter
and we resolved about 39, 26, and 22\% more stars (in $B$, $V$ and
$I$, respectively) than OGLE II. Coordinates were aligned to 
%rototranslated
%on 
OGLE II coordinate system and stars in common were counteridentified.
Over the total sample of 14734 common stars there
are 13688, 14483 and 14734 objects with $B$, $V$ and $I$
magnitude in the ranges 12.5$-$22.7, 12.6$-$23.1 and 12.3$-$21.6, 
respectively.
 Among these objects OGLE II reports 39 variable stars\footnote{\tiny from files:\\
ftp://sirius.astrouw.edu.pl/ogle/ogle2/var\_stars/lmc/rrlyr/lmc\_sc21/lmc\_sc21.tab
ftp://sirius.astrouw.edu.pl/ogle/ogle2/var\_stars/lmc/rrlyr/dmod.tab
ftp://sirius.astrouw.edu.pl/ogle/ogle2/var\_stars/lmc/rrlyr/other.tab
ftp://sirius.astrouw.edu.pl/ogle/ogle2/var\_stars/lmc/cep\\
/catalog/lmc\_sc21/lmc\_sc21.tab\\
ftp://bulge.princeton.edu/ogle/ogle2/var\_stars/lmc/cep/dmcep/tab2.txt
ftp://sirius.astrouw.edu.pl/ogle/ogle2/var\_stars/lmc/ecl/lmc\_sc21 %
\normalsize}. We recovered all of them.
Counteridentifications are provided in Table~14
along with average luminosities and classification in types in the two
photometries. There is general agreement in the type classification 
and in the derived periods that,
on average, agree within 2-3 decimal digits. %({\bf MARCELLA E' GIUSTO})
OGLE II classification does not match ours for 4 variable stars, namely 
the new candidate RRd, 2 candidate Anomalous Cepheids
and star \# 5148 that we classify as RRab while is classified 
RRc by OGLE II. A further object, 
star Id$_{SC\_21}$ =116626 is 
classified by OGLE II as CepFA; however, OGLE II light curves 
for this star are rather poor and the corresponding object in our 
photometry  (\# 22592) was not found to vary.  Finally, 
we have three additional variables in the area in 
 common that 
were apparently missed by OGLE II: an RRc, a binary system, and a
$\delta$ Scuti star, which are listed at the bottom of Table~14.
For 3 variables (namely stars \# 9604, 10320, and 25510) there is
a large discrepancy between OGLE II and our $V$ average magnitudes.
Two of these stars (\# 9604 and 10320) were discussed in Section 3.2.
Similarly to them, star \# 25510 has a very poor $V$ light curve
in OGLE II photometry and an average $V$ magnitude 0.62 mag fainter than ours, leading 
to unrealistic $<B-V>$=$-$0.11 and $<V-I>$=1.04 colours for an
RR Lyrae star. We suspect that these 3 stars may have been
wrongly counteridentified in the various photomeric bands.
 Figure~22 shows the point-to-point comparison of the $V$ light curves for 3
{\it ab-}type
RR Lyrae stars and one Cepheid representing respectively the best 
agreement (left panels) and the worst (right panels) comparison
between the two photometries (excluding the 3 above mentioned  discrepant 
stars).
Large discrepancies are also found among the $B$  magnitudes of
stars \# 4313 and 8723, that,  in the case of the first object, lead in 
OGLE II photometry to
a colour $<B-V>$=0.85 mag rather red for an RR Lyrae star. 
%We explicitly recall that the highest
%disagreement is indeed found for the $V$ photomerty of 
%stars \# 9604, 10320, and 25510.
\begin{figure*} 
\begin{center}
\includegraphics[width=15.0cm]{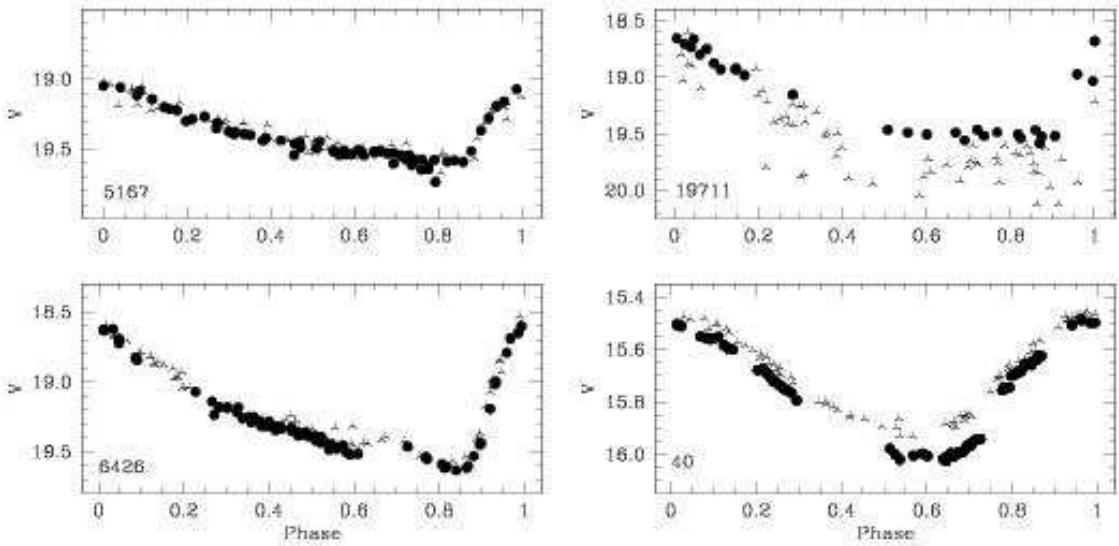}
\caption[]{Point-to-point comparison of the $V$ light curves for 3
{\it ab-}type 
RR Lyrae stars and a Classical Cepheid (lower right panel) in common with OGLE II. Filled dots: our photometry,
three arms crosses: OGLE II photometry. As in Figure 19, these represent 
the best (left) and worst (right) cases.}
%\label{f:fig5b}
\end{center}
\end{figure*}

\begin{figure} 
\includegraphics[width=8.8cm]{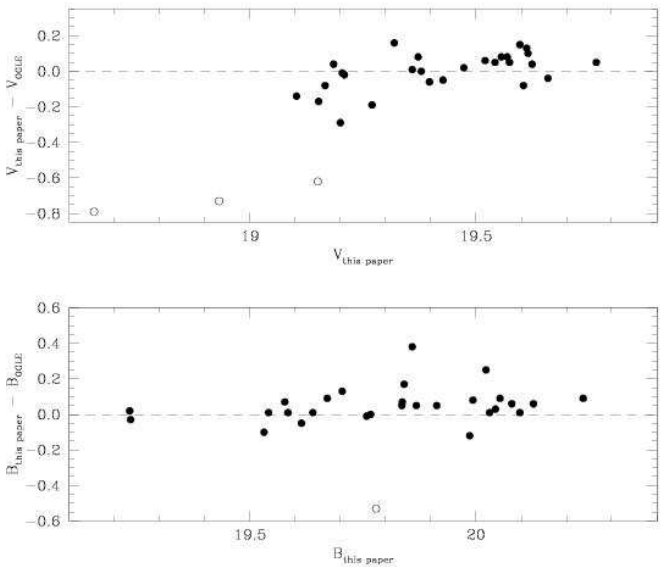}
\caption[]{Comparison between our and Ogle II 
mean $V$ and $B$ magnitudes for the variable stars in common.
Residuals are: this paper $-$ OGLE II. 
Open symbols are used for the most deviating stars (see text). 
%({\bf AGGIUNGERE IN QUESTA FIGURA TUTTE LE VARIABILI RECUPERATE DAL
%CONFRONTO CON MACHO})
%({\bf Questa figura va rifatta aggiundengo le stelle 28246, 28539 ed 
%eventualmente anche la 4313 e la 21007})
}
%%\label{f:fig1a}
\end{figure}

The comparison between our and OGLE II mean $V$,$B$ magnitudes 
for variable stars in common with 
complete light curves and no systematic shifts between our 1999 and 2001
photometry is shown in Figure~23. Average differences are 
$\Delta V$=0.01 mag ($\sigma=0.11$, 30 stars discarding stars
\# 9604, 10320 and 25510, open circles in Figure~23)
%({\bf PERCHE' SOLO 32 E NON 36, I.E. 39-3, LE ALTRE 4 CHE MANCANO
%HANNO CURVA DI LUCE NON COMPLETA?}) 
%({\bf CHECK QUESTI NUMERI MI SEMBRA MOLTO
%PICCOLA LA DIFFERENZA E PERCHE' SOLO 22 STELLE?
%QUI VA RIFATTO IL CONFRONTO E SI POSSONO USARE 33 STELLE})
and $\Delta B$=0.04 mag ($\sigma=0.15$, 29 stars, discarding also 
star \# 19711 that does not have $B$ magnitude in OGLE-II), respectively.
These average differences do not
change restricting the comparison only to the RR Lyrae stars.
Our photometry is on average slightly fainter than OGLE-II, 
again as expected on the
basis of the different reduction procedures used in the
two photometries (see Section 4.1).
The average $V$ magnitude of the RR Lyrae stars in common 
using objects with reliable photometry in both 
datasets is $<V_{RR}>$=19.444 mag ($\sigma=0.181$, 24 stars)
and  $<V_{RR}>$=19.427 mag ($\sigma=0.160$, 24 stars) in our and
OGLE-II photometry, respectively. These values are in good agreement with 
each other and with the average $V$ luminosity of our full sample
of RR Lyrae stars in field A (see end of Section 3.1 and C03),   
but about 0.06-0.08 mag fainter than 
the average $V$ magnitude from the total sample of OGLE II
LMC RRab's: $<V_{RR}>$=19.36$\pm$0.03 mag (and $<V_{RR}>$=19.31$\pm$0.021 
mag for the RRc's) by Soszy\'nsky et al. (2003).
Given the small sample of variable stars in common this systematic
shift might appear not very statistically significant, however it is
fully confirmed  by the comparison done on the much larger number of 
non variable stars at the
same magnitude level (see remaining part of this section and
Table~15).
%Restricting the comparison only to the RR Lyrae stars we find:
%$\Delta V=0.01$ mag ($\sigma=0.10$, 26 stars, discarding \# 25510), 
%and $\Delta B=0.03$ mag ($\sigma=0.14$, 26 stars). This difference
%found for the RR Lyrae stars is totally consistent
%({\bf IN REALTA' QUI LO SHIFT SEMBRA ESSERE LA
%META' MA VEDI MAIL DI MARCELLA, IN PROPOSITO}) with the
%systematic shift found for the non variable stars at the
%same magnitude level (see remaining part of this section and
%Table~15).  
%({\bf Calcoliamo e aggiungiamo qui qual'e' la differenza media e 
%il relativo sigma tra la 
%magnitudini medie V e B delle RR Lyrae tra noi e OGLE e su quante stelle
%sono calcolate queste differenze, ovviamente usando solo RR Lyrae con
%curve di luce complete. MARCELLA})
%Soszy\'nsky et al. (2003) derive $<V>$=19.36$\pm$0.03 mag and 
%19.31$\pm$0.021 mag from the total of OGLE II
%LMC RRab's and RRc's, respectively. The agreement is XXXXXX
%({\bf Qui si puo' aggiungere il confronto con le luminosita' medie 
%di questo sottocampione di RR, epoi confrontarsi con i nostri
%valori medi generali sui due campi e rimandare al C03 e alla sua
%tabella con i confronti tra le varie fotometrie}).

$B$, $V$ and $I$ residuals between our and OGLE II
photometry for the non variable stars in common are shown in Figure~24, 
while in Figure~25   
we plot the corresponding CMDs  (left panels: present paper; 
right panels: OGLE II photometry).
Our $B$,$V$ photometry is generally more accurate and deeper than
OGLE's. Objects falling off the main ridge lines of 
OGLE II~ $V,B-V$  CMD for $V> 20.0$ and $(B-V)<$0.2
%while are off  thus they 
are likely wrong measurements in OGLE II photometry  (e.g., blends, wrong
identifications, and wrong counteridentifications between $V$ and $B$)
since they fall very well on the main branches of our 
diagram.
In the $I$ band our photometry appears to be more 
uncertain.
However,
the objects that deviate most in our $I$ photometry 
($I> 20.0$ and $V-I <$ 0.0) 
have magnitudes generally
well below the magnitude level of RR Lyrae and clump stars, that 
are the luminosity levels we are mainly interested in.
%
%\begin{figure}
%%\includegraphics[%
%  clip,
%  scale=0.5]{last_fig.ps}
%\caption{The variable stars inside the common area. Filled  circles OGLE II stars
%open circles our variables.}
%\end{figure}

%
\begin{landscape}
\begin{table}
\caption{Variable stars inside the area in common with OGLE II field
LMC\_SC21. }
%({\bf Completare questa tabella con le informazioni sulle stelle aggiunte.})
%. Cep$=$ Cepheid, RR$=$RR Lyrae, Bin$=$ Binary, $\delta$S=Delta Scuti
%}
\begin{center}
\scriptsize
\begin{tabular}{clrcccclcccrr}
\hline
%\multicolumn{2}{c }{OGLE}& \multicolumn{8}{c }{ LMC\_SC21}&   \multicolumn{7}{c}{This work} \\ \hline
\multicolumn{1}{c}{Name}&\multicolumn{1}{c}{Type}&
\multicolumn{1}{c}{Id}&\multicolumn{1}{c}{$<V>$}&\multicolumn{1}{c}{$<B>$}&\multicolumn{1}{c}{$<I>$}&\multicolumn{1}{c}{Id}&\multicolumn{1}{c}{Type}&\multicolumn{1}{c}{$<V>$}&\multicolumn{1}{c}{$<B>$}&\multicolumn{1}{c}{$<I>$}&$\Delta V$&$\Delta B$  \\   
\multicolumn{1}{c}{$_{OGLE}$}&\multicolumn{1}{c}{$_{OGLE}$}&\multicolumn{1}{c}{$_{SC\_21}$}&\multicolumn{1}{c}{$_{OGLE}$}&\multicolumn{1}{c}{$_{OGLE}$}&\multicolumn{1}{c}{$_{OGLE}$}&\multicolumn{1}{c}{$_{this~paper}$}&\multicolumn{1}{c}{$_{this~paper}$}&\multicolumn{1}{c}{$_{this~paper}$}&\multicolumn{1}{c}{$_{this~paper}$}&\multicolumn{1}{c}{$_{this~paper}$}  & &\\   
\hline
OGLE052133.45-703951.6 & RRc   & 111870 & 19.29  & 19.57   & 18.90  & ~2249 & ~~~RRdm	 & 19.372 &  19.704 & 18.878 &   0.08 &  0.13 \\
OGLE052130.54-703711.5 & RRab  & 112191 & 19.48  & 20.01   & 18.86  & 15387 & ~~~RRab	 & 19.612 &  20.043 &	$-$  &   0.13 &  0.03 \\
OGLE052131.78-703646.5 & RRc   & 114344 & 19.48  & 19.77   & 18.99  & ~4388 & ~~~RRc	 & 19.427 &  19.758 &	$-$  & $-$0.05 & $-$0.01 \\
OGLE052148.39-703026.1 & CepFU & 116226 & 16.12  & 16.67   & 15.46  & ~~183 & ~~~Cep	 & 16.259 &  16.830 & 15.566 &   0.14 &  0.16 \\
OGLE052134.12-703024.8 & RRab  & 116880 & 19.72  & 20.15   & 19.02  & 25301 & ~~~RRab	 & 19.766 &  20.237 &	$-$  &   0.05 &  0.09 \\
OGLE052133.57-703157.5 & RRdm  & 117174 & 19.45  & 19.91   & 18.99  & 23032 & ~~~RRdm	 & 19.597 &  19.993 &	$-$  &   0.15 &  0.08 \\
OGLE052148.90-702801.2 & RRc   & 119123 & 19.44  & 19.27   & 17.91  & 10320 & ~~~AC      & 18.655 &  19.236 &	$-$  & $-$0.79 & $-$0.03 \\
OGLE052154.07-702917.8 & RRab  & 119283 & 19.58  & 20.09   & 18.92  & 26821 & ~~~RRab	 & 19.624 &  20.097 & 19.097 &   0.04 &  0.01 \\
OGLE052146.25-702813.5 & RRab  & 119434 & 19.46  & 19.96   & 18.76  & 28293 & ~~~RRab	 & 19.520 &  20.053 &	$-$  &   0.06 &  0.09 \\
OGLE052131.26-702812.2 & RRab  & 119439 & 19.21  & 19.63   & 18.59  & 10214 & ~~~RRab	 & 19.204 &  19.639 &	$-$  & $-$0.01 &  0.01 \\
OGLE052152.61-702929.0 & RRab  & 119754 & 19.45  & 19.86   & 18.93  & 26525 & ~~~RRab	 & 19.473 &  19.913 &	$-$  &   0.02 &  0.05 \\
OGLE052212.39-704010.0 & CepFO & 159587 & 15.69  & 16.18   & 15.07  & ~~~40 & ~~~Cep	 & 15.753 &  16.237 & 15.212 &   0.06 &  0.06 \\
OGLE052235.42-703828.6 & RRdm  & 160118 & 19.23  & 19.51   & 18.68  & ~3155 & ~~~RRdm	 & 19.209 &  19.577 & 18.792 & $-$0.02 &  0.07 \\
OGLE052157.21-703826.1 & RRe   & 160121 & 19.22  & 19.54   & 18.68  & ~3216 & ~~~RRc	 &  $-$   &    $-$  &	$-$  &    $-$ &   $-$ \\
OGLE052222.37-704000.1 & RRe   & 160409 & 19.70  & 20.11   & 19.32  & ~2119 & ~~~RRc	 & 19.659 &  19.986 & 19.407 & $-$0.04 & $-$0.12 \\
OGLE052216.65-703950.4 & RRc   & 160428 & 19.48  & 19.79   & 19.06  & ~2223 & ~~~RRc	 & 19.556 &  19.836 & 19.136 &   0.08 &  0.05 \\
OGLE052230.17-703553.8 & RRab  & 162785 & 19.24  & 19.63   & 18.60  & ~4933 & ~~~RRab	 & 19.103 &  19.531 & 18.542 & $-$0.14 & $-$0.10 \\
OGLE052159.82-703535.3 & RRab  & 162827 & 19.35  & 19.77   & 18.68  & ~5167 & ~~~RRab	 & 19.359 &  19.837 & 18.808 &   0.01 &  0.07 \\
OGLE052221.06-703534.2 & RRc   & 162831 & 19.02  & 19.41   & 18.40  & ~5148 & ~~~RRab	 &   $-$  &    $-$  &	$-$  &    $-$ &   $-$ \\
OGLE052209.71-703502.8 & RRab  & 162907 & 19.52  & 20.02   & 18.86  & ~5589 & ~~~RRab	 & 19.574 &  20.079 & 18.942 &   0.05 &  0.06 \\
OGLE052232.64-703348.9 & RRab  & 163060 & 19.15  & 19.57   & 18.57  & ~6426 & ~~~RRab	 & 19.185 &  19.584 & 18.555 &   0.04 &  0.01 \\
OGLE052208.39-703631.3 & RRab  & 163171 & 19.38  & 19.67   & 18.77  & 16249 & ~~~RRab	 & 19.379 &  19.671 &	$-$  &   0.00 &  0.09 \\
OGLE052238.32-703402.1 & RRab  & 163532 & 19.49  &  $-$    & 18.72  & 19711 & ~~~RRab	 & 19.200 &  19.535 & 18.607 & $-$0.29 &   $-$ \\
OGLE052210.67-703315.2 & CepFU & 165209 & 16.23  & 16.83   & 15.48  & ~~150 & ~~~Cep	 & 16.251 &  16.837 & 15.524 &   0.02 &  0.01 \\
OGLE052221.29-703244.2 & RRab  & 165596 & 19.37  & 19.84   & 18.83  & ~7211 & ~~~RRab	 &   $-$  &    $-$  &	$-$  &    $-$ &   $-$ \\
OGLE052230.18-703220.8 & RRab  & 165650 & 19.52  & 20.07   & 18.83  & ~7468 & ~~~RRab	 & 19.615 &  20.127 & 18.850 &   0.10 &  0.06 \\
OGLE052207.98-703200.1 & RRab  & 165710 & 19.43  & 20.06   & 18.79  & ~7734 & ~~~RRab	 &   $-$  &    $-$  &	$-$  &    $-$ &   $-$ \\
OGLE052226.55-703019.3 & RRc   & 165913 & 19.46  & 19.77   & 18.89  & ~8812 & ~~~RRc	 & 19.397 &  19.767 & 18.821 & $-$0.06 & $-$0.00 \\
OGLE052213.53-703011.8 & RRab  & 165930 & 19.77  & 19.66   & 18.73  & 25510 & ~~~RRab	 & 19.150 &  19.614 & 18.554 & $-$0.62 & $-$0.05 \\
OGLE052229.04-703036.1 & RRc   & 166393 & 19.49  & 19.82   & 19.01  & ~8622 & ~~~RRc	 & 19.542 &  19.868 & 19.103 &   0.05 &  0.05 \\
OGLE052207.18-702907.7 & RRab  & 168608 & 19.66  & 19.21   & 18.45  & ~9604 & ~~~AC      & 18.932 &  19.234 & 18.550 & $-$0.73 &  0.02 \\
OGLE052214.15-702835.3 & RRc   & 168696 & 19.25  & 19.53   & 18.71  & 27697 & ~~~RRc	 & 19.166 &  19.541 &	$-$  & $-$0.08 &  0.01 \\
OGLE052224.72-702740.8 & RRab  & 168833 & 19.49  & 19.77   & 18.89  & 10487 & ~~~RRab	 & 19.569 &  20.022 & 18.886 &   0.08 &  0.25 \\
OGLE052206.67-702755.9 & RRc   & 169354 & 19.79  & 20.01   & 19.23  & 28665 & ~~~RRc	 &   $-$  &  $-$    &	$-$  &    $-$  &   $-$  \\
OGLE052134.09-703652.8 & RRab  & 111805 & 19.46  & 20.31   & 18.77  &  ~4313 &~~~RRab	 & 19.270 &  19.779 & 18.451 & $-$0.19 & $-$0.53 \\
OGLE052213.85-700927.2 & EB    & 165895 & 19.32  & 19.48   & 18.81  &  ~8723 &~~~EB?	 & 19.152 &  19.859 &	$-$  & $-$0.17 &  0.38 \\
OGLE052219.01-703311.0 & RRab  & 166016 & 19.16  & 19.67   & 18.73  &  21007 &~~~RRab	 & 19.319 &  19.841 & 18.487 &   0.16 &  0.17 \\
OGLE052214.83-702807.0 & RRab  & 169311 & 19.68  & 20.02   & 19.06  &  28246 &~~~RRab	 & 19.605 &  20.030 & 19.144 & $-$0.08 &  0.01 \\
OGLE052155.09-703212.3 & CepFA & 116626 & 19.40  & 20.42   & 18.32  &  22592 &~~~(1)     &    $-$  &   $-$  &   $-$    $-$  &   $-$     \\
OGLE052235.68-702815.9 & $-$   & 169285 &   $-$   &   $-$  & $-$    & 28114  &~~~$\delta$S & 19.940 & 20.273 &	$-$  &  $-$  &   $-$     \\
OGLE052153.64-703635.4 & $-$   & 114367 &   $-$   &   $-$  & $-$    & ~4490  &~~~EB  &19.016 & 19.011 &  19.017 &   $-$  &   $-$   \\
OGLE052129.44-702923.6 & $-$   & 119268 &   $-$   &   $-$  & $-$    & 26715  &~~~RRc     &  19.378 & 19.725 &   $-$  &  $-$  &   $-$     \\
%OGLE052138.17-702726.0 & RRab  & 120039 &  19.368 & 19.895 & 18.837&200606 &  {\bf Non c'e'}    &    $-$  &   $-$  &   $-$   \\
%          $-$          & $-$   &   $-$  &   $-$   &   $-$  & $-$    &  2636  &  RRc	&  19.896 & 19.595 &  19.080 \\
\hline
\end{tabular}
\end{center}
(1) Star OGLE052155.09-703212.3, ( Id$_{SC\_21}$ =116626), is classified by OGLE II as CepFA,  
however OGLE II light curves for this star are rather
poor and the corresponding object in our  photometry  (\# 22592) is not was
found to vary.

%\scriptsize
\normalsize
\end{table}
\end{landscape}
%
%Among these objects OGLE II reports 114 ({\bf CHECK MARCELLA}) variable stars, 
%96 ({\bf CHECK MARCELLA}) of them have a 
%counterpart in 
%our database, but only 14 ({\bf CHECK MARCELLA}) are identified as variable stars in our photometry.
%On the other hand, in the same area we have 22 ({\bf CHECK MARCELLA}) further 
%variable stars all 
%having a
%counterpart, but not  recognized as variable objects 
%by OGLE II.
\noindent
\begin{table*}[ht]
%\vskip 4 cm
\begin{center}
\caption{Comparison of our and OGLE II photometry for the non variable 
stars in common. $<\Delta V>$,$<\Delta B>$,$<\Delta I>$ are: this 
paper $-$ OGLE II}
\scriptsize
\begin{tabular}{cccrcccccrcccccr}
\hline
\multicolumn{1}{c}{Bin}&\multicolumn{1}{c}{$<\Delta V>$}&
\multicolumn{1}{c}{$\sigma_{V}$}&\multicolumn{1}{c}{N}&&&
\multicolumn{1}{c}{Bin}&\multicolumn{1}{c}{$<\Delta B>$}&
\multicolumn{1}{c}{$\sigma_{B}$}&\multicolumn{1}{c}{N}&&&
\multicolumn{1}{c}{Bin}&\multicolumn{1}{c}{$<\Delta I>$}&
\multicolumn{1}{c}{$\sigma_{I}$}&\multicolumn{1}{c}{N}\\
\hline 
    15.00-17.00&     0.045&	0.017&   64&&& 16.00-17.00&	  0.014&     0.036&   16&&&  14.00-15.00&     0.063&     0.053&   35\\  
    17.00-17.25&     0.044&	0.027&   60&&& 17.00-17.50&  -0.006&     0.091&   22&&&  15.00-16.00&     0.070&     0.044&  129\\  
    17.25-17.50&     0.044&	0.021&   76&&& 17.50-18.00&   0.009&     0.027&   24&&&  16.00-17.00&     0.056&     0.040&  300\\
    17.50-17.75&     0.044&	0.023&   96&&& 18.00-18.25&   0.025&     0.036&   28&&&  17.00-17.50&     0.052&     0.042&  273\\    
    17.75-18.00&     0.043&	0.025&  107&&& 18.25-18.50&   0.011&     0.036&   51&&&  17.50-18.00&     0.042&     0.033&  368\\  
    18.00-18.25&     0.048&	0.028&  122&&& 18.50-18.75&   0.017&     0.032&   88&&&  18.00-18.25&     0.040&     0.037&  626\\   
    18.25-18.50&     0.048&	0.026&  174&&& 18.75-19.00&   0.022&     0.030&  149&&&  18.25-18.50&     0.055&     0.045& 1040\\    
    18.50-18.75&     0.043&	0.021&  194&&& 19.90-19.25&   0.026&     0.034&  161&&&  18.50-18.75&     0.052&     0.049&  479\\  
    18.75-19.00&     0.049&	0.026&  250&&& 19.25-19.50&   0.031&     0.038&  218&&&  18.75-19.00&     0.043&     0.054&  344\\   
    19.00-19.25&     0.049&	0.030&  628&&& 19.50-19.75&   0.034&     0.030&  330&&&  19.00-19.50&     0.045&     0.062&  966\\    
    19.25-19.50&     0.062&	0.032& 1105&&& 19.75-20.00&   0.032&     0.042&  524&&&  19.50-19.75&     0.027&     0.080&  723\\      
    19.50-19.75&     0.069&	0.036&  579&&& 20.00-20.25&   0.026&     0.049& 1115&&&  19.75-20.00&     0.031&     0.091& 1008\\     
    19.75-20.00&     0.068&	0.044&  510&&& 20.25-20.50&   0.044&     0.047& 1125&&&  20.00-20.25&     0.035&     0.113& 1340\\    
    20.00-20.25&     0.067&	0.051&  611&&& 20.50-20.75&   0.035&     0.063&  864&&&             &          &          &     \\
    20.25-20.50&     0.073&	0.063&  838&&& 20.75-21.00&   0.039&     0.066&  944&&&             &          &          &     \\   
               &          &          &     &&& 21.00-21.25&   0.038&     0.082& 1046&&&             &          &          &     \\	 
\hline 
\end{tabular}
\end{center}
\normalsize
%$<\Delta V>$,$<\Delta B>$,$<\Delta I>$ are: this paper $-$ OGLE II
%\label{t:tab6}
\end{table*}
%Open symbols mark the objects whose residuals in $V$ (circles),  $B$
%(squares) and $I$ (triangles) are larger than $\pm$0.75 mag, a total number of
%about 270XXXX ({\bf LUCA}) stars. 
In order to make a more meaningful comparison of the two photometries we 
restricted
the sample of the stars in common only to objects brighter 
than $V$=20.5, $B$=21.25, and $I$=20.25 mag. 
%and with photometric errors
%smaller than 0.1 mag ({\bf LUCA E' GIUSTO}) in both
%databases  (5414, 6705, and 7631 stars, respectively). 
Average residuals were computed
dividing the objects in magnitude bins and applying an 
iterative $\sigma$-rejection procedure
which discarded objects deviating more than 3$\sigma$ from the average
in the bin.
 Results are summarized 
in Table~15 (they are based on
5414, 6705, and 7631 stars in $V,B,I$ respectively). 
 At the magnitude level of RR Lyrae and clump stars ($V\sim19.4,B\sim19.8,I\sim18.8$;
and $V\sim19.3,B\sim20.2,I\sim18.3$, respectively) offsets are: $\Delta V$
= 0.06 ($\sigma_{V}$=0.03), $\Delta B$ = 0.03 ($\sigma_{B}$=0.04),
$\Delta I$ = 0.04 ($\sigma_{I}$=0.05), and $\Delta V$ = 0.06 ($\sigma_{V}$=0.03),
$\Delta B$ = 0.03-0.04 ($\sigma_{B}$=0.05), $\Delta I$ = 0.06 ($\sigma_{I}$=0.05),
Our photometry is systematically fainter than OGLE II photometry, again as
expected since DoPhot is reported to give systematically
brighter magnitudes for faint stars in crowded regions than DAOPHOT/ALLFRAME,
and since we resolve many more faint stars than OGLE II in the area
in common.
\begin{figure} 
\includegraphics[width=8.8cm]{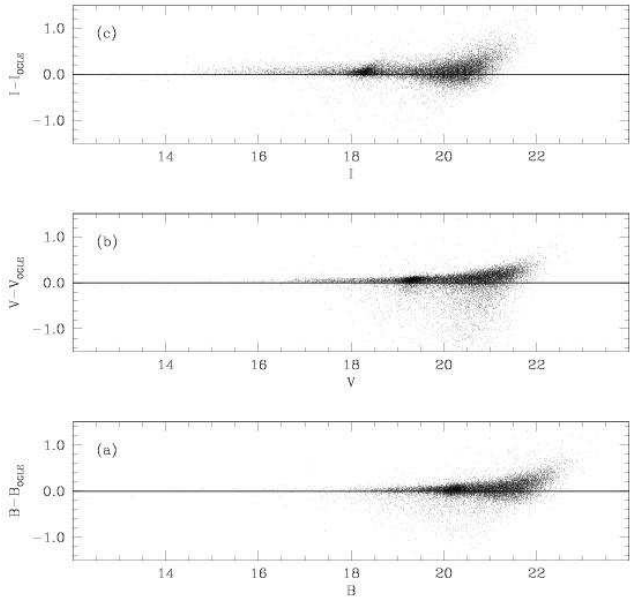}
\caption[]{Comparison between our and OGLE II 
photometry for 
 the about 14,000 stars in common. Residuals are this paper minus OGLE II.
%  Open symbols mark objects whose residuals 
%are larger than $\pm$0.75 mag in $V$ (circle), 
%$B$ (squares), and $I$ (triangles), respectively.
%({\bf Qui bisogna rifare l'I perche' le correzioni di apertura erano 
%sbagliate -  LUCA})
}
%\label{f:fig1a}
\end{figure}
%, respectively, with the most deviating
%objects marked as in Figure 12. 
%There are a number of deviating objects in the CMDs.
%In particular,  On the other hand  objects 
%with $I> 20.0$ and $V-I <$ 0.0 are likely to be . 
\begin{figure} 
\includegraphics[width=8.8cm]{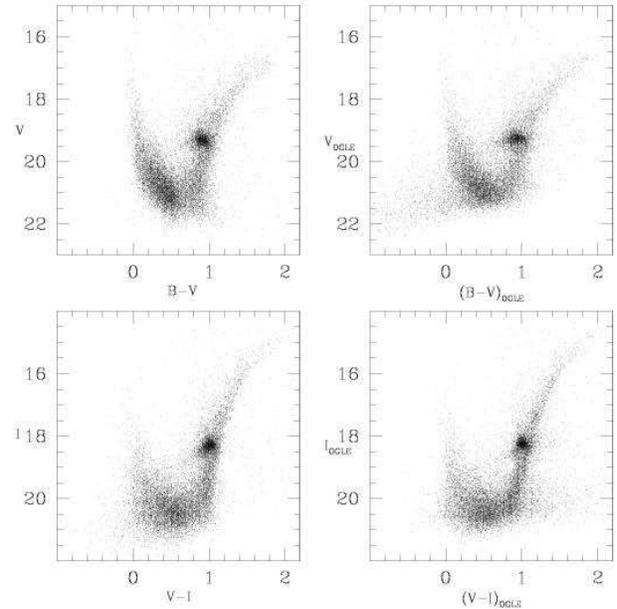}
\caption{$V\, vs\, (B-V)$ and $V\,\textit{vs\,}(V-I)$ CMDs for the stars in
common between our field A and OGLE II field LMC\_SC21. Left panels:
this paper; right panels: OGLE II.}
%\label{f:fig1a}
\end{figure}
Transformation equations between the two photometries 
were then computed as linear fits of the average 
residuals of all the bins:
$$B_{this~paper}-B_{OGLE}=0.00835 \times B_{this~paper}-0.13507$$
$$V_{this~paper}-V_{OGLE}=0.00751 \times V_{this~paper}-0.08626$$
$$I_{this~paper}-I_{OGLE}=-0.00622 \times I_{this~paper}+0.15914$$
Thus the transformation relations between OGLE II and our photometry
are:
$$B_{this~paper}=1.0084 \times B_{OGLE}-0.1362$$
$$V_{this~paper}=1.0076 \times V_{OGLE}-0.0869$$
$$I_{this~paper}=0.9938 \times I_{OGLE}+0.1582$$

\section{The pulsation characteristics of the RR Lyrae stars}
We have detected and derived periods for a total number of  135 RR Lyrae stars 
in our two fields (78 in field A, and 57 in field B). This number includes 
87 fundamental mode (RRab),
38 first overtone (RRc), and 10 double-mode (RRd) pulsators.
According to the completness of our photometry and the comparison
with MACHO and OGLE II catalogues 
(Sections 4.2 and 4.3) our sample of variables should be about 97\% complete.  
The two fields are found to contain about the same number of first overtone
RR Lyrae (20 in field A and 18 in field B) while the number of
fundamental mode pulsators is about 50\% larger in field A (52 RRab) 
than in field B (35 RRab).
We found that 17\% of the fundamental mode RR Lyraes in our two fields
are (or are suspected to be) affected by the Blazhko phase and amplitude 
modulation of the light curve (Blazhko 1907). 
This percentage is consistent with the 11.9\% and the 15\%
Blazhko incidence rates
among RRab's reported respectively by MACHO (Alcock et al. 2003b) and 
OGLE II (Soszy\'nsky et al. 2003), and maybe closer 
to
the 20\%-30\%  incidence rate commonly found for the Milky Way fundamental 
mode RR Lyrae (Szeidl 1988, Moskalik \& Poretti 2003). The 
first-overtone Blazhko variables are the 5.3\% of our RRc sample, again 
in agreement with both 
MACHO ($\sim$ 4\%, Alcock et al. 2003b), and OGLE II 
($\sim$ 6\%, Soszy\'nsky et al. 2003).
\begin{figure} 
\includegraphics[width=8.8cm]{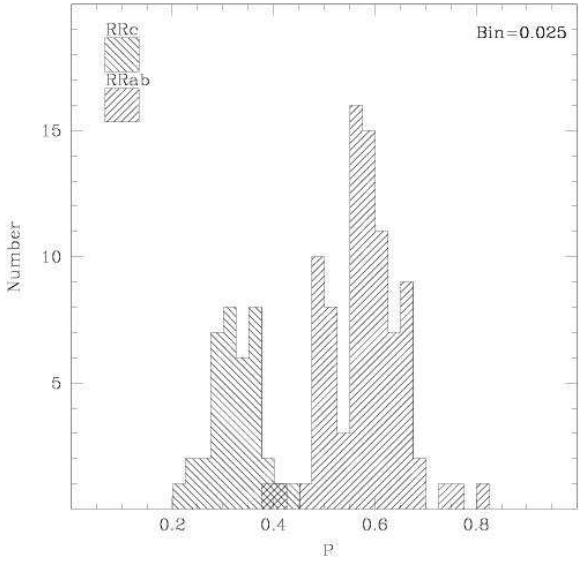}
\caption[]{ Number {\it vs} Period histogram of the single-mode RR Lyrae variables 
 in our sample (125 objects).} 
%\label{f:fig1a}
\end{figure}
Figure~26 shows the period distribution of the single-mode RR Lyrae's (125
objects). The two peaks correspond to the average period of the {\it c-} 
and {\it ab-}type 
pulsators, respectively:
%The mean period of the {\it c-} and {\it ab-}type RR Lyrae's is 
$<{\rm P_{RR_c}}>$=0.324 days ($\sigma$=0.048, 38 stars) and 
$<{\rm P_{RR_{\rm ab}}}>$=0.581 days ($\sigma$=0.071, 87 stars), 
%({\bf RICALCOLARE AGGIUNGENDO LE RRAB RECUPERATE DA MACHO}) 
to compare with 0.342 and 0.583 day of A96, and with 0.339 and 0.573 days by 
Soszy\'nsky et al. (2003).  
Our average periods are in good agreement with both A96 and Soszy\'nsky et al. 
(2003) results, which are based on much larger samples, and confirm 
%A96, on the basis of their $<{\rm P_{RR_{\rm ab}}}>$=0.583 day, 
%conclude that  
that the average period of the {\it ab}-type variables of the LMC 
is {\bf intermediate} 
between the periods of the  Galactic RR Lyrae stars of  
Oosterhoff type I (OoI) and II (OoII), 
 but  it is actually closer to the Oo I clusters  (being
$<$P$_{{\rm RR_{\rm ab}}}>$=0.55, and 0.65 days in Oo I and II
clusters, respectively; Oosterhoff 1939).
Our results also indicate that
the average pulsation properties of the RR Lyrae stars in the two fields 
are slightly different, with variables in field B being more  definitely of Oo
type I. Field B contains in fact  a larger number of {\it ab-}type RR Lyrae with
periods around half a day
(10 out of 35 RRab's in field B have P=0.50$\pm$0.02 days corresponding
to 28.6 \%, while only 5 out of 52 in field A, corresponding to
9.6 \%), as confirmed by the average periods computed  keeping
the variables in the two fields separate. These are: 
$<{\rm P_{RR_c}}>$=0.320$\pm 0.011$ days ($\sigma$=0.050, 20 stars),
$<{\rm P_{RR_{\rm ab}}}>$=0.593$\pm 0.010$ days ($\sigma$=0.065, 52 stars), 
and $<{\rm P_{RR_c}}>$=0.329$\pm 0.011$ days ($\sigma$=0.047, 18 stars),
$<{\rm P_{RR_{\rm ab}}}>$=0.562 $\pm 0.013$ days ($\sigma$=0.075, 35 stars),
in field A and B, respectively.

The $B$ and $V$ amplitudes (A$_B$, A$_V$, see Columns 14 and 15 of 
Tables 5 and 6) 
%were calculated for all  the RR Lyrae's
%with full coverage of the light curve as the difference between maximum  and 
%minimum of the best fitting models (see Di Fabrizio et al. 2002), and have been
were used together with the newly derived periods to build the period - amplitude
diagrams shown in Figure~27. The overlap in the transition region 
between {\it
ab} and {\it c-}type RR Lyrae is  small  (5 objects, see Figure~26).
Our shortest period {\it ab-}type RR Lyrae's are: 
star \#19450 in field A, P=0.398 days, A$_V$=1.344 and 
A$_B$=1.709 mag; and star \#19037 in field B, P=0.411 days, 
A$_V$=1.466 and A$_B$=1.821 mag. The longest period {\it c-}types
are: star \#6415 in 
field A, with P=0.443 days, 
A$_V$=0.438 and A$_B$=0.473 mag, and stars \# 6957 and \# 7064 in field B,
respectively with P=0.406 days, A$_V$=0.396, A$_B$=0.568 mag and 
P=0.401 days, A$_V$=0.474, A$_B$=0.607 mag.
These stars define the transition region between {\it
ab} and {\it c-}type RR Lyrae stars that, in our sample, occurs at
P$_{\rm tr} \sim$ 0.40 days, (P$_{\rm tr}$ = 0.457 days in A96).  
They are labelled in the period-amplitude distributions in 
Figure~27. 

\begin{figure*} 
\includegraphics[width=16cm]{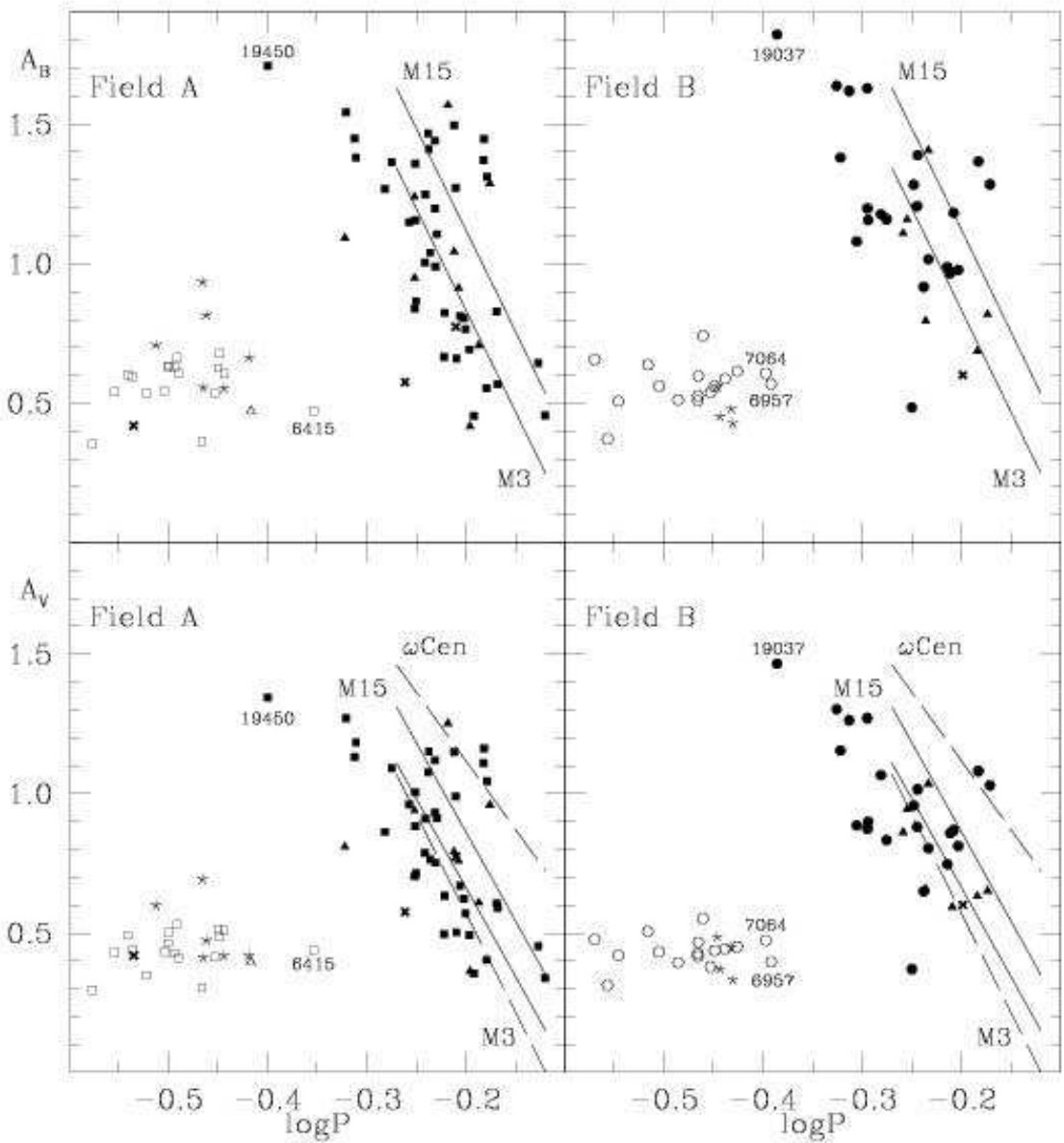}
\caption[]{A$_B {\it vs } \log {\rm P}$ 
and A$_V {\it vs } \log {\rm P}$ 
diagrams for the RR Lyrae's with
complete $B$ and $V$ light curves in field A and B, separately. Solid lines
show the distributions defined by the {\it ab-}type RR Lyrae
variables in the globular clusters M3 
from the photometry of 
Carretta et al. (1998), and M15 from Bingham 
et al. (1984).
Dashed lines in the lower panels are the 
A$_V vs \log P$
relations
derived for M3 and $\omega$ Cen by  
Clement (2000) only using RRab's with regular light curves.
Different symbols refer to {\it ab} (filled square and circles, in field A and B 
respectively), {\it c} 
(open square and circles in field A and B respectively),
{\it d-}type (asterisks) RR Lyrae variables, and candidate Anomalous 
Cepheids (crosses), respectively. Triangles mark the RR Lyrae variables that 
are found or sospected to be affected by Blazhko effect. Labels
identify RR Lyrae stars at the transition period between {\it ab-} and 
{\it c-}types 
%({\bf MARCELLA TOGLI LE LABELS della 27697, INOLTRE
%QUESTA FIGURA VA AGGIORNATA INCLUDENDO GLI OGGETTI RECUPERATI 
%DAL CONFRONTO CON OGLE II E MACHO PER CUI LE AMPIEZZE NON SONO INCERTE}).
}
%\label{f:fig1a}
\end{figure*}
%The shortest period {\it c-}type is star \#3216 in field A with P=0.$^d$218, 
%A$_V$=0.407 and A$_B$=0.515 mag, 
%while  
A96 discuss at some length the existence in their period and amplitude 
distributions (see figs. 1 and 6 of A96) of an extra-large number of 
variables with period around 0.28 days, 
which have asymmetric light curves, but low amplitudes. A96 
classify these variables as possible 
second-overtone  RR Lyrae's (type {\it e}), and see also the 
discussion in Soszy\'nsky et al. (2003).
Figure~26 does not show clear evidence for an extra peak 
around P$\sim$0.28 days. We have 8 objects in the 
period range from 
0.265 to 0.291 days (4 in each of the two fields). Only two of them show
asymmetric light curves, namely:  
star \#2223 in field A with P=0.288 days, 
A$_V$=0.493 mag, A$_B$=0.604 mag, and A$_I$=0.499 mag;
and star \#10585 in field B with P=0.270 days, A$_V$=0.478 mag, and 
A$_B$=0.657 mag.
Another RRc of slightly longer period has very asymmetric curves: 
star \#7490 in field B with P=0.305 days, 
A$_V$=0.505 and A$_B$=0.637.

The A$_V - \log {\rm P}$, A$_B - \log {\rm P}$ distributions of the 
variables in the two 
fields are similar (see Figure~27),
and resemble fig. 6 of A96, 
% with two major differences: (i) owing to our data
%sampling (observations in 4 consecutive nights) we do not have full 
%coverage of the light curve, hence we cannot derive A$_{\rm V}$ and 
%A$_{\rm B}$ values, 
%for variables with periods longer than 0.$^d$70 ($\log {\rm P} >-$0.16,
however our A$_V$ amplitudes range is slightly larger than in A96, with 
A$_V$ values from 0.29 to 1.47 mag in our sample to compare with 
0.35-1.35 in A96.

%A96 draw the attention to (i) the large number of Bailey {\it b-}type RR Lyrae
%with amplitude less than 1.0 mag in their period-amplitude distribution 
%(these variables
%outnumber by a factor 2 those with visual amplitudes larger than 1.0 mag); and 
%(ii) to the paucity of stars with V amplitude of 0.9$\pm$0.05 mag in the 
%period range $-0.3<\log{\rm P}<-0.2$. 
%Field A shows some evidence for a lack of {\it ab-}type variables with 
%amplitudes in the range 0.85$<A_V<$0.90 mag and 1.0$<A_B<$1.15 mag, 
%however the {\it a} and {\it b} variables seem to be equally 
%distributed around this gap, in agreement with 
%Walker (1992b, 1992c) finding for the LMC globular cluster variables. 
%On the other hand, the {\it ab-}type variables in field B seem to be 
%uniformally distributed in the amplitude range  0.60$<A_V<$1.20 mag, with 
%only very few objects (3 in total) overcoming these limits).

The period - amplitude distributions of the LMC variables were 
compared with the relations defined by the 
{\it ab-}type RR Lyrae's in the globular clusters M3, M15 and $\omega$ Cen, 
shown by lines in Figure~27.
Solid lines were derived from the photometry of 
Carretta et al. (1998) for M3, and Bingham 
et al. (1984) for M15, and were computed 
as follow: we first derived the 
period-amplitude relations using the M3 sample which is more extended; then 
we shifted the intercept of these relations while holding fixed the 
slopes, until a good fit (by eye) was obtained also for the variables in  
M15, which are too few in number to give a satisfactory best fit by 
themselves. Dashed lines in the lower panels of Figure~27 are the 
A$_V vs \log P$
relations
derived for M3 and $\omega$ Cen by  
Clement (2000) using only RRab's with regular light curves
(see also Clement \& Shelton 1999, and Clement \& Rowe 2000).

RR Lyrae's in field B seem to better follow the amplitude-period relations
of the variables in M3 and, as already noted, to belong to 
the OoI type. Variables in field A, instead, have pulsation properties 
more intermediate between the two Ooostheroff types.

\subsection{The double-mode pulsators in our sample}
According to A97, nine double-mode RR Lyrae stars were expected to fall in the
observed areas. We detected all of them and also found evidence  for
one possible additional RRd: star
\# 2249. This variable is tentatively classified as {\it d}-type mainly because
of the  large scatter of the observed V and B light curves (0.12 and 0.11 mag, 
respectively), which has no obvious  explanation since the object, although
rather faint, is not blended to other  stars on the frames. We fully covered the
light variation of all RRd's in  our sample; however our  sampling of their 
light curves is too coarse to allow a firm identification of the two 
periodicities, particularly for stars with fundamental  periods
around half a day. Periods from A97 have been adopted to phase  the data of
these variables, apart from  star \# 2249 for which we use our period.

%Intensity average $<B>$ and $<V>$ magnitudes of  the 10 RRd's are given in
%Table~5. For a comparison we also give the  $<V>$ values of A97 for these
%stars, and the values given in  the on line MACHO catalogue (see web site
%wwwmacho.mcmaster.ca/Data/MachoData.html) for 8 of  the RRd's in Table~5, since
%they are different than in A97. Variables are identified by our number, by the
%number in DF02  catalogue, and by MACHO/A97 identifier  preceded by two letters
%indicating the field the variable belongs  to and an ordering number according
%to tab.1 of A97, and used in Bragaglia et al. (2001). 

The average luminosity we derive from the 10 RRd's in our sample,
using the magnitude-averaged values in column 11 of Tables~5 and 6
($<V_{RRd}>$=19.335 $\pm$0.056, $\sigma$=0.176, 10 stars) is
%({\bf QUESTI NUMERI SONO STATI RICALCOLATI IL 31/05/04})is 
in very good
agreement with A00 average luminosity of the
LMC double mode RR Lyrae stars ($<V_{RRd}>$=19.327 $\pm$0.021).
%in very good agreement with 

%common with
%A97  is $<V>$=19.31$\pm 0.06$ ($\sigma$=0.17 -  9 stars), 
% and $<B>$=19.67$\pm 0.06$ ($\sigma$=0.18 - 9 stars), 
%to  compare with A97 $<V>$=19.13$\pm 0.04$ ($\sigma$=0.12 - 9  stars),  and 
%$<V>$=19.34$\pm 0.07$ ($\sigma$=0.186 - 8 stars)  from the MACHO on line
%catalogue.
%The average difference (this paper $-$ A97) 
%is $\Delta V=0.18 \pm$0.02  
%($\sigma$=0.07 - 9 stars) and 
%(this paper $-$ on line MACHO catalogue) 
%is $\Delta V=-0.08 \pm$0.03  
%($\sigma$=0.10 - 8 stars)

We may compare the average luminosity of the RRd variables  with the average
luminosity of the single-mode RR Lyrae stars in our two fields. Due to the difference in
reddening between the two fields (see C03) this comparison is better done keeping the
variables in the two  areas separated.
We found  
$<V(RRd)>_{Field A}$ = 19.378$\pm$0.055 ($\sigma$=0.135), and 
$<B(RRd)>_{Field A}$=19.731$\pm$0.058 ($\sigma$=0.141) from the 
average of the 6 RRd's in field A,
to compare with 
average values derived from the single-mode pulsators
 of 19.421$\pm$0.020 ($\sigma$=0.156, 61 stars), and 19.824$\pm$0.022 
 ($\sigma$=0.173, 61 stars), respectively;
and
$<V(RRd)>_{Field B}$ = 19.229$\pm$0.087 ($\sigma$=0.173) and 
$<B(RRd)>_{Field B}$=19.578$\pm$0.093 ($\sigma$=0.186) from the 
average of the 4 RRd's in field B,
to compare with 
average values derived from the single-mode pulsators
 of 19.326$\pm$0.023 ($\sigma$=0.155, 45 stars), and 19.687$\pm$0.023 
 ($\sigma$=0.156, 45 stars), respectively.
The RRd pulsators seem to be slightly brighter  than
the single-mode ones
 in the same field (by 0.043 mag in $V$ and 0.093 in $B$ in field A, and by 0.097 in $V$ and 0.109
in $B$ in field B), although the statistical significance of this result
might be weak  given the rather small number of objects. A similar conclusion was also
reached by G04.
%
%RRd's are on average slightly brighter 
% within the quoted errors and $<B>$ values for 
%the 
%RRd's in field A (and  
%are marginally brighter than  
% the while RRd's in 
%field B are on average $\sim$0.1 mag brighter (19.23$\pm$0.09, 
% and 19.58$\pm$0.09, average on 4 stars) 
%than the single-mode variables in 
%the same field (19.33$\pm$0.02 
% and 19.69$\pm$0.02, respectively).
%A similar trend is found in A97 $<{\rm V}>$ values.

\section{Fourier decomposition of the light curves}
%\subsection{Metallicities}
%
%({\bf QUI VA DESCRITTO COME SONO STATI CALCOLATI GLI ERRORI DELLE METALLICITA' 
%DAI PARAMETRI DI FOURIER VEDI PAPER SPETTROSCOPICO})
% 
%\subsection{Absolute magnitudes}
%
%\documentclass[a4paper,11pt]{report}   %% LaTex2e
%\usepackage[latin1]{inputenc}		%%
%\usepackage[italian]{babel}		%% "traduce" in italiano
%\usepackage[dvips]{graphicx} %%,colour,rotating}	%% per inserire grafici
%\pretolerance=10000 \tolerance=10000 \hyphenpenalty=10000 \hbadness =10000
%\textheight 240mm \textwidth 200mm
%\textheight 230mm \textwidth 130mm
%\topmargin -5mm %\oddsidemargin 9.6mm \evensidemargin 9.6mm
%\parindent 8mm
%\begin{document}
%\large
%\baselineskip 7mm
%\baselineskip 6.5mm
%\hoffset -0.5cm
%\hoffset -1.5cm \voffset -1cm
%\hsize 13cm               %% larghezza pagina
%\hsize 18cm
%
%%%%%%%%%%%%%%%%%%%%%%%%%%%%%%%%%%
\begin{table*}[ht]
\begin{center}
\caption{Fourier parameters of the light curves and corresponding estimate of the
star metallicity, absolute magnitude, intrinsic $(B-V)_0$ colour,  and effective temperature
 }
%\tiny
\scriptsize
%\label{stime}
%\vspace*{5mm}
%%%%%%%%%%%%%%%%%%%%%%%%%%%%%%%%%%%%%%%%%%%%%%%%%%%%%%
\begin{tabular}{cccccccccccc}
\hline
\hline
\multicolumn{1}{c}{Star}&
\multicolumn{1}{c}{P}&
\multicolumn{1}{c}{$<B-V>_{int}$}&
\multicolumn{1}{c}{A1}&
\multicolumn{1}{c}{A2}&
\multicolumn{1}{c}{A3}&
\multicolumn{1}{c}{A4}&
\multicolumn{1}{c}{A5}&
\multicolumn{1}{c}{A6}&
\multicolumn{1}{c}{A7}&
\multicolumn{1}{c}{A8}&
\multicolumn{1}{c}{  } \\
 7325 &0.48677& 0.410 & 0.34815  & 0.18973 &  0.13552  & 0.09432 &  0.05650 &  0.04252  & 0.01980  & 0.01321 &
 \hspace{0.8cm} \\
&&&&&&&&&&\\ 
\multicolumn{1}{c}{}&
\multicolumn{1}{c}{}&
\multicolumn{1}{c}{}&
\multicolumn{1}{c}{$\Phi_{21}$}&
\multicolumn{1}{c}{$\Phi_{31}$}&
\multicolumn{1}{c}{$\Phi_{41}$}&
\multicolumn{1}{c}{$\Phi_{51}$}&
\multicolumn{1}{c}{$\Phi_{61}$}&
\multicolumn{1}{c}{$\Phi_{71}$}&
\multicolumn{1}{c}{$\Phi_{81}$}&
\multicolumn{1}{c}{}\\
 &&&3.94169  & 1.90815  & 6.15817 &  4.23861 &  2.16828 &  0.40072 &  4.83247&\\
&&&&&&&&\\
\multicolumn{1}{c}{}&
\multicolumn{1}{c}{}&
\multicolumn{1}{c}{}&
\multicolumn{1}{c}{df1}&
\multicolumn{1}{c}{df2}&
\multicolumn{1}{c}{df3}&
\multicolumn{1}{c}{df4}&
\multicolumn{1}{c}{df5}&
\multicolumn{1}{c}{}\\
 &&&  1.414 & 0.909 & 0.352 & 1.388 & 2.824&  \\
&&&&&&&&\\
\multicolumn{1}{c}{}&
\multicolumn{1}{c}{}&
\multicolumn{1}{c}{}&
\multicolumn{1}{c}{df21}&
\multicolumn{1}{c}{df31}&
\multicolumn{1}{c}{df41}&
\multicolumn{1}{c}{df51}&
\multicolumn{1}{c}{}\\
 &&&0.661 & 0.757 & 0.894 & 0.501 &&\\
&&&&&&&&\\
\multicolumn{1}{c}{$[Fe/H]$}&
\multicolumn{1}{c}{M$_V$}& 
\multicolumn{1}{c}{$(B-V)_0$}&
\multicolumn{1}{c}{$ \log T_{eff}(B-V)$}\\
\multicolumn{1}{c}{}&
\multicolumn{1}{c}{}&
\multicolumn{1}{c}{}\\
 $-$0.874 &  0.684 & 0.332 &3.817&&&\\ 
\hline
\end{tabular}
\normalsize
\end{center}

Table 16 is presented in its entirety in the electronic edition
of the Journal. A portion is shown here for guidance regarding its form
and content.
\end{table*}

In recent years Jurcsik \& Kov\'acs (1996; hereinafter JK96),
%, A\&A, 312, 111), 
Kov\'acs \& Jurcsik (1996, 1997; hereinafter KJ96, KJ97),
% ApJ, 466, L17; 1997,  A\&A, 322, 218) 
and   
Kov\'acs  \& Walker (2001; hereinafter KW01)
%, A\& A, in press, astro-ph/0103188) 
have derived empirical relations between the parameters of the Fourier 
decomposition of the $V$ light curves of the fundamental mode RR Lyrae stars and 
their basic stellar quantities, namely: 
intrinsic magnitude and colours, effective temperature, gravity and 
metal abundance. These relationships were calibrated on Galactic field 
RR Lyrae (JK96) and on globular clusters variables (KJ96, KJ97, and KW01), 
and should allow to derive the physical parameters for any 
RRab provided that accurate Fourier parameters of $V$ light 
curve are available.  

Our sample of {\it ab-}type LMC RR Lyrae stars with high 
quality multiband light curves, metal abundances homogeneously 
derived and covering more than 1 dex metallicity range (G04),  
all at the same distance from us, and with reddening consistently derived
(C03), may be used to check these empirical relationships.

JK96 show that the 
light curves of the variable stars 
must satisfy completeness and regularity criteria, referred to by the
authors as {\it compatibility conditions}, 
for the Fourier parameters to predict reliable empirical quantities.  
 Namely, the deviations of the Fourier parameters  
 should not exceed the 
 maximum value (D$_{\rm m}$) of 3, with 
 maximum deviations  D$_{\rm m} > 3$ possibly 
 indicating that incompatibility with the empirical predictions can 
 be expected (Kov\'acs \& Kanbur 1998, hereinafter KK98).
The deviation parameters D$_{F}$ are defined as     
D$_{F}$ =$\mid F_{obs} - F_{calc} \mid / \sigma_F$, where  
 $F_{obs}$, $F_{calc}$ are respectively the observed value of a given
Fourier parameter and its predicted value from the other observed
parameters, and  $\sigma_F$  is the respective standard deviation 
(see eq. 6  and Table 6 of JK96). 
JK96 find that Blazhko stars
do not generally satisfy the {\it compatibility conditions}. 
However, Cacciari, Corwin, \& Carney (2004), in their extensive analysis of
the RR Lyrae stars in the globular cluster M3, based
on the large database of Corwin and Carney (2001),  
found that 40\% of the
variables with D$_{\rm m} <3$ were indeed Blazhko stars. 
%JK96 also caution that Blazhko stars
%do not generally satisfy the {\it compatibility conditions}.  

From our sample of 87 LMC RRab's we thus chose objects with fully 
covered $V$ 
light curves, no systematic shifts between 1999 and 2001 photometry, 
and not affected (or suspected to be affected) by Blazhko effect.
The selected variables were then tested against JK96 
{\it compatibility conditions}; 29 of them passed the test.
This sample includes 14 stars with D$_{\rm m} \leq 3$, and 
15 objects with  3$<$D$_{\rm m} \leq 5$, (D$_{\rm m} <5$ can
still provide acceptable results, cf. Cacciari et al. 2004).
Parameters from the Fourier decomposition of their $V$ light curves 
are provided in Table~16, while in column 3 of Table~17 
we report the highest maximum D$_{\rm m}$ value of each star.   
Metallicities ([Fe/H]), absolute magnitudes (M$_V$), intrinsic $(B-V)_0$ 
colours, and effectived temperatures (T$_{eff}$), 
were then computed from these parameters 
using the relationships by 
%of the
%Fourier decomposition of the $V$ light curves applying 
JK96, KW01 and Kov\'acs (2002; hereinafter K02).
They are provided in columns 4,6,8, and 10 of Table~17.
%to  those {\it ab-}type RR Lyrae's in our
%sample that satisfy the completness and regularity criteria defined by JK96, that is 
%to stars with . 
%There are 29 RRab's satisfying these 
%criteria in our sample.
\noindent
\begin{table*}[ht]
%\vskip 1 cm
\begin{center}
\caption{Metallicities, absolute magnitudes, $(B-V)_0$ colours, and 
effective temperatures from the Fourier 
parameters of the light curves for the subset of 29 RRab stars
%{\bf (Qui vanno aggiunti i sigma di Mv e Teff)}
}
%\vspace{0.5cm}
\scriptsize
\begin{tabular}{rclccccccc}
\hline
Id~~&Field&~~D$_m$    &[Fe/H]    &[Fe/H]&~~M$_V$  &~~M$_V$     &$(B-V)_0$&$(B-V)_0$   &$\log T_{eff}$\\ 
    &     &           &(Fourier) &G04   &(Fourier)&(this paper)&(Fourier)&(this paper)& (Fourier)  \\
\hline
 1408& B   &$\leq$4.159& $-0.588\pm0.172$&$-1.70\pm 0.11$&0.495$\pm$0.027&0.561&0.353&0.358&3.812\\
 2249& B   &$\leq$4.769& $-1.497\pm0.212$&$-1.56\pm 0.15$&0.520$\pm$0.027&0.564&0.352&0.366&3.806\\
 2525& A   &$\leq$3    & $-1.363\pm0.206$&$-2.06\pm 0.14$&0.513$\pm$0.027&0.465&0.356&0.334&3.806\\
 2884& B   &$\leq$3    & $-1.622\pm0.216$&$-1.90\pm 0.09$&0.507$\pm$0.027&0.435&0.355&0.354&3.804\\
 3054& B   &$\leq$4.158& $-0.944\pm0.208$&~~$-$ 	 &0.729$\pm$0.028&0.284&0.354&0.311&3.809\\
 3400& B   &$\leq$4.025& $-1.588\pm0.239$&$-1.45\pm 0.24$&0.624$\pm$0.028&0.687&0.315&0.305&3.817\\
 3412& B   &$\leq$3    & $-1.619\pm0.232$&~~$-$ 	 &0.572$\pm$0.028&0.643&0.324&0.356&3.814\\
 4540& B   &$\leq$3.111& $-1.403\pm0.216$&~$-$  	 &0.544$\pm$0.027&0.632&0.338&0.330&3.811\\
 4974& A   &$\leq$3    & $-1.085\pm0.200$&$-1.35\pm 0.09$&0.611$\pm$0.027&0.509&0.363&0.328&3.806\\
 5167& A   &$\leq$4.762& $-1.323\pm0.202$&~~$-$ 	 &0.553$\pm$0.027&0.484&0.369&0.375&3.802\\
 5902& B   &$\leq$3    & $-1.436\pm0.217$&$-2.12\pm 0.11$&0.531$\pm$0.027&0.339&0.337&0.320&3.811\\
 6398& A   &$\leq$3    & $-1.099\pm0.205$&$-1.40\pm 0.30$&0.587$\pm$0.027&0.442&0.346&0.339&3.811\\
 6426& A   &$\leq$3.667& $-1.707\pm0.212$&$-1.59\pm 0.09$&0.404$\pm$0.027&0.310&0.348&0.317&3.806\\
 7247& A   &$\leq$3    & $-1.262\pm0.211$&$-1.41\pm 0.10$&0.601$\pm$0.027&0.533&0.348&0.290&3.809\\
 7325& A   &$\leq$3    & $-0.874\pm0.209$&$-1.28\pm 0.09$&0.684$\pm$0.028&0.560&0.332&0.332&3.817\\
 7468& A   &$\leq$4.923& $-0.746\pm0.177$&~~$-$ 	 &0.626$\pm$0.026&0.740&0.392&0.404&3.799\\
 8094& A   &$\leq$4.088& $-1.188\pm0.178$&$-1.83\pm 0.12$&0.460$\pm$0.026&0.478&0.394&0.432&3.795\\
 8220& A   &$\leq$3.605& $-1.032\pm0.182$&~~$-$ 	 &0.496$\pm$0.026&0.594&0.375&0.350&3.802\\
 8720& A   &$\leq$3    & $-1.682\pm0.212$&$-1.76\pm 0.20$&0.365$\pm$0.028&0.254&0.336&0.283&3.810\\
 9494& A   &$\leq$3.889& $-1.526\pm0.219$&$-1.69\pm 0.28$&0.505$\pm$0.027&0.342&0.335&0.292&3.811\\
 9660& A   &$\leq$3.746& $-1.345\pm0.204$&~~$-$ 	 &0.580$\pm$0.026&0.517&0.373&0.367&3.800\\
10214& A   &$\leq$4.261& $-0.314\pm0.165$&$-1.48\pm 0.12$&0.621$\pm$0.027&0.329&0.374&0.332&3.807\\
12896& A   &$\leq$4.745& $-1.298\pm0.211$&$-1.53\pm 0.10$&0.564$\pm$0.027&0.714&0.347&0.351&3.809\\
14449& B   &$\leq$4.877& $-1.483\pm0.216$&$-1.70\pm 0.13$&0.585$\pm$0.027&0.732&0.357&0.370&3.805\\
18314& A   &$\leq$3    & $-1.324\pm0.209$&$-1.71\pm 0.12$&0.487$\pm$0.028&0.535&0.334&0.218&3.813\\
25301& A   &$\leq$3    & $-1.095\pm0.204$&$-1.40\pm 0.18$&0.547$\pm$0.027&0.891&0.336&0.396&3.814\\
25362& A   &$\leq$3    & $-1.279\pm0.209$&$-1.49\pm 0.10$&0.509$\pm$0.027&0.568&0.335&0.299&3.813\\
26525& A   &$\leq$3    & $-1.887\pm0.245$&$-1.63\pm 0.12$&0.602$\pm$0.028&0.598&0.329&0.368&3.811\\
26821& A   &$\leq$3    & $-1.235\pm0.205$&$-1.37\pm 0.13$&0.606$\pm$0.027&0.749&0.364&0.372&3.804\\
\hline
\end{tabular}
\end{center}
%\normalsize

Note: The $\log T_{eff}$'s from the Fourier parameters are from the 
Fourier $(B-V)_0$ colours.
\normalsize
%\label{t:tab7}
\end{table*}
These values were compared with the corresponding observed quantities
obtained in the present photometric study and in G04 spectroscopic
analysis. These comparisons are described in detail
in the following sections.

\subsection{Metallicities}
According to JK96 the [Fe/H] metal abundance of a fundamental mode 
RR Lyrae star is a linear function of the star's period $P$ and of the  
parameter $\phi$31 of the Fourier decomposition of the $V$ light curve.
We have estimated {\it photometric} metallicities for our
subsample of 29 {\it ab-}type RR Lyrae stars
using equation (3) of JK96 (see also K02).
Errors were calculated according to eq.s (4) and (5) of JK96, and
adopting for the Fourier parameters the standard deviations provided in 
Table~2 of KK98.
These {\it photometric} metallicities are based on Jurcsik (1995) 
metallicity scale. They span the range: $-0.31<$[Fe/H]$<-1.89$ with   
an average value of 
[Fe/H]=$-1.27$ ($\sigma= 0.27$, 29 stars), and mean uncertainty of 
about 0.21 dex (see column 4 of Table~17). 
\begin{figure*} 
\includegraphics[width=18cm]{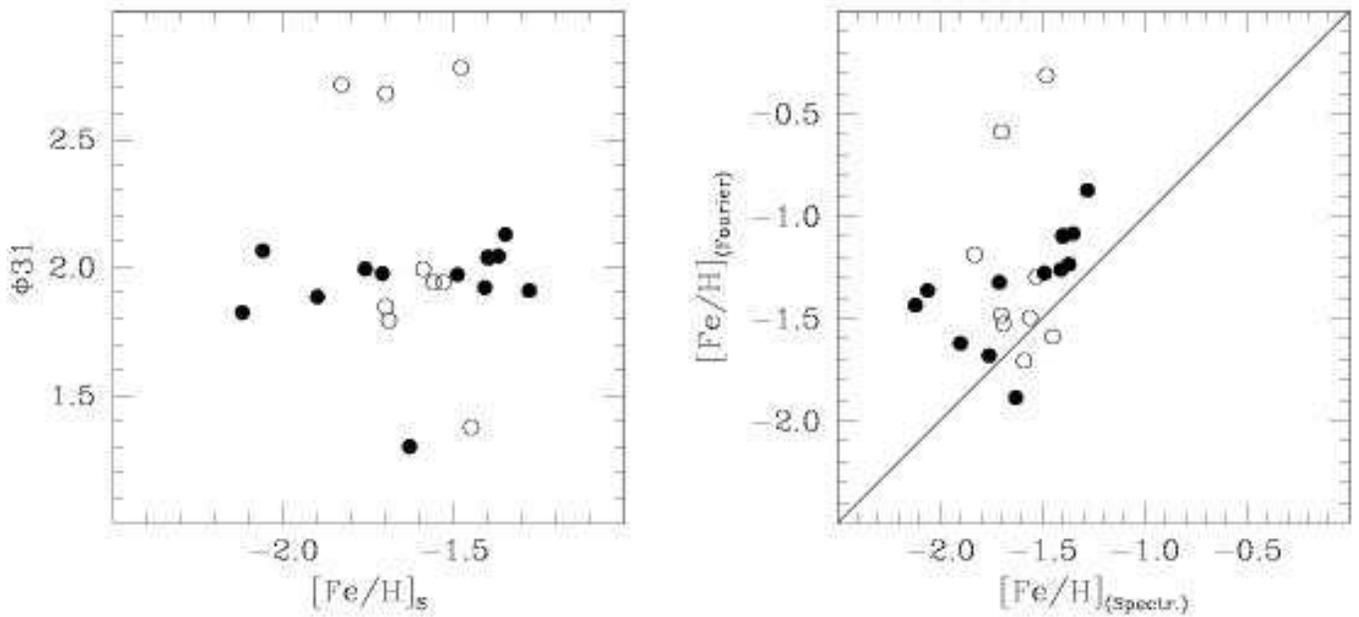}
\caption[]{Left panel: run of the $\phi$31 values with the spectroscopic 
metal abundance for the 22 stars analyzed by G04. 
Right panel: star-to-star comparison between {\it photometric} 
and G04 spectroscopic metallicities. For ease of comparison 
we show the 1:1 line. 
Filled and open symbols and are used for the
variables with D$_{\rm m} \leq 3$ and 3$<$D$_{\rm m} \leq 5$, respectively.
} 
%\label{f:fig1a}
\end{figure*}
G04 measured the metallicity for 22 of these stars using low
resolution spectroscopy obtained with the VLT. 
The spectroscopic abundances are listed in column 5 of Table~17.
They have average uncertainty of about 0.14 dex and 
span the metallicity range: $-2.12<$[Fe/H]$<-1.28$, in G04 metallicity
scale. This scale is 
% are tied to 
%Harris (1996) metal abundances for the Galactic
%globular clusters NGC~1851, NGC~3201, and M68 
on average
0.2 dex more metal poor than Jurcsik (1995) scale (see G04).

The average difference between photometric and spectroscopic metallicities
is 0.30$\pm$0.07 dex, with the photometric abundances being larger
as expected.  
In the left panel of Figure~28 we show the run of the $\phi$31 values with 
G04 metal abundances, 
%Filled and open symbols and are used for the
%variables with D$_{\rm m} \leq 3$ and 3$<$D$_{\rm m} \leq 5$, respectively.   
 and in the righ panel the star-to-star comparison between {\it photometric} 
and G04 spectroscopic metallicities for these 22 stars.

The correlations in both panels are not very strong, though, admittedly, some
of the most deviating objects have large D$_{\rm m}$ values.

%These metal abundances are on the metallicity scale defined by  Jurcsik
%(1995), where M3 is assumed to have [Fe/H]=$-$1.47 (Kraft et al. 1992, AJ 104,
%645). 

%It should be reminded that M3 in the Zinn \& West (1984) metallicity scale  has
%[Fe/H] =$-$1.66. Thus reddenings derived from the previous formula should be
%increased by about 0.01 mag  to account for the 0.2 dex difference between
%Jurcsik (1995) and Zinn \& West (1984)  scales.

%, the results are fully described in 
%Di Fabrizio et al (2002).

\subsection{Absolute magnitudes}
M$_V$ values were derived from the Fourier parameters A$_1$ and A$_3$ using 
equation (1) of K02 with the zero point set in agreement with the  
distance modulus: $\mu_{LMC}$=18.515$\pm$0.085 for the LMC and the dereddened
average visual magnitude of the LMC RR Lyrae stars: 
$<V(RR)>_0$=19.064$\pm$0.064 by C03, implying M$_{V}$=0.549 at [Fe/H]=$-$1.5.
These values are listed in column 6 of Table~17. Errors 
have been computed from 
equation A.2 in KW01, with standard deviations of the Fourier parameters 
A$_1$ and A$_3$ taken from Table~2 of KK98. The uncertainties
of the M$_V$ values appear surprisingly small. 
For comparison in column 7 we
list the M$_{V}$ values computed 
from the observed apparent intensity-averaged magnitudes (taken from column 8 
of Tables~5 and 6) dereddened for $E(B-V)$=0.116 and 0.086 in field A and B
respectively (C03) and the standard estinction law 
A$_V$=3.1$\times E(B-V)$, on the assumption of  
$\mu_{LMC}$=18.515$\pm$0.085 (C03).  
%\clearpage
\begin{figure*} 
\includegraphics[width=18cm]{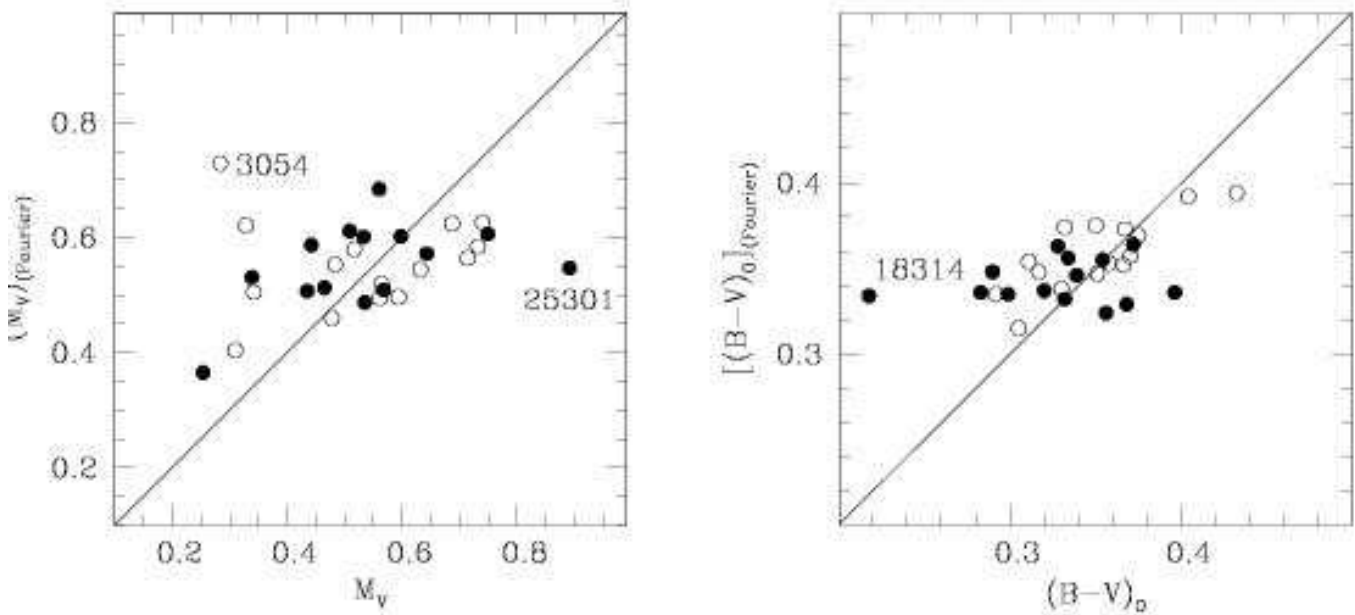}
\caption[]{Star-to-star comparison between  M$_{V}$ values (left panel) and
$(B-V)_0$ colours (right panel)  
derived from the Fourier parameters of the light curves and the 
corresponding observed quantities. For ease of comparison 
we show the 1:1 lines. 
% observed
%apparent magnitudes, on the assumption of $\mu_{LMC}$=18.815$\pm$0.085 (C03).
%Right panel: star-to-star comparison between 
Filled and open symbols mark variables with D$_{\rm m} \leq 3$ and 
3$<$D$_{\rm m} \leq 5$, respectively.
} 
%\label{f:fig1a}
\end{figure*}
The star-to-star comparison between M$_{V}$ values is shown in the
left panel of Figure~29.
%The zero point shift by about 0.15-0.2 mag 
%and 
The
 reduced range of the (M$_{V}$)$_{Fourier}$ values compared to the
observed M$_{V}$'s is quite surprising. If we remove the two major outliers
(stars \# 3054 and 25301) the (M$_{V}$)$_{Fourier}$ 
values still span only 60\% of the range spanned by the observed M$_{V}$'s. 
The larger range of the observed M$_{V}$'s can be only partially justified by
the actual intrinsic depth of our LMC observed fields
(see discussion in Section 3.1 of C03).  
%We have used Sturch's method as described in Walker (1990, 1998), where
%the reddening zero point has been adjusted to give E(B$-$V)=0.0 mag at the
%Galactic poles, and the [Fe/H] is that of Zinn \& West (1984) metallicity
%scale, whereby 
%$$ E(B-V) = (B-V)_{min} -0.24P -0.056[Fe/H]_{ZN}-0.336$$ 
%
%Derived $E(B-V)'s$ exhibit a rather large spread; they are given in column 6 of
%Table 7.

\subsection{Intrinsic $(B-V)_0$ colours and effective temperatures}
$(B-V)_0$ intrinsic colours were computed from 
the Fourier parameters A$_1$ and A$_3$ using
equation (6) of KW01 which is based on the zero points established
by KJ97 for magnitude-averaged magnitudes. The values of 
$\log T_{eff}$ were then computed from these $(B-V)_0$ colours
 using equation (11) of KW01
and adopting $\log g$=2.75 for the average  
gravity as suggested by several Baade-Wesselink studies.
Derived values are listed in columns 9 and 10 of Table~17, respectively. 
Observed $(B-V)_0$ colours were computed from the magnitude-averaged
values in column 11 of Tables 5 and 6 and dereddened according to the
$E(B-V)$ values in C03, these colours are provided in 
column 8 of Table~17.
The comparison between derived and observed colours is shown in the
right panel of Figure~29. As with the absolute magnitudes, the 
[$(B-V)_0$]$_{Fourier}$ colours cover an interval about 40\% smaller 
than that spanned by the observed $(B-V)_0$'s. 

In conclusion, the comparison between empirical determinations from the
Fourier parameters of the light curves and corresponding observed
quantities for the 29 {\it ab-}type RR Lyrae stars in the LMC has 
revealed a number of discrepancies, in particular 
between the derived and observed M$_{V}$ and $(B-V)_0$ values, deserving 
deeper investigation based on larger samples of stars than available here.
In this respect, we notice that similar discrepancies in 
the M$_{V}$ and $(B-V)_0$ values have been found by Cacciari et al. (2004), 
from the analysis of the RR Lyrae stars in M3, and in the M$_{V}$ values 
of the variables in $\omega$ Cen (Clement \& Rowe 2000) and 
M15 (Kaluzny et al. 2000). 

%\begin{figure} 
%%\includegraphics[width=8.8cm]{fig11_MACHO.ps}
%\caption[]{Comparison of our and MACHO 
%photometry for the variable stars in common.
%Residuals are: this paper $-$ MACHO. 
%Filled and open symbols are used for variable stars in field A and 
%B, respectively. Triangles are the double mode RR Lyrae stars.}
%%%\label{f:fig1a}
%\end{figure}

%\section{Summary and conclusions}

%({\bf Mettiamo questa sezione?})
                                                                                                                              
\begin{acknowledgements}                                                                                                      
                                                                                                                              
% Special thanks go to P. Montegriffo for his expert advice during 
%the reduction
%phase and for the development of the specific software used for the 
%detection of the variable stars and in the study of their periodicities. 
We are indebted to G. Rodighiero for her advice on the use of  
SExtractor, to 
M. Bellazzini and M. Messineo 
for their help in setting the DoPHOT reductions, 
%to
%M. Zoccali for her precious contribution in setting up the reductions with
%DAOPHOT and ALLFRAME, 
and to R. Merighi for 
help in the layout of
some of the figures of the paper. 
It is a pleasure to thank D. Alves, C. Clement and G. Kov\'acs for
 providing
some of the data on which the comparison with MACHO 
photometry is based.
%G.C. whishes to thank M. Marconi and
Special thanks go to 
C. Cacciari for many valuable discussions 
%about the classification
%of some of the RR Lyrae variables, and 
on the parameters of the  Fourier decomposition of the light curves, and 
for lending us her macros to compute the basic stellar quantities from the
Fourier parameters.
%, and the reddening of M3.
We thank the anonymous referee for useful suggestions.

This paper utilizes public domain data obtained by the MACHO Project, jointly
funded by the US Department of Energy through the University of California,
Lawrence Livermore National Laboratory under contract No. W-7405-Eng-48, by the
National Science Foundation through the Center for Particle Astrophysics of the
University of California under cooperative agreement AST-8809616, and by the
Mount Stromlo and Siding Spring Observatory, part of the Australian National
University.

This work was partially supported by MIUR - Cofin98 under the project 
''Stellar
Evolution", by MIUR - Cofin00 under the project ''Stellar observables 
of cosmological relevance", and by MIUR - Cofin02 under the project
''Stellar populations, distances and star formation histories in Local 
Group galaxies of all morphological types".

\end{acknowledgements}

\appendix
                               
\section{Atlas of the light curves} 
                                                                   
Atlas of the light curves for the 162 short period variables stars in our two LMC fields.
The photometric data 
are folded with the ephemerides given 
in Table~5 and 6.
Variables stars are divided 
per field and grouped by type: RR Lyrae stars ({\it ab-, c-, d-}type 
separately), $\delta$ Scuti, candidate Anomalous Cepheids, Cepheids, eclipsing
binaries, and within each 
group are ordered by increasing period. 

%\newpage

\begin{figure*} 
\includegraphics[width=18cm]{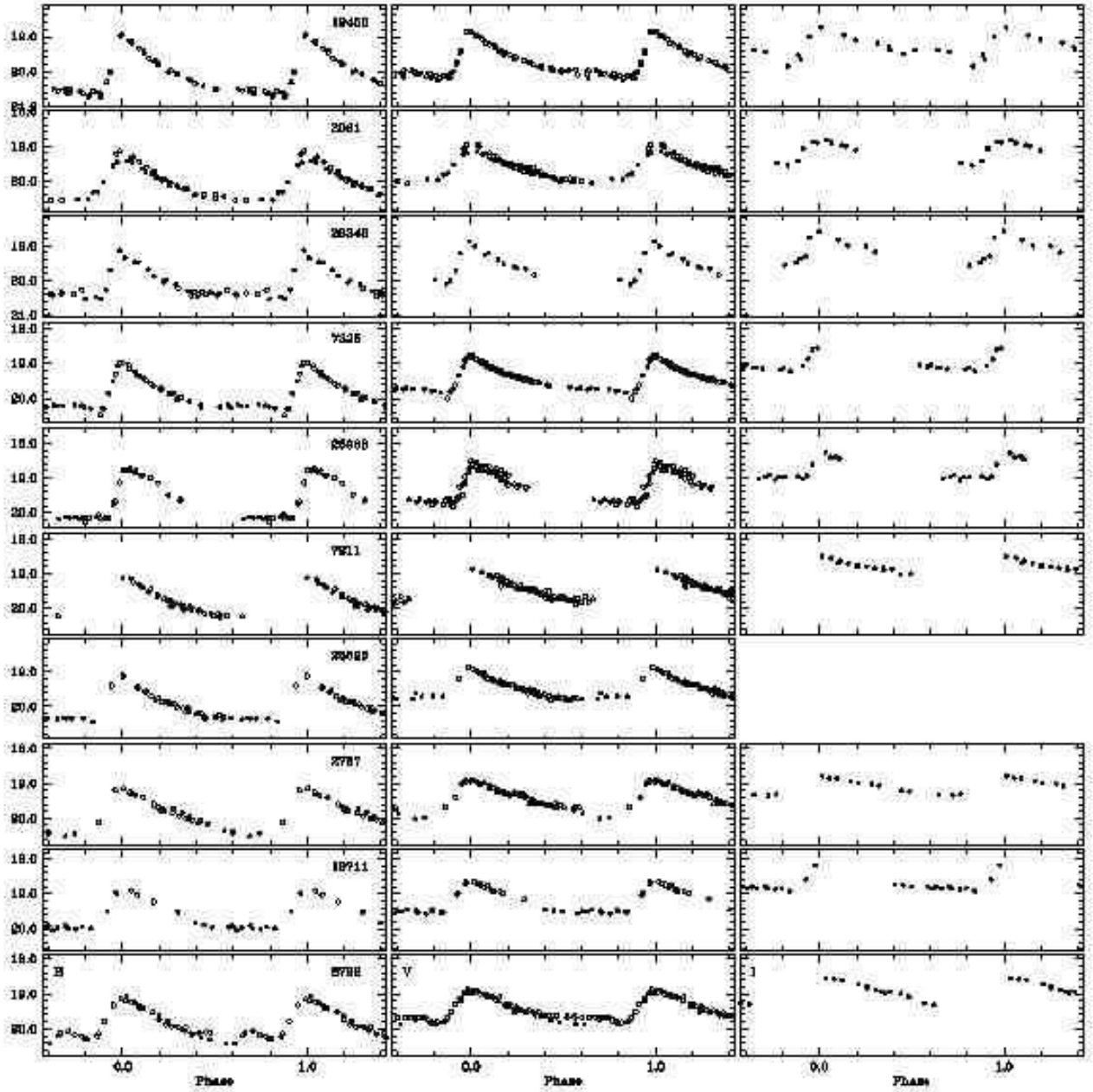}
\caption[]{$B,V,I$ light curves of the {\it ab-}type RR Lyrae stars
in field A, variables are ordered by increasing period.
 Open and filled symbols are used for the 1999 and 2001 data, respectively.}
%\label{f:fig1a}
\end{figure*}

\begin{figure*} 
\includegraphics[width=18cm]{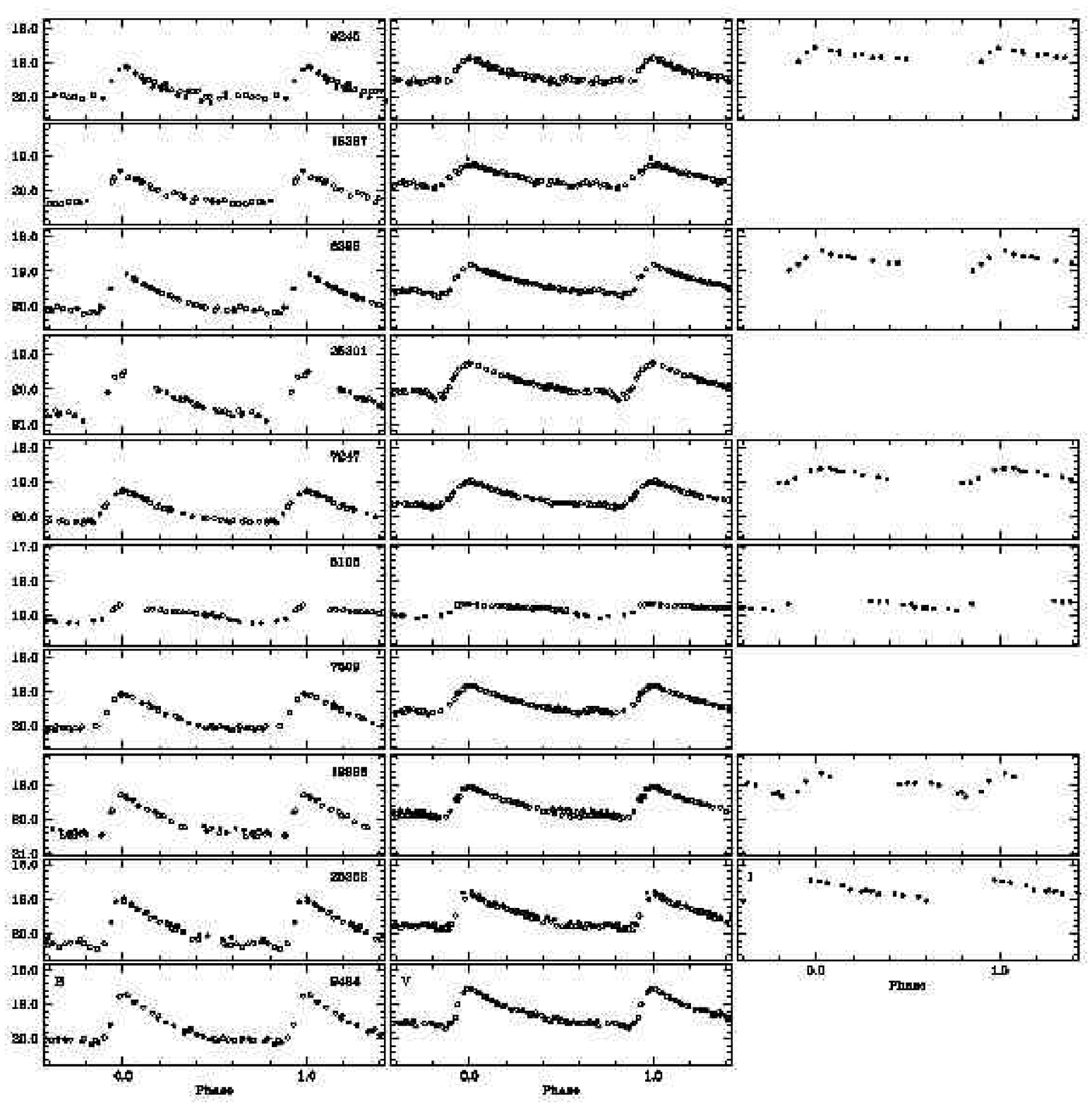}
{Fig. {\bf A.1.} -- continued --}
%\label{f:fig1b}
\end{figure*}

\begin{figure*} 
\includegraphics[width=18cm]{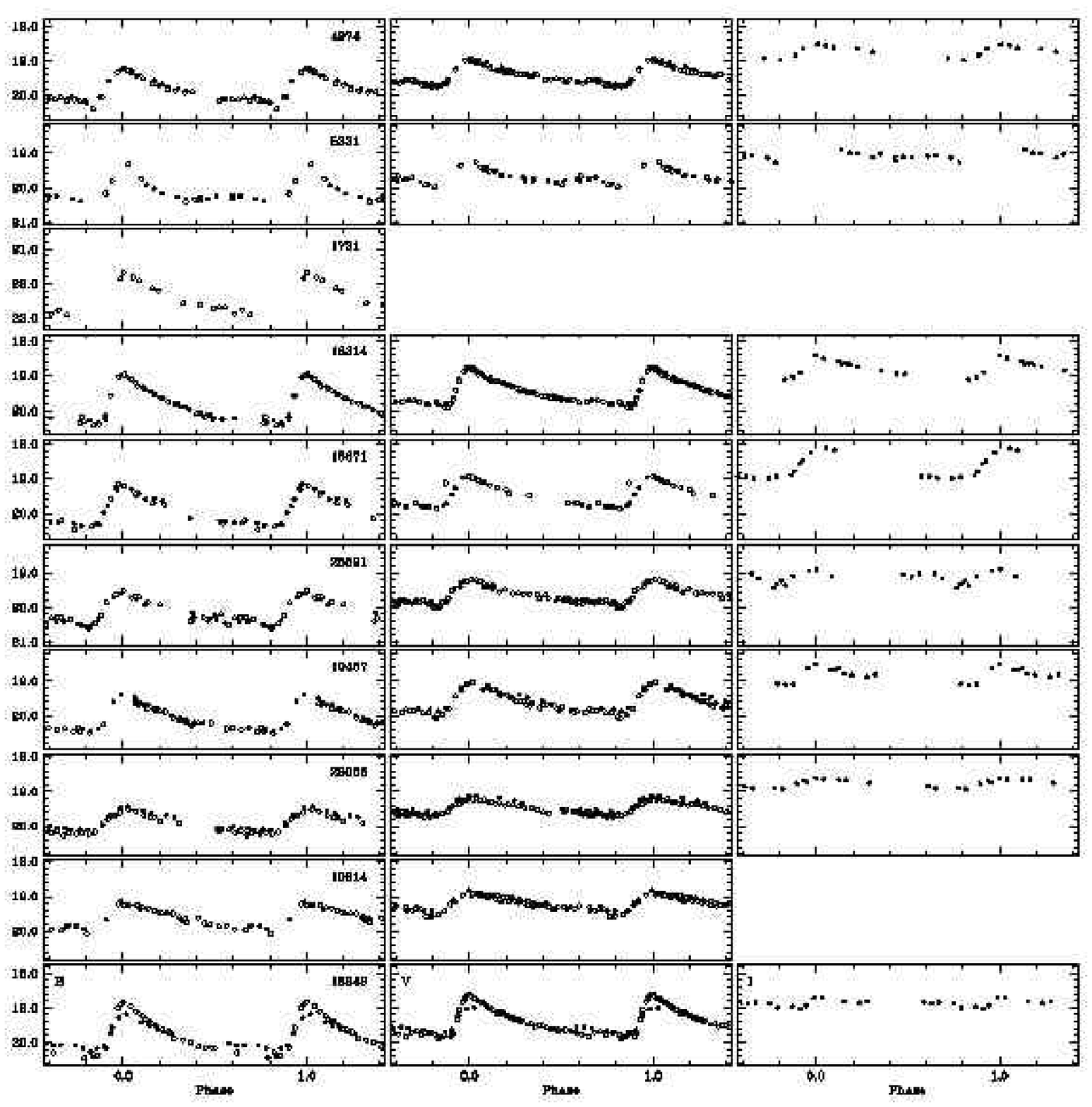}
{Fig. {\bf A.1.} -- continued --}
%\label{f:fig1c}
\end{figure*}
 
\begin{figure*} 
\includegraphics[width=18cm]{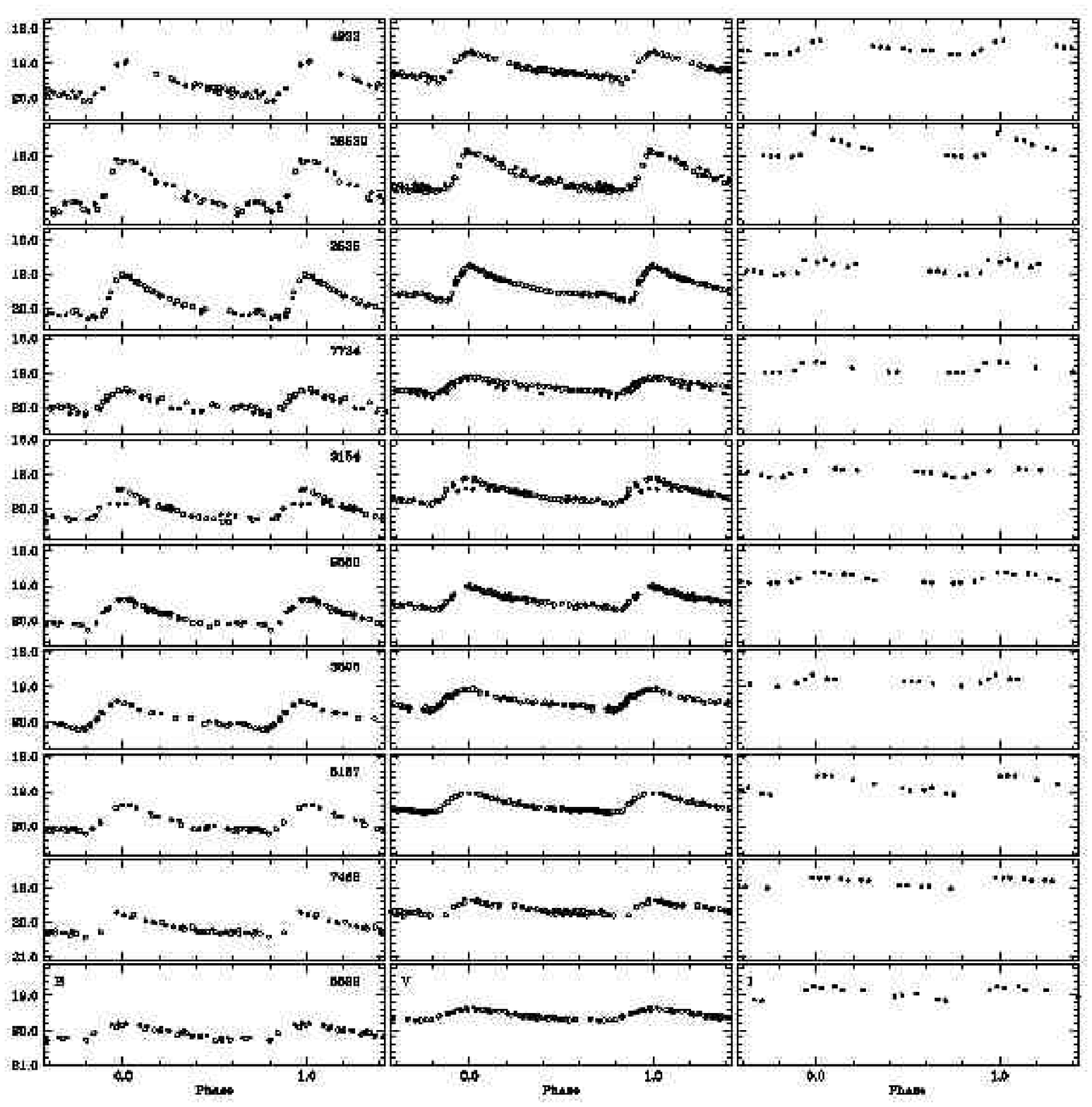}
{Fig. {\bf A.1.} -- continued --}
%\label{f:fig1d}
\end{figure*}

\begin{figure*} 
\includegraphics[width=18cm]{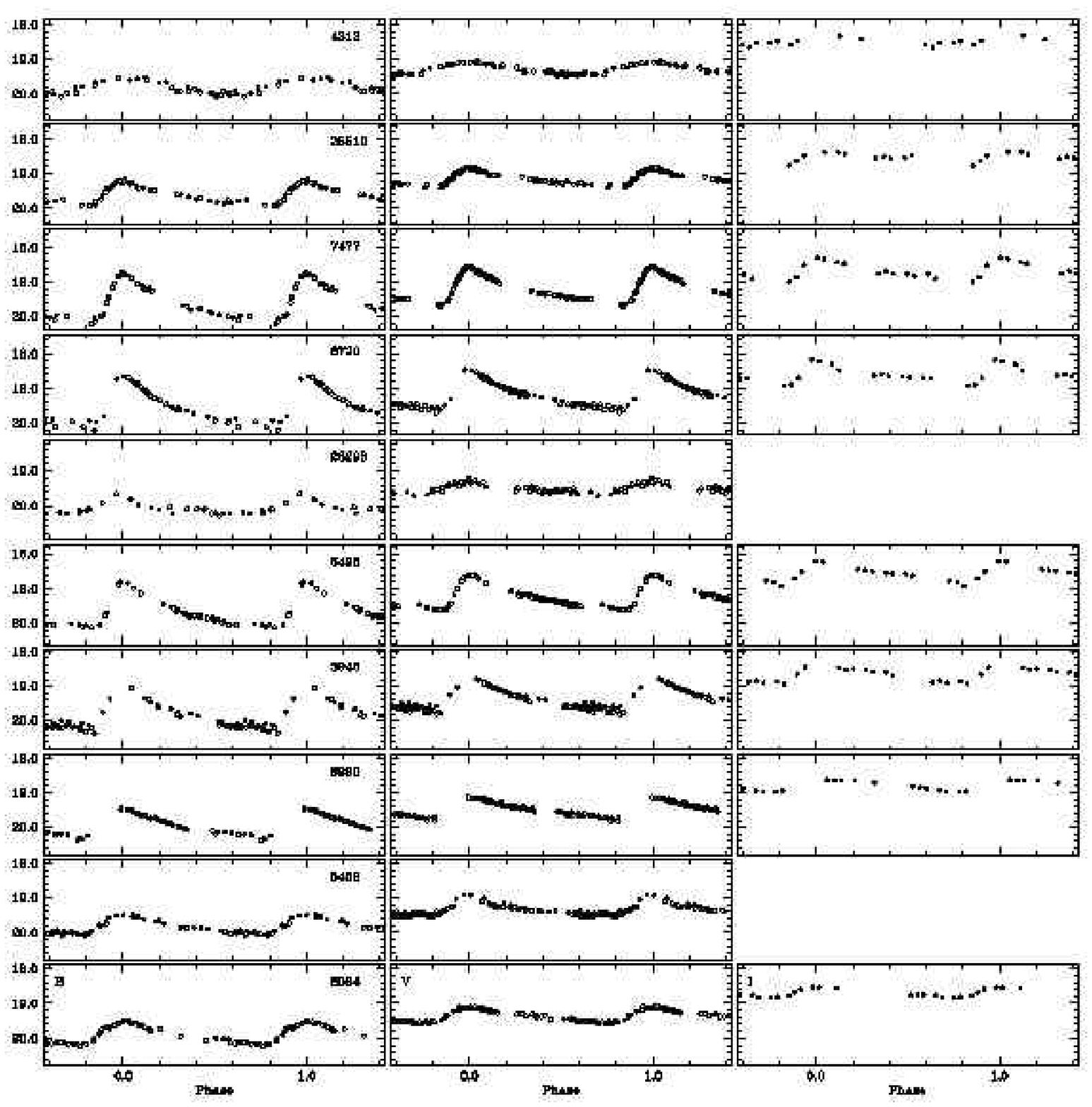}
{Fig. {\bf A.1.} -- continued --}
%\label{f:fig1e}
\end{figure*}

\begin{figure*} 
\includegraphics[width=18cm]{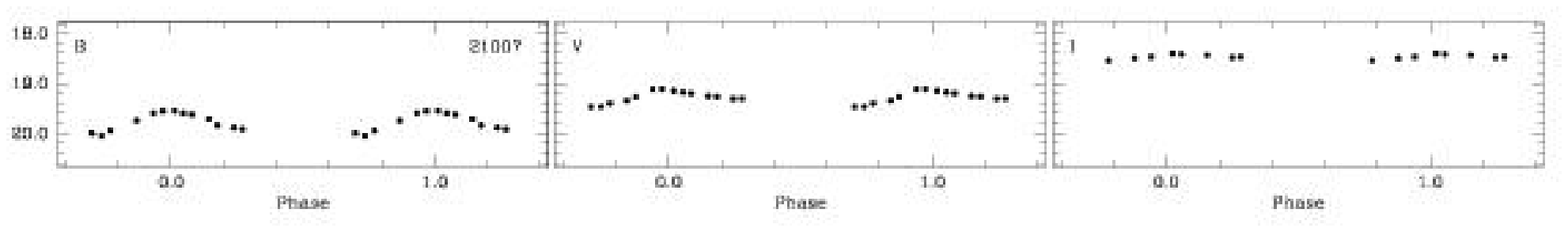}
{Fig. {\bf A.1.} -- continued --}
%\label{f:fig1f}
\end{figure*}

\begin{figure*} 
\includegraphics[width=18cm]{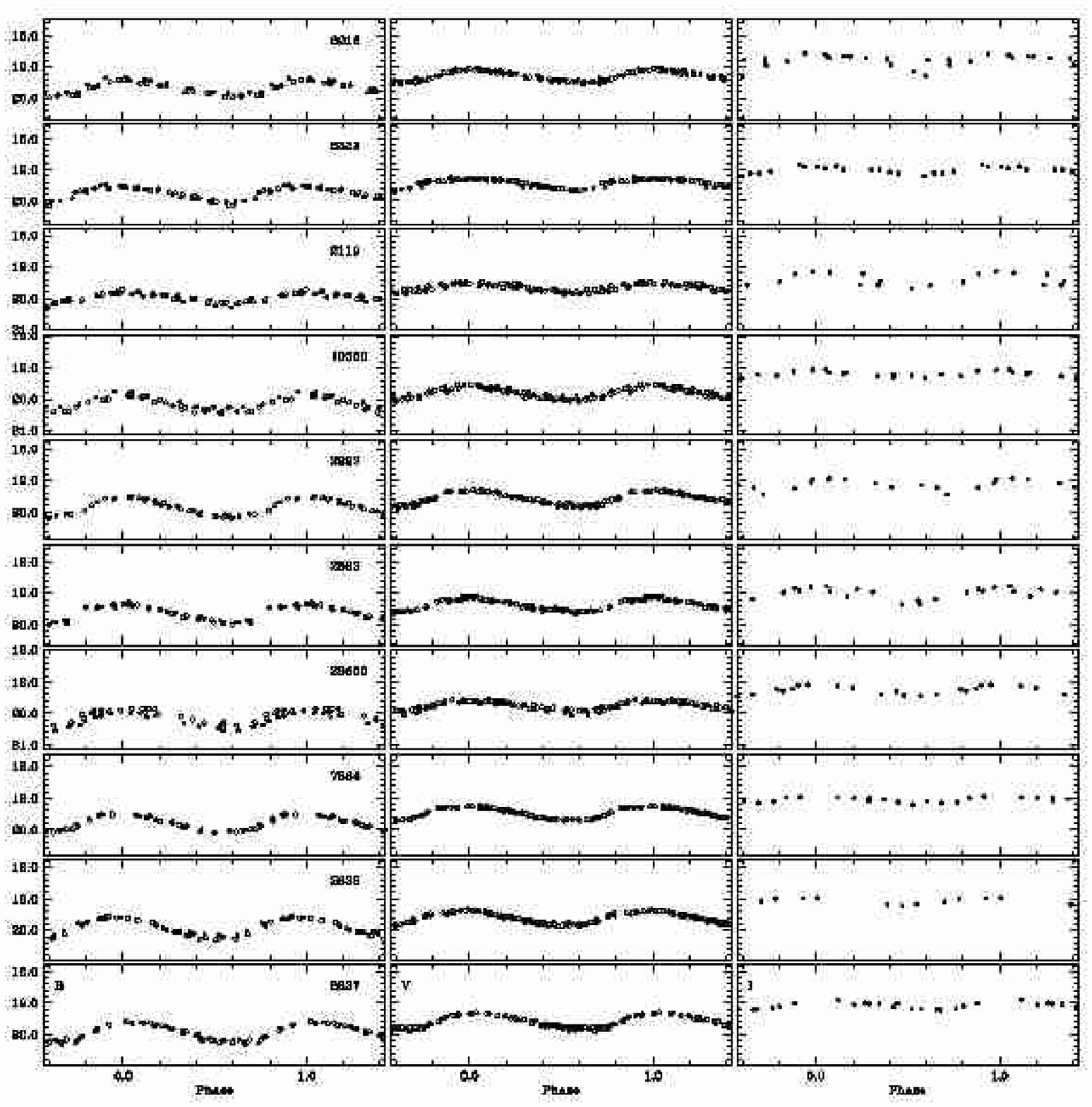}
\caption[]{$B,V,I$ light curves of the {\it c-}type RR Lyrae stars
in field A, variables are ordered by increasing period.
Open and filled symbols are used for the 1999 and 2001 data, respectively.}
%\label{f:fig2a}
\end{figure*}

\begin{figure*} 
\includegraphics[width=18cm]{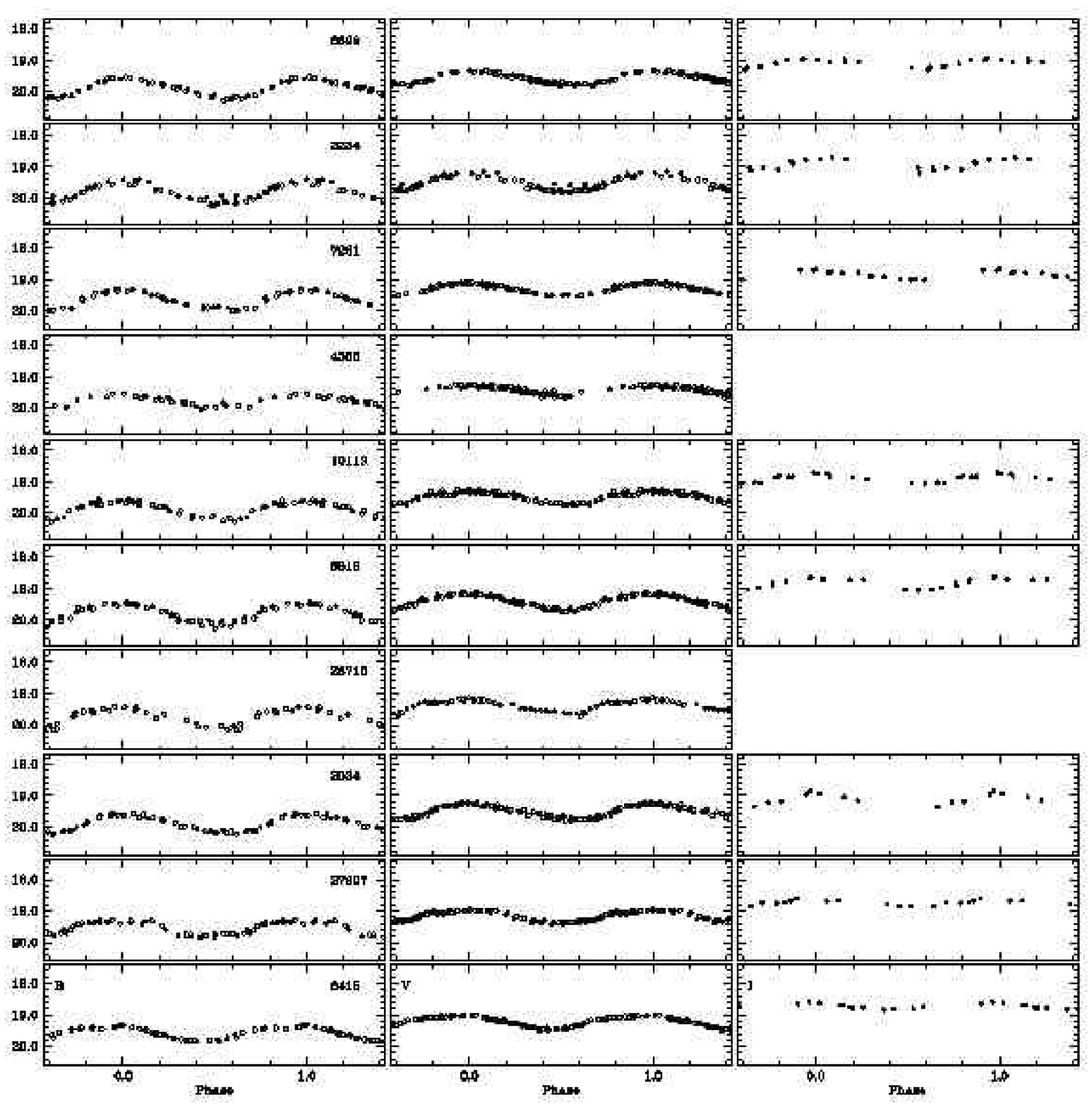}
{Fig. {\bf A.2.} -- continued --}
%\label{f:fig2b}
\end{figure*}

%\begin{figure*} 
%%\includegraphics[width=18cm]{RR_c_ca_page_3.ps}
%%\includegraphics[width=18cm]{Cef_ca_page_1.ps}
%{Fig. {\bf A.2.} -- continued --}
%\label{f:fig2c}
%\end{figure*}

\begin{figure*} 
\includegraphics[width=18cm]{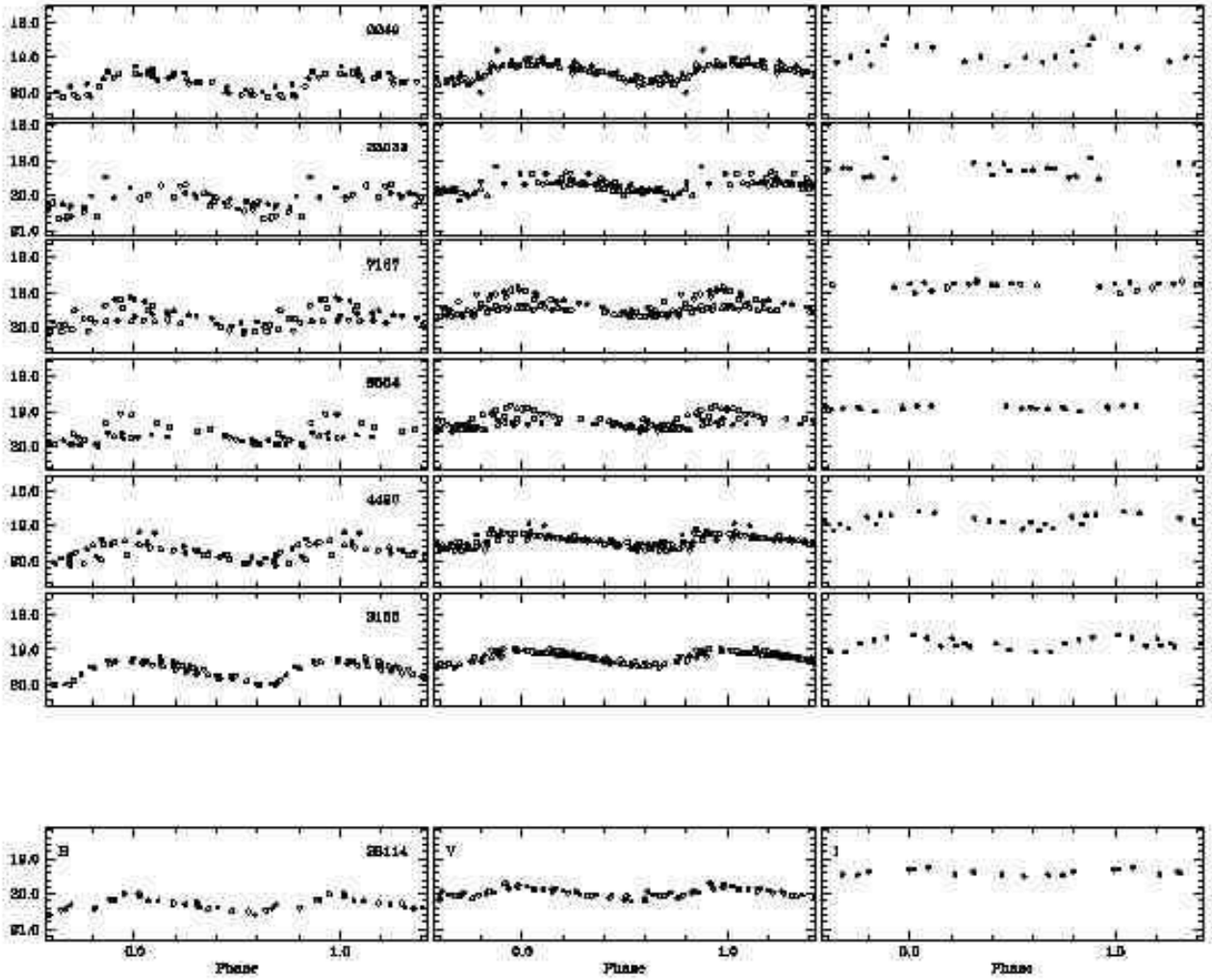}
\caption[]{$B,V,I$ light curves of the {\it d-}type RR Lyrae stars and
of the $\delta$ Scuti star (bottom panel)
in field A, variables are ordered by increasing period. 
Open and filled symbols are used for the the 1999 and 2001 data, respectively.}
%\label{f:fig3a}
\end{figure*}

\begin{figure*} 
\includegraphics[width=18cm]{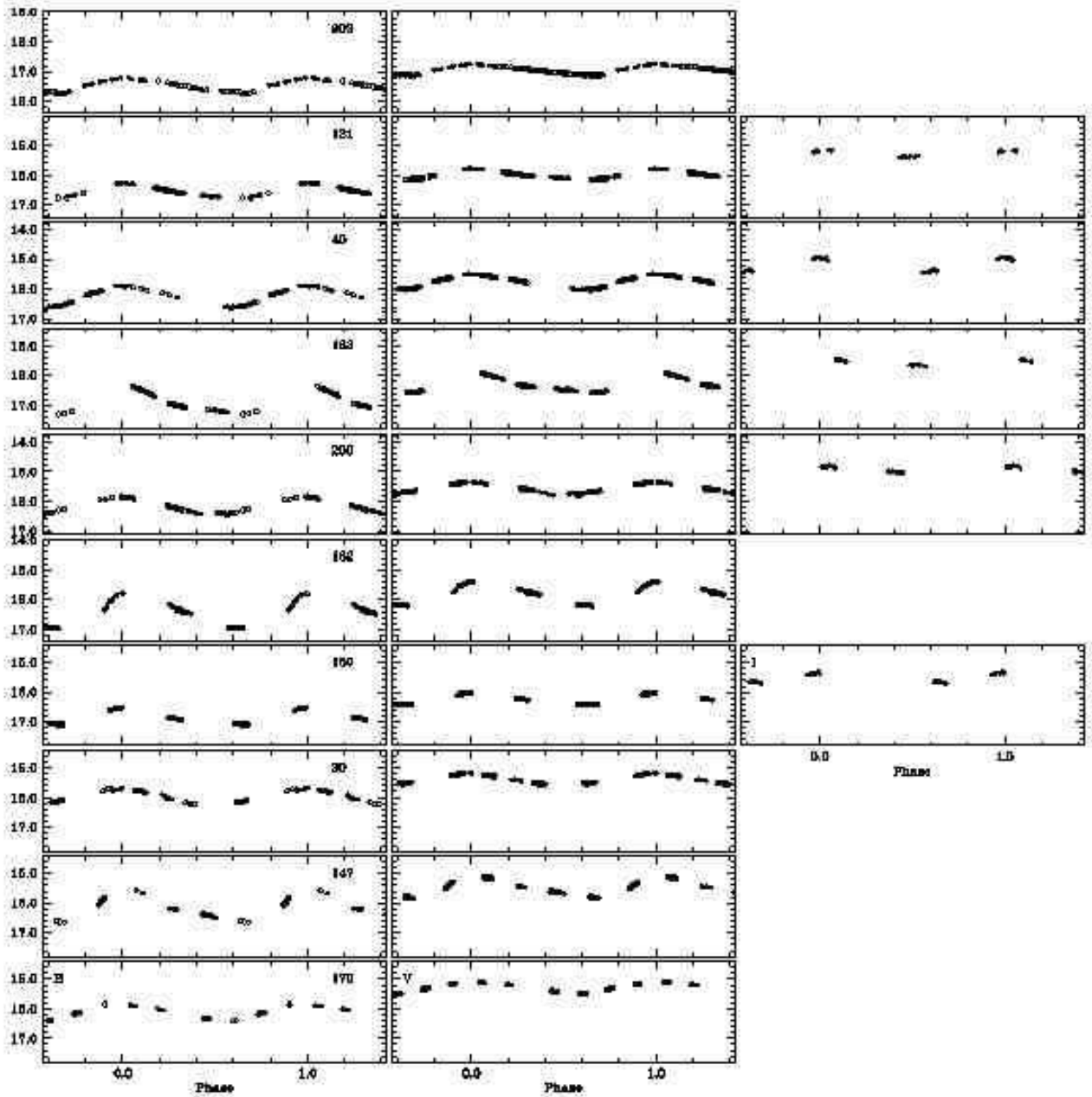}
\caption[]{$B,V,I$ light curves of Classical Cepheids 
in field A, variables are ordered by increasing period.
Open and filled symbols are used for the 1999 and 2001 data, respectively.}
%\label{f:fig4a}
\end{figure*}

%\begin{figure*} 
%%\includegraphics[width=18cm]{Cef_ca_page_2.ps}
%{Fig. {\bf A.4.} -- continued --}
%\label{f:fig4b}
%\end{figure*}

\begin{figure*} 
\includegraphics[width=18cm]{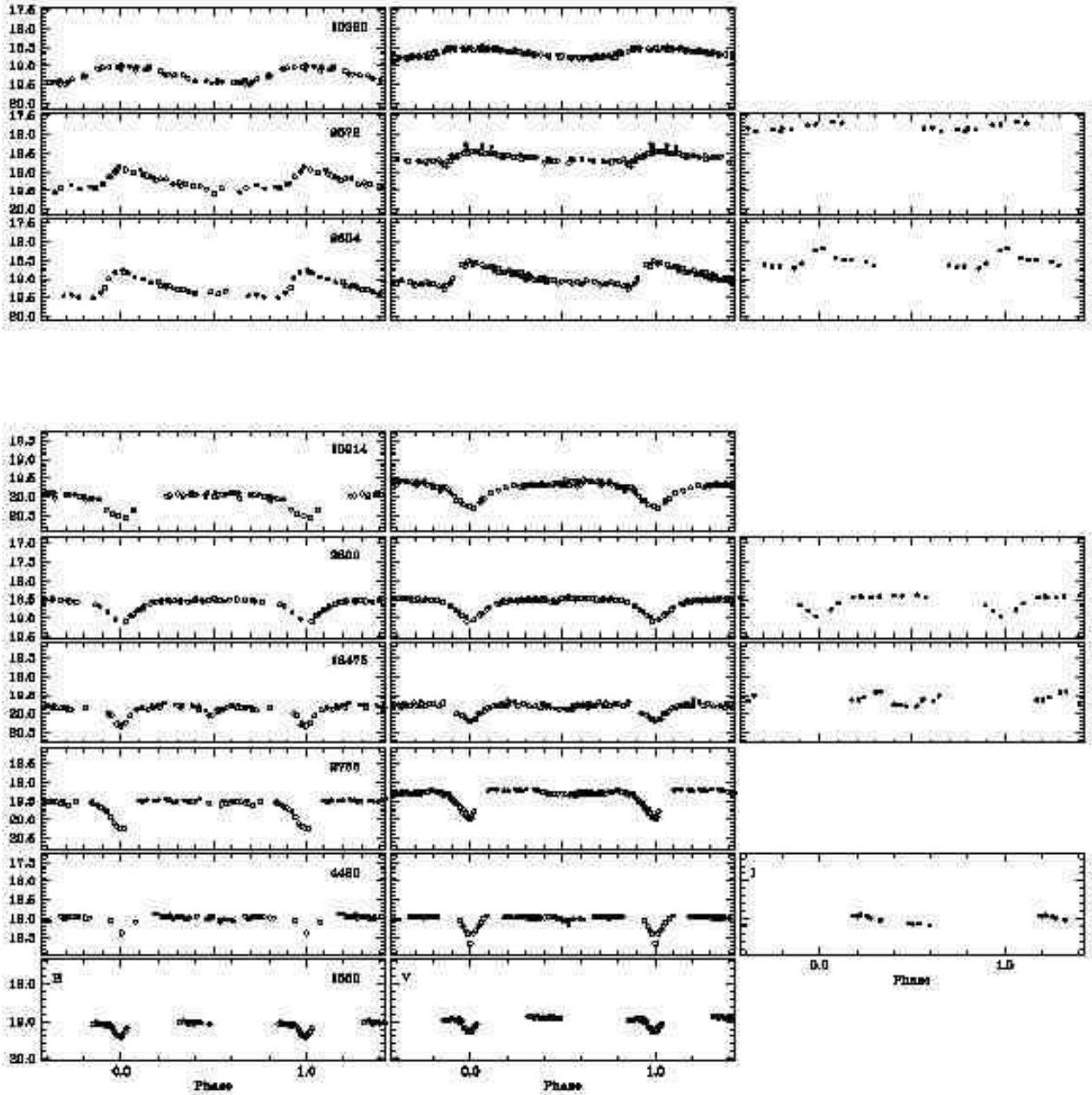}
\caption[]{$B,V,I$ light curves of candidate Anomalous Cepheids (first three top panels)
and binaries in field A, variables are ordered by increasing period. 
Open and filled symbols are used for the 1999 and 2001 data, respectively.}
%\label{f:fig5a}
\end{figure*}

\begin{figure*} 
\includegraphics[width=18cm]{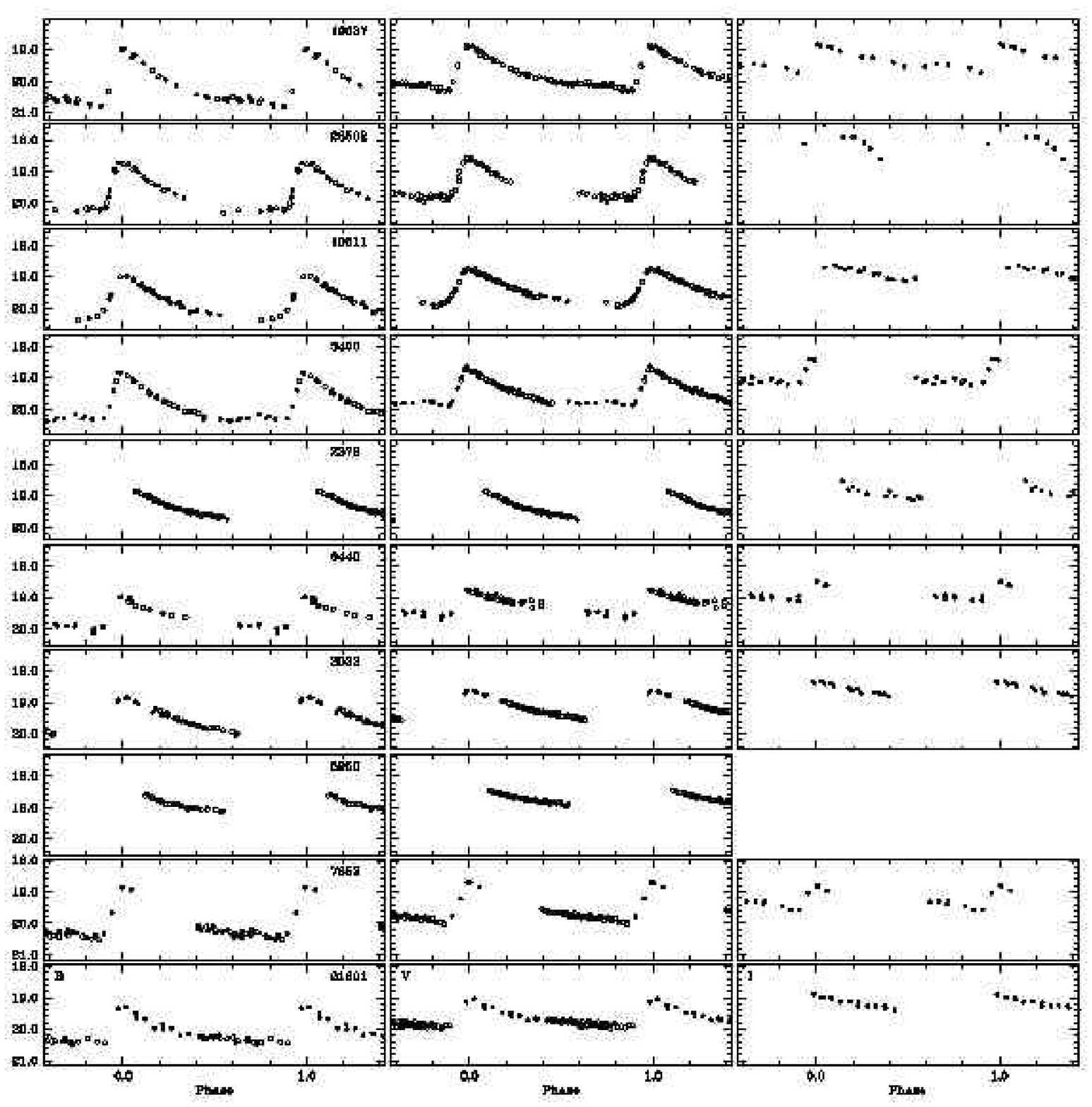}
\caption[]{$B,V,I$ light curves of the {\it ab-}type RR Lyrae stars
in field B, variables are ordered by increasing period.
Open and filled symbols are used for the 1999 and 2001 data, respectively.}
%\label{f:fig1a}
\end{figure*}

\begin{figure*} 
\includegraphics[width=18cm]{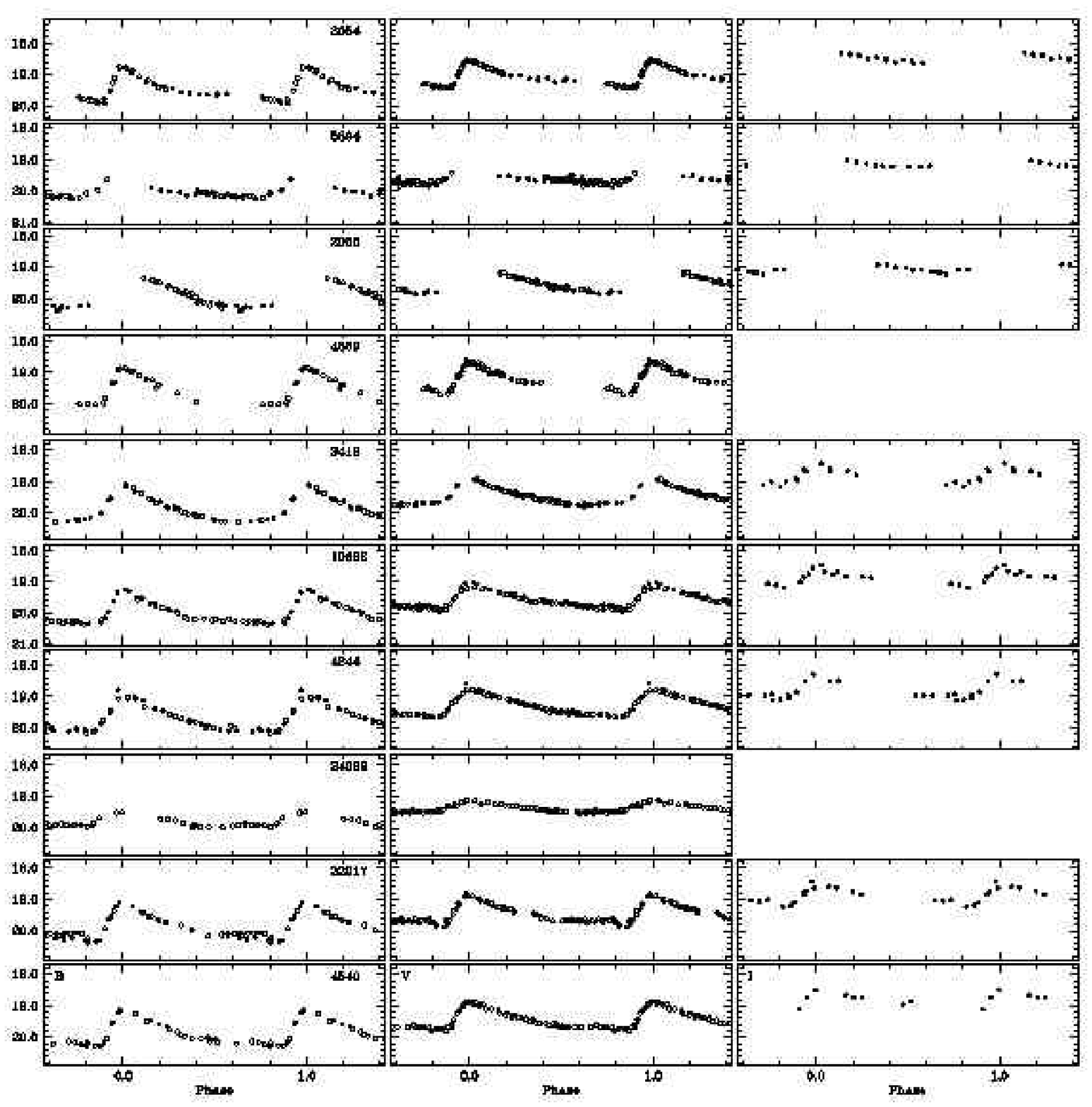}
{Fig. {\bf A.6.} -- continued --}
%\label{f:fig1b}
\end{figure*}

\begin{figure*} 
\includegraphics[width=18cm]{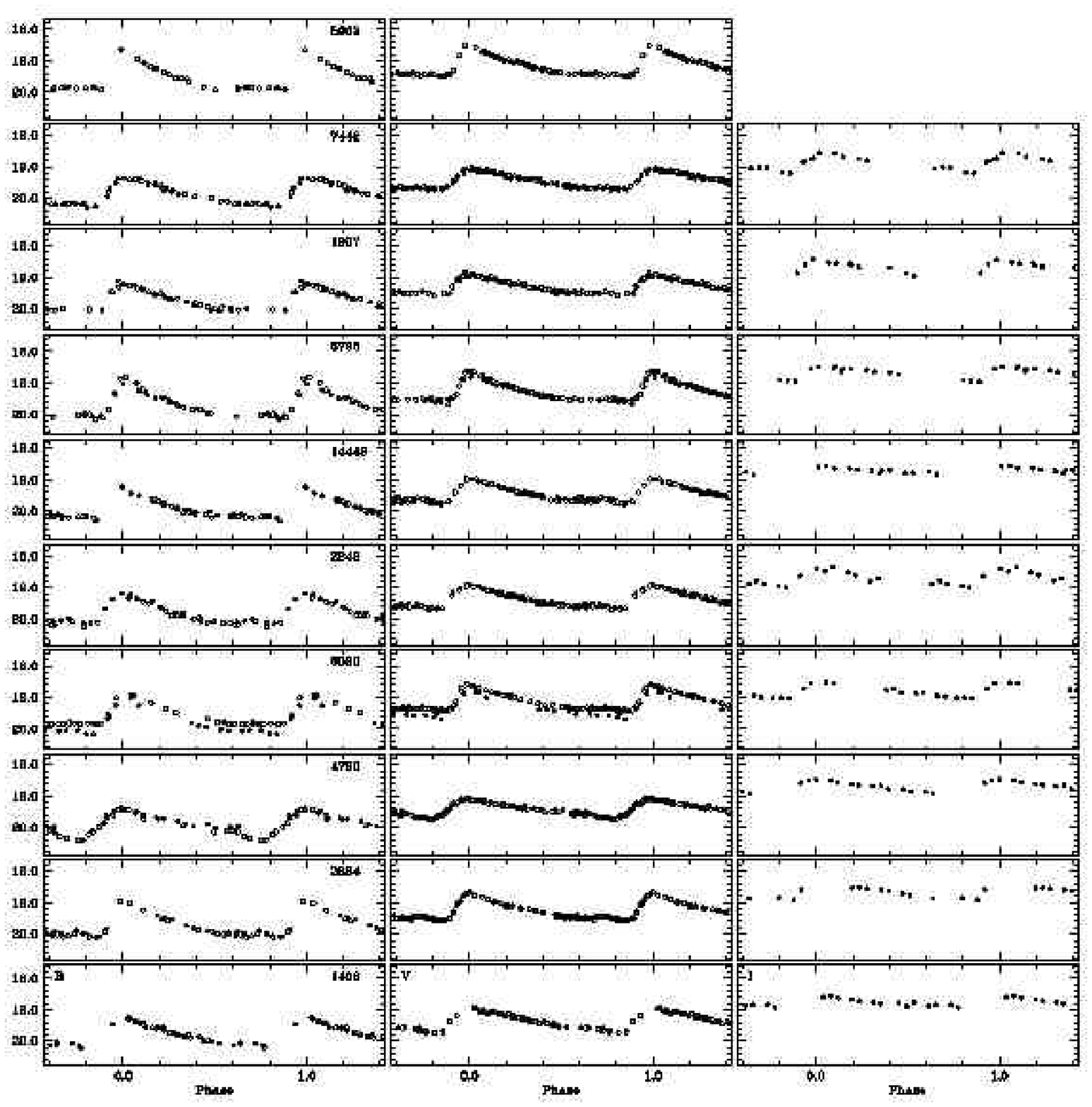}
{Fig. {\bf A.6.} -- continued --}
%\label{f:fig1c}
\end{figure*}
 
\begin{figure*} 
\includegraphics[width=18cm]{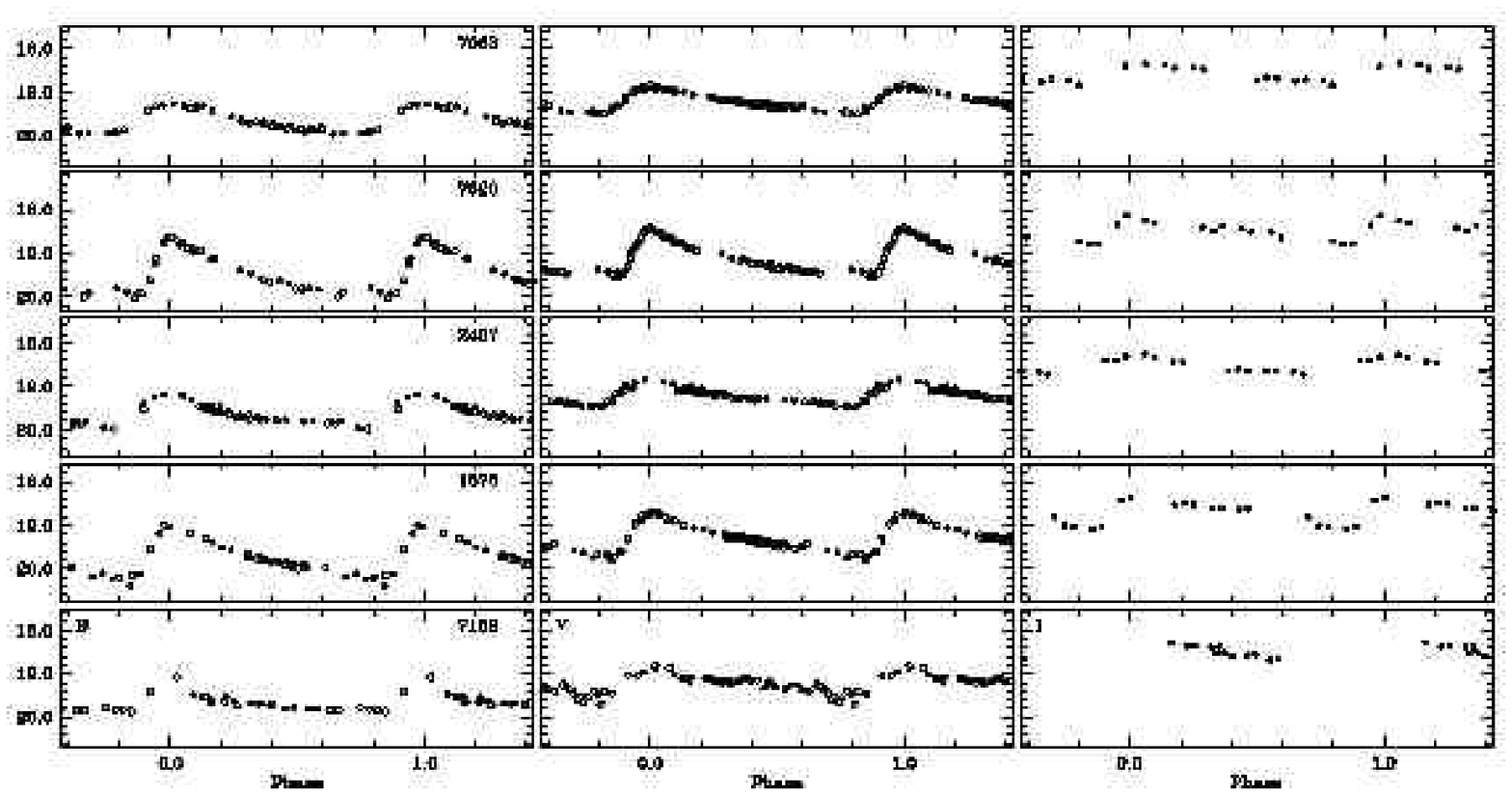}
{Fig. {\bf A.6.} -- continued --}
%\label{f:fig1d}
\end{figure*}

\clearpage
\begin{figure*} 
\includegraphics[width=18cm]{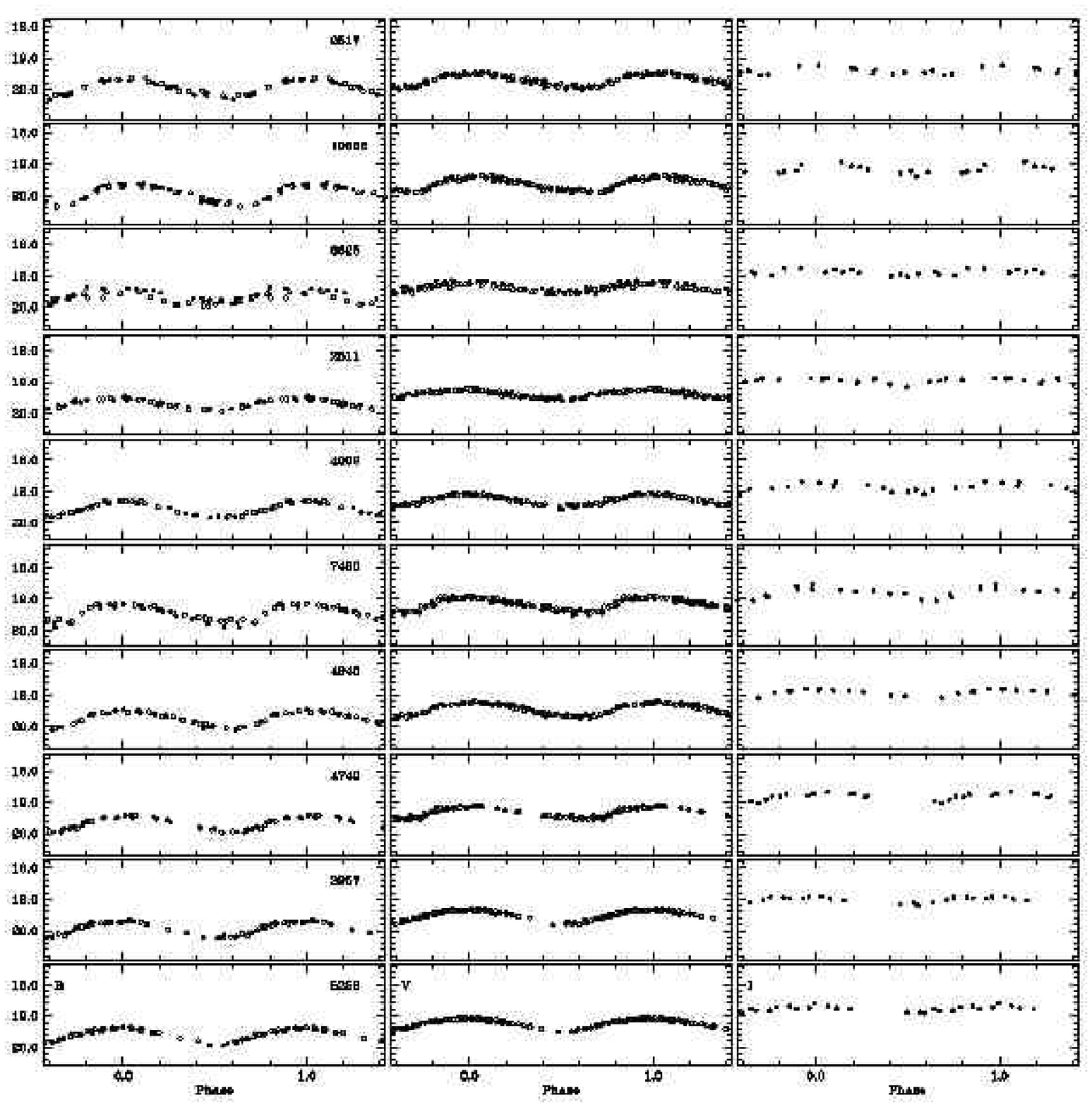}
\caption[]{$B,V,I$ light curves of the {\it c-}type RR Lyrae stars
in field B, variables are ordered by increasing period. 
Open and filled symbols are used for the 1999 and 2001 data, respectively.}
%\label{f:fig2a}
\end{figure*}

\begin{figure*} 
\includegraphics[width=18cm]{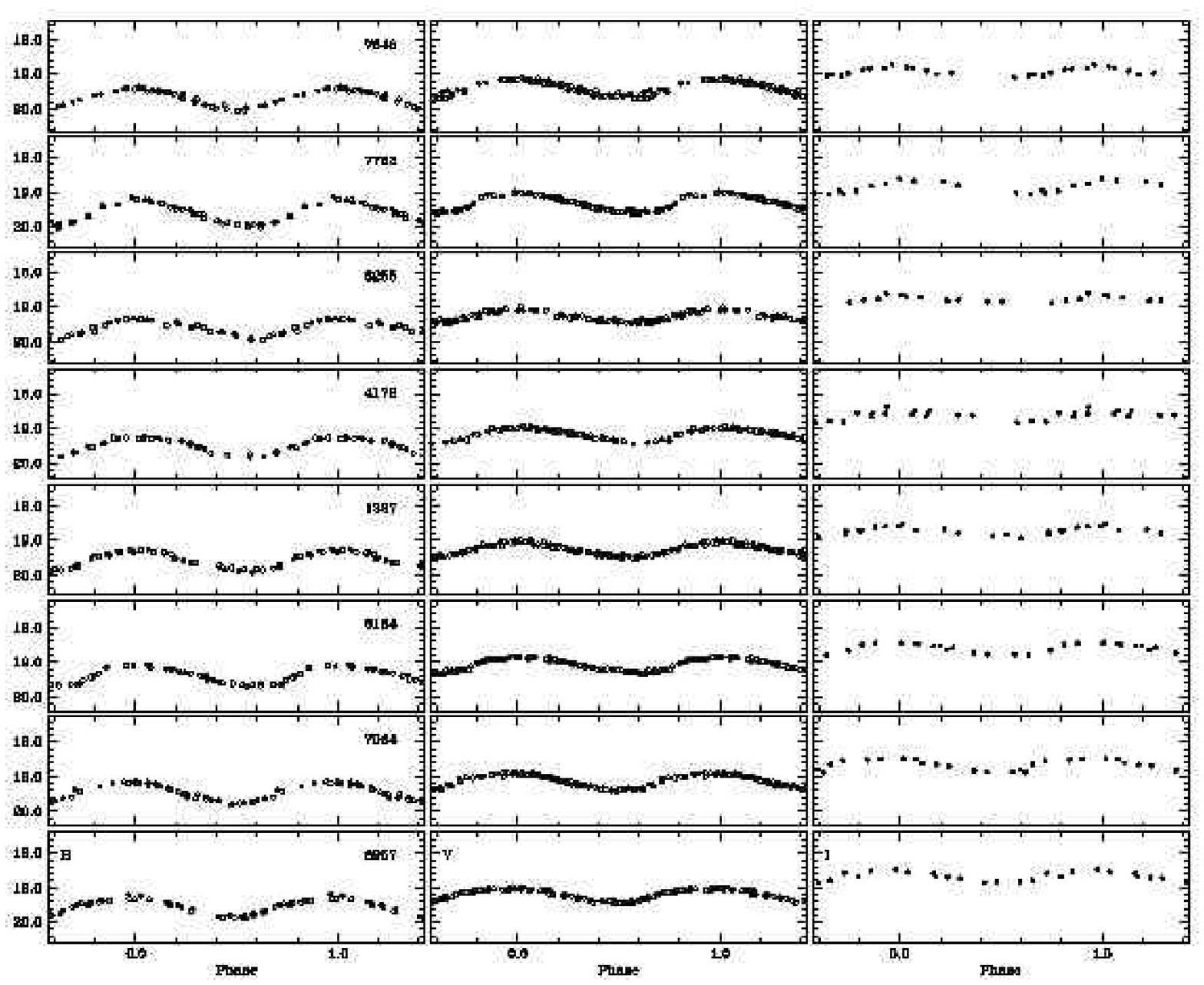}
{Fig. {\bf A.7.} -- continued --}
%\label{f:fig2b}
\end{figure*}

%\begin{figure*} 
%%\includegraphics[width=18cm]{RR_c_cb_page_3.ps}
%{Fig. {\bf A.7.} -- continued --}
%\label{f:fig2c}
%\end{figure*}
\clearpage
\begin{figure*} 
\includegraphics[width=18cm]{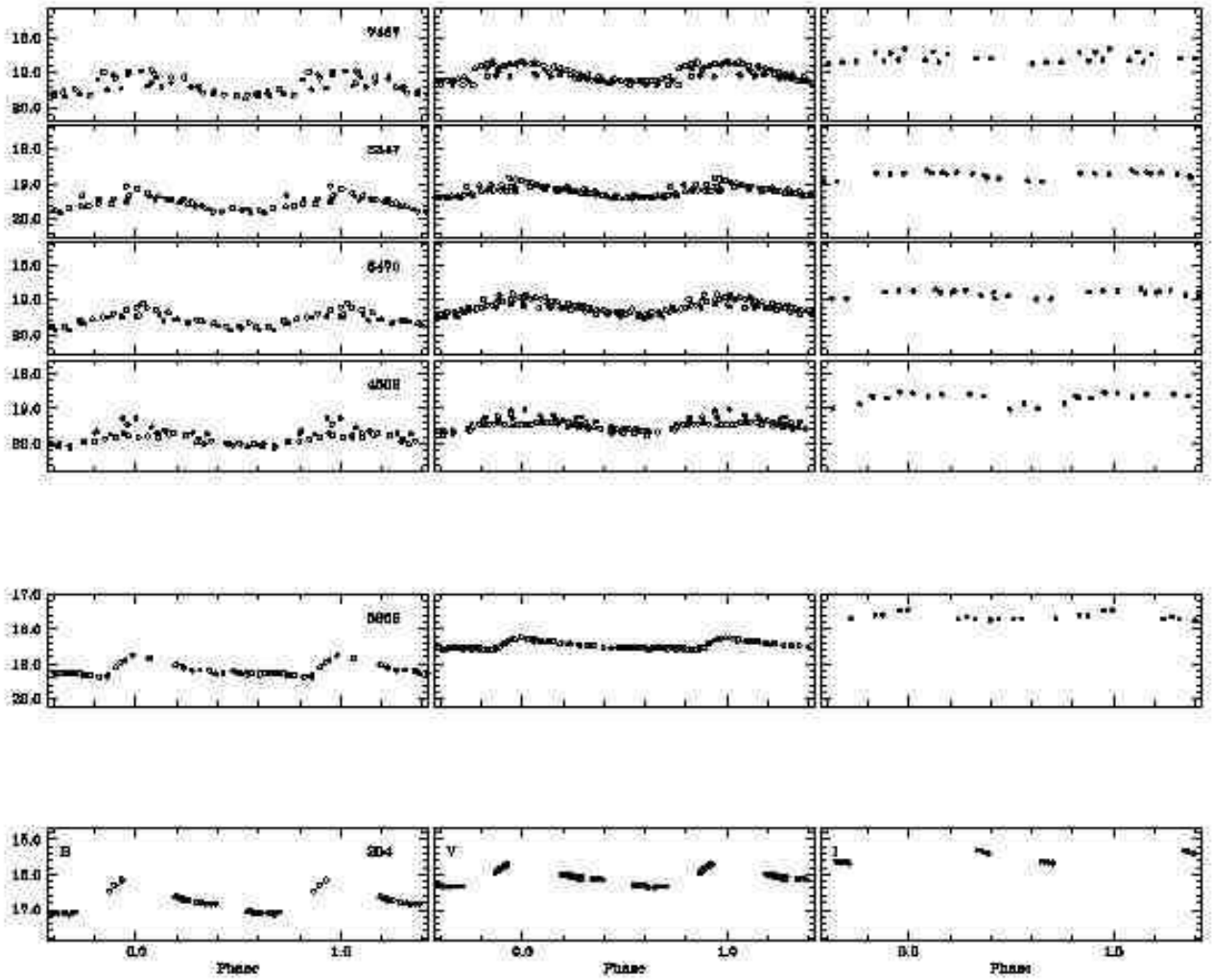}
\caption[]{$B,V,I$ light curves of the {\it d-}type RR Lyrae stars
(first four top panels),
Anomalous (fifth panel) and Classical Cepheids (bottom panel)
in field B, variables are ordered by increasing period.
Open and filled symbols are used for the 1999 and 2001 data, respectively.}
%\label{f:fig3a}
\end{figure*}

%\begin{figure*} 
%%\includegraphics[width=18cm]{Cef_cb_page_1.ps}
%\caption[]{$B,V,I$ light curves of Anomalous (top
%panel) and Classical Cepheids 
%(bottom panel)  
%in field B, variables are ordered by increasing period.}
%\label{f:fig4a}
%\end{figure*}

\clearpage
\begin{figure*} 
\includegraphics[width=18cm]{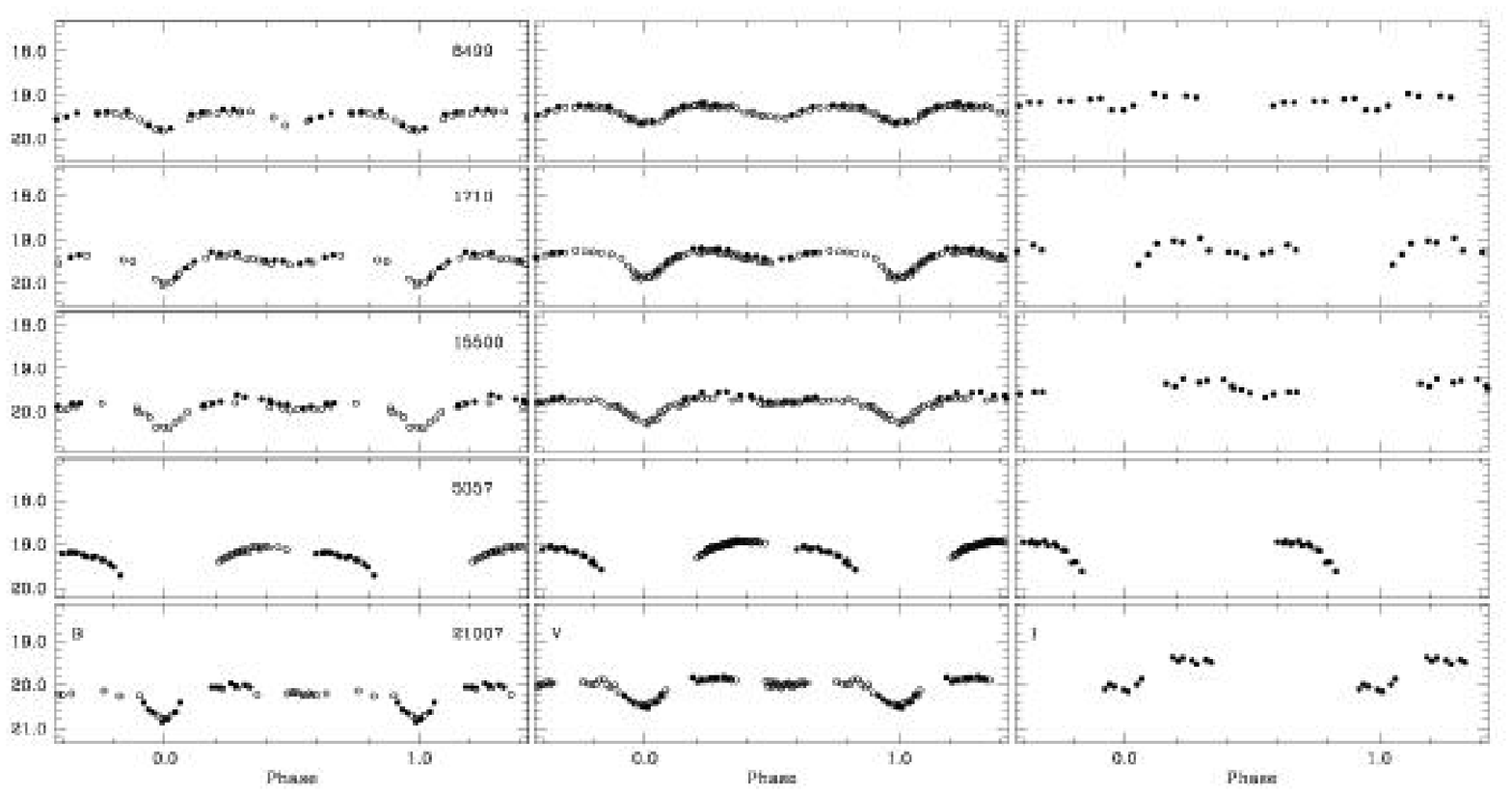}
\caption[]{$B,V,I$ light curves of  binaries 
in field B, variables are ordered by increasing period. 
Open and filled symbols are used for the 1999 and 2001 data, respectively.}
%\label{f:fig5a}
\end{figure*}


\begin{thebibliography}{}      
                        
\bibitem[]{} Alcock, C. et al.  1996, AJ, 111, 1146 (A96)
\bibitem[]{} Alcock, C. et al.  1997, ApJ, 482, 89 (A97)					     
\bibitem[]{} Alcock, C. et al. (the MACHO collaboration) 1999, PASP, 111, 1539 (A99)
\bibitem[]{} Alcock, C. et al. 2000, AJ, 119, 2194 (A00)					 
\bibitem[]{} Alcock, C. et al. 2003a, VizieR On-line Data Catalog: II/247
\bibitem[]{} Alcock, C. et al. 2004, AJ, 127, 334 (astro-ph/0310281)
\bibitem[]{} Alcock, C. al. (the MACHO collaboration) 2003b, ApJ,  598, 597 
\bibitem[]{} Alves, D.R.  2004,  Proc. of JD 13 on "Extragalactic Binaries", 
IAU XXV General Assembly, eds. A. Gimenez and I. Ribas, in
press (astro-ph/0310673) 
\bibitem[]{}Baldacci, L., Clementini, G., Held, E.V., \& Rizzi, L. 2004, in
 Communications in Asteroseismology, Asteroseismology and Stellar Evolution, 
 eds. Z. Kollath \& G. Handler, Austrian Academy of Sciences,  Vol. 145,
 p.32 (astro-ph/0311170)
\bibitem[]{} Barning, F.J.M. 1963, Bull. Astron. Inst. Netherlands, 17, 22
\bibitem[]{} Bingham, E.A., Cacciari, C., Dickens, R.J., \& Fusi Pecci,
F. 1984, MNRAS, 209, 765  
\bibitem[]{} Blazhko, S. 1907, Astron. Nachr., 175, 325
\bibitem[]{}Bono, G., Caputo, F., Santolamazza, P., Cassisi, S., \&
 Piersimoni, A. 1997, AJ, 113, 2209
\bibitem[]{}Bono, G., Caputo, F., \& Stellingwerf, R.F. 1995, ApJS, 99, 263
\bibitem[]{} Bragaglia, A., Gratton, R.G., Carretta, E., Clementini, G., 
 Di Fabrizio, L., \& Marconi, M. 2001, AJ, 122, 207
\bibitem[]{} Cacciari, C., Corwin, T.M., \& Carney, B.W. 2004, AJ, in press
 (astro-ph/0409567)
\bibitem[]{} Carretta, E., Cacciari, C., Ferraro, F.R., Fusi Pecci, F., 
\& Tessicini, G. 1998, MNRAS, 298, 1005
\bibitem[]{} Caputo, F. 1998, A\&ARv, 9, 33 
\bibitem[]{} Carney, B.W., Storm, J., \& Jones, R.V. 1992, ApJ, 386, 684
\bibitem[]{} Clement, C.M. 2000, in The Impact of Large-Scale Surveys on Pulsating 
 Star Research, ASP Conference Series, ed. L. Szabados and D. Kurtz,  Vol. 203,
 p.266
\bibitem[]{} Clement, C.M., \& Rowe, J. 2000, AJ, 120, 2579
\bibitem[]{} Clement, C.M., \& Shelton, I. 1999, ApJ, 515, L85 
\bibitem[]{} Clementini, G., Gratton, R.G., Bragaglia, A., Carretta, E., 
 Di Fabrizio, L., \& Maio, M. 2003a, AJ, 125, 1309 (C03)
\bibitem[]{} Clementini, G., Held, E.V., Baldacci, L., \& Rizzi, L. 2003b, 
 ApJ, 588, L85   
\bibitem[]{} Corwin, T.M., \& Carney, B.W. 2001, AJ, 122, 3183
\bibitem[]{} Dolphin, A.E., Saha, A., Claver, J., Skillman, E.D., Cole, A.A., 
 Gallagher, J.S., Tolstoy, E., Dohm-Palmer, R.C., \& Mateo, M. 2002, AJ, 123, 3154   
\bibitem[]{} Freedman, W.L., et al. 2001, ApJ, 553, 47
\bibitem[]{} Gallart, C., Freedman, W.L., Aparicio, A., Bertelli, G., \& Chiosi, C. 
 1999, AJ, 118, 2245
\bibitem[]{} Gallart, C., Aparicio, A., Freedman, W.L., Madore, B.F., 
 Martinez-Delgado, D., \& Stetson, P.B. 2004, AJ, 127, 1486 
\bibitem[]{} Gratton, R.G., Bragaglia, A., Clementini, G.,Carretta, E., 
 Di Fabrizio, L., Maio, M., \& Taribello, E. 2004, A\&A, 421, 937 (G04)	    
\bibitem[]{} Harris, W.E. 1996, AJ, 112, 1487
\bibitem[]{} Jurcsik, J. 1995, AcA, 45, 653
\bibitem[]{} Jurcsik, J., \& Kov\'acs, G. 1996, A\&A, 312, 111 (JK96)
\bibitem[]{} Kaluzny, J., Kubiak, M., Szyma\'nski, A., Udalski, A., 
 Krzemi\'nski, W., \& Mateo, C. 1997, A\&AS, 125, 343
\bibitem[]{} Kaluzny, J., Olech, A., Thompson, I., Pych, W., Krzeminski, W.,
 \& Schwarzenberg-Czerny, A. 2000, A\&AS, 143, 215
\bibitem[]{} Kov\'acs, G., 2002, in $\omega$ Centauri: A Unique Window into
Astrophysics,  ASP Conference Series, ed.s F. van Leeuwen, G. Piotto and J. Hughes, 
Vol. 265, 163 (K02)
\bibitem[]{} Kov\'acs, G., \& Jurcsik, J. 1996, ApJ, 466, L17 (KJ96)
\bibitem[]{} Kov\'acs, G., \& Jurcsik, J. 1997, A\&A, 322, 218 (KJ97) 
\bibitem[]{} Kov\'acs, G., \& Kanbur, S.M., 1998, MNRAS, 295, 834 (KK98)
\bibitem[]{} Kov\'acs, G., \& Walker, A.R, 2001, A\&A, 371, 579 (KW01)
\bibitem[]{} Landolt, A.U. 1992, AJ, 104, 340
\bibitem[]{} Lomb, N.R.1976, Ap\&SS, 39, 447
\bibitem[]{} Marconi, M., \& Clementini, G. 2004, in preparation
\bibitem[]{} Marconi, M., Fiorentino, G., \& Caputo, F. 2004, A\&A, 417,
1101
\bibitem[]{} Moskalik, P., \& Poretti, E. 2003, A\&A, 398, 213 
\bibitem[]{} Nemec, J.M., Nemec, A.F.L., \& Lutz, T.E. 1994, AJ, 108, 222
\bibitem[]{} Norris, J.E., Freeman, K.C., \& Mighell, K.L. 1996, ApJ, 462,
 241 
\bibitem[]{} Oosterhoff, P.T. 1939, Observatory, 62, 104 
\bibitem[]{} Pancino, E., Pasquini, L., Hill, V., Ferraro, F., 
 \& Bellazzini, M. 2002, ApJ, 568, 101         
\bibitem[]{} Pritzl, B.J., Armandroff, T.E., Jacoby, G.H., \& Da Costa, G.
 S. 2002, AJ, 124, 1464
\bibitem[]{} Sandage, A., 1990, ApJ, 350, 603
\bibitem[]{} Sandage, A., 1993, AJ, 106, 703
\bibitem[]{} Scargle, J.D. 1982, ApJ, 263, 835  
\bibitem[]{} Schechter, P.L., Mateo, M., Saha, A. 1993, PASP, 105, 
 1342
\bibitem[]{}Smith, H.A., Silbermann, N.A., Baird, S.R., \& Graham, J.A. 1992, AJ, 104
 1430
\bibitem[]{} Soszy\'nsky, A., Udalski, A., Szyma\'nski, M., 
 Kubiak, M., Pietrzy\'nski, G., Wo\'zniak, P., Zebru\'n, K., 
 Szewczyk, O., Wyrzykowski, L.
 2003, Acta Astron., 53, 93
\bibitem[]{} Stetson, P.B. 2000, PASP, 112, 925
\bibitem[]{} Stetson, P.B. 1994, PASP, 196, 250
\bibitem[]{} Stetson, P.B. 1996, "Users manual for DAOPHOT II"
%\bibitem[]{} Sturch, C. 1966, ApJ, 143, 774 
\bibitem[]{} Suntzeff, N.B., \& Kraft, R.P. 1996, AJ, 111, 1913
\bibitem[]{} Szeidl, B. 1988, in Multimode Stellar Pulsations, ed. G. Kov\'acs, 
      L. Szabados, \& B. Szeidl (Budapest: Konkoly Obs.), 45 	
\bibitem[]{} Udalski, A., Kubiak, M., \& Szyma\'nski, M. 1997, Acta Astron., 47, 319
\bibitem[]{} Udalski, A. et al. 2000, Acta Astron. 50, 307																  
\bibitem[]{} Walker, A.R. 1992, ApJ 390, L81
\bibitem[]{} Wallerstein, G., \& Cox, A.N. 1984, PASP, 96, 677
\bibitem[]{}Zinn, R., \& West, M.J. 1984, ApJS, 55, 45
   
\end{thebibliography}
\end{document}